Università degli Studi di Padova

# Università degli Studi di Padova

Centro di Ateneo di Studi e Attività Spaziali
"Giuseppe Colombo" — CISAS

---

PhD Course in "Scienze Tecnologie e Misure Spaziali"
Curriculum in Sciences and Technologies for Aeronautics and Satellite Applications — Series XXXV

# Optical and Opto-mechanical Analysis and Design of the Telescope for the Ariel Mission

**Coordinator**
Prof. Francesco Picano

**PhD Candidate**
Paolo Chioetto

**Supervisor**
Dr. Paola Zuppella

**Co-supervisor**
Dr. Vania Da Deppo

To mountains and climbers,
and to whom made me love them.

# Abstract

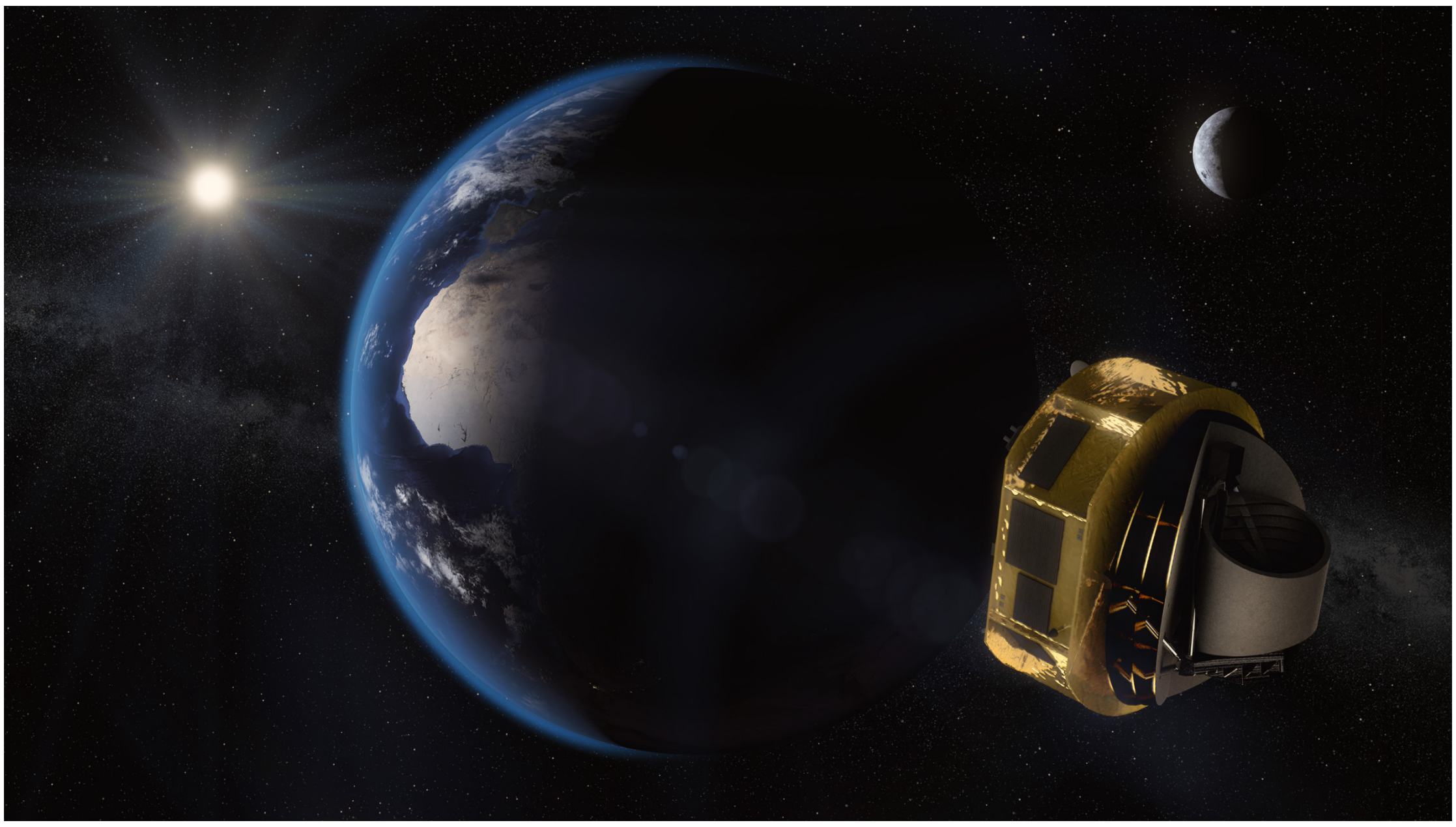

**Figure 1:** Artist's impression of Ariel on its way to Lagrange Point 2 (L2). Here, the spacecraft is shielded from the Sun and has a clear view of the whole sky. Image Credit: ESA/STFC RAL Space/UCL/Europlanet-Science Office.

THE ATMOSPHERIC REMOTE-SENSING INFRARED EXOPLANET LARGE-SURVEY (Ariel) is the first space mission dedicated to measuring the chemical composition and thermal structures of thousands of transiting exoplanets, enabling planetary science far beyond the boundaries of the Solar System. Ariel was officially adopted in 2020 as the fourth medium size (M4) mission in the scope of ESA "Cosmic Vision" program, with launch expected in 2029. The mission will operate from the Sun-Earth Lagrangian point L2.

The scientific payload consists of two instruments: a high resolution spectrometer covering the waveband 1.95–7.8 µm, and a multi-purpose fine guidance system / visible photometer / low resolution near-infrared spectrometer with wavelength coverage between 0.5 µm and 1.95 µm. The instruments are fed a collimated beam from an unobscured, off-axis Cassegrain telescope. Instruments and telescope will operate at a temperature below 50 K.

The mirrors and supporting structures of the telescope will be realized in aerospace-grade aluminum. Given the large aperture of the primary mirror ($0.6\,m^2$), it is a choice of mate-rial that requires careful optical and opto-mechanical design, and technological advances in the three areas of mirror substrate thermal stabilization, optical surface polishing and optical coating.

This thesis presents the work done by the author in these areas, as member of the team responsible for designing and manufacturing the telescope and mirrors.

The dissertation starts with a systematic review of the optical and opto-mechanical requirements and design choices of the Ariel telescope, in the context of the previous development work and the scientific goals and requirements of the mission.

The review then progresses with the opto-mechanical design, examining the most important choices in terms of structural and thermal design. This will serve as an introduction to a statistical analysis of the deformations of the optical surface of the telescope mirrors and of their alignment in terms of rigid body motions.

The qualification work on thermal stabilization, polishing and coating is then presented. The three procedures have been set up and tested to demonstrate the readiness level of the technological processes employed to fabricate the mirrors.

The first process, substrate thermal stabilization, is employed to minimize deformations of the optical surface during cool down of the telescope to the operating temperature below 50 K. Purpose of the process is to release internal stress in the substrate that can cause such shape deformations.

Then a combined optical surface figuring/polishing process is applied to reduce residual surface shape errors and bring surface roughness to below 10 nm RMS. Polishing of large aluminum surfaces to optical quality is notoriously difficult due to softness of the material, so a dedicated polishing recipe was set up and tested.

Finally, an optical coating recipe based on protected silver was characterized in terms of reflectivity and qualified for environmental stability, particularly at cryogenic temperatures, and for uniformity. Some of the coated samples are also being monitored and measured periodically for any sign of performance degradation while they age.

All tests were performed on samples of the same aluminum alloy chosen as baseline for the mirror substrates and on a full-scale prototype of the Ariel primary mirror.

Results from the coating characterization were also used to prepare an estimation of the various components contributing to the expected throughput of the telescope at the end of the scientific lifetime of the mission.

# Preface

THE RESEARCH WORK described in this doctoral dissertation covers specific themes in the context of the optical and opto-mechanical design and technological development of the telescope for the ESA Ariel Mission. The work was performed by the author during the preliminary definition phases of the mission (B1 and B2) as member of the Italian team in charge of the design and manufacturing of the telescope and mirrors, under the supervision of Dr. Paola Zuppella and Dr. Vania da Deppo of CNR-IFN Padova[1]. The most relevant results from this work are captured in a series of papers: a peer-reviewed journal article and five conference proceedings contributions.

The dissertation is divided in two parts. The first introduces the Ariel mission, its scientific goals, history and main design characteristics and then reports the author's work in the areas of optical and opto-mechanical analysis and design and mirrors coating qualification. Each chapter provides the necessary introductory background to put each of the papers in the context of the development work for the mission, and provides further literary references and clarifications of relevant aspects of the research.

The second part consists of the six papers, one per chapter, reformatted for consistency, but otherwise presenting the same unaltered content as compared with their published counterparts. The list of papers is included also here below for convenience, together with a brief clarification of the author's contribution to each of them.

## Paper I

P. Chioetto et al. "Preliminary Analysis of Ground-to-Flight Mechanical Tolerances of the Ariel Mission Telescope". In: *Proc. SPIE 12180, Space Telescopes and Instrumentation 2022: Optical, Infrared, and Millimeter Wave.* Aug. 27, 2022, 121804R. DOI: 10.1117/12.2628900

The author performed all analyses and computations introduced in Section 3.2.2 and presented in the paper, and devised most of the simulation steps, with the help of the rest of the Italian optical team. Some of the techniques employed in the analysis were developed for a previous work, described in Section 3.2.1, to which the author contributed participating in all the preparatory discussions and by independently verifying the results.

[1] CNR-Istituto di Fotonica e Nanotecnologie di Padova, Via Trasea 7, 35131 Padova, Italy

### Paper 2

P. Chioetto et al. “Qualification of the Thermal Stabilization, Polishing and Coating Procedures for the Aluminum Telescope Mirrors of the ARIEL Mission”. In: *Experimental Astronomy* 53 (Apr. 19, 2022), pp. 885–904. DOI: 10.1007/s10686-022-09852-x

The paper presents the development and characterization of the thermal treatments, polishing and coating processes devised for the manufacturing of the primary mirror for the Ariel telescope. The processes were designed in collaboration with two industrial partners (MediaLario[2] and CILAS[3]), who performed the actual manufacturing steps. The author, as member of the team who supervised the partners, followed closely each of the various phases and contributed to all planning, designing and characterization discussions. He also reviewed all of the analyses described in the paper, and actually performed or verified a large number of them.

### Paper 3

P. Chioetto et al. “The Primary Mirror of the Ariel Mission: Cryotesting of Aluminum Mirror Samples with Protected Silver Coating”. In: *Proc. SPIE 11451, Advances in Optical and Mechanical Technologies for Telescopes and Instrumentation IV*. Dec. 13, 2020, 114511A. DOI: 10.1117/12.2562548

This paper and the following one present the coating qualification activities carried out at CILAS under direct coordination and supervision of Dr. Da Deppo, Dr. Zuppella and the author, who followed closely the planning and implementation of optical characterization measurements, some of which he performed directly, and the quality inspections.

### Paper 4

P. Chioetto et al. “Test of Protected Silver Coating on Aluminum Samples of ARIEL Main Telescope Mirror Substrate Material”. In: *Proc. SPIE 11852, International Conference on Space Optics — ICSO 2020*. June 11, 2021, p. 118524L. DOI: 10.1117/12.2599794

See Paper 3 above.

### Paper 5

P. Chioetto et al. “Long Term Durability of Protected Silver Coating for the Mirrors of Ariel Mission Telescope”. In: *International Conference on Space Optics — ICSO 2022, in press*. Oct. 3, 2022

The entire activity, including measurements and analyses, was planned and is being implemented by the author under supervision of Dr. Zuppella.

[2]MediaLario Srl, Via al Pascolo, 10, 23842 Bosisio Parini (Lecco), Italy.
[3]CILAS-ArianeGroup, 8 avenue Buffon, CS16319, 45,063 Orleans CEDEX 2, France.

Paper 6

P. Chioetto et al. “Initial Estimation of the Effects of Coating Dishomogeneities, Surface Roughness and Contamination on the Mirrors of Ariel Mission Telescope”. In: *Proc. SPIE 11871, Optical Design and Engineering VIII*. Oct. 4, 2021, 118710N. doi: 10.1117/12.2603768

The author developed all models, simulations and analyses reported in the paper with the supervision of Dr. Zuppella.

# Contents

# List of Figures

# List of Tables

# List of Acronyms

**AFM** .......... Atomic Force Microscopy

**AIRS** .......... Ariel Infrared Spectrograph

**AIV** ........... Assembly, Integration and Verification

**AOI** ........... Angle Of Incidence

**BSDF** .......... Bidirectional Scattering Distribution Function

**CAD** .......... Computer-Aided Design

**CFRP** ......... Carbon Fiber Reinforced Polymer

**CMM** ......... Coordinate Measuring Machine

**CNR** .......... Consiglio Nazionale delle Ricerche (National Research Council)

**CO** ............ Common Optics

**CSL** ........... Centre Spatial de Liège (Liege Space Center)

**CTE** ........... Coefficients of Thermal Expansion

**DT** ............ Diamond Turing

**ECSS** .......... European Cooperation for Space Standardization

**EE** ............. Enclosed Energy

**EOL** ........... End Of Life

**ESA** ........... European Space Agency

**EXP** ........... Exit Pupil

**EChO** ......... Exoplanet Characterisation Observatory

**FEA** ........... Finite Element Analysis

**FGS** ........... Fine Guidance Sensor

**FoV** ........... Field of View

**GEO** ........... Geosynchronous Orbit

**INAF** .......... Istituto Nazionale di AstroFisica (National Institute for Astrophysics)

**IR** ............. Infrared

**IRMOS** ........ Infrared Multi-Object Spectrograph

**ISO** ............ International Organization for Standardization

**JWST** .......... James Webb Space Telescope

**LEO** ........... Low Earth Orbit

**MF** ............ Merit Function

**MIRI** .......... Mid-Infrared Instrument

**NIR** ........... Near-Infrared

**NIRSpec** ....... Near-IR Spectrometer

**OPD** .......... Optical Path Difference

**PDR** .......... Preliminary Design Review

**PI** ............. Principal Investigator

**PSF** ............ Point Spread Function

**PTM** .......... Pathfinder Telescope Mirror

**PVD** ........... Physical Vapor Deposition

**QE** ............ Quantum Efficiency

**RH** ............ Relative Humidity

**RMS** .......... Root Mean Squared

**RSA** ........... Rapid Solidifying Aluminum

**RSS** ........... Root Square Sum

**SEM-EDX** ..... Scanning Electron Microscopy–Energy-dispersive X-ray spectroscopy

**SFE** ............ Surface Error

**SNR** ........... Signal to Noise Ratio

**SRR** ........... System Requirements Review

**STOP** .......... Structural, Thermal, Optical Performance analysis

**TA** ............ Telescope Assembly

**TDA** .......... Technology Development Assessment

**TIS** ............ Total Integrated Scatter

**TMA** .......... Three Mirror Anastigmat

**TRL** ........... Technology Readiness Level

**UV** ............ Ultra Violet

**VI** ............. Visual Inspection

**WFE** .......... Wavefront Error

**WLI** ........... White Light Interferometer

# Part I

# Introduction and Overview

*Four things do not come back: the spoken word, the sped arrow, the past life, and the neglected opportunity.*

Ted Chiang, *Exhalation* (2019)

# 1 Introduction

Space exploration missions are probably one of the most complex scientific and engineering human endeavors, involving large teams and resources for long periods of time. Alongside the scientific results and technological fallout, they are also a great tribute to the curiosity and passion of the hundreds of individuals who contribute to each mission, since the very beginnings of the space age in the late '50s, to the present day.

It is fascinating to see how an idea takes shape in the mind of few driven scientists and slowly gains momentum and support, and eventually blooms into a large project. I hope this introduction will convey all of this while presenting the work that went into the early successful proposal through the various design and review phases of the Ariel mission.

This chapter serves also as an introduction to the work performed by the author within the Italian contribution to the mission. It begins with an overview of ESA's "Cosmic Vision", the program to which Ariel belongs, it then summarizes the science and scientific objectives of the mission and the scientific requirements that will guide the engineering design and development work.

It will then provide a high level description of the mission payload, review the operational and management aspects of the entire mission development and the contribution of the Italian team, and conclude with the author's role.

Note that the project is still in a design phase, so all specific details provided in this and the coming chapters are provisional and may change as the mission progresses towards the implementation phase.

## 1.1 ESA Cosmic Vision Program

The vision of the European Space Agency is that "Europe must have the ambition to have a space programme and a space agency that is world-class and is leading" [2]. This vision is articulated into a series of commercial, technological and scientific/exploration programs.

Science missions involve a large number of individuals and institutions, both in the public and private sectors, that rely on the stability of a long term plan to invest time and resources in projects that can take over two decades to go from initial concept to the production of scientific results.

"Cosmic Vision" is the current cycle of ESA's long-term planning for space science missions. It follows "Horizon 2000" and "Horizon 2000 Plus", that in the span of 20 years saw the successful launch of Herschel and Planck in 2009, Gaia in 2013, BepiColombo in 2018 and the European contribution to the James Webb Space Telescope (launched in 2021).

"Cosmic Vision 2015-2025", created in 2005 [11], is the logical continuation into the next decade of the ESA science planning cycle. The program already saw the successful launch of its first missions, CHEOPS and Solar Orbiter 2019 and 2020, respectively.

ESA manages the program by periodically issuing a call for proposals for new science missions. The call includes descriptions of the scientific goals, size and cost of the mission, and details the programmatic and implementation details. Calls are addressed to the scientific community.

Missions fall into three categories: small (S-class), medium (M-class) and large (L-class), based on the planned size and breadth of scientific goals addressed, and therefore reflecting on the cost and development time required. A fourth category was introduced in 2018 for a fast (F-class) mission with reduced development time and modest size.

There are four broad scientific themes at the heart of the "Cosmic Vision" program: *what are the conditions for planet formation and the emergence of life? How does the Solar System work? What are the fundamental physical laws of the Universe? How did the Universe originate and what is it made of?*

The study of exoplanets is therefore a stronghold of the program, with a lineup of missions that will keep Europe at the forefront of this growing field by addressing each a unique aspect of exoplanet research (Figure 1.1).

Cheops (CHaracterizing ExOPlanet Satellite, launched in 2019) is the first S-class mission, dedicated to searching for exoplanetary transits by performing ultra-high precision photometry on bright stars already known to host planets.

Plato (PLAnetary Transits and Oscillations of stars, launch planned for 2026) is the third M-class mission with the purpose of finding and studying a large number of extrasolar planetary systems, and in particular of investigating the properties of terrestrial planets in the habitable zone around solar-like stars.

And finally Ariel (Atmospheric Remote-sensing Infrared Exoplanet Large-survey), the 4th adopted M-class mission, is the first dedicated to measuring the chemical composition and thermal structures of the atmospheres of thousands of transiting exoplanets, enabling planetary science far beyond the boundaries of the Solar System.

The three other missions in the M-class lineup are: Solar Orbiter (M1), launched in 2020, Euclid (M2), planned for 2023 and EnVision (M5), the Venus orbiter selected in 2021 [1].

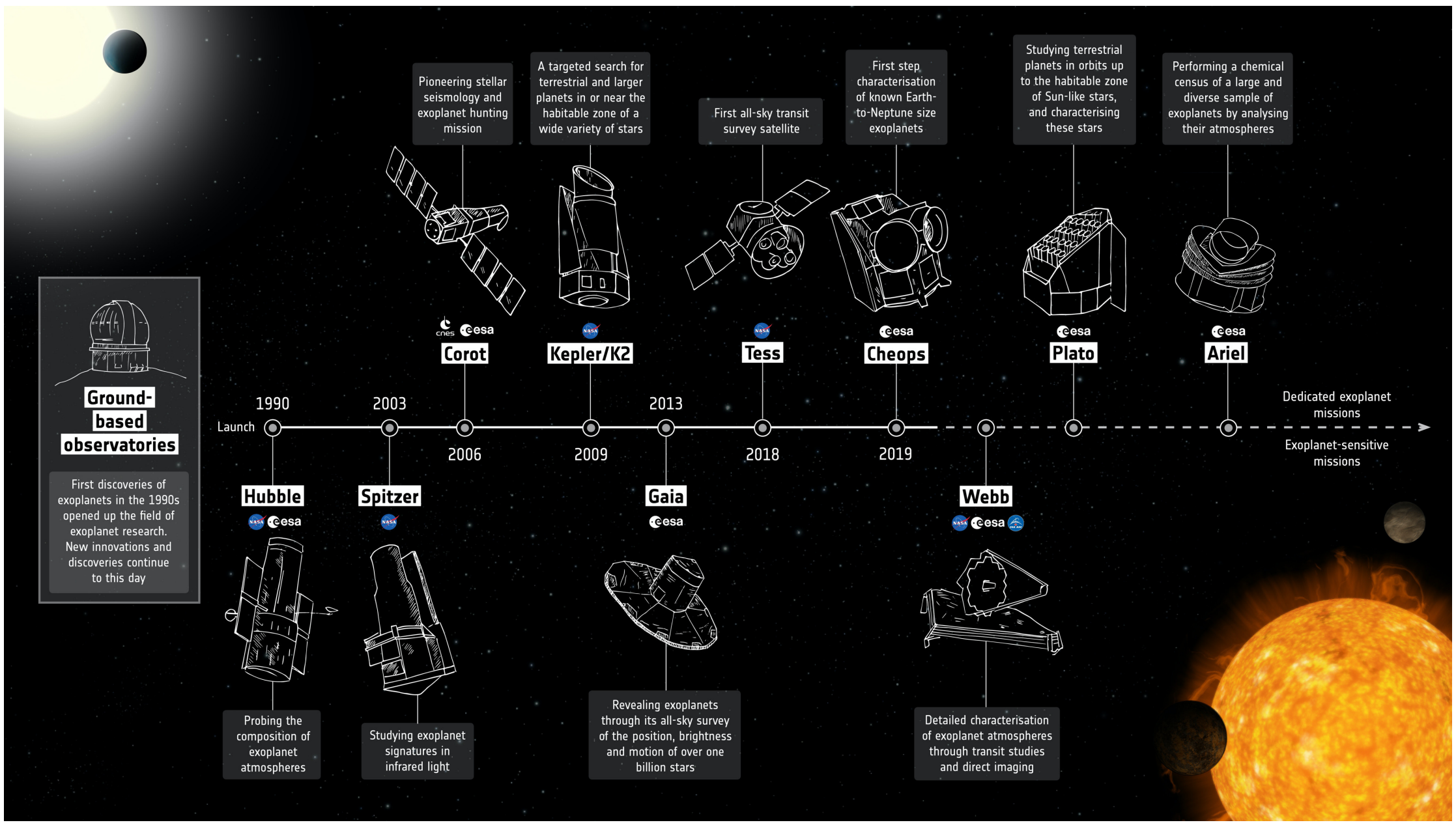

**Figure 1.1:** Infographic of the main space missions dedicated in part or in total to the field of exoplanet research, already flown or in development. Credit ESA.

## 1.2 The Science of Ariel

### 1.2.1 History and scientific objectives

Just a couple of years before the ground-breaking (and Nobel Prize worthy) discovery of the first exoplanet by Mayor and Queloz in 1995 [7], ESA was considering a very ambitious mission, named Darwin, to find Earth-like planets and analyze their atmospheres to detect chemical signatures of life. The mission was however considered too ambitious from a technological point of view and the science not sufficiently developed, so the proposal was eventually rejected in 2007 after the phase-A study [5].

ESA and the exoplanetary community therefore discussed the need to introduce an intermediate step in the research, instead of aiming at the study of Earth-like planets directly. This led to the proposal of the Exoplanet Characterisation Observatory (EChO) as M3 mission in the "Cosmic Vision" program, dedicated to the detection and characterization of exoplanets atmospheres through transit spectroscopy. Giovanna Tinetti was the P.I. of the proposal [9].

The M3 slot was eventually given to Plato, but the technological and scientific research work led in 2015 to a renewed proposal for a transit spectroscopy survey mission for the "Cosmic Vision" M4 slot: Atmospheric Remote-sensing InfraRed Large-survey (ARIEL). The mission was then green-lighted for the preliminary study phase the same year, selected as M4 in 2018 for the definition study Phase B1 (see paragraph 1.4.1 for a description of ESA mission phases), and officially adopted in 2020. The name of the mission was changed to Ariel after selection.

The purpose of the mission is to address the first of the four themes of the "Cosmic Vision" program, reported in Section 1.1: *"What are the conditions for planet formation and the emergence of life?"* by investigating the atmospheres of a large sample of planets of different types, including gas giants, Neptunes, super-Earths and Earthsize planets, orbiting stars of different types.

The reason for observing a sample of a thousand planets is to make a statistical study of the link (if any arises) between the composition and evolution of a planet's atmosphere, its orbital parameters and the nature of the hosting star.

Ariel will employ transit and eclipse spectroscopy in the 1.1–7.8 μm spectral range and photometry in multiple narrow bands covering the optical and near-infrared (NIR). The target is to detect elemental composition of atmospheres of warm and hot planets (eg. C, O, N, S, Si), as well as all the expected major atmospheric gases from e.g. $H_2O$, $CO_2$, $CH_4$, $NH_3$, HCN, $H_2S$ through to the more exotic metallic compounds, such as TiO, VO, and condensed species, and probe the thermal structure, identify clouds and monitor the stellar activity.

The key performance parameters to achieve this target are a high stability of the observations and high noise rejection, in order to detect spectroscopic signals at the level of 10–50 part per million (ppm) relative to the star [10].

### 1.2.2 Exoplanet Atmospheres Detection Techniques

As mentioned in the previous section, the atmosphere of an exoplanet can be studied by discerning the spectroscopic signatures of its component chemical species from the light coming from its hosting star. We review here the techniques that will be employed by Ariel (illustrated in Figure 1.2), since they are relevant to understand mission requirements.

**Transit** When a planet passes in front of its host star, it can be detected by measuring the dimming of the star light flux. This small reduction (a few percentage points) is related to the ratio of the planet-to-star projected area (transit depth). If we additionally measure the wavelength dependence of the dimming, we can observe the spectral signatures of the additional absorption of atomic or molecular species present in the terminator region of the exoplanet's atmosphere. This measurement can also provide an estimation of the radius at the surface of the planet and of its atmosphere.

**Eclipse** When a planet passes behind its host star, its emission/reflection can be obtained directly by measuring the difference between the combined star+planet signal before and after the eclipse, and the stellar flux alone, measured during the eclipse. Recording the flux during the ingress and egress phases of the eclipse additionally allows the creation of a spatially-resolved map of the day-side hemisphere.

**Phase Curve** For the reminder of the time in between transit and eclipse, emission from the planet at different phase angles can be measured by recording the flux of the star+planet system. Such observations can only be performed from space, as they typically span a time interval of more than a day.

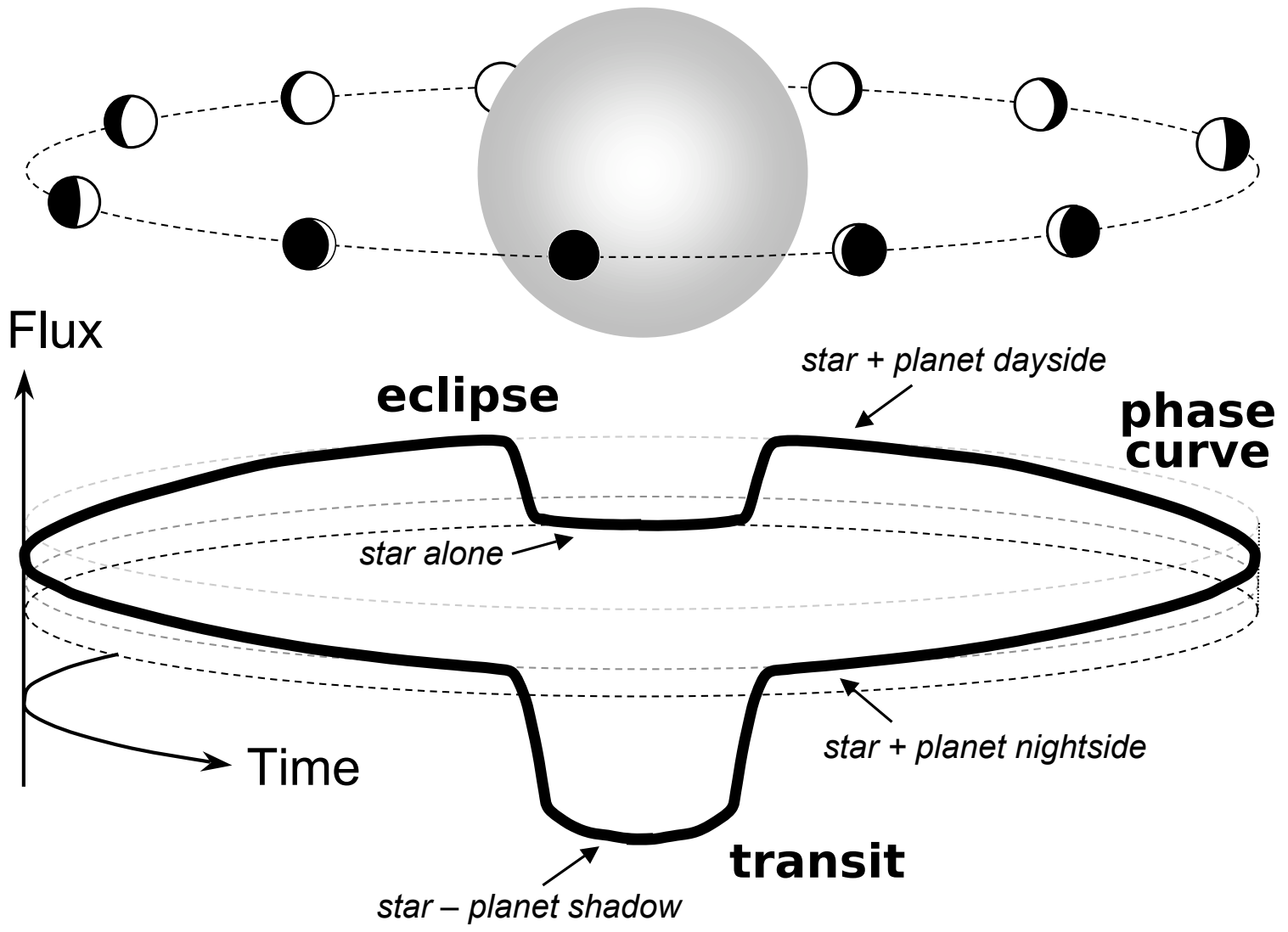

**Figure 1.2:** Depiction of the relative dimming of a star flux when a planet is in transit or eclipse. Credit J. N. Winn (2010) [12].

For a subset of the target sample of exoplanets, Ariel will perform repeated observations employing the three methods described above. This will permit the creation of a record of time series representing the meteorological variations in the planetary atmospheres.

### 1.2.3 Tiered Observational Approach

The detailed and repeated observations described in the previous section could not be carried out during the four years of nominal science operations of Ariel. The observations plan has been therefore structured into the four tiers below.

1. **Reconnaissance Survey**. This tier will cover the largest sample possible (approximately a thousand exoplanets already known at the time of planning) with observations spanning no more than a couple of transits/eclipses, for a rapid characterization of a large population of objects.

2. **Deep Survey**. A subset of Tier 1 planets will be observed at a higher spectral resolution and Signal to Noise Ratio (SNR). The goal is the detection of the atmospheric composition and structure, and to investigate potential correlations between atmospheric chemistry and basic parameters such as planetary radius, density, temperature, stellar type and metallicity.

3. **Benchmark Planets**. The most promising targets in terms of SNR and spectral resolving power will be selected and observations will be repeated through time to assess atmospheric variability, due for example to variations in the cloud coverage or patterns in the global circulation.

4. **Bespoke and Phase-curves observations**. This fourth tier will be dedicated to the observation of interesting targets that do not fit the tiered approach, and to detailed measurement of phase curves. As a dedicated mission, Ariel has the capability and the time for a detailed study of these identified interesting objects.

## 1.3 The Ariel Mission

Now that we reviewed the history and scientific requirements of Ariel, we can proceed to a brief description of the mission profile and the main modules that comprise the payload.

### 1.3.1 Mission Profile

ESA is planning to launch Ariel with an Ariane 6.2 launcher from the Guiana Space Center, in Kourou, French Guiana, in 2029. After launch, the spacecraft will be injected into a direct transfer trajectory towards the Sun-Earth Lagrange L2 point, approximately 1.5 million kilometers away from the Earth. The separation sequence will be optimized taking into account the specific constraints of the cryogenic payload, avoiding as much as possible direct Sun illumination, to minimize any risk of damage.

The final configuration will consist in a large halo orbit of approximately 1 million km around L2. This orbit, similar to the one followed by the James Webb Space Telescope (JWST) [6] and several other missions, has been selected because it offers a very stable thermal environment, important to achieve the level of performance stability required by science specifications, and it provides a large unobscured field of view.

The orbit offers also a constant distance from the Earth (and Sun) and a fixed Sun-Spacecraft angle, simplifying the design of the communications and power subsystems [10].

An additional advantage of the L2 orbit is that it lies far away from Earth orbits crossing the radiation belts, and mostly outside the Magnetosheath, the area of higher energy plasma caused by the acceleration of the solar wind by the Earth's magnetosphere. This provides a relative benign plasma environment [8].

After the mission is completed, the spacecraft will be moved away from the halo orbit with a manoeuvre that will minimize the risk of returning to Earth and polluting the protected LEO and GEO regions.

### 1.3.2 Payload description

The Ariel payload is divided into two broad sections, based on their operating temperature. The "warm" section is located inside the main body of the spacecraft, and consists in the con-trol units of the telescope and the instruments, and the active cooler control electronics and compressors.

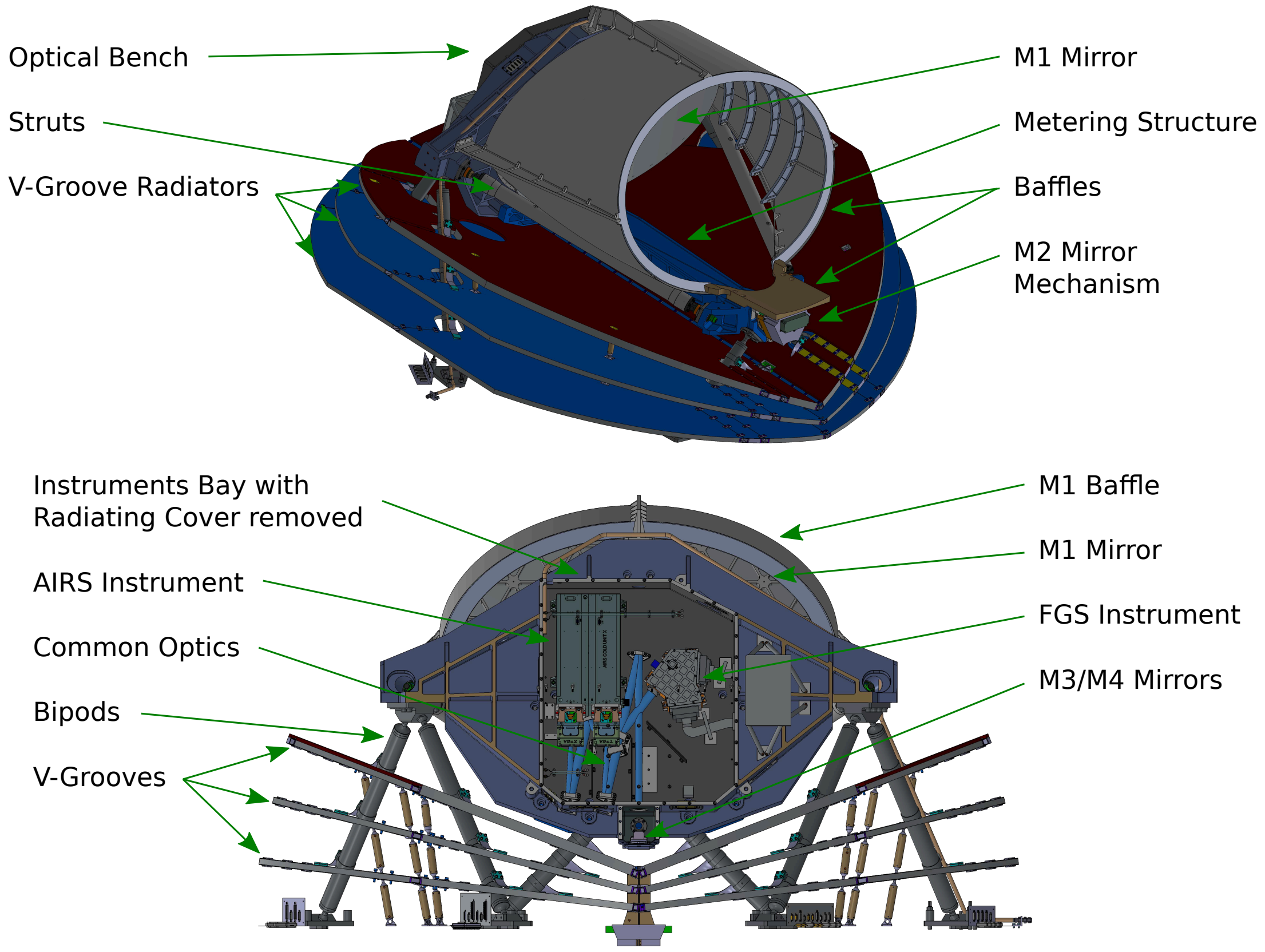

**Figure 1.3:** Ariel Payload mechanical configuration and layout.

The "cold" section includes all the items operating at cryogenic temperatures, shielded from the heat emanating from the spacecraft by the "V-Grooves". The units of the cold section are listed below and illustrated in Figure 1.3.

- The telescope system, consisting of four mirrors (M1–M4), a re-focusing mechanism (M2M) attached to M2, the supporting structures and baffles. The optical and opto-mechanical design of the TA are described in Sections 2.2.2 and 3.1.1.

- An optical bench and metering structure serving as support to the mirrors and instruments. The two items, together with the telescope, are referred to as the Telescope Assembly (TA).

- The Common Optics (CO), a set of fold mirrors and dichroics, relaying the collimated output beam from the telescope to the various instruments.

- The Ariel IR Spectrometer (AIRS), the main scientific instrument of the Payload.
- The combined Fine Guidance Sensor (FGS) / Visible Photometer (VisPhot) / Near-IR Spectrometer (NIRSpec).
- The thermal hardware consisting of the coolers, radiators, V-Grooves and supporting structures to insulate the cold payload from the warmer spacecraft structures.

The telescope is designed to feed a collimated beam into the Common Optics, which in turn split the wavelength bands by dichroics into the two separate sets of instruments with coincident fields of view.

The FGS/VisPhot/NIRSpec instrument contains three photometric channels (VisPhot 0.50–0.60 μm, FGS1 0.60–0.80 μm and FGS2 0.80–1.10 μm) and a coarse resolution spectrometer (NIRSpec 1.10–1.95 μm with resolving power[1] R ≥ 15). FGS1 and FGS2 will also be used as a redundant system for providing guidance and closed-loop feedback to the high stability pointing of the spacecraft. The FGS provides simultaneous information on the photometric stability of the target stars, and NIRSpec is optimized for cloud characterisation.

AIRS contains two channels (AIRS-Ch0 1.95–3.90 μm with resolving power R ≥ 100 and AIRS-Ch1 3.90–7.80 μm with R ≥ 30).

The payload module is passively cooled to a temperature below 50 K by isolation from the spacecraft bus via a series of three V-Groove radiators; the detectors for the AIRS are the only items that require active cooling to <42 K via an active Ne JT cooler.

## 1.4 Ariel Program Management

We present here a brief overview of the mission life cycle and project management philosophy followed by ESA, since its concepts and terminology are used throughout this dissertation, and they are in any case a useful knowledge base for anyone involved in research and development for a space mission.

Most of the concepts presented here are described in a set of standards prepared by the European Cooperation for Space Standardization (ECSS), an "initiative established to develop a coherent, single set of user-friendly standards for use in all European space activities" [4].

### 1.4.1 Mission Development Phases

A space project is typically divided into a series of formal phases, from the initial mission proposal to decommissioning [3]:

- Phase 0 – Mission analysis / needs identification
- Phase A – Feasibility

[1] Defined as $R = \lambda/\Delta\lambda$

- Phase B – Preliminary Definition
- Phase C – Detailed Definition
- Phase D – Qualification and Production
- Phase E – Utilization
- Phase F – Disposal

Each project phase is defined by the activities on system and product level, and progress from one phase to the next is subjected to a formal review process.

Phases 0, A, and B are focused on the elaboration and definition of the main functional and technical requirements at system level, the identification of the activities and resources to be employed to develop the mission and the initial design of the high level program and project management plan.

An important aspect of the initial phases of the project is to identify the level of technological readiness of each process and component employed by the mission, and to plan a proper set of technology development activities to ensure the appropriate level is reached (see for example Section 4.3 for Ariel development program for telescope mirrors).

In the framework of ESA "Cosmic Vision" program, Phases 0 and A usually see various candidate missions competing to be selected for a specific slot, and only the selected mission is then made to progress to Phase B.

In the case of Ariel, Phase B has been further divided into a first definition phase (B1), followed by formal adoption of the mission by ESA, and the start of the implementation phase (B2), when industrial contracts for the development of the service module of the spacecraft and the various instruments and modules could be finally issued.

After Phase B, Phases C and D consist in the development and qualification of the space and ground segments and their products.

Phase E consists in the activities related to the launch, commission, utilization and maintenance of the mission, and finally Phase F comprises all activities to be performed in order to safely dispose all products launched into space as well as ground segment.

As we saw in Section 1.2.1, Ariel initial proposal was submitted to ESA in January 2015. It was then selected, among other two, in June 2015 and entered the study Phase 0/A.

The ESA Science Programme Committee (SPC) eventually selected Ariel as the M4 mission in 2018, kickstarting the Definition Phase B1 study, leading to formal adoption in November 2020, as ESA "Cosmic Vision" M4 mission, and marking the beginning of the Implementation Phase (B2/C), the selection of the industrial partners and kick-off of the co-engineering activities.

### 1.4.2 Management and Italian Role

As explained above, ESA space missions are proposed by a consortium of several research institutes belonging to different European and non-European countries with a focus on experience and diversity since one of the goals of ESA is to promote the development of space related technologies in emerging countries.

During the study phases and after a proposal has been officially adopted, several industrial contracts are also issued for further technological development and manufacturing of the components, and for the Assembly, Integration and Verification (AIV) of the spacecraft.

The Ariel Consortium is led by Prof. Giovanna Tinetti[2], Principal Investigator and Paul Eccleston[3], Consortium Project Manager, and comprises teams from 11 European countries, the USA and Japan.

Each team is responsible for specific modules or design/analysis functions. The UK team is responsible for AIV of the scientific payload and for managing the entire program (Figure 1.4).

The Italian team is responsible for developing the technologies required for the manufacturing of the telescope mirrors, and since Phase B2, also for the Telescope Assembly Work Package. This package consists in several deliverables, some requiring coordination with teams from other counties:

- manufacturing of the optical bench, metering structure and connecting struts;
- design and manufacturing of the four mirrors and supports;
- design and manufacturing of the thermal control and decontamination hardware (eg. heaters, cryo-harnesses and sensors);
- the M2 mirror refocusing mechanism (M2M), provided by SENER[4] under the supervision of IEEC[5];
- the telescope baffles, provided by AST[6].

Other transport and testing equipment and facilities will be provided by Admatis[7] and CSL[8]. A team at UPM[9] is in charge of the overall mechanical design of the TA, with input from the various teams involved.

[2] Department of Physics and Astronomy, University College London, Gower Street, London WC1E 6BT, UK
[3] RAL Space, STFC Rutherford Appleton Laboratory, Didcot, Oxon, OX11 0QX, UK
[4] SENER Grupo de Ingeniería, Cervantes, 8, 48930 Getxo, Spain
[5] Institut d'Estudis Espacials de Catalunya, Gran Capità, 2-4, Edifici Nexus, Desp. 201, 08034 Barcelona, Spain
[6] Active Space Technologies, S.A., Parque Indústrial de Taveiro, Lote 12, 3045-508 Coimbra, Portugal
[7] Admatis, 5. Kandó Kálmán Street, 3534 Miskolc, Hungary
[8] Centre Spatial de Liège, Avenue du Pré-Aily, 4031 Angleur, Belgium
[9] Escuela Técnica Superior de Ingeniería Aeronáutica y del Espacio (ETSIAE), Universidad Politécnica de Madrid (UPM), Pza.Cardenal Cisneros 3, 28040 Madrid, Spain

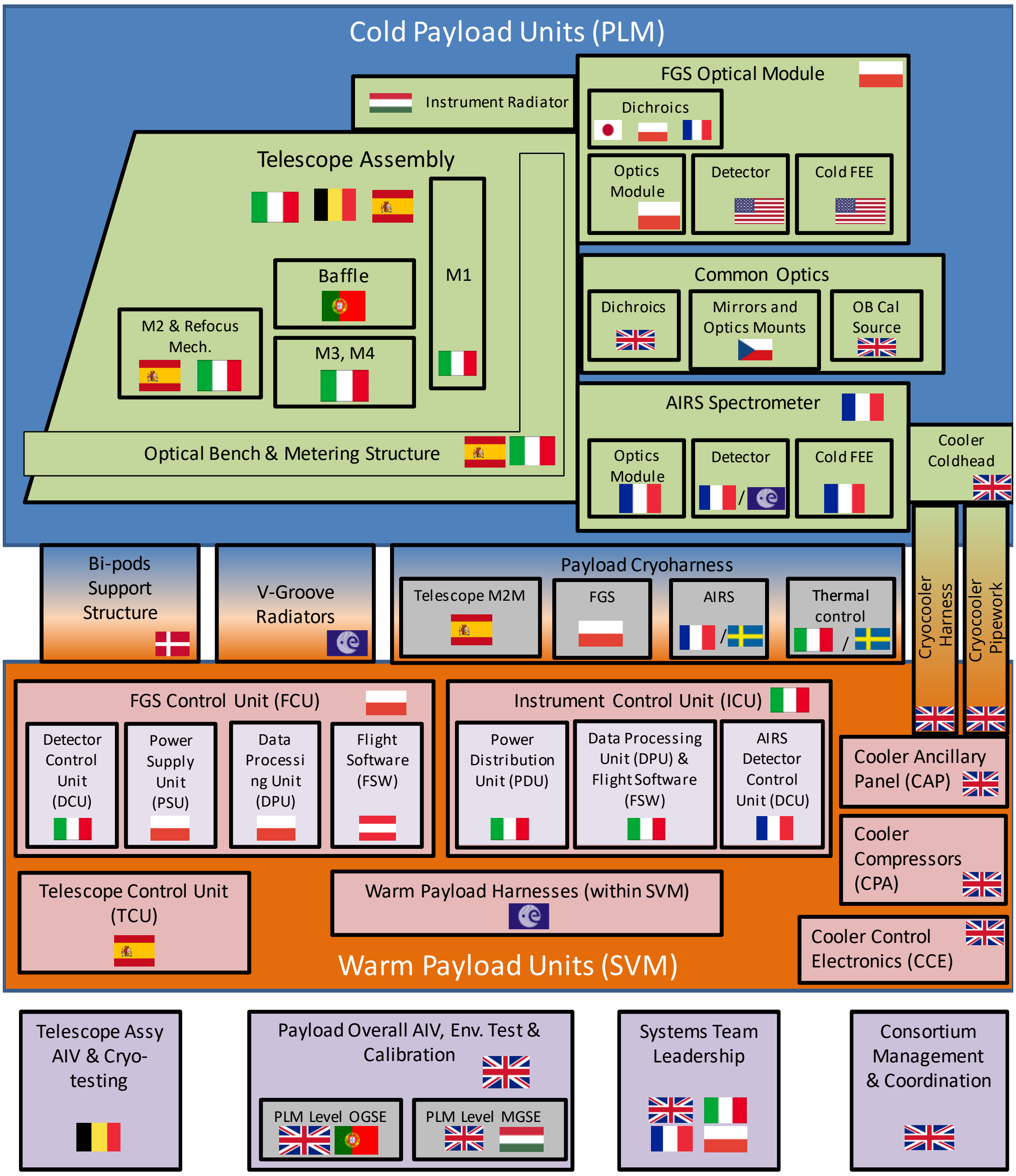

**Figure 1.4:** Infographic of Payload architecture and responsibilities of the countries in the Ariel Consortium. Credit: Ariel Consortium.

The contract for manufacturing the Italian TA components and for the AIV relative to the current project phase has been awarded to Leonardo S.p.A.[10].

The Italian team is led by the co-PIs Dr. Giuseppina Micela[11] and Dr. Giuseppe Malaguti[12], and the Project Manager Prof. Emanuele Pace[13].

Within the Italian contribution to the mission, the team at CNR-IFN in Padova was in charge, during Phases A and B1, of the Telescope optical design and the opto-mechanical development of the Telescope mirrors (including optical coating), under the lead of Dr. Vania Da Deppo. For Phases B2 and C, and with the lead of Dr. Paola Zuppella, it is currently responsible for the PA/QA Management of the Telescope Assembly.

### 1.4.3 Role of the author

The author's involvement with the Ariel mission started during Phase B1, within the work package concerned with the development of the telescope mirrors.

In particular, the activities carried out with the author's support concerned the fabrication and characterization of a full size demonstrator of the primary mirror of the telescope, with the purpose of qualifying the technological readiness of the manufacturing technologies. Specifically, the activities consisted of:

- the optical and tolerance analysis of the design of the telescope and mirrors (Chapter 3);
- the design and analysis of the procedures employed to qualify for technological readiness the telescope mirrors manufacturing processes, and in particular substrate thermal stabilization, optical surface shaping and polishing (Chapter 4);
- the analysis of surface shape and surface roughness measurements (Chapter 4);
- the support to optical simulations of the performance degradation induced by thermoelastic effects;
- the qualification of the optical coating process, the simulations and characterization of coating performance and the assessment of its lifetime durability by environmental testing (Chapter 5).

After successful completion of the study phase and official mission adoption in November 2020, involvement of the author continued as member of the Italian Optical Engineering and PA/QA teams, formally within the work package for the telescope PA/QA Management.

The work consisted on further refinement of the analyses and design of the telescope and in particular on the following areas:

[10] Leonardo S.p.A. Space Division, Via delle Officine Galileo, 1, 50013 Campi Bisenzio (FI), Italy

[11] INAF-Osservatorio Astronomico di Palermo, Piazza del Parlamento 1, 90134 Palermo, Italy

[12] INAF-Osservatorio di Astrofisica e Scienza dello spazio di Bologna, Via Piero Gobetti 93/3, 40129 Bologna, Italy

[13] Dipartimento di Fisica ed Astronomia-Università degli Studi di Firenze, Largo E. Fermi 2, 50125 Firenze, Italy

- mirrors surface error tolerancing (Chapter 3);
- tolerancing of mirrors mechanical alignment both on ground and in flight (Chapter 3);
- Structural, Thermal, Optical Performance (STOP) analysis (Chapter 3);
- analysis of the telescope throughput performance (Chapter 5).

## References

[1] *ESA Science & Technology - Policy for Missions of Opportunity in the ESA Science Directorate*. URL: https://sci.esa.int/web/cosmic-vision/-/59977-missions-of-opportunity (visited on 06/22/2022).

[2] *ESA Vision*. URL: https://vision.esa.int/ (visited on 06/22/2022).

[3] European Cooperation for Space Standardization. *Space Project Management – Project Planning and Implementation*. ECSS-M-ST-10C Rev. 1. ECSS Secretariat, ESA-ESTEC, Mar. 6, 2009.

[4] *European Cooperation for Space Standardization (ECSS)*. URL: https://ecss.nl/ (visited on 06/24/2022).

[5] A. Leger and T. Herbst. *DARWIN Mission Proposal to ESA*. arXiv:0707.3385. July 23, 2007. arXiv: 0707.3385. URL: http://arxiv.org/abs/0707.3385 (visited on 06/23/2022).

[6] P. A. Lightsey. "James Webb Space Telescope: A Large Deployable Cryogenic Telescope in Space". In: *Boulder Damage Symposium XXXIX: Annual Symposium on Optical Materials for High Power Lasers*. Oct. 10, 2007, 67200E. DOI: 10.1117/12.754010.

[7] M. Mayor and D. Queloz. "A Jupiter-mass Companion to a Solar-Type Star". In: *Nature* 378.6555 (Nov. 23, 1995), pp. 355–359. DOI: 10.1038/378355a0.

[8] J. Minow, W. Blackwell, and A. Diekmann. "Plasma Environment and Models for L2". In: *42nd AIAA Aerospace Sciences Meeting and Exhibit*. Jan. 5, 2004. DOI: 10.2514/6.2004-1079.

[9] G. Tinetti et al. "EChO: Exoplanet Characterisation Observatory". In: *Experimental Astronomy* 34.2 (Oct. 2012), pp. 311–353. DOI: 10.1007/s10686-012-9303-4.

[10] G. Tinetti et al. *Ariel: Enabling Planetary Science across Light-Years, Ariel Definition Study Report*. Version 1. arXiv, 2021. arXiv: 2104.04824.

[11] A. Wilson et al., eds. *Cosmic Vision: Space Science for Europe 2015-2025*. ISBN 978-92-9092-489-0 247. Noordwijk, The Netherlands: ESA Publications Division, ESTEC, 2005. 109 pp. ISBN: 978-92-9092-489-0.

[12] J. N. Winn. *Transits and Occultations*. arXiv:1001.2010. arXiv, Sept. 24, 2014. arXiv: 1001.2010.

*Music is the arithmetic of sounds as optics is the geometry of light.*

Claude Debussy, quoted in Simonton, D. K., *Greatness: Who Makes History and Why* (1994)

# 2
# Ariel Telescope Optical Analysis and Design

This chapter presents the optical design of the Ariel telescope, providing the necessary background to the research activities described in the reminder of this dissertation.

The design choices are explained in terms of the optical performance requirements, showing how they derive from the higher level scientific requirement of the mission.

## 2.1 Ariel Scientific Requirements

Ariel will employ transit and eclipse spectroscopy and photometry to study exoplanets atmospheres by detecting many molecular species, probing the thermal structure, identifying and characterizing clouds, and monitoring host stars activity (Section 1.2.2). These techniques require temporal resolution of the signal, but no significant angular resolution: this is the governing principle upon which the design of Ariel's telescope is based, driving the need of a very stable and low noise system.

So far, exoplanet atmospheres have been detected using general-purpose space- or ground-based instrumentation, but these measurements suffer from limited and patchy spectral coverage, and from systematic errors due to pointing jitter, thermal and opto-mechanical instability or other shortcomings linked to detectors. The Ariel payload was designed considering the lessons learned in measuring exoplanetary atmospheres using Spitzer, the Hubble Space Telescope and ground-based instruments [9].

The wavelength range was specifically optimized to cover all the expected major atmospheric gases from, e.g. $H_2O$, $CO_2$, $CH_4$, $NH_3$, HCN, $H_2S$, through to the more exotic metallic compounds, such as TiO, VO, and condensed species.

Comprehensive radiometric (*ArielRad* [6]) and transit spectroscopy (*ExoSim* [7]) simula-tions of the Ariel payload have been performed since early on in the project to predict performance and guide the optimization of the payload design.

**Table 2.1:** Summary of Ariel required spectral coverage by instrument and resolving power. For an overview of the instruments see Section 1.3.2.

| Instrument | Wavelenght Range | Resolving Power |
|---|---|---|
| VisPhot | 0.50–0.60 μm | Integrated band |
| FGS1 | 0.60–0.80 μm | Integrated band |
| FGS2 | 0.80–1.10 μm | Integrated band |
| NIRSpec | 1.10–1.95 μm | R ≥ 20 |
| AIRSCh0 | 1.95–3.90 μm | R ≥ 100 |
| AIRSCh1 | 3.90–7.80 μm | R ≥ 30 |

**Table 2.2:** Summary of the high level optical requirements of the Ariel Mission.

| Parameter | Value |
|---|---|
| Spectral SNR | average ≥7 (≥10 for Tier 4 observations) |
| Photometric stability | 20–100 ppm over an individual transit/occultation, depending on target brightness |
| Effective collecting area | ≥0.6 $m^2$ |
| System throughput | Spectral channels: ≥40 % |
| | Photometric channels: ≥50 % |

These studies show that Ariel will obtain spectroscopic and photometric time series of transiting exoplanets with better than 20 to 100 ppm stability over a single transit observation, depending on the target brightness and provided a sufficient margin is guaranteed on system throughput, noise, stability and pointing jitter.

The resulting set of requirements are summarized in Tables 2.1 and 2.2.

## 2.2 Ariel Telescope Design

### 2.2.1 Telescope Performance Requirements

The set of requirements imposed on the telescope of the Ariel mission, derived from the higher level science requirements (Section 2.1) and from the characteristics of the instruments downstream from the telescope, are summarized in Table 2.3.

The design is driven by the requirement that the telescope in operational conditions has to be diffraction limited at the wavelength of 3 µm over a Field of View (FoV) of 30″, which corresponds approximately to an RMS Wavefront Error (WFE) of 200 nm according to the Marechal Criterion (see Section 2.3.2). This requirement was derived from the performance simulations cited in Section 2.1.

**Table 2.3:** Summary of Ariel Telescope optical requirements.

| Parameter | Value |
|---|---|
| Collecting area | ≥0.6 $m^2$ |
| FoV | 30″ with diffraction limited performance |
| | 41″ with optical quality for centroiding |
| | 50″ unvignetted |
| Wavefront Error (WFE) | Diffraction limited at 3 µm |
| Wavelenght range | 0.5–7.8 µm |
| Throughput | Minimum >0.78 |
| | Average >0.82 |
| Output beam dimension | 20.0 mm × 13.3 mm (ellipse axes) |
| Angular magnification | 55 |

The telescope throughput requirement is derived instead from a break-down of the overall system throughput budget between telescope, common optics and instruments. More details on the budget are provided in Chapter 5.

The Field of View (FoV) of the telescope is derived from the specific requirements from the instruments, and from the analysis of the alignment between instruments and telescope, with appropriate margins.

Figure 2.1 shows the FoV requirements in the case of the maximum mis-alignment of the FGS and AIRS instruments. The main requirement is to have a well resolved PSF at the center of the AIRS spectrometer FoV of 6.4″ × 26.4″ (orange rectangle). Considering an additional offset of ±10″ and a measurement margin of ±2″, the total diffraction limited FoV radius becomes ±15.2″, approximated in the requirements to a total FoV angle of 30″.

A larger 41″ telescope FoV is given by the 17″ FoV of FGS (blue transparent circle). The image over this 41″ annulus can be of lower quality, but must be sufficient to allow the star centroid to be well enough resolved to be initially located and then brought to the centre of the FGS FoV, where the telescope image quality is better.

A still larger telescope FoV, extending to 50″, is required to capture the slit background. There are no image quality requirements over this additional FoV, the only real requirement is for the FoV to be unvignetted so that background photons reach the slit.

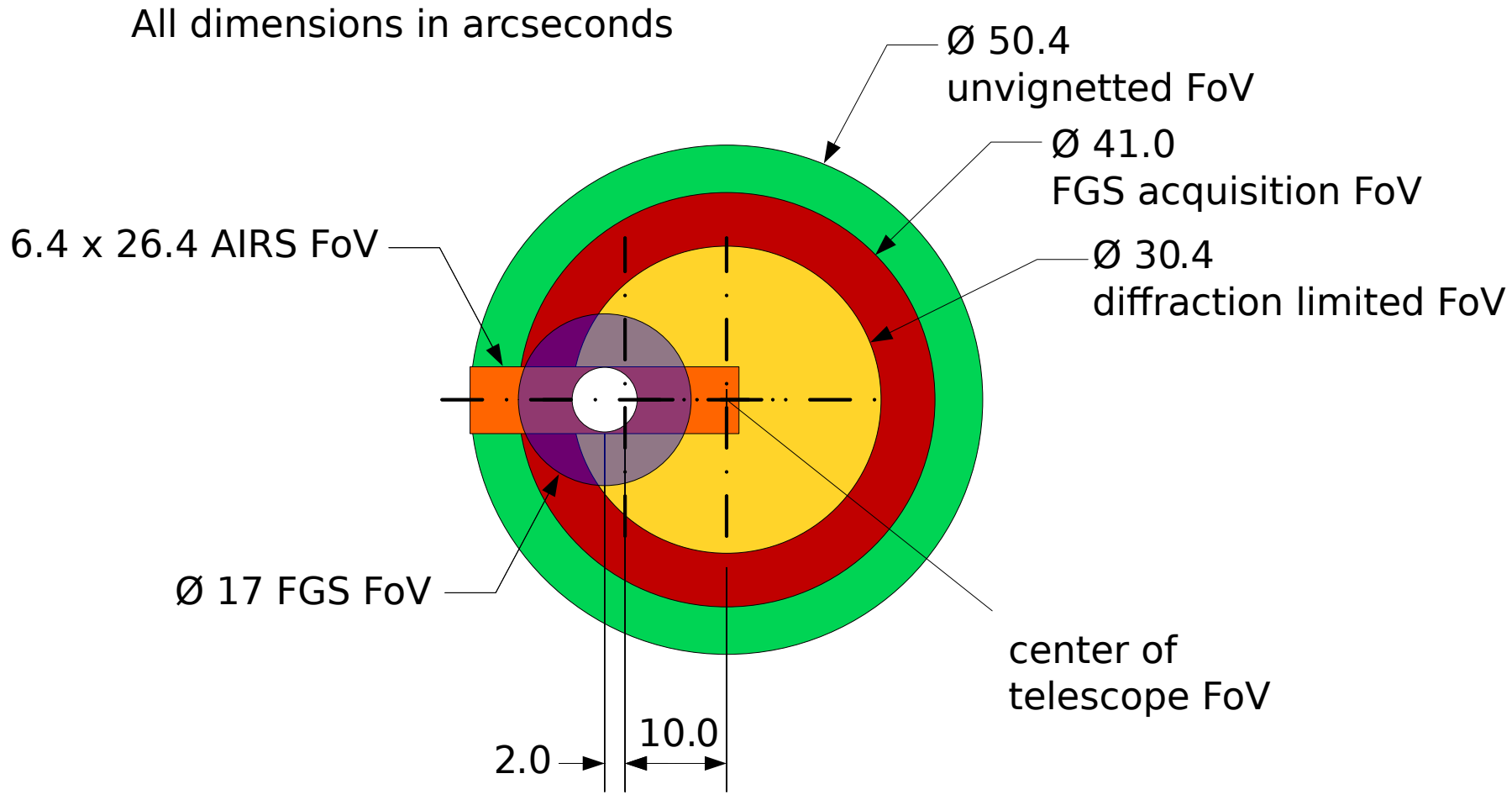

**Figure 2.1:** Telescope FoV, showing the offset of $\pm 10''$ and a measurement margin of $\pm 2''$.

### 2.2.2 Telescope Optical Design

The main requirements driving the design of Ariel's telescope derive from the goal of observing exoplanets atmospheres by discriminating the contribution of the light reflected or absorbed by the exoplanet from that emitted by the host star system. The two components will be separated by following the temporal evolution of the spectrogram of the incoming light (i.e. studying the transits and eclipses of the exoplanet), rather than trying to achieve a spatial separation. This methodology relaxes the requirements on the optical quality and spatial resolution of the system (setting aside the requirements for centroiding and pointing done by the FGS subsystem).

Moreover, exoplanetary systems will be observed one at a time, requiring a very small Field of View (FoV). This small FoV and relaxed optical quality requirements allow emphasis to be placed on optimising other aspects of the telescope such as stability, rejection of stray light and minimization of the thermal background.

As we saw in the introduction (Section 1.2.1), the Ariel Mission concept descends from the preliminary study done for EChO, and in particular the design of Ariel's telescope derives directly from it [2], with a series of straightforward adjustments to reduce mass and size. In particular, to maintain the overall payload design with minimum adaptations, the telescope collecting area was reduced to 0.6 $m^2$ , almost half of the one originally envisioned for EChO. This implies an entrance pupil of the order of 1 m in diameter.

To select a design for the EChO telescope, the study team realized a trade-off between 10 different proposals, based on performance, technological readiness and overall manufacturability. The choice for the baseline telescope concept then fell on an afocal Three Mirror Anastigmat (TMA) of the Korsch type, consisting of an elliptical primary mirror in an off-axis, unobscured configuration in order to maximize throughput without increasing the collecting area. Contrary to a typical Korsch system with a large FoV, the EChO concept had been optimised to produce a very small exit pupil, to simplify procurement and optimizing volume and cost of the optics for the instruments [2].

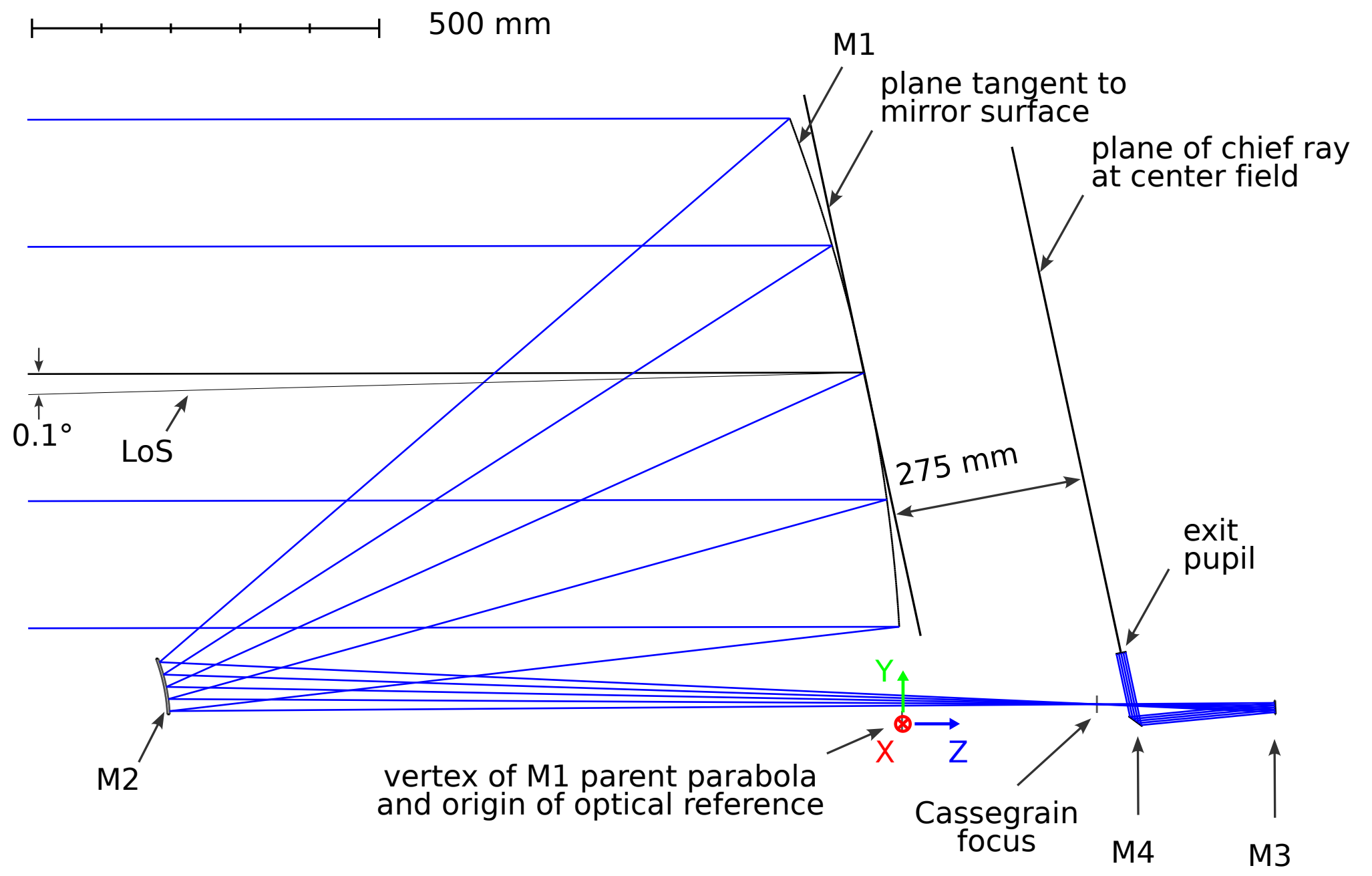

**Figure 2.2:** Scale drawing of the telescope optical design on the Y-Z optical axis plane. The 0.1° offset is exaggerated for clarity.

The Ariel design team eventually opted for a Cassegrain design (parabolic primary M1 and hyperbolic secondary M2) with the third mirror M3 used to recollimate the beam to its desired elliptical footprint with 20.0 mm × 13.3 mm axes (Figure 2.2). The three mirrors share the same optical axis.

In the following paragraphs we will describe the main details of the optical design of Ariel's telescope. Please note that all design dimensions in this section refer to the operational telescope temperature of 50 K.

After M3, a plane mirror (M4) is redirecting the exit beam parallel to the back of M1 at a distance of 275 mm, where the OB and instruments are located.

The aperture stop, located at the primary mirror defines the elliptical entrance pupil with major and minor axes dimensions of 1100 mm × 730 mm.

The center of the FoV of the telescope is inclined of 0.1° in the Y-Z plane with respect to the optical axis of the telescope defined by the mirrors common optical axis, to give an accessible return beam from M3.

M2 has a refocus mechanism with three degrees of freedom: focus and tip/tilt. The purpose is to correct for one-off movements due to launch loads and cool-down and to make occasional adjustments, for example to compensate for any long term drifts in structural stability. The use of M2 to correct the specific misalignments that may arise due to launch loads

**Table 2.4:** Summary of the characteristics of the four telescope mirrors at operating temperature (50 K). Apertures are defined on a plane orthogonal to the optical axis, except for M4, where it is defined on the plane of the mirror itself.

| Mirror | M1 | M2 | M3 | M4 |
|---|---|---|---|---|
| Radius of curvature (mm) | 2319.432 | 239.141 | 509.697 | – |
| Type | Concave | Convex | Concave | Plane |
| Conic constant $k$ | -1 | -1.392 | -1 | – |
| Aperture decenter (y direction, mm) | 500 | 50 | 20 | – |
| Clear aperture shape | Elliptical | Elliptical | Elliptical | Circular |
| Aperture $x, y$ semi-axes (mm) | 550 × 365 | 56 × 40 | 15 × 11 | 12 × 12 |

is analyzed in details in Paper 1, introduced in Section 3.2.2.

Table 2.4 summarizes the characteristics of the optical elements.

## 2.3 Optical Performance

Before proceeding with a brief overview of the performance of the nominal design of the telescope, we will review two basic optical analysis concepts that will be used throughout this dissertation, and in particular for the tolerance analysis described in Section 3.2.2: wavefront error and encircled energy.

For a more thorough treatment of the concepts, please refer for example to [3, Chapter 4] and [4, Chapter 10].

### 2.3.1 Wavefront Error

The wavefront error (WFE) is a very commonly used performance parameter for optical systems since it can be easily and quickly modeled with optical analysis software and measured with precision using interferometers. In the Geometric Optics approximation, and considering the wavefront of the light wave progressing through the various elements of an optical system, the wavefront error at a specific section of the system is defined as the optical path difference (OPD) between the actual wavefront and an ideal one. It therefore gives a measure of the degree of aberration of an optical system. In the case of a focal system, the ideal wavefront is a sphere, while for an afocal system it is a plane.

The WFE is usually expressed in terms of the difference between the maximum and minimum OPD, or Peak to Valley (PtV) using an obvious geographical analogy, or as the square root of the mean of the squares of the OPDs, sampled regularly on the wavefront area. This is indicated as RMS (Root Mean Squared), and is calculated with the formula below, where $n$ stands for the number of samples.

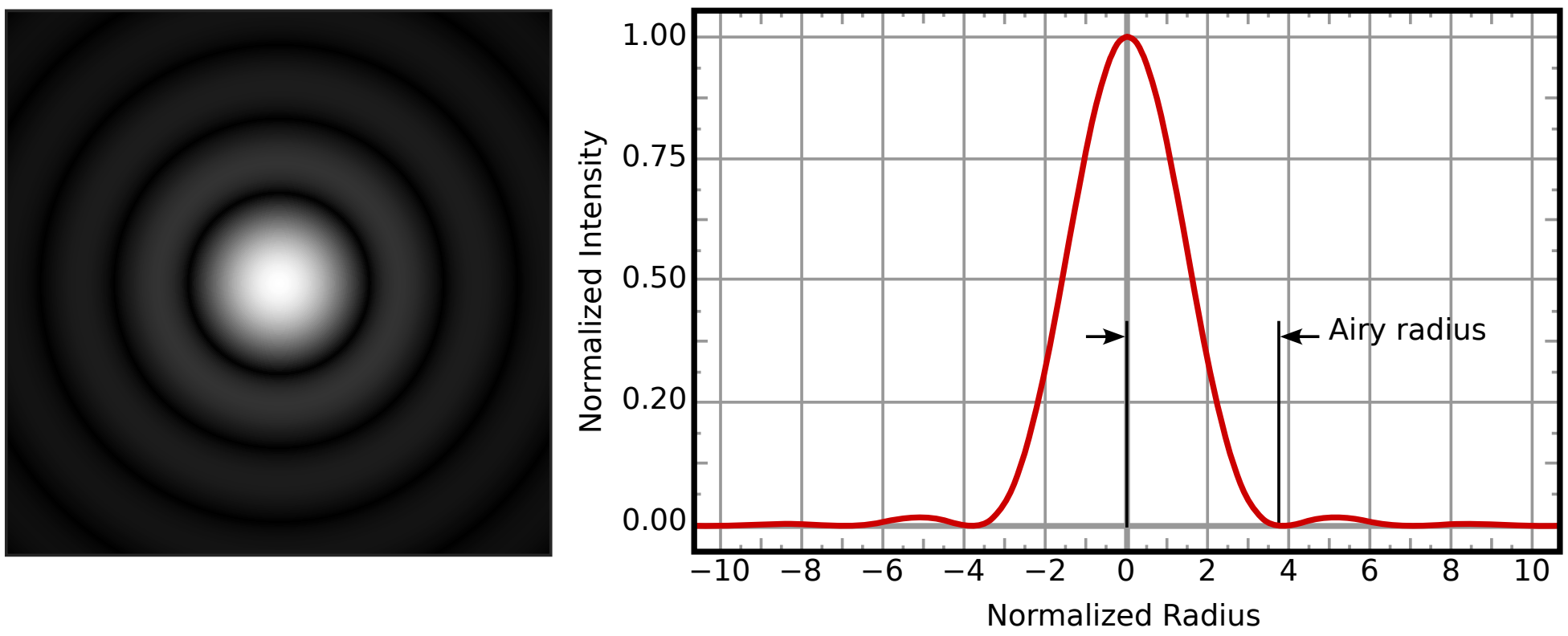

**Figure 2.3:** 2-dimensional representation of the Airy disk (left) and cross-section (right), with the radius of the first ring highlighted.

$$\text{RMS WFE} = \sqrt{\frac{1}{n}\sum_{i=1}^{n}\text{OPD}_i^2}$$

In the case of a plane or spherical mirror at normal incidence, the WFE of the light reflected by it is simply twice the surface error (SFE), intended as the difference in height between the actual surface of the mirror and the nominal one [5].

### 2.3.2 Encircled Energy

The Encircled Energy is instead defined in terms of the so called Point Spread Function (PSF), which is the response of an optical system to a point source or point object.

This response is generally expressed as the irradiance distribution at the image plane, as a function of linear spatial coordinates for a stigmatic system, or angular coordinates for an afocal system such as Ariel's telescope.

For a "perfect" system, that is a system that generates an image that is perfectly in focus, with no geometrical aberrations, the PSF is determined solely by the diffraction of the incoming wavefront through the aperture, and when the aperture is circular, takes the well-known shape of the Airy disk (Figure 2.3).

The energy contained in the central circular area of the disk (first ring) is approximately 83.8 % of the total. The radius of the central area, for a system with focal length $f$ and aperture $D$, at the wavelength $\lambda$, is:

$$r_1 \approx 1.22\frac{f\lambda}{D}.$$

Such systems, where aberrations are negligible and the PSF shape is only determined by diffraction, are called *diffraction limited*. A. Maréchal in 1947 provided an approximate criterion, still in use today, to determine the maximum RMS WFE for a diffraction limited system using the ratio $\lambda/14$. In the case of Ariel telescope, this leads to

$$3\ \mu\text{m}/14 = 214\ \text{nm RMS},$$

which was then approximated to 200 nm RMS in the preliminary telescope requirements reported in Section 2.2.1.

Equivalently, considering the ratio $S$ (Strehl ratio) between the image intensity of the aberrated system and the maximum attainable intensity of an ideal optical system limited only by diffraction and with the same aperture, a system can be considered approximately *diffraction limited* if $S \geq 0.8$ [1, Chapter 9.2].

In an aberrated system, the energy in the PSF is spread out and a way to quantify this is to calculate the radius of the circle that encloses a specific amount of energy, hence the term *Encircled Energy*. In the case of Ariel, the elliptical aperture produces an "elliptical" PSF, so we talk instead of energy enclosed in an ellipse.

### 2.3.3 Telescope Optical Performance

The optical design and analysis of the Ariel telescope has been carried out with the optical design software Zemax[1] OpticStudio®. Optimization of the design is based on minimal average wavefront error for nine point sources (eigth equally spaced on a circle, and one on the center) placed at an infinite distance from the telescope entrance pupil (fields), and at an angle of 30″ from the telescope Line of Sight and center of the FoV, and for the three most important observational wavelengths (1.95 µm, 3 µm and 7.8 µm), to guarantee diffraction limited performance according to the requirement specified in Section 2.2.1.

While the actual design was optimized considering also the additional optical paths through the Common Optics for the FGS and AIRS channels, since they are outside the scope of this dissertation, we report only the optical performance analysis at the telescope exit pupil.

Since the telescope is afocal, the spot diagrams are reported with the *afocal image space* option of OpticStudio turned on, appropriate for systems with collimated output.

In Figure 2.4 we report the Spot Diagrams corresponding to the eight fields at the edge of the 30″ FoV and at the wavelength of 3 µm, where the telescope is required to have diffraction limited performance. A spot diagram shows where a uniformly spaced set of rays, traced from a specific field, are intersecting the image surface. It is a visual representation of the light distribution over the image plane and provides information on the magnitudes of the various geometrical aberrations in the optical system [8].

The RMS of the spot (i.e. the RMS of the distances of the points of intersection from the centroid) can be compared with the Airy disk radius to give an idea of how far the system is from the diffraction limit. In our case, the largest RMS spot radius is 2.3″, much smaller than the Airy disk radii, respectively 38″ and 57″ in the $x$ and $y$ directions.

[1]ANSYS, Inc., 2600 Ansys Drive, Canonsburg, PA 15317, United States

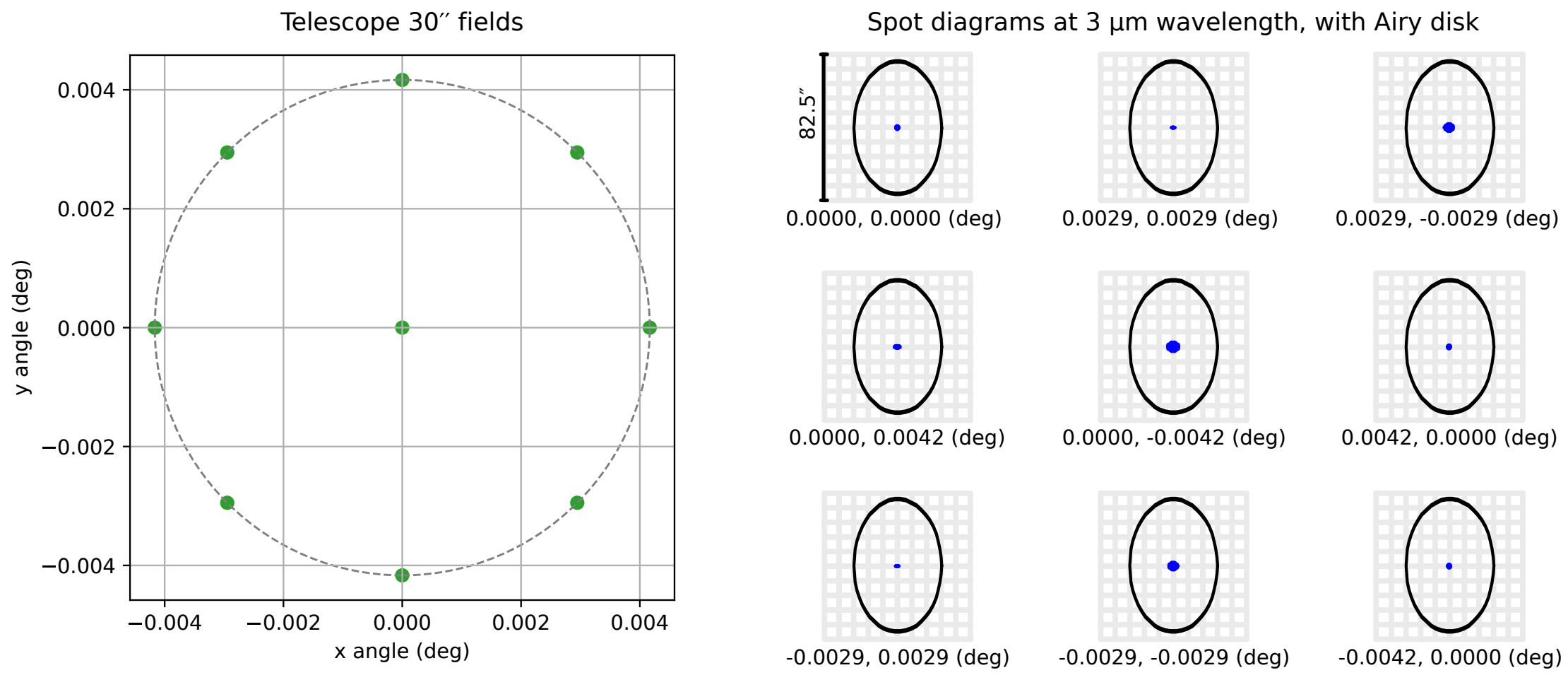

**Figure 2.4:** Position of the fields on a 30″ circle used for the optical analysis (left), and Spot Diagrams for each of the fields shown with the Airy disk (black ellipse) corresponding to the major and minor axes of the telescope aperture, at the wavelength of 3 µm (right).

Figure 2.5 shows instead the RMS wavefront error and the Strehl Ratio as functions of the angles in $x$ and $y$ of the field position with respect to the telescope FoV. The maximum value of the wavefront is

$$9.27 \times 10^{-3} \text{waves} \cdot 3\ \mu\text{m} = 27.8\ \text{nm RMS},$$

to be compared with the requirement of 200 nm RMS. Note however that the requirement is given for the as-built, operational telescope, so compliance can only be verified through additional analyses, as we will see in Chapter 3.

Finally, Figure 2.6 is the PSF of one of the radial fields at the 30″ FoV, showing no visual sign of distortion as expected since the telescope at this wavelength is diffraction limited.

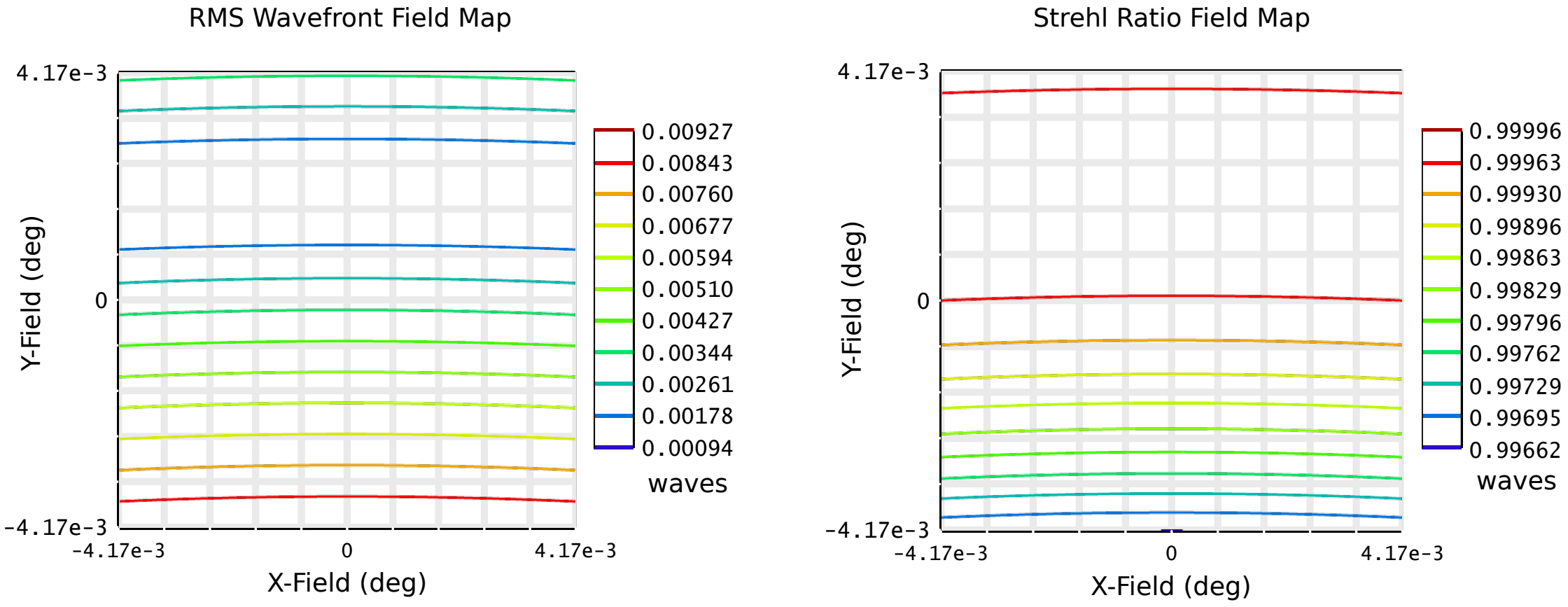

**Figure 2.5:** Chart of the RMS Wavefront error (left) and Strehl ratio (right) of the nominal telescope as functions of the field angle over 30″ in the $x$ and $y$ directions, at the wavelength of 3 µm.

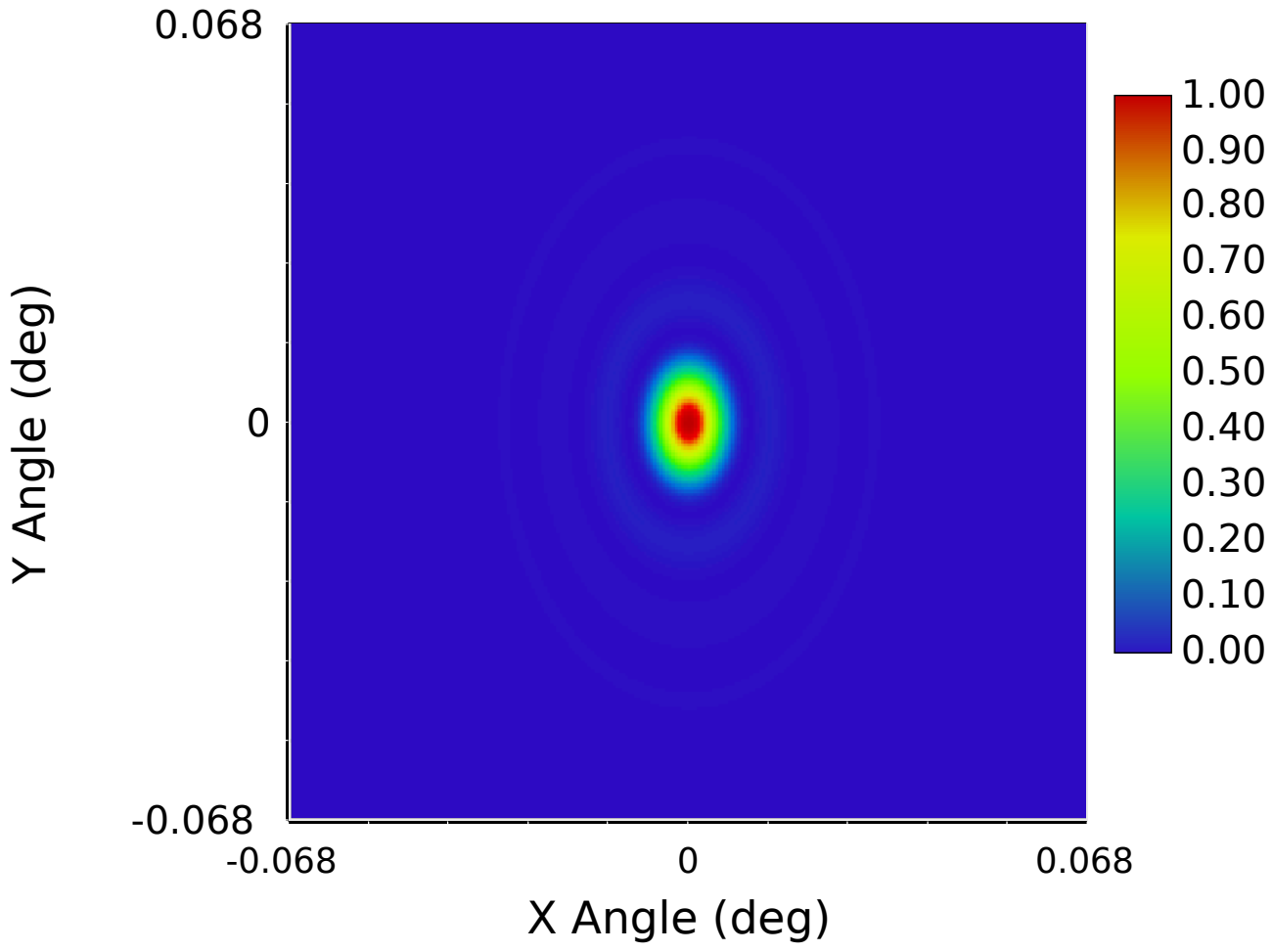

**Figure 2.6:** Telescope normalized PSF for the radial field with the highest aberration (−30″ $y$ field), at 3 µm wavelength.

## References

[1] M. Born and E. Wolf. *Principles of Optics: Electromagnetic Theory of Propagation, Interference and Diffraction of Light*. Cambridge; New York: Cambridge University Press, 1999. ISBN: 978-0-521-64222-4.

[2] EChO Study Science Team. *Exploring the Atmospheres of Diverse Worlds beyond Our Solar System, EChO Assessment Study Report (Yellow Book)*. Dec. 16, 2013. URL: https://sci.esa.int/s/89za2QA.

[3] R. E. Fischer, B. Tadic-Galeb, and P. R. Yoder. *Optical System Design*. New York, London: SPIE Press, 2008. ISBN: 978-0-8194-6785-0.

[4] E. Hecht. *Optics*. 5 ed. Boston: Pearson Education, Inc, 2017. 714 pp. ISBN: 978-0-13-397722-6.

[5] D. Malacara, ed. *Optical Shop Testing*. 3rd ed. Wiley Series in Pure and Applied Optics. Hoboken, N.J: Wiley-Interscience, 2007. 862 pp. ISBN: 978-0-471-48404-2.

[6] L. V. Mugnai et al. "ArielRad: The Ariel Radiometric Model". In: *Experimental Astronomy* 50.2 (Dec. 1, 2020), pp. 303–328. DOI: 10.1007/s10686-020-09676-7.

[7] S. Sarkar et al. "ExoSim: The Exoplanet Observation Simulator". In: *Experimental Astronomy* 51.2 (Apr. 1, 2021), pp. 287–317. DOI: 10.1007/s10686-020-09690-9.

[8] O. N. Stavroudis and D. P. Feder. "Automatic Computation of Spot Diagrams*". In: *Journal of the Optical Society of America* 44.2 (Feb. 1, 1954), p. 163. DOI: 10.1364/JOSA.44.000163.

[9] G. Tinetti et al. *Ariel: Enabling Planetary Science across Light-Years, Ariel Definition Study Report*. Version 1. arXiv, 2021. arXiv: 2104.04824.

*Is it possible to arrange the lenses in telescopes in such a way that the deficiency in the one will correct the deficiency in the other [...] ?*

Baruch Spinoza, Letter 30A (1665)

# 3
# Ariel Telescope Opto-mechanical Analysis and Design

After a review of the science goals and requirements of the Ariel mission, and the analysis of the optical design chosen for the Ariel telescope, this and the following chapters present one of the core themes of the research work described in this thesis: the development of the technologies involved in the manufacturing of the telescope mirrors and their integration into the mechanical support structure to form the complete Telescope Assembly, with the objective of guaranteeing the required optical performance.

The discipline that is concerned with the mechanical implementation of optical systems is named "Optomechanics". More precisely, as appropriately defined by D. Vukobratovich in [1, Chapter 2], "Optomechanics is defined as that part of optical engineering concerned with maintaining the shape and position of the surfaces of an optical system".

This chapter introduces the opto-mechanical design and analysis of the Telescope Assembly (TA) in order to provide the context of the author's work presented in Paper 1:

> P. Chioetto et al. "Preliminary Analysis of Ground-to-Flight Mechanical Tolerances of the Ariel Mission Telescope". In: *Proc. SPIE 12180, Space Telescopes and Instrumentation 2022: Optical, Infrared, and Millimeter Wave*. Aug. 27, 2022, 121804R. doi: 10.1117/12.2628900

Chapter 4 will instead be devoted specifically to the the development of the technologies that are required for the manufacturing of the telescope mirrors.

## 3.1 Opto-mechanical analysis and design

The optical design of a telescope consists of ideal surfaces that are defined and located in space with infinite accuracy (or at least within the limitations of the computation engine used for the simulations, which are usually much smaller than those of mechanical systems). Real telescopes, however, are subjected to the specific characteristics and limitations of materials, manufacturing processes and environmental conditions both during construction and operations. These limitations impose deviations from the ideal shapes and positions of the elements, which need to be determined or estimated in order to assess the impact on the final optical performance, and eventually budgeted in the design, or counteracted.

At the core of the opto-mechanical design of a telescope is therefore the study of the sources of deviations and the design techniques that can overcome them.

The first source of deviations to consider is the physical limitations of manufacturing technologies. In the case of mirrors, especially large ones such as the primary mirror of the Ariel telescope, the main issue is the deviation of the manufactured shape of the optical surface from the ideal one. This causes aberrations in the wavefront that may result in a loss of image quality and a reduction of signal, as part of the light is dispersed outside of the useful area of the detector, be it an imaging system such as FGS, or a spectrometer as AIRS. This topic will be covered in more details in Chapter 4.

There are then a series of issues inherent to the assembly of an optical system: alignment procedures and measurement techniques have limitations that set a hard floor to the level of accuracy that can be reached when mounting the mirrors in the supporting structure of the telescope.

The exact nature of these two types of deviations are generally not predictable in advance, eg. determining what specific aberration will be present in a manufactured mirror or what kind of misalignment will be present in the system is difficult, at least until prototypes are made. For this reason, the initial approach is to make estimations from statistical simulations based on what is known of the technological processes employed, in what is termed "Tolerance Analysis". We will see this in Section 3.2 below.

Finally, three other sources of deviations are usually considered together in the so called "STOP Analysis", by the initials of Structural, Thermal and Optical Performance analysis, as described below.

**Structural** This area covers the mechanical analysis of the telescope and mirrors, with the goal of determining the level of deformations induced on single elements or assemblies by mechanical means, primarily stresses due to external static or dynamic loads (eg. gravity and vibrations), stresses induced in optical elements by their holding structures and stresses that are intrinsic to the materials. The latter topic is specially crucial for the stability of metallic mirrors, and will be treated in details in the next chapter.

**Thermal** Temperature changes, either permanent such as the cool-down of a space payload, or fluctuations, induce deformations to structural elements and distortions of optical elements. All the effects involved need to be foreseen and accounted for.

**Optical** This area covers the analysis of the way in which deformations and misalignments affect the final optical performance of the system.

These three types of analyses, contrary to tolerance analysis, are amenable to simulation to predict the expected effects. They are usually treated together by means of an iterative process in which, at each phase in the project, the telescope design is refined towards a specific set of interim requirements; the upgraded design is then subjected to Structural, Thermal and Optical analyses to determine the optical performance, and the results of these analyses are then used as the basis for the following process iteration until the desired performance is reached.

The results of the tolerance analysis and STOP analysis are then combined to provide an overall budget that is checked against the requirements, to confirm compliance of the design.

In the following sections we will first provide a brief overview of the most important concepts concerning the structural and thermal analysis and design of the Ariel telescope, in the context of the STOP analysis, and then introduce the paper on ground-to-flight tolerances.

### 3.1.1 Structural Analysis and Design

The optical elements of a telescope, once mounted and aligned, must maintain their shape and their relative position within design tolerances for the entire duration of the mission. This is especially challenging for space telescopes that are subjected to heavy mechanical loads and are assembled on ground under gravity, but then operate in a weightless environment.

One of the key performance parameter to be considered for the optomechanical stability is stiffness, more precisely the stiffness-to-mass ratio or specific stiffness, since materials and structures with a high ratio lead to the construction of lighter, less deformable telescopes.

The selection of the appropriate material for the optical elements and support structures is one of the aspects to take into consideration to achieve a high specific stiffness. In the case of the Ariel Telescope, as we will see in Section 4.1, the choice was more heavily influenced by other criteria, such as thermal properties and machinability, among others. The choice of a material with relatively lower specific stiffness led therefore to more stringent requirements on the stiffness of supporting structures.

A very important parameter commonly used to evaluate the specific stiffness of structures and mirrors with a complex geometry, such as heavily lightweighted mirrors, is the fundamental frequency or frequency of the first resonant mode. For a system with one degree of freedom, it is in fact directly related to the deflection of the structure under its own weight, according to the relation [1]:

$$f_n = \frac{1}{2\pi}\sqrt{\frac{g}{\delta}},$$

where:

- $f_n$ is the fundamental frequency, in Hz;
- $g$ is the gravitational acceleration;

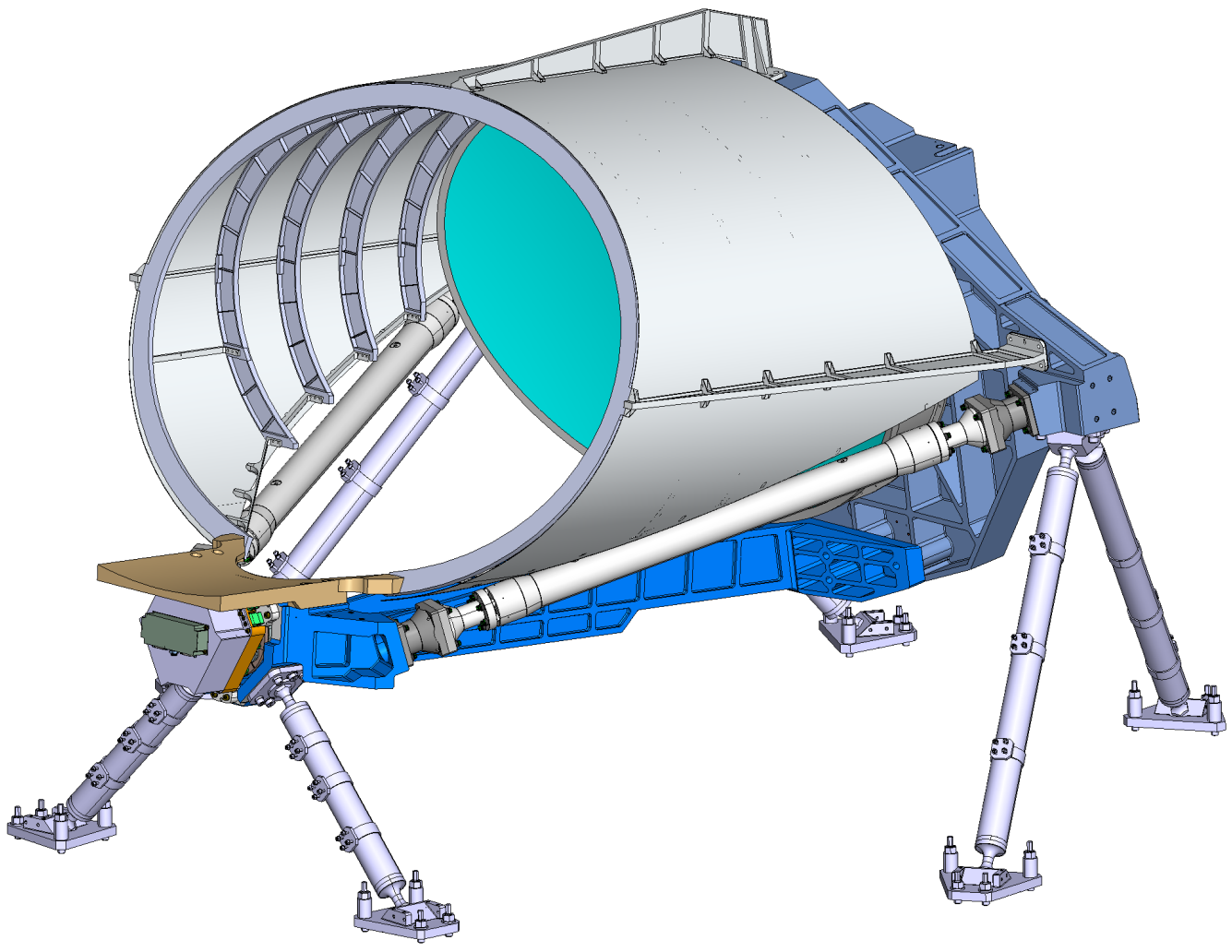

**Figure 3.1:** Telescope assembly mechanical design.

- $\delta$ is the self-weight deflection of the structure.

Deflection caused by gravity is obviously not an issue per-se for spaceborne telescopes, but it needs to be taken into consideration since manufacturing, assembly and alignment of the mirrors is performed on ground, where all productions and measurement steps are affected by gravity. Because of this, as we will see in the next chapter, special care needs to be taken in the last phases of manufacturing of the main telescope mirror, where the optical surface is brought close to the nominal shape within few tens of nanometers, and so a deflection of even tenths of microns can have an important effect on optical performance.

In terms of integration and alignment, the deformations of the mirrors support structure must be either minimized by employing an appropriate integration setup, or compensated with the addition of adequate shims or other fine adjustment means that can then be removed before flight.

In the case of space-borne telescopes, a specific minimum frequency value is also required to avoid excessive stresses caused by resonance with the low frequency vibrations associated with launch.

### Ariel Telescope Mechanical Design

The Ariel Telescope Assembly (TA) [10] is an all aluminum structure consists of the following elements (Figures 3.1 and 3.2):

- the M1, M2, M3 and M4 mirrors and mounts;
- the optical bench, supporting M1 and enclosing the common optics and the instruments;

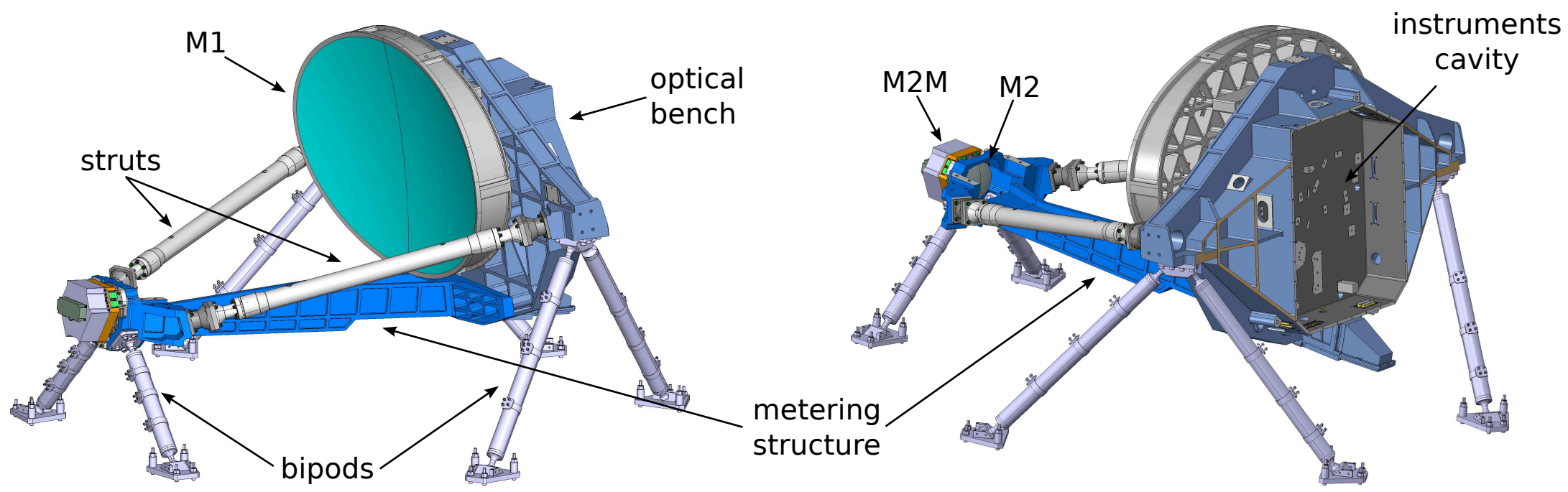

**Figure 3.2:** Overview of the telescope assembly mechanical design mounted on the carbon fiber bipods, with the primary mirror baffle removed.

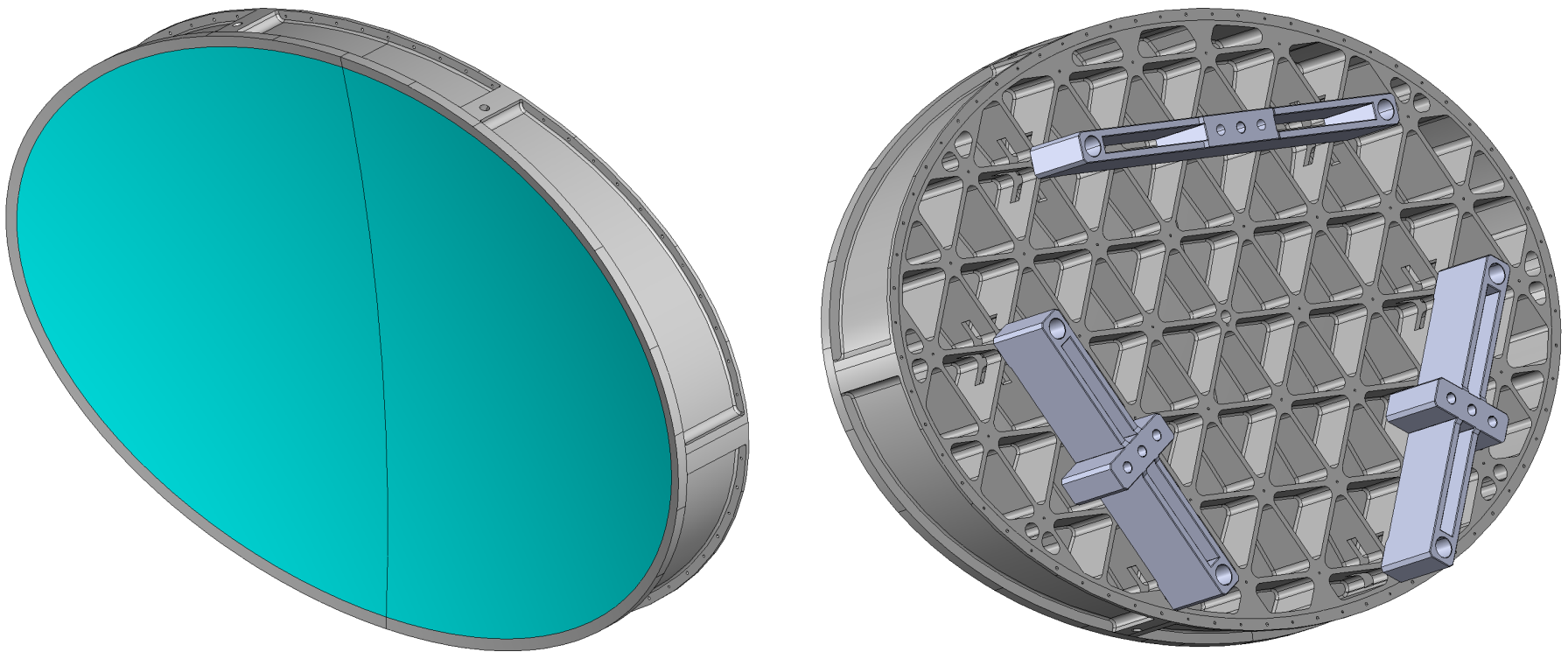

**Figure 3.3:** M1 mirror showing the lightweigthing pattern and flexible connectors to the optical bench on the back.

- the metering structure, connecting the optical bench to the M2M mechanism;
- the struts, connecting optical bench to metering structure and acting as structural reinforcement;
- the M1 baffle.

M1 is possibly the most critical component of the TA, with stringent requirements both at the mechanical and optical level, so its design is crucial to achieve the required level of optical performance.

Being the heaviest item in the telescope assembly (it represents almost a third of the total TA mass of approximately 230 kg), it is aggressively lightweighted (Figure 3.3) with a uniform pattern of triangular pockets.

M1 is going to be mounted on the optical bench through three flexible hinges bolted on the back of the mirror, leaving it relatively free to contract or expand with respect to the support structure.

M1 is also the largest cryogenic aluminum mirror ever to fly in a space missions, so crucial technologies are being developed in order to support its manufacturing, as we will see in Chapter 4.

M2 is mounted at the front of the telescope structure and is equipped with a refocus mechanism (M2M). The details of the type and range of motion supported by M2M are given in Paper 1.

The telescope is supported by three carbon fiber reinforced polymer (CFRP) bipods with titanium flexures, attached to the service module in a semi-kinematic configuration [1, Chapter 2]. The front bipods support the telescope metering structure near the M2M, while the other two bipods are at the rear of the telescope, connected to the optical bench. The positions were chosen to maximize vibration modes of the payload, while the material provides thermal insulation.

Finally, a baffle will restrict view of the open sky from M1 and the top of M2. The baffle will also serve as heat radiator, aiding thermalization of the payload.

### 3.1.2 Thermal Analysis and Design

The purpose of the thermal design of an optical system is to minimize the effects on optical performance of variations in temperature, a task referred to as athermalization. There are mainly three types of effects to be taken into consideration, as described below.

1. Distortions in the optical elements due to intrinsic properties of the material, for example a change in shape of a mirror during cool-down, caused by the release of residual stresses in the substrate. These effects are difficult to predict and simulate, so they can only be prevented, as we will see in more details in the next chapter with specific regard to the primary mirror of the Ariel telescope.

2. Distortions in the optical elements caused by temperature gradients or stresses induced by their mounts. They are mainly caused by a difference in CTE or in temperature between optical element and mount.

3. Distortions in the optical support structures, causing a misalignment of the optical elements.

The latter two effects can be both predicted by thermal and thermo-elastic Finite Element Analysis (FEA) simulations.

In the case of the Ariel telescope, the main strategy to achieve athermalization is by using the same material both for mirrors and supporting structures, thus ensuring the same nominal CTE for the entire telescope. After reaching the operational temperature, and with proper thermal control to avoid temperature gradients, ideally the final shape of the telescope shall be uniformly scaled down and maintain its optical prescription.

There will be deformations caused by residual thermal gradients and stresses induced by the bipods connecting the TA to the spacecraft Service Module: these are modelled and analysed as part of the STOP Analysis.

### 3.1.3 STOP Analysis

We report here a very brief description of the STOP Analysis procedure applied to the the Ariel telescope, to serve as context in which the tolerance analysis was performed. It consists of a set of cases defined to model specific boundary conditions during the life-cycle of the mission:

- on-ground alignment and test cases, to study the effect of gravity on optical quality considering different orientations of the TA, both at ambient and cryogenic temperatures;
- cool-down cases, to study the stress incurred by the TA during the controlled transition to the operating temperature of 50 K;
- operational cases, to assess the optical performance of the telescope from the beginning of scientific observations.

The study of each case follows the same sequence of steps: first the boundary conditions are used to generate a thermal map of the system, then the map is applied to the CAD model in a thermo-mechanical simulation to calculate a map of mechanical displacements of the various elements. This output is then converted to an optical model, separating the displacement map into a set of rigid body motions to be applied to the location and orientation of the optical elements, and a deformation map to be applied to to the surface of the optical elements.

The resulting "distorted" telescope model is then analyzed with an optical simulation software to determine what level of performance can be achieved after minimizing aberrations by means of the M2M refocusing mechanism and repointing of the entire telescope (see Paper 1 for a detailed explanation of how this optimization is performed).

This is the last step of each cycle in the STOP analysis, and it is carried out to verify compliance with requirements using the same indicators and tools employed for the optical performance analysis of the nominal design seen in Section 2.2.2.

It is worth noting that the optical performance of the telescope is defined at operating temperature, while mechanical models employed for manufacturing are at room temperature. Therefore, the optical design of the telescope must fulfill both goals of ensuring compliance at operating temperature, but must also used as template for the mechanical design of mirrors and supporting structures at room temperature.

There are therefore two Ariel TA optical models, one the exact scaled up/down version of the other, using the scale factor of 1.004 07 calculated form the integrated CTE of the specific aluminum alloy[1] between 50 K and 293 K, using data from NIST [3].

The model at 293 K is conventionally treated as the "master", and serves as template for the mechanical model. It is also used to analyze the performance of the room temperature STOP Analysis cases. The "as built" telescope manufactured using this model, as one may correctly assume, will not perform as the ideal telescope neither at room temperature, where the effect of gravity on the various elements induces distortions that affect optical quality, nor at operating temperature, where thermal effects also cause (hopefully smaller) deformations. These effects are all quantified in the STOP Analysis, and compensated using appropriate shims to be applied both for alignment and verification on ground and during operation in flight (a separate set of shims for each case).

The model at 50 K is instead used to verify theoretical compliance of the undeformed telescope with optical performance, and predicted compliance of the distorted telescope in the flight STOP Analysis cases at operating temperature.

## 3.2 Tolerance Analysis

As we mentioned in the previous sections, there are two types of deviations of optical elements that cannot be treated by deterministic simulation and must therefore be the subject of a statistical study: they are deviation of the manufactured optical surfaces from their nominal shape, and the accuracy of the alignment.

The study is generally conducted in two steps: in the first, each deviation is taken separately, and its effects are compared with performance requirements to determine the range of admissible deviations; in the second the effects of all possible deviations are combined statistically to assess their effect on performance.

The second step is usually performed using Monte Carlo methods, a set of computing algorithms that assign a probability distribution to a set of variables, in our cases the deviations, and then repeatedly make a random sampling of the variables and calculate the resulting performance

An initial set of tolerances for the Ariel telescope was devised in Phases A and B1, together with a preliminary budget breaking down the total WFE for the telescope to its single components and assembly processes [5].

The main components of the budget are the tolerances of:

1. manufacturing
2. alignment
3. launch

[1] As the CTE of an alloy may vary even between production batches from the same provider, it has already been planned to update the simulations with actual measurements of the coefficient from samples of the supply that will be used for manufacturing.

4. cool-down and gravity release

5. in-flight stability

Budget components are expressed as WFE RMS (defined in Section 2.3.1), and being in fact variances, and under the generally used assumption that they are not correlated, they are combined with the Root Square Sum (RSS) operator as defined below:

$$\mathrm{WFE}_{total} = \sqrt{\sum_i \mathrm{WFE}_i^2}$$

STOP Analysis, discussed briefly in Section 3.1 is mostly concerned with cool-down and gravity release (Item 4).

Here we will review the work done during Phase B2 to further the analysis of the manufacturing tolerances and the misalignments that may happen during launch (ground-to-flight tolerances).

### 3.2.1 Mirrors Manufacturing Tolerance Analysis

As we anticipated in the Introduction, one of the most innovative technological element of the Ariel telescope is its primary mirror. Its large aperture, and the fact that it will be realized in aluminum without an overcoating of a material that is easier to shape through deterministic polishing, such as an electroless nickel or nickel-phosphorus plating, make achieving the required surface shape and finish particularly challenging, as we will see in Chapter 4. For this reason, it is important to provide to the manufacturing process a requirement on optical surface shape as relaxed as possible, within the constraints of the scientific goals of the mission.

The initial requirement on telescope optical quality was expressed in terms of WFE RMS, as we saw in Section 2.2.1, and this requirement can be directly allocated by setting an upper limit to the surface error of each mirror. It was derived from the condition of having a diffraction limited telescope at the wavelength of 3 µm, employing the Maréchal criterion (Section 2.3.2).

This requirement is perfectly able to guarantee the optical quality close to 3 µm and above, where the shape of the PSF is dictated by diffraction effects rather than WFE. The requirement is then apt for Ariel primary instrument AIRS, operating between 1.95 µm and 7.8 µm. At lower wavelenghts however, the requirement isn't enough to restrain the shape of the PSF in order to ensure precise photometry for science through the VisNIR instrument, and accurate pointing with the FGS.

To better understand this point, it is worth remembering that while there is a precise relation between the wavefront of an optical system and its PSF[2], there is no univocal relation between the RMS of a WFE and the shape of the PSF. By reducing the WFE to its RMS

[2]It can be shown that the PSF is in fact related to the wavefront function via a Fourier transform [7, Chapter 11]

value, information on the distribution of spatial frequencies is lost, and so the possibility of univocally determining the PSF, and ultimately the radius of encircled energy.

In our specific case, it was shown that there are in fact families of WFE functions with the same RMS that do not produce PSFs compliant with scientific requirements [11].

**Table 3.1:** Enclosed Energy criterion for telescope optical performance, expressing the dimensions of the ellipse that shall enclose at least 83.8 % of the energy of the PSF.

| Instrument | Wavelength | Ellipse axes (83.8 % EE) | |
|---|---|---|---|
| | (nm) | $y$ (arcsec) | $x$ (arcsec) |
| VisPhot | 0.55 | 41 | 27 |
| FGS1 | 0.70 | 40 | 26 |
| FGS2 | 0.90 | 41 | 27 |
| | 1.00 | 42 | 28 |
| | 1.24 | 45 | 30 |
| NIRSpec | 1.48 | 48 | 32 |
| | 1.71 | 49 | 33 |
| | 1.95 | 53 | 35 |
| AIRS Ch0 | 1.95 | 53 | 35 |
| | 3.00 | 73 | 49 |
| | 3.90 | 91 | 61 |
| AIRS Ch1 | 3.90 | 91 | 61 |
| | 5.90 | 132 | 88 |
| | 7.80 | 169 | 113 |

A new requirement was therefore devised based on Enclosed Energy (EE for short, see Section 2.3.2), so as to constrain the size of the PSF regardless of its shape. The criterion indicates, for the wavelength range of each instrument, what is the ellipse that should contain the specific amount of energy corresponding to a diffraction limited system (83.8 %, Table 3.1).

The Airy disk for each wavelength is fully contained within the corresponding specified ellipse, so the new criterion is generally more relaxed than requesting to be diffraction limited, but being specified in terms of PSF, it is guaranteed to satisfy scientific requirements. Specifying instead a larger amount of WFE RMS, as we just saw, would not achieved the same purpose.

While well defined in mathematical terms, an EE-based criterion is less practical in terms of analysis and manufacturing practices since there is no direct relation to the shape of the optical surfaces. As we saw in Section 2.3.1, the WFE is directly related to the surface error of an optical element, so its measurement can guide directly the manufacturing process indicating how the element deviates from the perfect shape, and as we will see in Chapter 4, it is directly used by deterministic polishing machines such as the one being designed for the fabrication of M1.

It became therefore necessary to devise a way to translate the EE criterion to a set of criteria on surface errors that are of easier application by the mirrors manufacturer, both in terms of implementation and verification. This led, following the suggestion of Prof. E. Pascale (Ariel Consortium Payload Scientist), to the work described in [6].

The idea behind the work is to decompose the surface error of each mirror into a set of components, and then provide an RMS requirement on each of the components. For the decomposition, we opted for the Zernike polynomials, a set of orthogonal polynomials defined in polar coordinates $\rho$ and $\theta$ on circular apertures, and widely used for the definition of optical surfaces [2]. The decomposition consists in determining the set of coefficients $c_{nm}$ that satisfy the following equation:

$$W(\rho, \theta) = \sum_{n=0}^{\infty} \sum_{m=0}^{n} c_{nm} Z_n^m(\rho, \theta)$$

where $W(\rho, \theta)$ is the wavefront error or surface error function, and $Z_n^m$ are the Zernike polynomials indexed by $n$ and $m$ positive indexes.

Zernike polynomials are used to model low order aberration (characterized by low spatial frequencies), and so are particularly suitable to represent the smooth manufacturing deviations of an optical surface that contribute to determine the shape of the corresponding PSF [9]. In our case, following a common practice, the first 36 terms are used to model the surface errors of the mirrors (Figure 3.4 illustrates the first 36 polynomials on elliptical apertures).

It is worth nothing that Zernike polynomials are orthogonal and orthonormal only on circular pupils and as continuous functions. When the aberration function is known only at a discrete array of points, for example when determined by ray tracing or by measurement, then Zernike polynomials are not orthogonal over the data set, and, therefore, the coefficients are not independent of each other or the number of terms used in the expansion. Similarly, Zernike polynomials are not orthogonal when the pupil is elliptical as in our case, regardless of the number of points [8, Chapter 13]. In such cases, the Zernike coefficients that match the data values depend on the calculation procedure used, so this must be thoroughly specified.

Using the Zernike decomposition, the surface of each mirror can then be expressed as the sum of the nominal conic function (parabolic or hyperbolic) and a deformation term:

$$z = \frac{cr^2}{1 + \sqrt{1 - (1+k)c^2 r^2}} + \sum_{n=1}^{N} c_n Z_n(\rho, \phi),$$

where $z$ is the surface “sag” (z-coordinate, along the optical axis), $c$ is the surface curvature, $r = \sqrt{x^2 + y^2}$ is the radial coordinate and $k$ the conic constant (equal to $-1$ for M1 and M3, to $-1.392$ for M2); $c_n$ and $Z_n$ are the Zernike coefficients and polynomials respectively, using a single index notation such as Noll's [9].

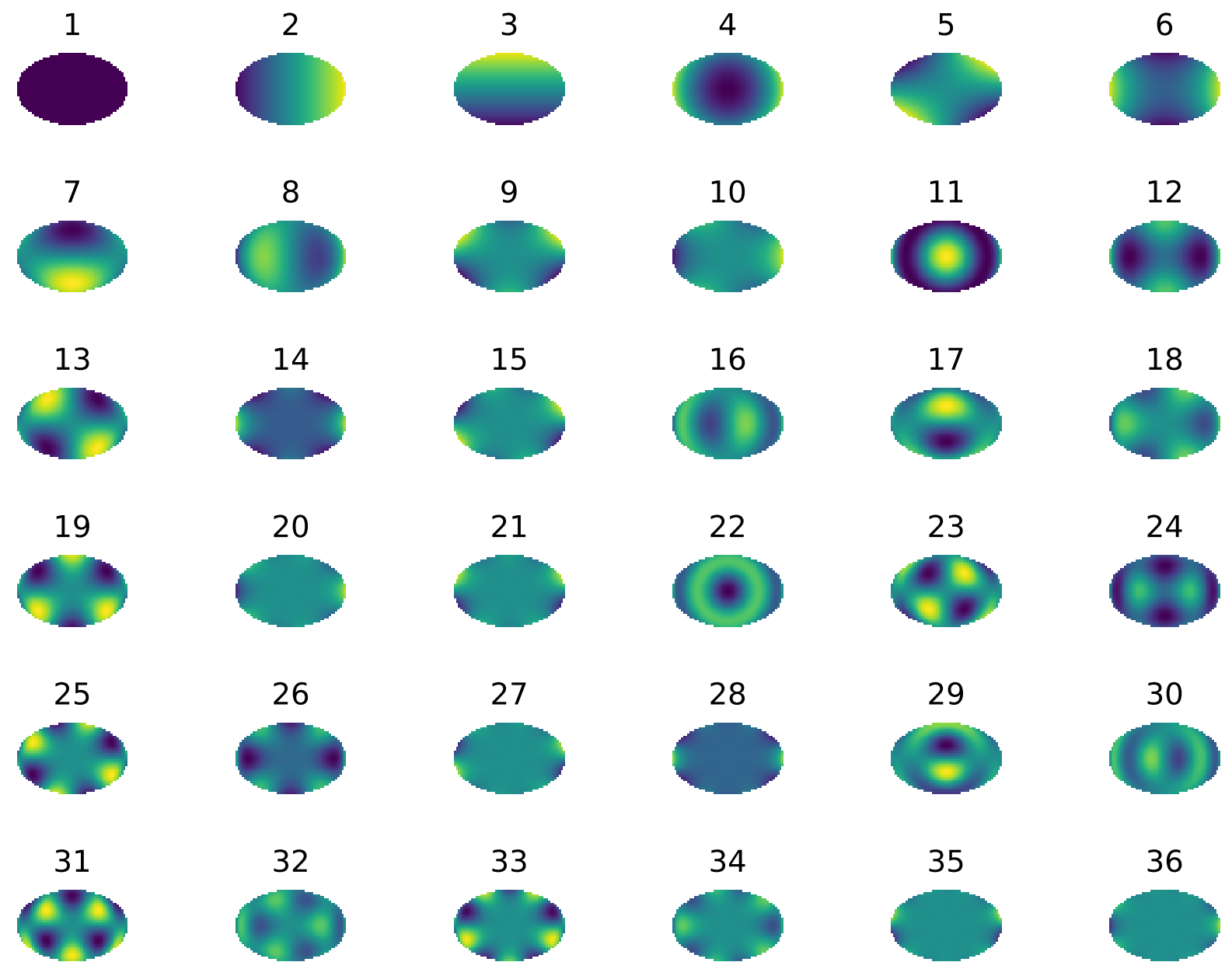

**Figure 3.4:** The first 36 Zernike polynomials with unit amplitude, represented on the elliptical aperture of mirror M1, indexed following the Noll sequential notation.

Each mirror can then be modeled as the nominal surface with a set of deformations added on top: it is sufficient to sample the Zernike parameters from appropriate random variables. A set of Monte Carlo cases can then be created and analyzed to determine how many configurations satisfy the EE requirement.

At this point we have all the ingredients to perform the analysis. The method is iterative and consists of the following steps:

1. determine the minimum and maximum value of each Zernike coefficient of each mirror that produces a telescope with the mimimum acceptable Enclosed Energy. This is done by assigning an increasing value to each Zernike coefficient $c_n$ individually (and leaving the rest to zero) until the requirement is no longer satisfied (a methodology called *inverse sensitivity analysis*);

2. prepare a set of telescope realizations with deformations in all Zernike modes of the four mirrors, by sampling all the $c_n$ coefficients within the ranges found in Step 1, and calculate the percentage of realizations that comply with the EE requirement;

3. reduce the ranges of Step 1 until the vast majority of cases (eg. 95 %) are compliant.

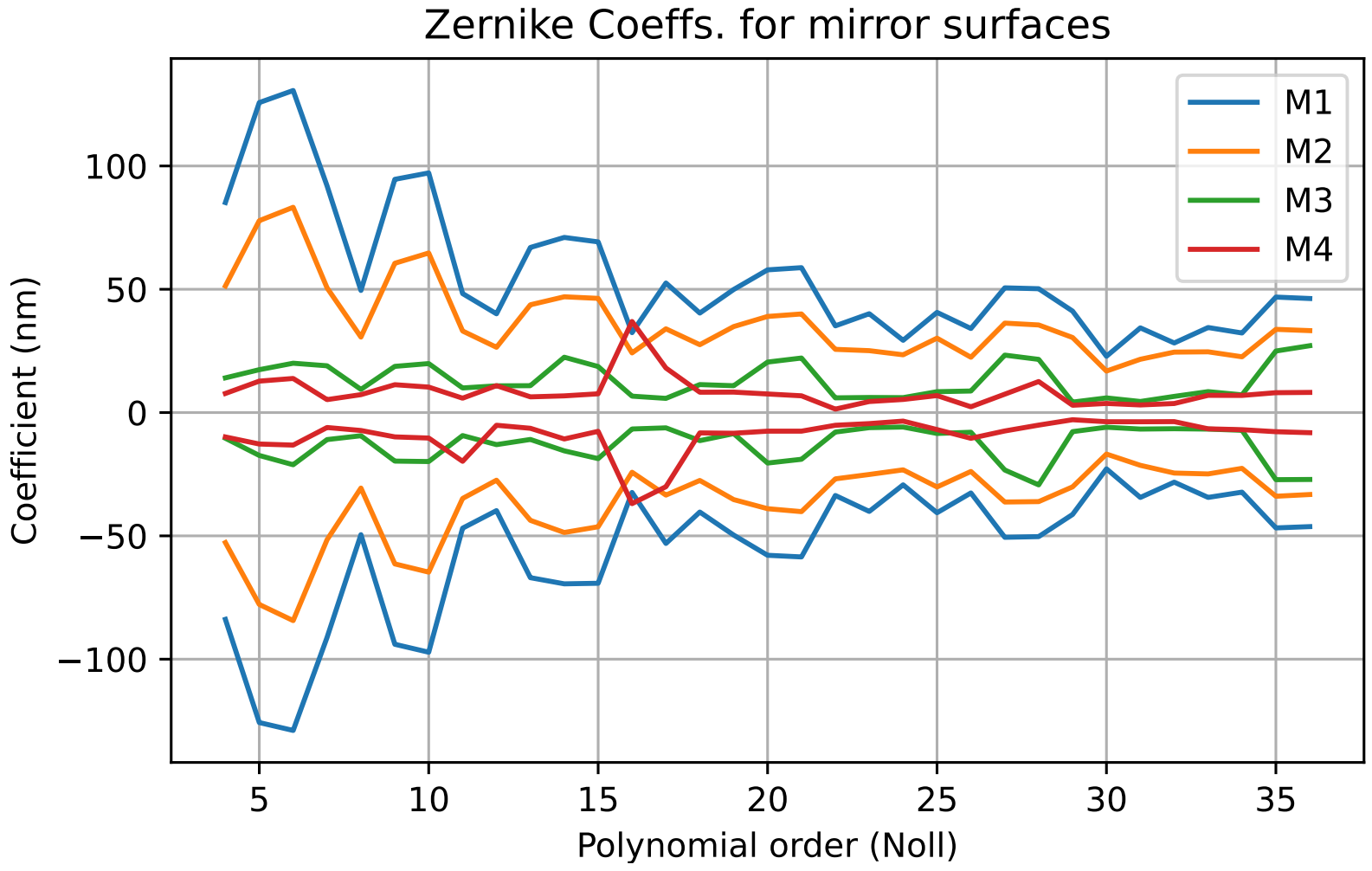

**Figure 3.5:** Set of ranges for the Zernike coefficients $c_n$ for the four mirrors that produce telescope instances within EE requirements.

The resulting set of ranges (as for example the one charted in Figure 3.5) can then be used as guidance to the mirrors manufacturer to assess progress in the figuring process: if all measured Zernike modes of the mirror is within the ranges, then it is sufficiently probable that the mirror will satisfy the requirement on Enclosed Energy.

### 3.2.2 Ground-to-flight Tolerance Analysis

The overall concept and methods developed for the calculation of mirror manufacturing tolerances of the previous section served as starting point for the work on *ground-to-flight* analysis, i.e. the determination of the effects and sensitivity of displacements of the mirrors on telescope performance, assuming nominal optical surfaces.

The analysis starts with the nominal telescope design, and considers all the ways that its mirrors can be displaced considering only rigid body motions, i.e. translations and rotations of the mirrors, but excluding any deformation of the mirrors surfaces.

The *ground-to-flight* case considers a simplified model of the behavior of the telescope after it has been aligned on ground. The hypotheses considered for the case are presented below.

1. The starting point is the telescope nominal design at the operating temperature of 50 K, under the assumption that the telescope has been aligned perfectly, that mirror optical surfaces are nominal and that there are no deformations due to stresses introduced by cool-down or temperature gradients. This is obviously not realistic, but this case serves as a base model for further analyses: the results from this case are going to be included in the larger framework of the STOP Analysis, where all thermal and mechanical deformations will be taken into consideration.

2. Mirrors can only rigidly translate (shift) or rotate (tilt), without deformations. The center of curvature (CoR) for rotation was chosen arbitrarily for M1, M3 and M4, since a final design for the mounting structures was not available at the time of performing the analysis. For M2, the CoR of the M2M is considered.

3. After a specific set of misalignments is applied to the nominal design, the telescope is re-optimized to achieve maximum optical performance. The optimization variables are the M2 refocusing mechanism that shifts the mirror along the optical axis and tilts it around the two axes orthogonal to the optical axis, and the line of sight of the telescope, operating in fact a repointing.

Performance is evaluated in terms of Enclosed Energy, as for the mirrors manufacturing analysis (Section 3.2.1), and also determining the shift and tilt of the collimated beam exiting the telescope, that must be restricted for the downstream instruments to work properly.

The purpose of this analysis is twofold. The first is to establish a set of tolerances, i.e. the maximum admissible displacement ranges that can happen to the telescope, after it is aligned on ground, without compromising performance. This part of the analysis is the subject of the publication presented this section.

The second purpose, to which this analysis will serves as starting point, is to guide the definition of the TA alignment procedure, and to estimate the level of alignment that can be achieved.

The analysis and results are presented in Paper 1.

## References

[1] A. Ahmad, ed. *Handbook of Optomechanical Engineering*. Boca Raton, Fla: CRC Press, 1997. 396 pp. ISBN: 978-0-8493-0133-9.

[2] M. Born and E. Wolf. *Principles of Optics: Electromagnetic Theory of Propagation, Interference and Diffraction of Light*. Cambridge; New York: Cambridge University Press, 1999. ISBN: 978-0-521-64222-4.

[3] P. Bradley, R. Radebaugh, and M. Lewis. "Cryogenic Material Properties Database, Update 2006". In: *Proceedings of ICMC '06 Twenty First International Cryogenic Engineering Conference and 9th Cryogenics*. 2006, pp. 13–21.

[4] P. Chioetto et al. "Preliminary Analysis of Ground-to-Flight Mechanical Tolerances of the Ariel Mission Telescope". In: *Proc. SPIE 12180, Space Telescopes and Instrumentation 2022: Optical, Infrared, and Millimeter Wave*. Aug. 27, 2022, 121804R. DOI: 10.1117/12.2628900.

[5] V. Da Deppo et al. "The Afocal Telescope Optical Design and Tolerance Analysis for the ESA ARIEL Mission". In: *Proc. SPIE 10590, International Optical Design Conference 2017*. Nov. 27, 2017, 105901P. DOI: 10.1117/12.2292299.

[6] E. Diolaiti et al. *Telescope Assembly Zemax OpticStudio® Tools for Statistical Analysis of Optical Tolerances*. Ariel Payload Consortium Phase B2 Study ARIEL-INAF-PL-TN-015 Issue 1.0. 2021.

[7] E. Hecht. *Optics*. 5 ed. Boston: Pearson Education, Inc, 2017. 714 pp. ISBN: 978-0-13-397722-6.

[8] D. Malacara, ed. *Optical Shop Testing*. 3rd ed. Wiley Series in Pure and Applied Optics. Hoboken, N.J: Wiley-Interscience, 2007. 862 pp. ISBN: 978-0-471-48404-2.

[9] R. J. Noll. "Effect of Mid- and High-Spatial Frequencies on Optical Performance". In: *Optical Engineering* 18.2 (Apr. 1, 1979). DOI: 10.1117/12.7972338.

[10] E. Pace et al. "The Telescope Assembly of the Ariel Space Mission". In: *Proc. SPIE 12180, Space Telescopes and Instrumentation 2022: Optical, Infrared, and Millimeter Wave*. Vol. 12180. Aug. 27, 2022, p. 36. DOI: 10.1117/12.2629432.

[11] E. Pascale. *ARIEL Wavefront Distorsion Analysis*. ARIEL Payload Consortium Phase B1 Payload Study ARIEL-SAP-PL-TN-005 Issue 3.0. 2020.

*I think that 'perfectionism' is a term that's thrown about mostly by people who are lazy. We're not trying to do something perfect [...] you're looking for stuff to be imperfect in exactly the right way.*

David Fincher, interview to *MovieMaker podcast* (2021)

# 4
# Telescope Mirrors Technological Development and Prototyping

The primary mirror of the Ariel Telescope, with its optical aperture of 0.6 m$^2$, stretches the current technological limits on aluminum machining and polishing for optical space application. For this reason, Phases A and B1 of the project included a specific program to demonstrate the level of technological development of all processes required for the manufacturing of M1, and that culminated in the realization of a full size prototype mirror.

This chapter, after making an overview of the program and discussing the main mirror fabrication and measurement technologies employed, introduces the work done by the author following up the companies involved in the program for the two key areas of thermal stabilization of the mirrors substrate material and polishing of the optical surface, constituting the first two section of Paper 2:

> P. Chioetto et al. "Qualification of the Thermal Stabilization, Polishing and Coating Procedures for the Aluminum Telescope Mirrors of the ARIEL Mission". In: *Experimental Astronomy* 53 (Apr. 19, 2022), pp. 885–904. doi: 10.1007/s10686-022-09852-x

The third topic treated in the paper, optical coating qualification, will be covered in Chapter 5 together with optical throughput estimations.

## 4.1 Choice of substrate material

One of the key steps in the process of translating the optical blueprint of a space telescope into an opto-mechanical design is the choice of materials for the mirrors substrates and supporting structures.

Large mirror substrates for cryogenic space telescopes are realized in various materials. Recent examples include the 6.5 m primary mirror of the James Webb Space Telescope (launched in 2021), composed of 18 hexagonal segments made from sintered beryllium powder and having each a diameter of 1.32 m flat-to-flat [6, 9], and the 3.5 m silicon carbide (SiC) primary mirror of the Herschel Space Observatory, launched in 2009 [18].

Aerospace-grade Aluminum alloy 6061 has cryogenic space heritage for smaller mirrors, with apertures below 0.5 m of diameter [15], but it is considered a promising material for its thermal conductivity and diffusivity properties and for its excellent manufacturability since it is easy to machine and it can be polished and heat-treated [21].

Aluminum is also an excellent structural material, so it can be used easily and inexpensively for the construction of the entire Telescope Assembly (TA), providing athermal design by matching the CTE of the mirror substrates with that of their supports [24], as explained also in Section 3.1.2.

During the early stages of the Ariel feasibility study in 2017, the Italian telescope team conducted a preliminary recollection of information from literature and industry opinions, and then compiled a detailed trade-off analysis [7] among the following materials:

1. aluminum alloy RSA443 with a nickel-phosphor overcoating on the optical surface for polishing, and matching supporting structures;
2. aluminum alloy 6061 for mirror substrates and supporting structures (with specific reference to the heritage of the JWST MIRI instrument, although referring to a much smaller 200 mm mirror [13]);
3. Zerodur® glass-ceramics for substrates with supporting structures in carbon fiber reinforced polymers (CFRP) with similar CTE;
4. silicon carbide (SiC) for both mirrors substrates and supporting structures;
5. beryllium mirrors with support structures either also in beryllium, or in a CFRP with similar CTE.

The materials were evaluated based on a mix of quantitative and qualitative criteria:

1. availability, in terms of procurement lead time and cost;
2. optical polishability, in terms of resulting surface shape and roughness;
3. thermo-mechanical properties that facilitate the design of a cryogenic telescope, a qualitative assessment mostly based on literature of success cases;
4. machinability for light-weighting;

5. the ratio of thermal conductivity over CTE, as a measure of thermo-optical stability, since a high thermal conductivity leads to reaching thermal equilibrium faster while a low CTE reduces strains from cooling down and thermal gradients (see also Section 3.1.2);

6. specific stiffness (Young's modulus over density), a parameter to evaluate how well the material lends itself to a stiff and light structure;

7. Consortium ability to produce the mirror, based on an evaluation of capabilities and heritage of member institutions.

Aluminum alloy 6061 received the highest overall score, and so it was selected as the baseline material for both mirror substrates and supporting structures.

Al 6061 is a precipitation-hardened aluminum alloy, containing magnesium, silicon and copper as its major alloying elements. The specific temper and forge selected for Ariel is designated "T651 rolled plate" and it refers to the following hardening and stress relieving processes: *solution heat-treated*, *artificially aged* and *stress relieved by stretching* [2].

The two main drawbacks of aluminum, identified in the trade-off study, are the level of micro-roughness achievable for the optical surface and dimensional stability.

One of the key objectives of Phase A of the Ariel mission study was therefore to assess the level of readiness of the technologies required for the fabrication of the telescope mirrors, and especially M1, since it would be the first aluminum mirror of this size to fly in space.

## 4.2 Technology Readiness Levels

ESA employs ISO Standard 16290:2013 [12] for the definition of Technology Readiness Levels (TRLs) and assessment criteria. The classification, employed mostly for space hardware and devised originally by NASA [20], divides the technical maturity of technologies into nine levels, as indicated in Table 4.1.

As anticipated in Section 1.4.1, one of the requirements for mission adoption is that all technologies employed reach at least TRL 6: "System/subsystem model or prototype demonstration in a relevant environment (ground or space)". This was the goal of the the specific development program designed for the telescope mirrors and described in the next section.

## 4.3 Technology Development Program

As we saw in Section 2.2.1, when the development program was designed, the main requirements on the Ariel telescope were a total wavefront error of 200 nm RMS at operational temperature and *zero-g*, and a minimum throughput of 82 %.

These values were then broken down for the four mirrors, resulting in a requirement of a surface error of 80 nm RMS for the optical surface of M1 as built, and a surface roughness of 10 nm RMS (the latter is explained in more details in Section 5.4).

**Table 4.1:** The basic Technology Readiness Levels according to ISO 16290:2013

| Readiness Level | Definition | Explanation |
|---|---|---|
| TRL 1 | Basic principles observed and reported | Lowest level of technology readiness. Scientific research begins to be translated into applied research and development. |
| TRL 2 | Technology concept and/or application formulated | Once basic principles are observed, practical applications can be invented and R&D started. Applications are speculative and may be unproven. |
| TRL 3 | Analytical and experimental critical function and/or characteristic proof-of concept | Active research and development is initiated, including analytical / laboratory studies to validate predictions regarding the technology. |
| TRL 4 | Component and/or breadboard validation in laboratory environment | Basic technological components are integrated to establish that they will work together. |
| TRL 5 | Component and/or breadboard validation in relevant environment | The basic technological components are integrated with reasonably realistic supporting elements so it can be tested in a simulated environment. |
| TRL 6 | System/subsystem model or prototype demonstration in a relevant environment (ground or space) | A representative model or prototype system is tested in a relevant environment. |
| TRL 7 | System prototype demonstration in a space environment | A prototype system that is near, or at, the planned operational system. |
| TRL 8 | Actual system completed and "flight qualified" through test and demonstration (ground or space) | In an actual system, the technology has been proven to work in its final form and under expected conditions. |
| TRL 9 | Actual system "flight proven" through successful mission operations | The system incorporating the new technology in its final form has been used under actual mission conditions. |

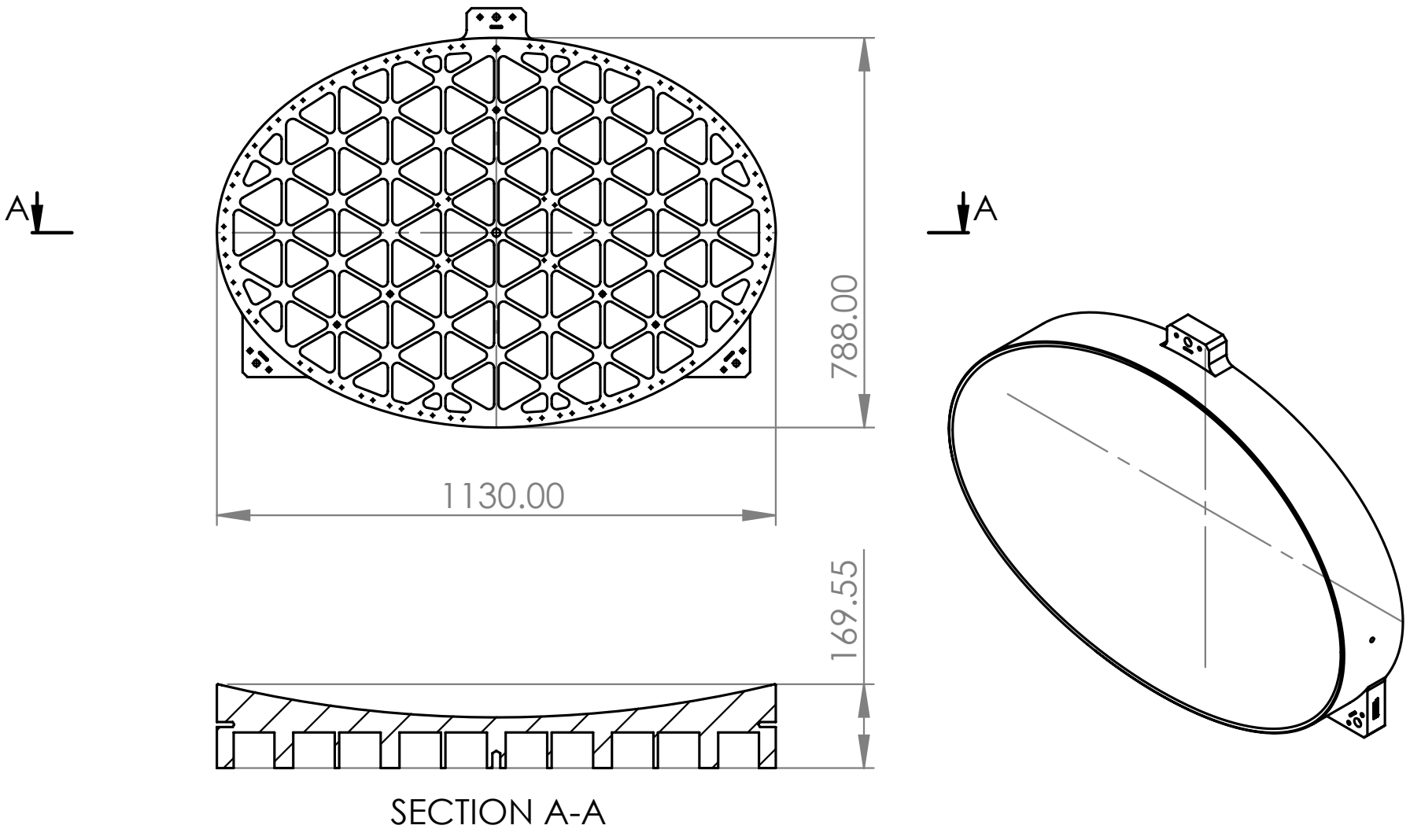

**Figure 4.1:** Simplified mechanical drawing of the PTM prototype. Quotes in mm.

The goals of the development program were to demonstrate that the processes and technologies selected for the fabrication of M1 could be employed to manufacture a prototype fulfilling the TRL 6 mandate of being "tested in a relevant environment" simulated on ground in a cryogenic vacuum chamber.

The Technology Development Assessment (TDA) program was then proposed by the Telescope team and agreed with ESA and the Ariel Consortium. It commenced in Phase A of mission development with the material selection process described in section 4.1 and the plan of designing and manufacturing a prototype mirror: the Pathfinder Telescope Mirror, or PTM.

The mechanical design of the mirror (Figure 4.1) followed the structural and thermal design principles described in Sections 3.1.1 and 3.1.2. The main characteristics of the prototype are:

- spherical optical surface with radius of curvature of ~2401 mm, derived from fitting of the paraboloid surface of M1;
- simplified 30 % lightweighting (the back pockets have a flat bottom, and do not follow the optical surface curvature), to a total mass of ~157 kg;
- shape accuracy of <100 nm RMS;
- surface roughness of <10 nm RMS.

The relaxed requirements of the PTM had the purpose of reducing the development time by simplifying those processes were a sufficient level of technological maturity was present, while focusing on the following key processes:

- thermal treatments for improving cryogenic opto-mechanical stability by minimizing residual substrate stresses;
- optical surface polishing to give the mirror its final shape and surface roughness according to specifications;
- optical surface coating for protection and reflectivity improvement.

The development program was carried out in two steps: in the first one, the processes were set up and tested on a series of samples of the mirrors substrate material, and in the second, they were applied to the prototype mirror itself. Working on samples permitted to conduct several tests in parallel, with faster machining and treatment times.

The samples consisted of three sets of discs with varying dimensions, cut from the same aluminum plate used to manufacture the PTM:

1. small, 6 mm thick samples with a diameter of 25 mm, used primarily for coating qualification tests;
2. 19 mm thick samples with a diameter of 50 mm, used to set up the polishing process;
3. 19 mm thick samples with a diameter of 150 mm, used to test the thermal treatments.

The following two Sections 4.3.1 and 4.3.2 describe the development of the thermal treatments recipe and surface polishing, while the work done to qualify the optical coating will be described in Chapter 5.

### 4.3.1 Thermal Treatments

As anticipated in Section 4.1, one of the potential issues of aluminum as a substrate material for high precision mirrors with large apertures is dimensional stability, defined as "the dimensional change that occurs in response to internal or external influences" [1, Chapter 4].

Here we are not referring the expected homotetic contraction of the mirror during cooling down, but rather the generally smaller variations in dimensions that are generally not predictable and often irreversible.

There are several types of dimensional instabilities that may affect metal mirrors, generally happening after a temperature change or in time. Most are caused by relaxation of residual stresses harbored in the material, and can be mitigated by an appropriate stress release procedure.

The Ariel telescope will be manufactured at room temperature, will experience thermal variations en route to the L2 orbit and a final cool down phase when it arrives. If mirror substrates hold residual stress, we expect the greater part of it to be released during the final cool down, and envision only smaller changes (if any) in the relatively short mission nominal duration (four years).

**Table 4.2:** Steps of the thermal stress release procedure tested under the mirror Technology Development Assessment program. Step 8 was not part of the original procedure, but was added after reviewing the outcomes of Step 7.

| Stress Release Recipe Steps | |
|---|---|
| 1. | Age at 175 °C for 8 hours. |
| 2. | Finish machining, leaving 1 mm of margin for SPDT/polishing. |
| 3. | Age again at 175 °C for 8 hours. |
| 4. | Perform three thermal cycles from −190 °C to 150 °C with rates not to exceed 1.7 °C / minute. |
| 5. | Repeat three thermal cycles as in Step 4. |
| 6. | Diamond turning/polishing. |
| 7. | Repeat three thermal cycles as in Step 4. |
| 8. | Repeat three thermal cycles as in Step 4 (validation cycle). |

Developing a thermal treatment procedure for stress release is a very complex and time consuming activity, requiring specific metallurgical expertise. Fortunately a procedure for the same alloy and temper chosen for the Ariel telescope existed in the literature, developed by R. G. Ohl *et al.* from NASA/Goddard Space Flight Center and successfully applied to the Infrared Multi-Object Spectrograph (IRMOS) cryogenic instrument [17].

The Ariel telescope team decided to adopt this procedure with small variations dictated by the available thermal cycling and testing facilities, and to subject it to a qualification procedure, first applying it to the 150 mm samples and then to the PTM itself. The modified procedure, detailed in Table 4.2, consisted of several thermal cycles interspersed with the machining and polishing fabrication steps. After every step of the procedure, surface error was measured and compared with the previous steps to assess surface form variations. After Step 5, the mirror substrate was supposed to have reached stability, so it was machined to its final specifications and underwent a final verification thermal cycle.

The first test of the procedure was performed on one of the 150 mm samples. The results are described by the author in [3] and were deemed successful, although an additional thermal cycle was added at the end to serve the purpose of the verification cycle, since cycle number 7 did cause a measurable change in optical surface shape, likely caused by additional stress induced by the diamond turning/polishing process. The final cycle instead did not introduce any measurable change, so the stress release procedure was updated to include the 8$^{th}$ step as verification cycle. Results of the sequence of measurements taken after each step are reported in Figure 4.2.

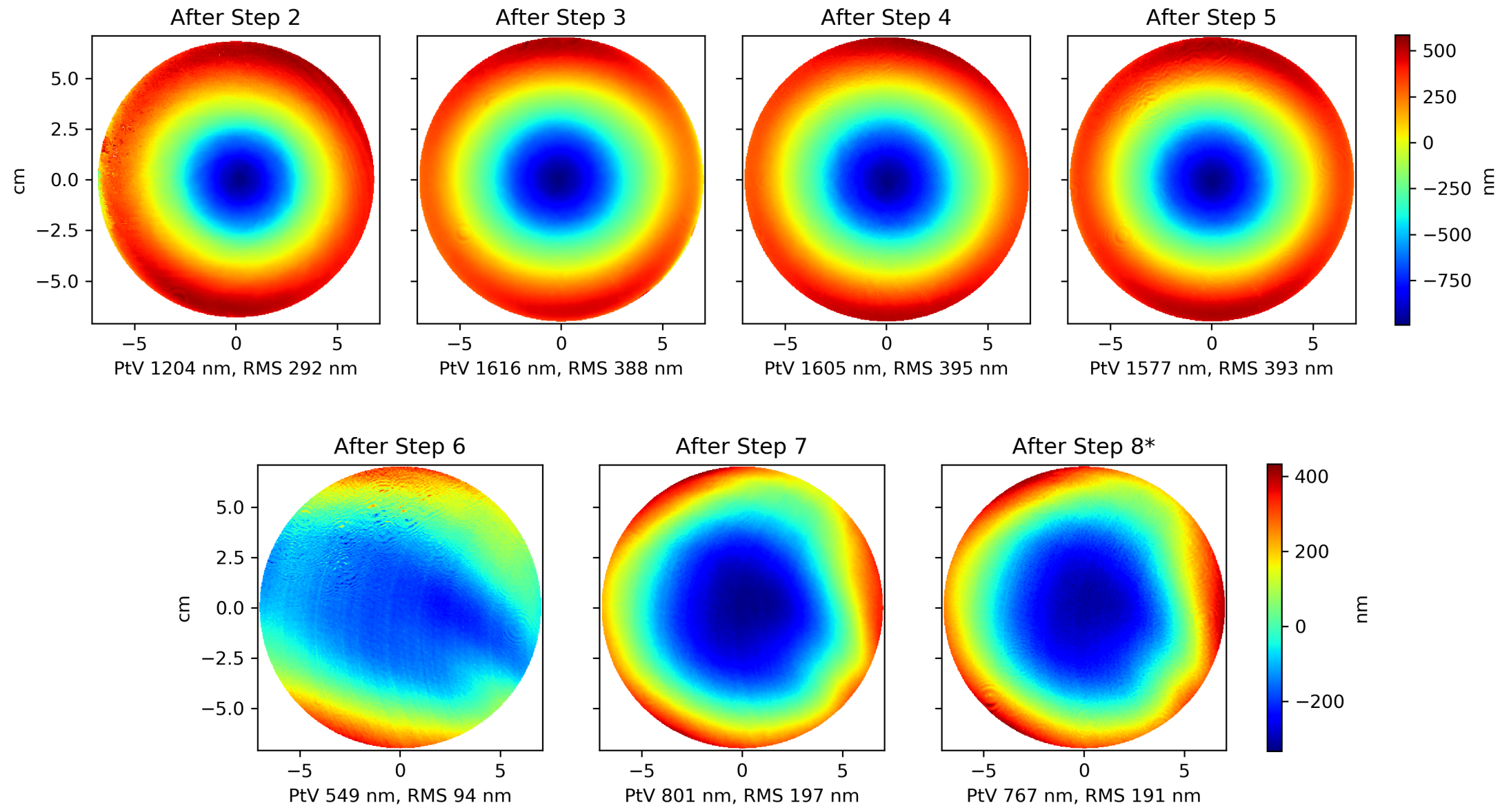

**Figure 4.2:** Surface error maps of the first 150 mm sample after each step of the stress release procedure. Step 7 introduced a significant surface form variation, so the additional Step 8* was introduced to assess if the substrate was finally stabilized.

After the successful results from the first test, the procedure was applied to three more 150 mm samples to have statistical relevance. The tests, reported in Section 2.2 of Paper 2, confirmed the viability of the recipe for two out of the three samples. For the third one the results were not considered significant since the polished optical surface presented a higher than average roughness that increased the error of the interferometric measurements to levels comparable with the measurement itself.

### 4.3.2 Optical Surface Polishing

Polishing processes are employed to give the optical surface of a mirror its final shape and smoothness in order to comply with the requirements of surface error and roughness (80 nm RMS and 10 nm RMS respectively).

As we saw in Section 2.2.1, surface error $\sigma_w$ and roughness $\sigma_s$ are impacting directly the performance of the telescope in terms of Strehl's ratio $S$ and Total Integrated Scatter TIS through the exponential relations:

$$S = \exp\left[-\left(\frac{2\pi\sigma_w}{\lambda}\right)^2\right]$$

$$\mathrm{TIS} = 1 - \exp\left[-\left(\frac{4\pi\cos\theta_i\sigma_s}{\lambda}\right)^2\right]$$

where $\lambda$ is the wavelength and $\theta_i$ the angle of incidence of the incoming light. The polishing process is therefore of paramount importance for requirements compliance.

Optical polishing is generally achieved through controlled removal of small quantities of material from a surface, usually by mechanical means such as abrasion by microscopic particles suspended in a liquid medium, called slurry.

The process is fundamentally different than the ones employed in previous phases of mirrors manufacturing, such as grinding, cutting or milling. A single point diamond turning machine for example, such as the ones employed for the high precision optics of the Ariel telescope mirrors, work by removing material through a cutting tool passing over the rotating surface. Ideally, the removal is determined solely by the shape and path of the tool on the surface, while of course the imperfections in the shape of the tool, stiffness of the machine or limited precision in the position control will result in deviations from the required surface shape.

Polishing instead is fundamentally a statistical process mediated by the microscopic abrasive particles present in the slurry. In polishing, the amount of material removed does not depend on the precise position of the tool that is "rubbing" the surface with the slurry, but it is directly proportional to the elapsed time, applied pressure and relative speed between tool and surface, in what is known as Preston's law [19].

The technique allows a great control over the removal rate, potentially achieving very precise figuring of surfaces when the removal rate is high, and smoothing of microscopical surface defects with a low rate.

In the case of aluminum as substrate material for telescope mirrors, there are specific issues that make it difficult to polish the optical surface to a very high degree of smoothness. The result is a level of surface scatter that makes aluminum unsuitable for applications in the visible and ultraviolet wavelengths: until recently, state of the art diamond turned aluminum mirrors were achieving a surface finish between 4 and 10 nm RMS [5], depending also on the kind of alloy used [16].

As an alternative, an electroless nickel-phosphorus or pure aluminum plating onto the optical surface can be polished further to just a few nanometers [24], but the process has only been tested to smaller apertures and may present issues at the cryogenic operating temperature of Ariel.

Lately, improved results for direct polishing of Al 6061-T6 mirror substrate have been reported, obtaining a surface roughness of 2–3 nm RMS [22, 14], although with smaller diameter mirrors than Ariel M1.

The issues with polishing bare aluminum appear to be caused by its softness and the presence of the alloying elements Si, Fe, Cu, and Mg in Al 6061, creating grains, inclusions and voids and presenting different hardness levels to the material removal process. Oxidation may also be induced or accelerated by the polishing process [25].

The development of a polishing process for the Ariel mirrors was therefore expected to be a challenging task, so it was divided into steps involving various samples and requiring a certain degree of experimentation with different combinations of slurries, pressions and speeds, before transferring the process to the PTM.

A detailed description of the development of the polishing procedure on samples and of the initial results are reported in Section 2.3 of Paper 2.

### 4.3.3 Results on the PTM Prototype

#### Thermal Treatments

After successful setup and verification of the thermal stabilization procedure of Table 4.2 on samples, the same was applied to the full size PTM prototype.

The mirror was subjected to three thermal cycles and then its optical surface was machined with a prototype diamond cutter.

Measurements of the surface error to assess the effect of each of the thermal cycles performed before optical surface polishing were carried out with a Coordinate Measuring Machine (CMM), a device that measures the geometry of physical objects by sensing discrete points on the surface of the object by means of a mechanical probe.

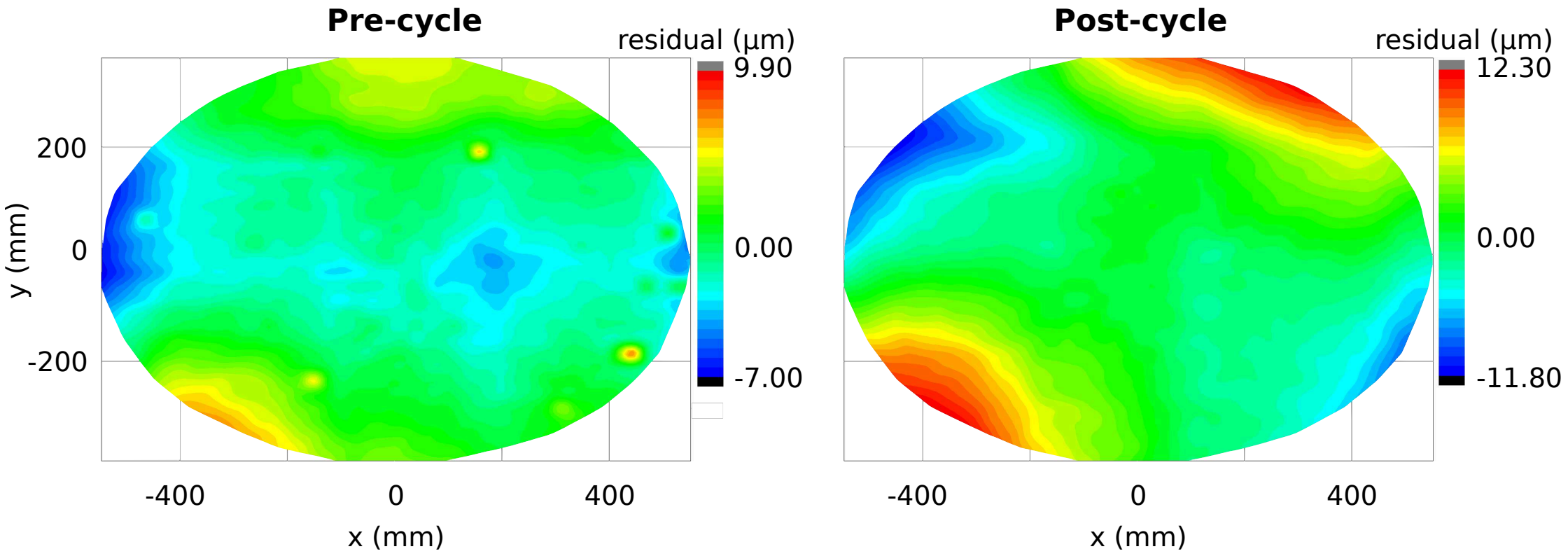

**Figure 4.3:** M1 surface shape before (left) and after (right) the third thermal stabilization cycle (Step 4 in the recipe of Table 4.2), after removing the best fit sphere of radius 2402.9 mm. The SFE RMS changed from 2.1 µm to 2.4 µm.

Figures 4.3 and 4.4 give a qualitative overview of the result of the two thermal cycles done before and after the diamond machining, in terms of surface shape variation. Although the accuracy of CMMs is limited to approximately 1 µm over the entire measurement range (as reported by their datasheets), feedback from the operator is that he used them consistently for measuring the shape of mirrors during early manufacturing stages and reported confidence in being able to detect shape variations in the order of hundreds of micrometers.

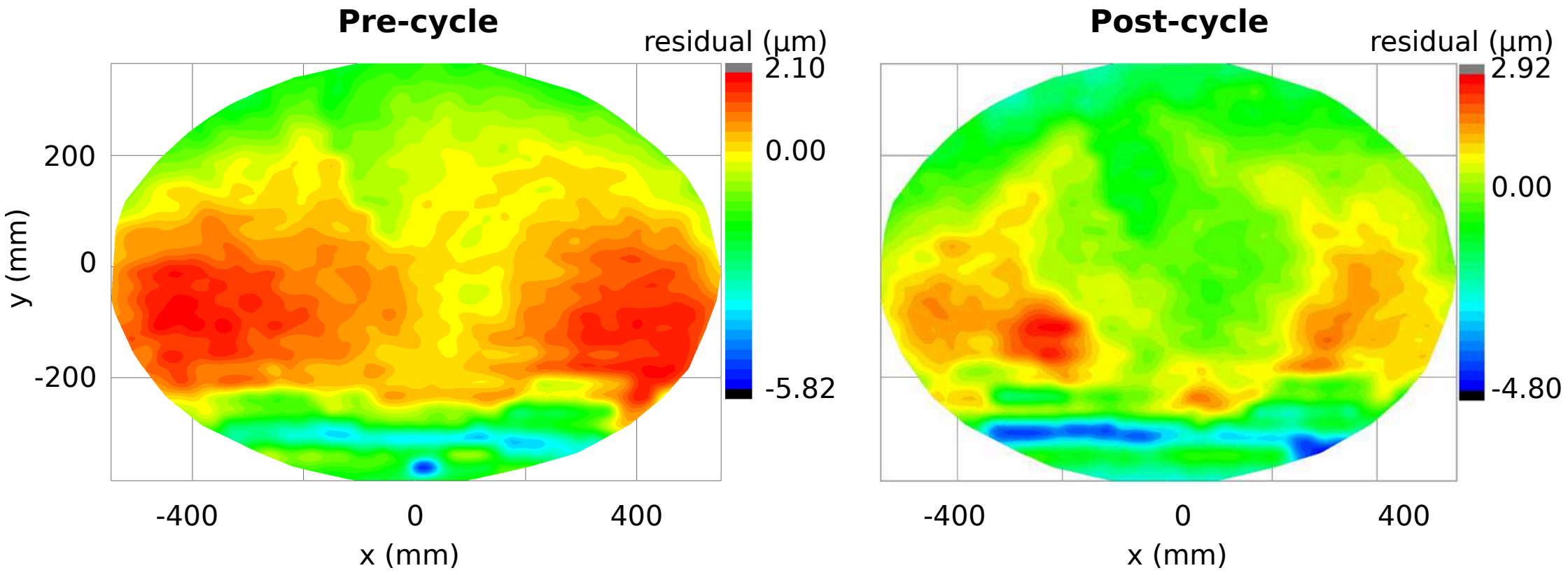

**Figure 4.4:** M1 surface shape before (left) and after(right) the thermal stabilization cycle after diamond machining (Step 7 in the recipe of Table 4.2), after removing the best fit sphere of radius 2402.9 mm. The SFE of both maps is 1.2 μm RMS.

Before diamond cutting, the thermal cycle brought a change in RMS surface shape, after removing the best fit sphere of radius 2402.9 mm, from 2.1 μm to 2.4 μm. After diamond cutting, the two maps of Figure 4.4 appear to highlight a reduction in astigmatism, however the surface error measurements have the same value of 1.2 μm RMS.

After polishing, before and after the last thermal cycle, the mirror was measured with a fizeau-type interferometer operating at the wavelength of 632.8 nm, with an F/2.2 transmission sphere, which covers almost entirely the optical surface with a pixel size of 0.95 mm. The mirror was held vertically on the structure shown in Figure 4.5 on the left.

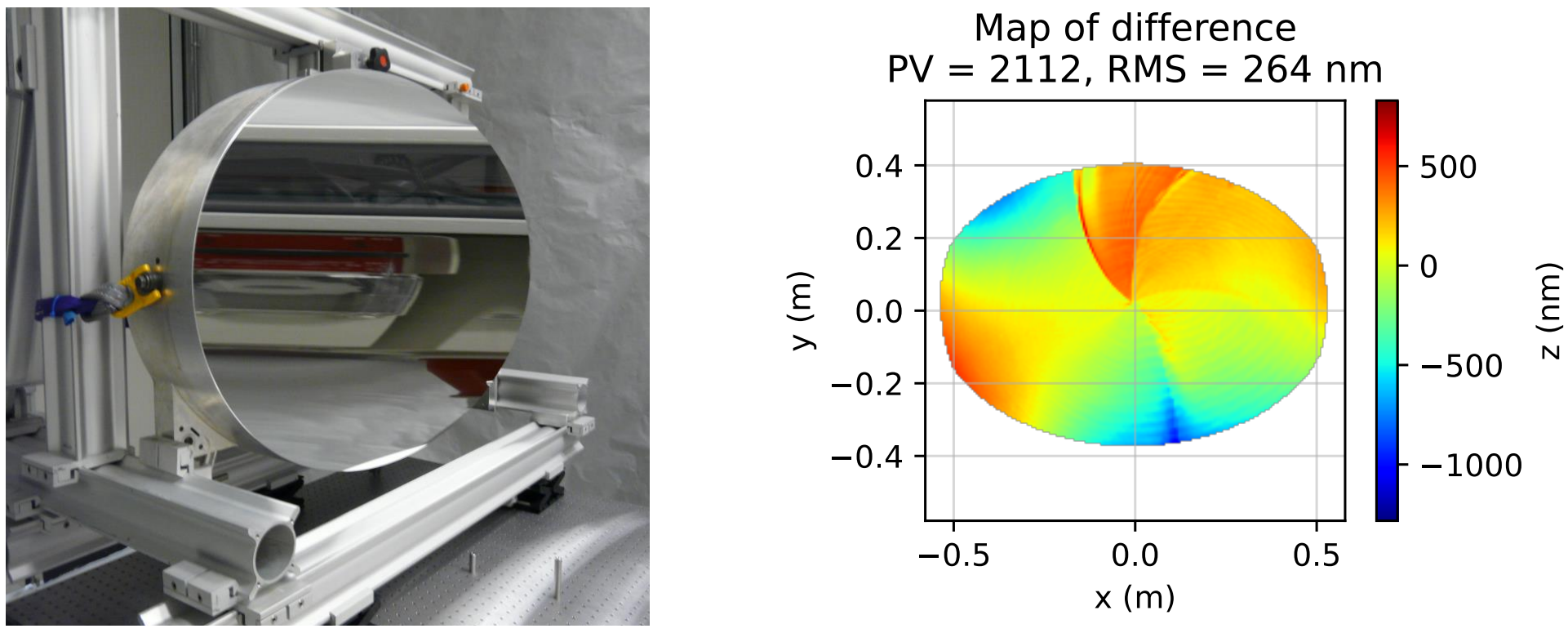

**Figure 4.5:** Image of the interferometric measurement setup with M1 (left, credit Media-Lario Srl) and map of the difference of the surface errors measured after and before the last verification thermal cycle, Step 8 in the recipe of Table 4.2 (right).

The Peak to Valley error resulting from the measurements on 95 % of the surface is 3612 nm with a corresponding RMS value of 996 nm (only piston and tip/tilt were subtracted by the interferometer software). Strong astigmatism was found, aligned with the elliptical aperture major semi-axis.

The comparison before and after the heat treatment shows (Figure 4.5, on the right) a significant map-to-map difference, in the shape of astigmatism. The measured surface errors were 999 vs. 1048 nm RMS, but the measurement was affected by a systematic error, discovered only during the second measurement instance, so no direct conclusion can be drawn about the impact that the cycle itself had on the part.

The metrology setup in fact included a screw to hold the mirror in position, since the centre of gravity of the mirror is located outside the lower pads used to support it, and the mirror would otherwise flip over. The force acting on the pad when the screw is in place, although minimal, was not measured in the first instance, but was later found to increase the SFE up to 1300 nm RMS.

### Optical Surface Polishing

Polishing of the PTM consisted in three phases: a first figuring phase characterized by a high and constant rate of material removal, a second figuring phase with a lower and variable rate of removal and finally a polishing phase to obtain the final surface roughness.

The first phase consists of a series of runs with a constant removal rate in the order of 1 μm per run, to take away small surface defects and subsurface stress left by the diamond turning. The process utilizes a silicon carbide slurry and hard polyurethane pad. These runs allow addressing large shape error correction, in the order of few tens of microns PtV. The surface roughness produced by this operation is of the order of 150 nm RMS.

In the second phase the time spent by the tool on each area is determined by a shape measurement made after each run. This allows a finer control of the amount of material removed, permitting to address features with higher spatial frequency. For this phase, the same polyurethane pad is used in combination with an aluminum oxide slurry. The removal rate is approximately 5 times lower, but surface roughness is improved to about 50 nm RMS.

The final polishing step was developed specifically to achieve the surface roughness specifications requirements for Ariel primary mirror. Extensive experimentation was carried out to optimize machine parameters by testing many different slurries and clothes. As for the second phase, the process in each run is guided by the input of shape metrology data, to control positioning and dwell time across the surface.

The results appeared promising, with some areas of the mirror reaching the required 10 nm RMS roughness. The process had however to be interrupted because of schedule constraints.

### Further Developments

Despite the satisfactory outcomes of the qualification activities on samples, the results of the thermal treatments and especially the polishing process applied on the PTM were not con-sidered sufficient to fulfill the TRL 6 requirements as such, but showed nonetheless to be sufficiently promising by ESA review board.

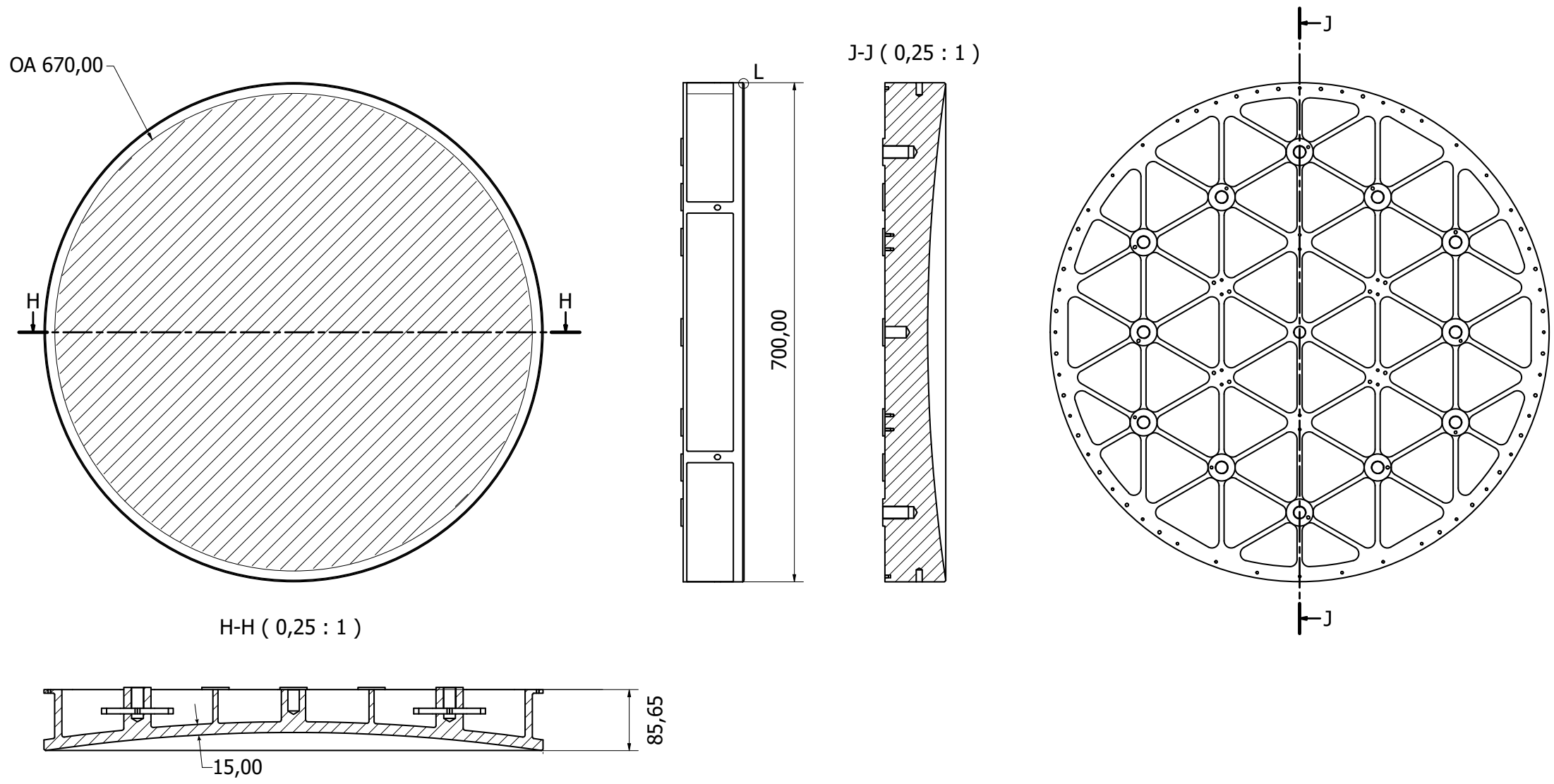

**Figure 4.6:** Simplified mechanical drawing of the M1 breadboard model used in Phase B2 for further development of the thermal treatments, machining and polishing procedures. All quotes in millimeters.

Upon consideration also of the very tight schedule, the entire set of payload studies conducted as part of Phase B1 of the project was deemed sufficient to grant adoption and transition to Phase B2, under condition that further qualification studies would be carried out.

For the Ariel telescope M1 mirror, it was therefore agreed to start a new qualification program to consolidate and further develop the know-how and results obtained in Phase B1 in terms of thermal treatments [11] and polishing process. The program, still ongoing at the time of writing, consists in the manufacturing of two new prototype mirrors [23] with slightly smaller dimensions (a diameter of 700 mm, Figure 4.6) but a mechanical design equivalent to the updated M1 specifications in terms of lightweighting and mounting interfaces [8, 10].

One of the prototype will follow the entire procedure of thermal cycling, diamond turning and polishing, and shall then be tested for shape variations in cryogenic conditions. The other will be used to obtain earlier results on the diamond turning and polishing technologies.

## References

[1] A. Ahmad, ed. *Handbook of Optomechanical Engineering*. Boca Raton, Fla: CRC Press, 1997. 396 pp. ISBN: 978-0-8493-0133-9.

[2] ASM Handbook Committee, ed. *ASM Handbook, Volume 2 – Properties and Selection: Nonferrous Alloys and Special-Purpose Materials*. ASM International, Jan. 1, 1990. DOI: 10.31399/asm.hb.v02.9781627081627.

[3] P. Chioetto et al. "The Primary Mirror of the ARIEL Mission: Testing of a Modified Stress-Release Procedure for Al 6061 Cryogenic Opto-Mechanical Stability". In: *EPSC Abstracts*. Vol. 13. Sept. 15–20, 2019, p. 2. URL: https://meetingorganizer.copernicus.org/EPSC-DPS2019/EPSC-DPS2019-1625.pdf.

[4] P. Chioetto et al. "Qualification of the Thermal Stabilization, Polishing and Coating Procedures for the Aluminum Telescope Mirrors of the ARIEL Mission". In: *Experimental Astronomy* 53 (Apr. 19, 2022), pp. 885–904. DOI: 10.1007/s10686-022-09852-x.

[5] E. L. Church and P. Z. Takacs. "Survey Of The Finish Characteristics Of Machined Optical Surfaces". In: *Optical Engineering* 24.3 (June 1, 1985). DOI: 10.1117/12.7973495.

[6] G. C. Cole et al. "An Overview of Optical Fabrication of the JWST Mirror Segments at Tinsley". In: *Proc. SPIE 6265, Space Telescopes and Instrumentation I: Optical, Infrared, and Millimeter, 62650V*. June 10, 2006. DOI: 10.1117/12.672926.

[7] V. Da Deppo et al. *ARIEL Telescope Material Trade-Off*. Technical Paper ARIEL-INAF-PL-TN-004. 2017.

[8] C. Del Vecchio et al. "Optimization of the Ariel Primary Mirror". In: *Proc. SPIE 12180, Space Telescopes and Instrumentation 2022: Optical, Infrared, and Millimeter Wave*. Vol. 12180. Aug. 27, 2022, p. 194. DOI: 10.1117/12.2629819.

[9] L. D. Feinberg et al. "James Webb Space Telescope Optical Telescope Element Mirror Development History and Results". In: *Proc. SPIE 8442, Space Telescopes and Instrumentation 2012: Optical, Infrared, and Millimeter Wave, 84422B*. Sept. 21, 2012. DOI: 10.1117/12.924271.

[10] D. Gottini et al. "FEA Testing the Pre-Flight Ariel Primary Mirror". In: *Proc. SPIE 12180, Space Telescopes and Instrumentation 2022: Optical, Infrared, and Millimeter Wave*. Vol. 12180. Aug. 27, 2022, p. 195. DOI: 10.1117/12.2629815.

[11] E. Guerriero et al. "Heat Treatment Procedure of the Aluminium 6061-T651 for the Ariel Telescope Mirrors." In: *Proc. SPIE 12180, Space Telescopes and Instrumentation 2022: Optical, Infrared, and Millimeter Wave*. Vol. 12180. Aug. 27, 2022, p. 39. DOI: 10.1117/12.2628178.

[12] International Organization for Standardization. *Space Systems — Definition of the Technology Readiness Levels (TRLs) and Their Criteria of Assessment*. ISO 16290:2013. International Organization for Standardization, 2013.

[13] G. Kroes et al. "MIRI-JWST Spectrometer Main Optics Opto-Mechanical Design and Prototyping". In: *Proc. SPIE 5877, Optomechanics 2005*. Aug. 18, 2005, 58770P. DOI: 10.1117/12.614784.

[14] J. Lyons III and J. Zaniewski. "Process for Polishing Bare Aluminum to High Optical Quality". In: *NASA tech briefs* 23 (2001), pp. 58–59.

[15] T. Newswander et al. "Aluminum Alloy AA-6061 and RSA-6061 Heat Treatment for Large Mirror Applications". In: *Proc. SPIE 8837, Material Technologies and Applications to Optics, Structures, Components, and Sub-Systems, 883704*. Sept. 30, 2013, p. 883704. DOI: 10.1117/12.2024369.

[16] A. A. Ogloza et al. "Optical Properties And Thermal Stability Of Single-Point Diamond-Machined Aluminum Alloys". In: *Proc. SPIE 0966, Advances in Fabrication and Metrology for Optics and Large Optics*. Jan. 29, 1989, p. 228. DOI: 10.1117/12.948069.

[17] R. G. Ohl IV et al. "Comparison of Stress Relief Procedures for Cryogenic Aluminum Mirrors". In: *Proc. SPIE 4822, Cryogenic Optical Systems and Instruments IX*. Nov. 1, 2002, p. 51. DOI: 10.1117/12.451762.

[18] G. L. Pilbratt et al. "*Herschel* Space Observatory: An ESA Facility for Far-Infrared and Submillimetre Astronomy". In: *Astronomy and Astrophysics* 518 (July 2010), p. L1. DOI: 10.1051/0004-6361/201014759.

[19] F. W. Preston. "The Theory and Design of Plate Glass Polishing Machines". In: *Journal of the Society of Glass Technology* 11 (1927), pp. 214–256.

[20] S. Sadin et al. "NASA Technology Push towards Future Space Mission Systems Space and Humanity Conference". In: *Selected Proceedings of the 39th International Astronautical Federation Congress, Acta Astronautica*. 1989.

[21] C. R. Sandin et al. "Materials Evaluation for the Origins Space Telescope". In: *Journal of Astronomical Telescopes, Instruments, and Systems* 7.01 (Feb. 10, 2021). DOI: 10.1117/1.JATIS.7.1.011011.

[22] R. ter Horst et al. "Directly Polished Lightweight Aluminum Mirror". In: *Proc. SPIE 7018, Advanced Optical and Mechanical Technologies in Telescopes and Instrumentation, 701808*. July 23, 2008. DOI: 10.1117/12.788657.

[23] A. Tozzi et al. "Toward ARIEL's Primary Mirror". In: *Proc. SPIE 12180, Space Telescopes and Instrumentation 2022: Optical, Infrared, and Millimeter Wave*. Vol. 12180. Aug. 27, 2022, p. 193. DOI: 10.1117/12.2628906.

[24] D. Vukobratovich and J. P. Schaefer. "Large Stable Aluminum Optics for Aerospace Applications". In: *Proc. SPIE 8125, Optomechanics 2011: Innovations and Solutions, 81250T*. Sept. 8, 2011. DOI: 10.1117/12.892039.

[25] Z. Yin and Z. Yi. “Direct Polishing of Aluminum Mirrors with Higher Quality and Accuracy”. In: *Applied Optics* 54.26 (Sept. 10, 2015), p. 7835. DOI: 10.1364/AO.54.007835.

*One does not see with the eyes; one sees with the brain, which has dozens of different systems for analyzing the input from the eyes.*

Oliver Sacks, *Hallucinations* (2012)

# 5 Mirrors Optical Coating Qualification and Performance Assessment

THIS CHAPTER describes the research work on the qualification of a suitable protected silver coating for the Ariel telescope mirrors. The activities detailed here are part of the Technology Development Program introduced in Chapter 4.

The qualification was carried out in two steps. The first step consisted in a series of performance tests of the coating on samples of the baseline aluminum alloy selected for the telescope mirrors, including some of the samples employed for the development of the thermal treatments and polishing procedure described in Section 4.3. This work is the subject of Paper 3:

> P. Chioetto et al. "The Primary Mirror of the Ariel Mission: Cryotesting of Aluminum Mirror Samples with Protected Silver Coating". In: *Proc. SPIE 11451, Advances in Optical and Mechanical Technologies for Telescopes and Instrumentation IV*. Dec. 13, 2020, 114511A. DOI: 10.1117/12.2562548

and of Section 2.4 of Paper 2:

> P. Chioetto et al. "Qualification of the Thermal Stabilization, Polishing and Coating Procedures for the Aluminum Telescope Mirrors of the ARIEL Mission". In: *Experimental Astronomy* 53 (Apr. 19, 2022), pp. 885–904. DOI: 10.1007/s10686-022-09852-x

The second step concerned the verification of coating performance on the prototype of the primary mirror, as described in Paper 4:

> P. Chioetto et al. “Test of Protected Silver Coating on Aluminum Samples of ARIEL Main Telescope Mirror Substrate Material”. In: *Proc. SPIE 11852, International Conference on Space Optics — ICSO 2020*. June 11, 2021, p. 118524L. DOI: 10.1117/12.2599794

After the successful conclusion of the qualification process, the samples have been kept in storage in cleanroom conditions and are periodically tested for morphological and/or performance variations, for the characterization of ageing process of the coating. A description of this work, with interim results, is described in Paper 5:

> P. Chioetto et al. “Long Term Durability of Protected Silver Coating for the Mirrors of Ariel Mission Telescope”. In: *International Conference on Space Optics — ICSO 2022, in press*. Oct. 3, 2022

Finally, the measurements from the coating qualification campaign were used to update the estimations of the telescope throughput, one of the key optical performance requirements described in Section 2.3.

The details of the work are described in Paper 6:

> P. Chioetto et al. “Initial Estimation of the Effects of Coating Dishomogeneities, Surface Roughness and Contamination on the Mirrors of Ariel Mission Telescope”. In: *Proc. SPIE 11871, Optical Design and Engineering VIII*. Oct. 4, 2021, 118710N. DOI: 10.1117/12.2603768

## 5.1 Selection of a Coating Material

The optical surface of telescope mirrors is usually protected and rendered highly reflective in the target observational wavebands by application of a coating thin film.

For the mirrors of the Ariel telescope, the Consortium made a trade-off between the common metal coating options with good reflectance in the IR and visible wavelengths: aluminum, gold and silver. Each has abundant references in literature and space heritage.

Aluminum has the best overall reflectance from the UV through the far-IR, is less susceptible to atmosphere tarnishing than silver and is generally more durable with regard to handling than both the other two options, but its reflectance drops in the IR at wavelenghts greater than 1 μm [12].

Gold is a popular choice for infrared space telescopes: JWST [16], NEOWISE [8] and the Cassini/CIRS instrument [12], all use protected gold, however reflectance of such coatings drops entering visible wavelengths at about 600 nm, making it unsuitable for Ariel since the operating waveband starts at 500 nm.

Silver is a popular choice for visible and infrared applications since it has the highest reflectance of the three metals in the waveband 500–8000 nm, as can be seen in Figure 5.1 and therefore represented the optimal choice also for Ariel mirrors.

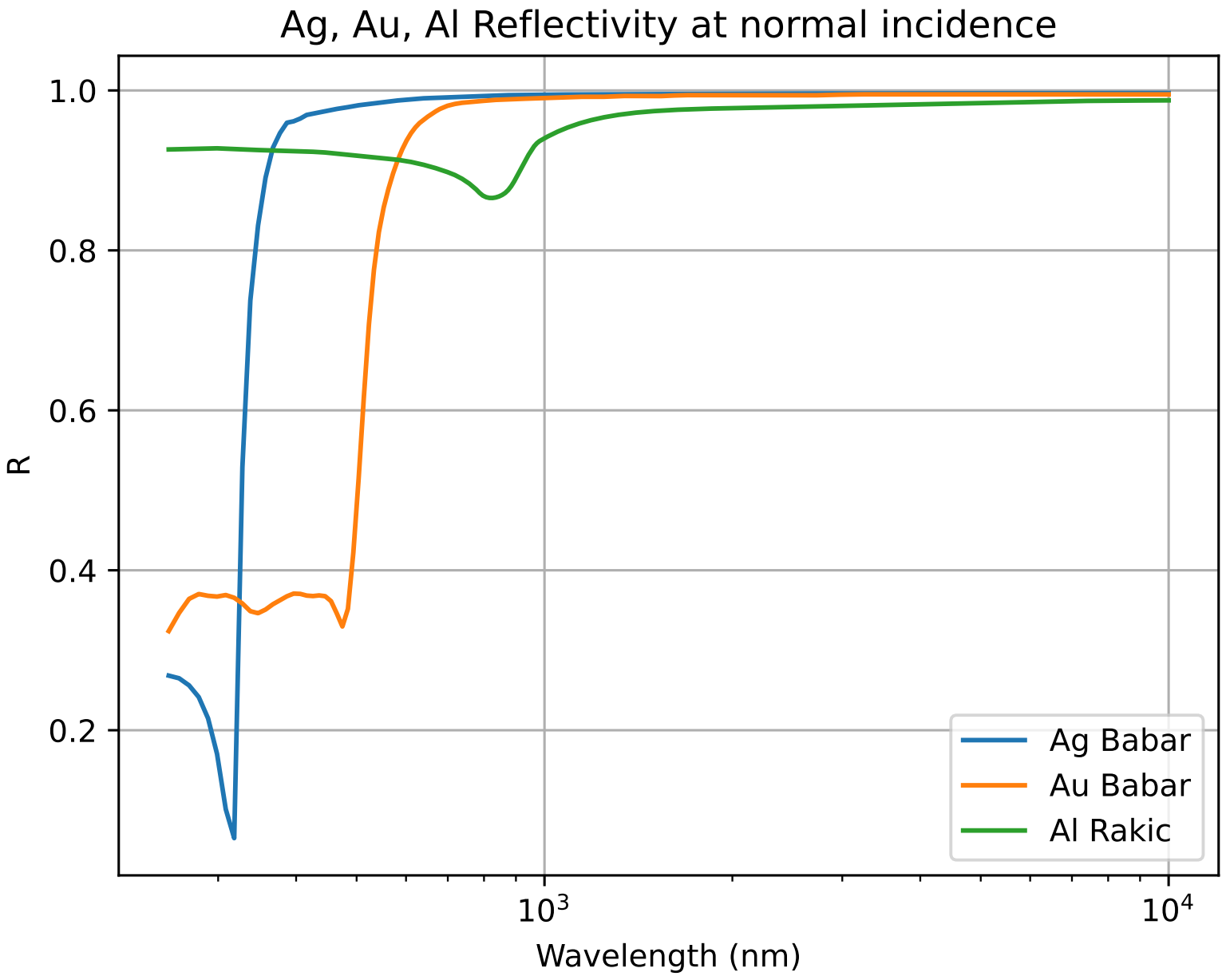

**Figure 5.1:** Simulated reflectance of unprotected Au, Ag and Al metal coatings from optical constants (Babar and Weaver for Au and Ag [1], Rakić for Al [18]).

It is however particularly sensitive to damaging from contact with humidity, sulfur and chlorine pollutants normally present even in the controlled atmosphere of a cleanroom, so design and test of an appropriate protection overcoat is of paramount importance to avoid loss of performance over time or even catastrophic damage [10].

Adhesion of the silver to the aluminum substrate is also an important characteristic to consider to avoid delamination due to mechanical or chemical stress, so an appropriate adhesion layer is generally introduced [12].

A protected reflective coating therefore usually consists of a series of layers: an adhesion layer, the bulk reflecting layer, optionally another intermediate adhesion layer, and finally one or more dielectric layers serving as protection and possibly also to improve reflectance at specific wavebands (Figure 5.2).

For Ariel telescope mirrors, a survey of existing providers in Europe led to the choice of a protected silver coating from CILAS, based on their capabilities of coating objects of up to 2 m of diameter, and space heritage [20, 11].

The coating consists of a NiCr adhesion layer followed by the Ag layer of enough thickness to result opaque to incoming visible light (at least 150 nm) and a final dielectric protection layer, whose adhesion may be possibly improved by an intermediate layer. The total coating thickness amounts to approximately 350 nm, but the exact composition and thicknesses cannot be disclosed at the time of writing.

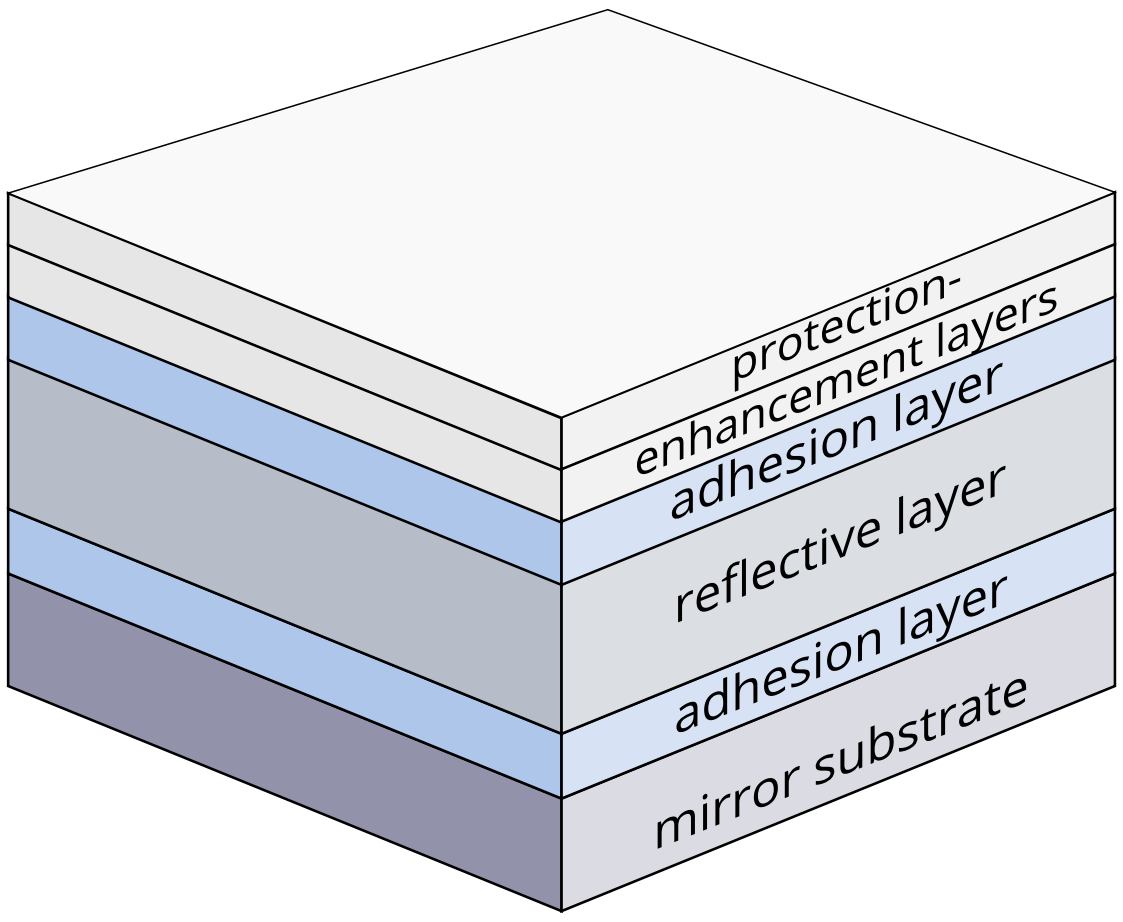

**Figure 5.2:** Example of a protected reflective coating stack, not to scale.

The layers are applied to the substrate through magnetron sputtering, a Physical Vapor Deposition (PVD) technique in which atoms of the coating material are ejected in vacuum from a target. The ejection happens through cathodic sputtering by bombardment with positive ions derived from an electrical discharge in a gas.

The ejected atoms then travel towards the object to be coated and adhere to its surface. The magnetron is used to generate a magnetic field that confines the sputtering plasma in the vicinity of the target, improving atoms extraction rate and reducing unwanted collisions of plasma ions with the substrate to be coated [23].

## 5.2 Qualification Campaign

The coating recipe proposed by CILAS had flown already in space missions, but in order for the process to achieve TRL 6, it needed to undergo an additional qualification "on a representative model or prototype, tested in a relevant environment" (see Table 4.1 for a definition of TRL levels).

In particular, the coating needed to be tested at the cryogenic operating temperature of Ariel (50 K) and on a surface of at least 0.6 $m^2$ for the qualification to be as representative as possible. The purpose of the test was to ensure that the difference in CTE between the aluminum substrate of the mirrors and the various coating layers, or other unforeseen effects, did not cause any delamination or any performance degradation of the coating.

A qualification campaign of thin film coatings for space applications consists of a series of environmental tests to ensure that the coating will reach the end of life of the instrument without damages and with the intended performance.

The European Cooperation for Space Standardization maintains a specific set of test methods, conditions and measurements in the ECSS Q-ST-70-17C standard [9]. The qualification campaign for an optical coating consists of the following tests:

1. performance
2. adhesion
3. cleanability
4. moderate abrasion
5. humidity
6. thermal vacuum and cycling
7. particle and UV radiation
8. other mission-specific tests (eg. contamination effects or solar illumination)

Acceptance criteria after each test, according to the standard, require:

1. no visual degradation of the coating;
2. no delamination or adherence loss;
3. thickness confirms to requirements;
4. performance measurements comply with coating specifications.

For the specific case of Ariel telescope mirrors, since the coating had already been qualified for space use, the initial campaign to reach TRL 6 for Phase B1 required only a delta qualification, and focused on the first six test types. Performance measurements were also limited to spectral reflectance between 500 and 1100 nm of wavelength and near normal incidence, although some samples have been measured in the full observational wavelength range of Ariel, and at a wider range of angles of incidence.

Particle exposure testing was also deferred to a later phase, on the premise that tests of similar coatings in comparable or harsher radiation environments than the one expected at L2 did not show significant degradation [12, 21, 15], while no significant UV exposure is expected since exposure to solar light will be avoided.

### 5.2.1 Measurement Methods

#### reflectance Measurements

For the measurements of reflectance as specified in the ECSS Q-ST-70-17C standard, alongside the commercial spectrophotometers employed by CILAS (see Section 2.4 of Paper 2 for details), the author developed a simple setup for in-house measurements of specular reflectance at variable angle [24].

The setup (Figure 5.3) consists of:

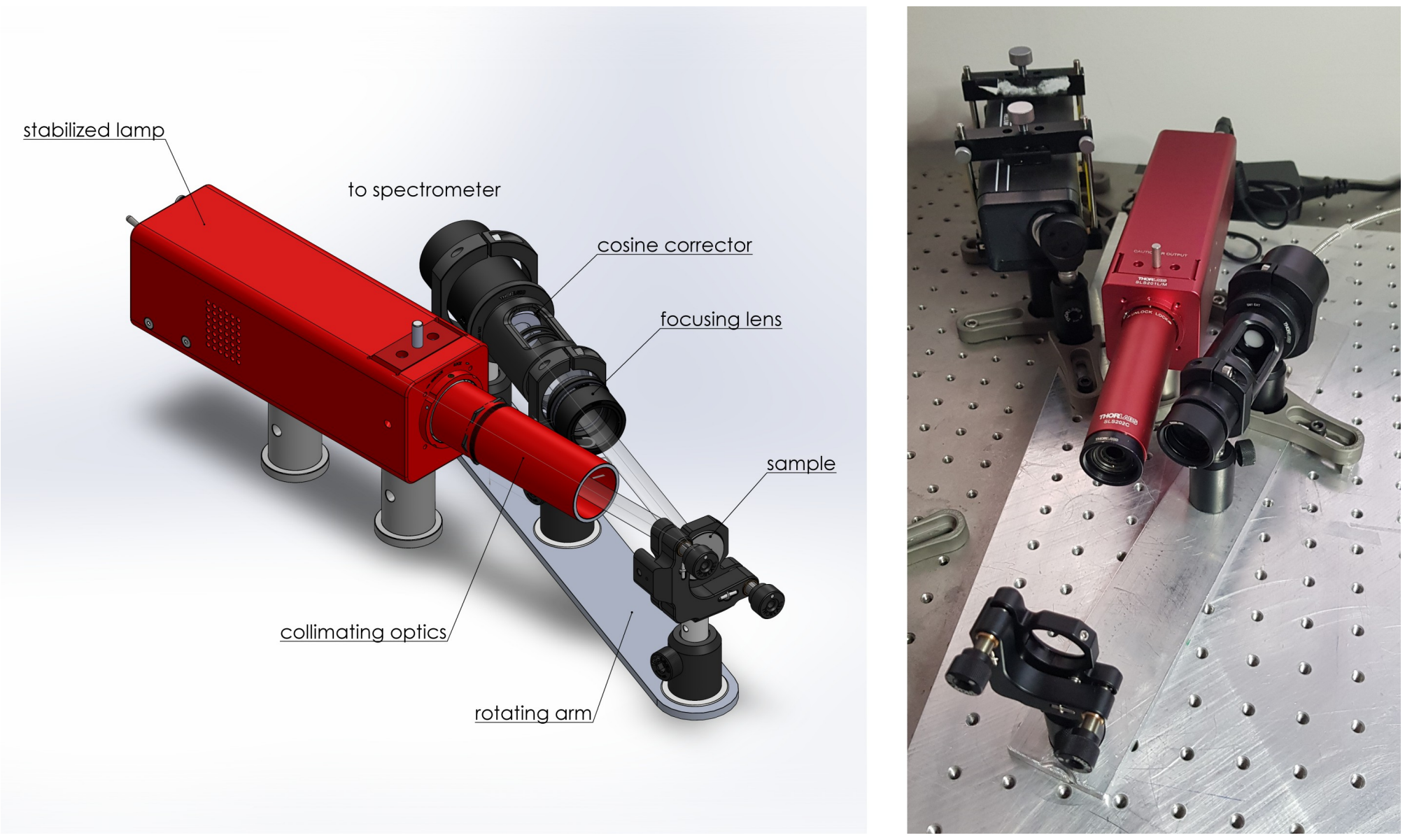

**Figure 5.3:** Drawing (left) and photograph (right) of the specular reflectance setup at variable angle used for part of the coating qualification test measurements.

- a stabilized halogen lamp with collimating optics;
- sample holder on a rotating platform;
- a rotating arm centered on the sample holder and hosting a cosine corrector block with focusing optics;
- a silicon-based spectrophotometer connected to the cosine corrector through an optical fiber.

Thanks to the rotating arm, the setup can perform both relative and absolute measurements: *relative measurements* are performed by measuring under the same conditions the spectral irradiance of the beam reflected by the sample $I_{sample}(\lambda)$ and reflected from a reference $I_{ref}(\lambda)$ with known reflectance $R_{ref}(\lambda)$, and applying the formula:

$$R_{sample}(\lambda) = \frac{I_{sample}(\lambda)}{I_{ref}(\lambda)} R_{ref}(\lambda).$$

To avoid slight variations of the angle of incidence caused by differences in planarity of samples and reference, resulting also in a variation in the centration of the beam on the cosine corrector entrance screen, a laser is shone on the sample at a larger angle than the one used for the measurement, and the sample is adjusted in tilt and tip until the reflection is centered on a far target.

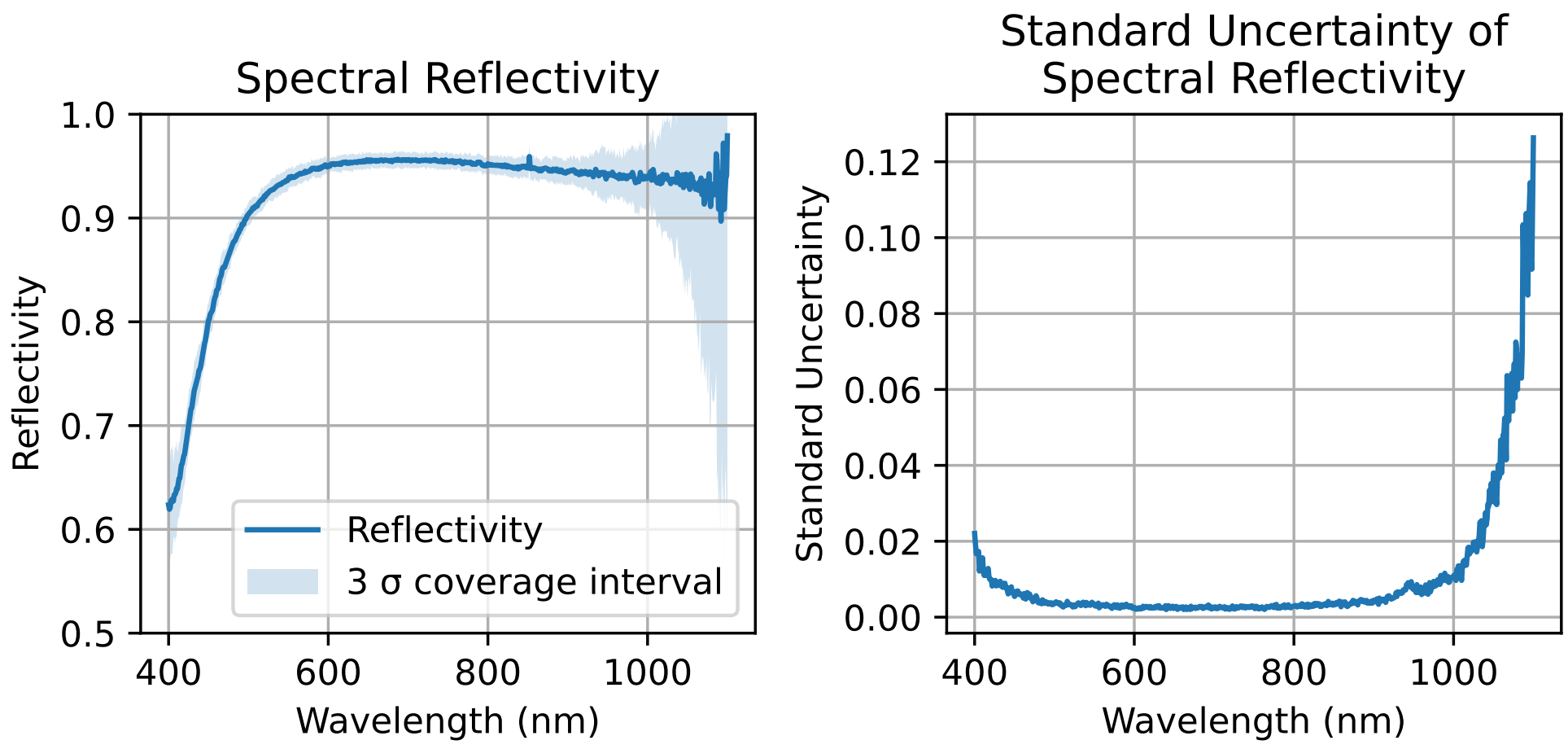

**Figure 5.4:** Plots of the expectation value of a reflectance measurement with the $3\sigma$ coverage interval (left), and standard uncertainty (right).

*Absolute measurements* instead require measuring the spectral irradiances of the beam reflected from the sample $I_{ref}(\lambda)$ and the beam directly from the lamp $I_{dir}(\lambda)$ through the empty sample holder by rotating the arm to 180°, and dividing the two:

$$R(\lambda) = \frac{I_{ref}(\lambda)}{I_{dir}(\lambda)}.$$

As the beam is not perfectly collimated, for absolute measurements the setup is designed to have the same distance in both the direct and reflected optical paths in order to limit the effect of potential spatial inhomogeneities of the samples or the cosine corrector. A visual check also shows that the beam is well within the aperture size of the cosine corrector.

The light source has been tested for stability on a time scale comparable to the measurement time, after a suitable warm-up period.

An estimation of the measurement error in absolute measurement mode has also been made according to the ISO Guide to the expression of Uncertainty in Measurement (ISO GUM [14]). The expectation value of the measured reflectance is plotted in Figure 5.4, together with the $3\sigma$ coverage interval calculated from the standard uncertainty of the measurement.

### Atomic Force Microscopy

The ECSS Q-ST-70-17C standard on durability testing of optical coatings requires visual inspections before and after each test to spot any macroscopic degradation appearing on the surface, such as a change of color, cracks or pits.

The standard suggests following Annex C of ISO 9211-4:2022, that mandates the use of two cool white 15 W lamps positioned directly above the sample, and to look at the sample against a black matte background at a distance of ≤45 cm and at a near grazing angle [13]. If however degradation is suspected, microscopic images may be taken to further qualify it.

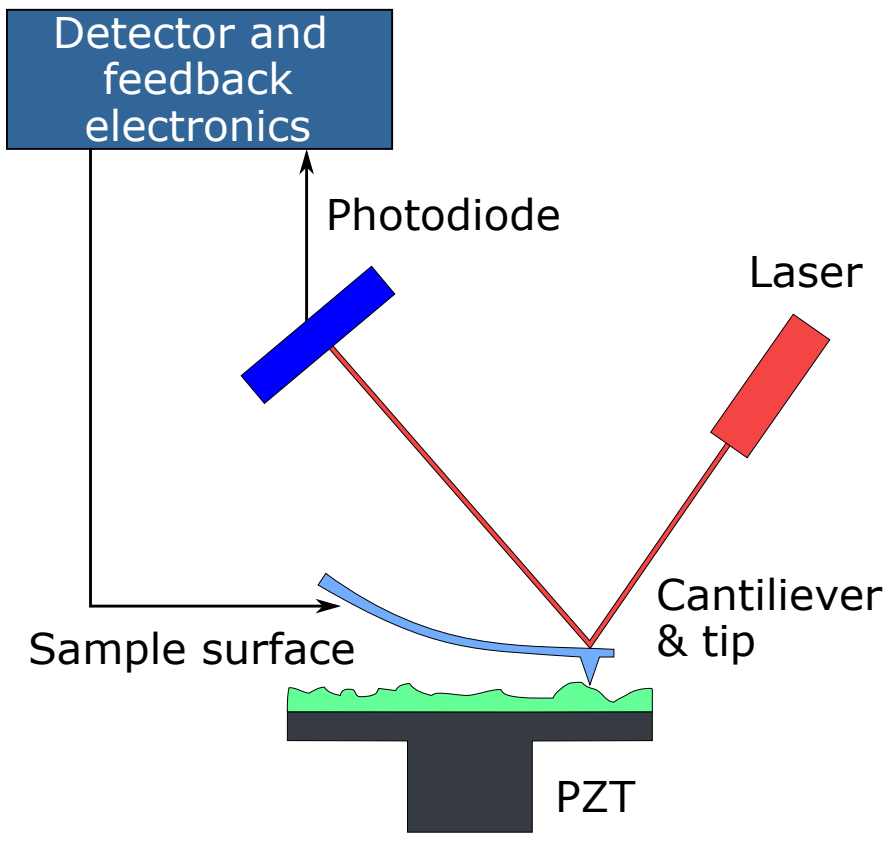

**Figure 5.5:** Schematic view of the principle of operation of an Atomic Force Microscope.

Thanks to the availability in-house of an atomic force microscope (a Park Systems[1] XE-Series 70) [24], for the specific tests described in Papers 3 and 4 we opted to take microscopic images of the surface before and after the cryogenic tests and during the ageing period to detect and possibly qualify any difference, and to measure surface roughness.

Atomic Force Microscopy (AFM) is a high-resolution non-optical imaging technique first demonstrated by Binnig, Quate and Gerber in 1985 [2], and further developed into a powerful measurement tool for surface analysis.

AFM allows the accurate and non-destructive characterization of topographical and morphological properties of the surface of samples at the atomic level, without requiring any specific sample preparation nor vacuum.

The basic operation principle of a standard AFM system with optical feedback (Figure 5.5), involves scanning the surface of a sample in a raster pattern with a probe, consisting of a microscopic tip positioned near the free end of a flexible cantilever. The AFM tip is usually made of silicon or silicon nitride.

The probe scans the surface, moved by a piezoelectric ceramic driver that controls the lateral and the vertical position of the probe. As the tip moves over surface features of different height, the deflection of the cantilever is captured by a laser beam reflected from its back side and directed into a photodetector.

The AFM can operate in contact or non-contact modes. In contact mode, the AFM operates similarly to a profilometer, but with a much smaller loading force on the tip, of the order of $1 \times 10^{-7}$ to $1 \times 10^{-11}$ N, making the contact area between tip and surface extremely small and making atomic resolution measurements possible. When the tip is in contact with the sample, repulsive short-range interatomic forces come into play, and cause a deflection of the cantilever.

[1]Park Systems Inc., KANC 15F, Gwanggyo-ro 109, Suwon 16229, Korea

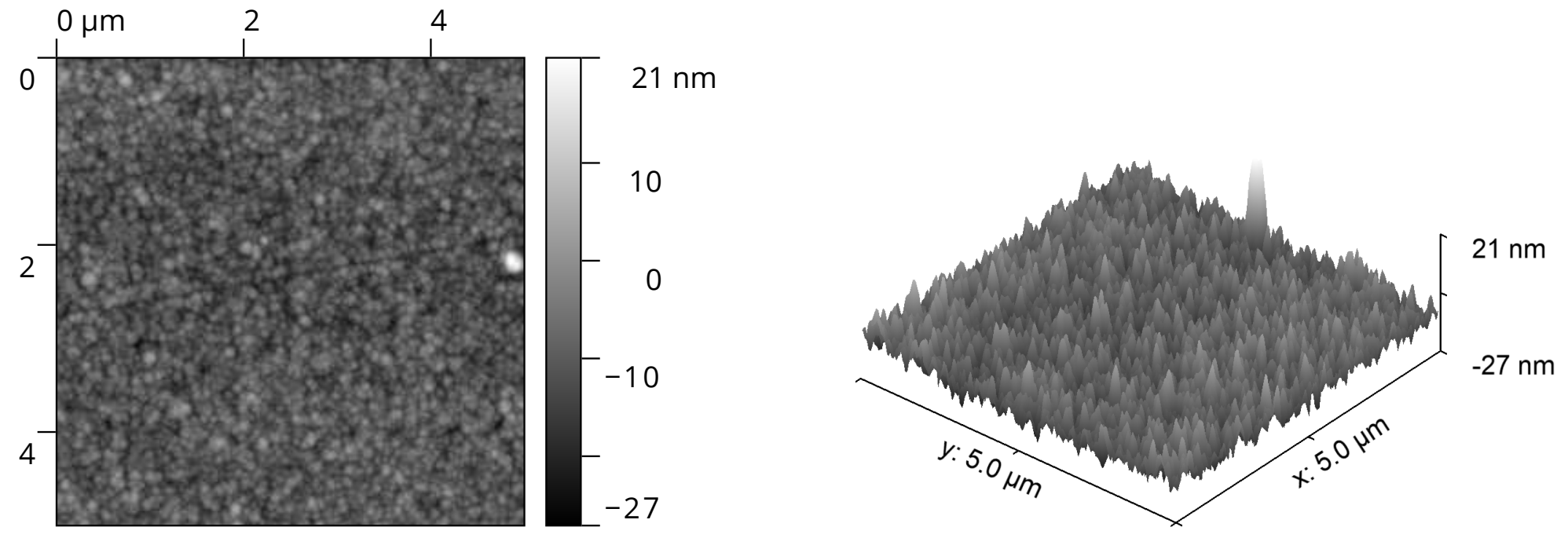

**Figure 5.6:** Example of an AFM measurement, taken by the author, of an aluminum sample with the baseline protected silver coating. On the left, the two-dimensional representation of the measurement and on the right, the three-dimensional view. Images processed with Gwyddion [17].

A feedback loop controls the vertical extension of the scanner in order to maintain near-constant cantilever deflection and hence a constant interaction force. The coordinates that the AFM tip tracks during the scan are combined to generate a three-dimensional topographic image of the surface.

In non-contact mode instead, the cantilever is driven to vibrate with a small amplitude of 1 nm or less, at or near its resonance frequency, and the probe tip is kept at a distance of several nanometers to several tens of nanometers away from the surface in the region of attractive interaction forces.

When the probe senses the presence of a force (or rather the gradient of a force), the oscillating frequency changes since the spring constant of the system is altered. This change in frequency or phase is then measured [19].

A sample measurement taken by the author is shown in Figure 5.6.

### 5.2.2 Campaign Phases

The campaign was planned by the team at CNR-IFN together with the Italian Telescope Team and the industrial partner MediaLario S.r.l., and scheduled to be run in two phases. The first one consisted in the actual qualification tests on aluminum and glass samples of various sizes, while the second consisted in the coating of the PTM primary mirror demonstrator alongside smaller samples, to verify repeatability of the coating process and to perform additional adhesion tests on the PTM itself.

The number of samples was determined such that in case of degradation happening during or after a test, it would be possible to trace it back to the specific process likely to have caused it, and to have at least three samples undergoing each test, for statistical significance.

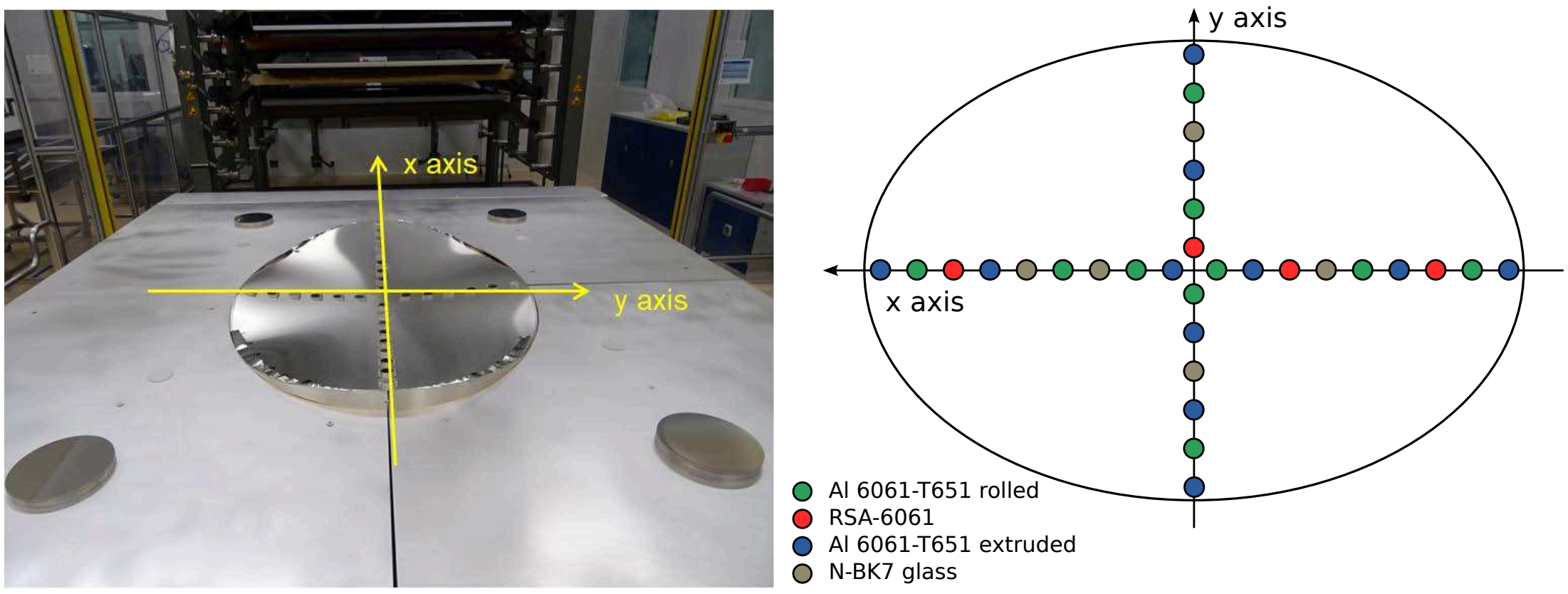

**Figure 5.7:** Picture of the sample holder on the coating chamber loading tray, with four additional 150 mm samples (left) and schematic representation of samples positions on the holder (right). Photo courtesy of CILAS.

### Qualification

The first phase of the campaign covered two coating batches. Samples of the first batch were coated while laying on a flat surface and then subjected to a reduced set of tests. The purpose of this batch was to set up both the coating and testing processes.

Two of the samples from this first batch, since they fully represented both the substrates and the coating process, were later kept in-house for a more detailed analysis of the effects of the cryogenic tests, including further spectral reflectance measurements and surface scans with the AFM. The satisfactory results of the tests and measurement were reported in Paper 3.

For the second coating batch, an aluminum holder was designed to approximate the shape of Ariel telescope M1 mirror in size and curvature, and the 25 mm samples were positioned along the major and minor axes of the elliptical aperture (Figure 5.7).

A total of 30 samples were employed for the tests of different materials:

- the baseline aluminum 6061-T651 in rolled plate forge;
- two alternatives to the baseline (RSA-6061 and Al 6061-T651 in extruded forge), in case of future needs;
- glass samples, used to measure coating thickness and as a reflectance reference.

The samples were then subjected to the entire battery of tests planned for the qualification, which passed successfully as described in Section 2.4 of Paper 2.

The use of glass samples is of particular importance since their surface roughness (declared to be below 1 nm RMS by the coating manufacturer) can be considered negligible at the operating wavelengths of Ariel in terms of scattering light, so they were used as reflectance references as we will see in Section 5.4.

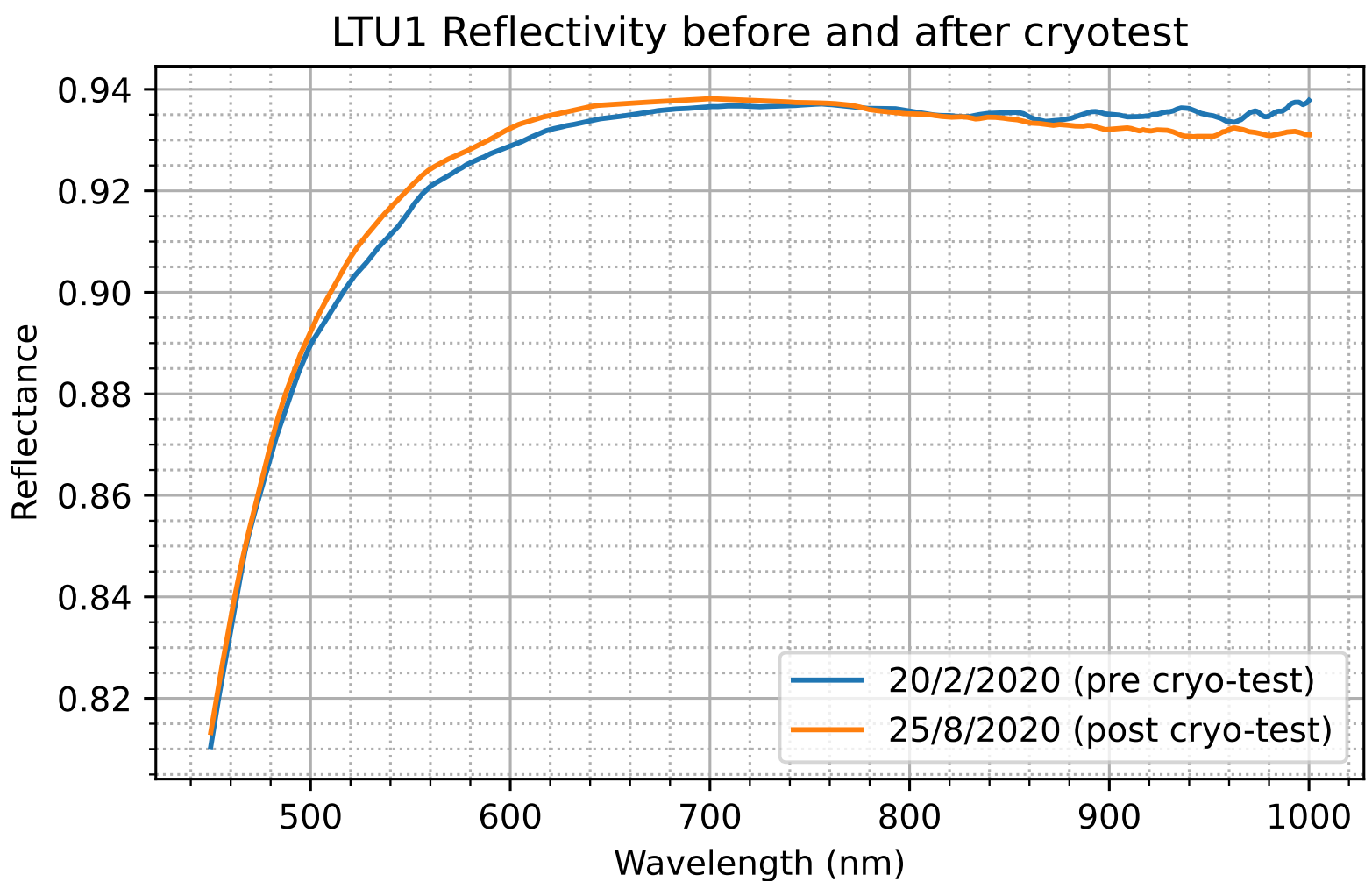

**Figure 5.8:** Measurements of one of the 150 mm diameter sample before and after the cryotest.

Additionally, four 150 mm samples were also coated alongside the sample holder. These samples were not, strictly speaking, part of the qualification process but represented important verification items since they had been subjected to the entire baseline stress releasing thermal treatments recipe (Section 4.3.1), so they could be considered representative of the telescope mirrors substrates. After coating, they were subjected to visual inspection and adhesion tests to verify compliance of the coating, and then to a single thermal cycle in vacuum with temperature range 54–293 K, a temperature change rate ≤5 K and a dwell time at 54 K of 15 min. Visual inspection before and after the cycle did not highlight any significant change.

Reflectance for one of them was measured by the author before and after the cryotest with a version of the setup described in Section 5.2.1 modified to hold the much larger and heavier sample, with the results shown in Figure 5.8. No change in reflectance was detected within the repeatability of the setup.

### Verification

After the qualification phase was successfully completed, the next step was to verify the results and to further confirm repeatability of the coating process by coating and testing the PTM primary mirror prototype itself, together with a set of 6 aluminum and several glass witness samples. The aluminum samples, as for the previous phases of the campaign, had been obtained from the same plate of the PTM, and underwent the same polishing process.

After coating, the samples were tested under the same conditions described in the previous section (except the cryogenic tests that were not performed on the samples), with equivalent results. The PTM, due to its size, could not be tested for reflectance, but only for adhesion in two different areas of this surface.

Results of the verification tests were also satisfactory, as reported by the author in Paper 4.

## 5.3 Coating Ageing Monitoring

Ageing performance of coated samples is verified, according to ECSS Q-ST-70-17C standard, with an thermal ageing test and a humidity test, as degradation caused by chemical reactions will be accelerated at a higher temperature.

Since the successful termination of the qualification tests, the samples have been stored in cleanroom conditions and checked periodically to detect any performance degradation by visual inspection and reflectance measurements.

Interim results are presented in Paper 3.

## 5.4 Throughput Estimation

One of the key performance parameters for the Ariel telescope, as described in Section 2.3, is the optical throughput.

Throughput in this case is defined as the efficiency of the optical system represented by the telescope, and is used for the estimation of the overall photometric efficiency of Ariel. More formally, let's consider the *optical extent* $G$, defined as:

$$G = n^2 A\Omega,$$

where $A$ is the area of the entrance pupil of the telescope, $\Omega$ the solid angle subtended at this pupil by the object and $n$ is the refractive index of the medium between object and pupil. $G$ is a purely geometrical quantity that can be shown to be invariant for the system, eg. if calculated at the entrance or exit pupil of a telescope.

The optical extent relates directly to the the radiant flux $\Phi$ transmitted by the telescope and the radiance $L$ of a source by the relation:

$$\Phi = \tau G L,$$

where $\tau$ is the transmittance of the optics. We can then define *throughput* as the quantity $\tau G$ [22].

In our case, since all sources are at infinity and therefore the quantity $G$ is always constant, we are specifically interested transmittance of the the telescope system $\tau$, that can be derived from the reflectance $R_{\mathrm{M}}$ of each mirror as:

$$\tau = R_{\mathrm{M1}} R_{\mathrm{M2}} R_{\mathrm{M3}} R_{\mathrm{M4}}.$$

The requirement on throughput is given at the end of the expected operating life of the mission (End of Life, or EOL), so the estimation must consider all the physical effects that are expected to contribute to the reflectance of the mirrors from coating to launch and though the entire life of the mission:

- nominal coating reflectance, i.e. reflectance of the coating when deposited on a surface with negligible scattering;
- variations in reflectance caused by coating dishomogeneities;
- scattering caused by surface roughness, surface defects and particulate contamination;
- absorption caused by molecular and ice contamination.

A first high level estimation was compiled in the early study phases of the mission, using reflectance values from coating catalog data and including the most probable causes of throughput loss. After Phase B1 was completed and actual measurements of the baseline coating were available, the author updated the estimation with more detailed modelling and analysis of additional scattering sources. The results of the estimation are presented in Paper 6.

## References

[1] S. Babar and J. H. Weaver. "Optical Constants of Cu, Ag, and Au Revisited". In: *Applied Optics* 54.3 (Jan. 20, 2015), p. 477. DOI: 10.1364/AO.54.000477.

[2] G. Binnig, C. F. Quate, and C. Gerber. "Atomic Force Microscope". In: *Physical Review Letters* 56.9 (Mar. 3, 1986), pp. 930–933. DOI: 10.1103/PhysRevLett.56.930.

[3] P. Chioetto et al. "Long Term Durability of Protected Silver Coating for the Mirrors of Ariel Mission Telescope". In: *International Conference on Space Optics — ICSO 2022, in press*. Oct. 3, 2022.

[4] P. Chioetto et al. "Initial Estimation of the Effects of Coating Dishomogeneities, Surface Roughness and Contamination on the Mirrors of Ariel Mission Telescope". In: *Proc. SPIE 11871, Optical Design and Engineering VIII*. Oct. 4, 2021, 118710N. DOI: 10.1117/12.2603768.

[5] P. Chioetto et al. "Qualification of the Thermal Stabilization, Polishing and Coating Procedures for the Aluminum Telescope Mirrors of the ARIEL Mission". In: *Experimental Astronomy* 53 (Apr. 19, 2022), pp. 885–904. DOI: 10.1007/s10686-022-09852-x.

[6] P. Chioetto et al. "Test of Protected Silver Coating on Aluminum Samples of ARIEL Main Telescope Mirror Substrate Material". In: *Proc. SPIE 11852, International Conference on Space Optics — ICSO 2020*. June 11, 2021, p. 118524L. DOI: 10.1117/12.2599794.

[7] P. Chioetto et al. "The Primary Mirror of the Ariel Mission: Cryotesting of Aluminum Mirror Samples with Protected Silver Coating". In: *Proc. SPIE 11451, Advances in Optical and Mechanical Technologies for Telescopes and Instrumentation IV*. Dec. 13, 2020, 114511A. DOI: 10.1117/12.2562548.

[8] R. M. Cutri. *Explanatory Supplement to the WISE All-Sky Data Release Products*. 2015. URL: https://wise2.ipac.caltech.edu/docs/release/allsky/expsup/sec3_2.html (visited on 05/29/2022).

[9] European Cooperation for Space Standardization. *ECSS-Q-ST-70-17C – Durability Testing of Coatings (1 February 2018)*. European Space Agency, Feb. 1, 2018.

[10] K. A. Folgner et al. "Environmental Durability of Protected Silver Mirrors Prepared by Plasma Beam Sputtering". In: *Applied Optics* 56.4 (Feb. 1, 2017), p. C75. DOI: 10.1364/AO.56.000C75.

[11] C. Grèzes-Besset et al. "High Performance Silver Coating with PACA2M Magnetron Sputtering". In: *Proc. SPIE 11180, International Conference on Space Optics — ICSO 2018*. July 12, 2019, p. 1118083. DOI: 10.1117/12.2536210.

[12] J. B. Heaney et al. "Preferred Mirror Coatings for UV, Visible, and IR Space Optical Instruments". In: *Proc. SPIE 8510, Earth Observing Systems XVII, 85100F*. Oct. 12, 2012. DOI: 10.1117/12.931005.

[13] International Organization for Standardization. *ISO 9211-4:2022 Optics and Photonics — Optical Coatings — Part 4: Specific Test Methods: Abrasion, Adhesion and Resistance to Water*. Vernier, Geneva, Switzerland: International Organization for Standardization, 2022.

[14] International Organization for Standardization. *Uncertainty of Measurement — Part 3: Guide to the Expression of Uncertainty in Measurement (GUM:1995)*. ISO/IEC Guide 98-3:2008. International Organization for Standardization, 2008.

[15] R. A. Keski-Kuha et al. "James Webb Space Telescope Optical Telescope Element Mirror Coatings". In: *Proc. SPIE 8442, Space Telescopes and Instrumentation 2012: Optical, Infrared, and Millimeter Wave, 84422J*. Sept. 21, 2012. DOI: 10.1117/12.925470.

[16] P. A. Lightsey et al. "Optical Transmission for the James Webb Space Telescope". In: *Proc. SPIE 8442, Space Telescopes and Instrumentation 2012: Optical, Infrared, and Millimeter Wave*. Aug. 22, 2012, 84423A. DOI: 10.1117/12.924841.

[17] D. Nečas and P. Klapetek. "Gwyddion: An Open-Source Software for SPM Data Analysis". In: *Open Physics* 10.1 (Jan. 1, 2012). DOI: 10.2478/s11534-011-0096-2.

[18] A. D. Rakić. "Algorithm for the Determination of Intrinsic Optical Constants of Metal Films: Application to Aluminum". In: *Applied Optics* 34.22 (Aug. 1995), pp. 4755–4767. DOI: 10.1364/AO.34.004755.

[19] D. Rugar and P. Hansma. “Atomic Force Microscopy”. In: *Physics today* 43.10 (1990), pp. 23–30.

[20] I. Savin de Larclause et al. “PACA2m Magnetron Sputtering Silver Coating: A Solution for Very Big Mirror Dimensions”. In: *Proc. SPIE 10563, International Conference on Space Optics — ICSO 2014*. Jan. 5, 2018, p. 1056308. DOI: 10.1117/12.2304237.

[21] D. A. Sheikh. “Improved Silver Mirror Coating for Ground and Space-Based Astronomy”. In: *Proc. SPIE 9912, Advances in Optical and Mechanical Technologies for Telescopes and Instrumentation II*. July 22, 2016, p. 991239. DOI: 10.1117/12.2234380.

[22] W. H. Steel. “Luminosity, Throughput, or Etendue?” In: *Applied Optics* 13.4 (Apr. 1974), pp. 704–705. DOI: 10.1364/AO.13.000704.

[23] S. Swann. “Magnetron Sputtering”. In: *Physics in Technology* 19.2 (Mar. 1988), pp. 67–75. DOI: 10.1088/0305-4624/19/2/304.

[24] P. Zuppella et al. “Optical Coatings: Applications and Metrology”. In: *The 2nd International Electronic Conference on Applied Sciences*. Oct. 15, 2021, p. 50. DOI: 10.3390/ASEC2021-11137.

# 6
# Conclusion

Following a brief introduction on the scope and purpose of the Ariel mission, the research work presented in this dissertation covers several topics in the field of the opto-mechanical analysis and design of Ariel's telescope and mirrors.

Many of the analysis and design methodologies described here present several novel aspects and, although developed and applied to a specific use case, are of broader applicability for the development of space telescopes in general.

Of particular interest are the manufacturing technologies and processes employed for the realization of large lightweighted aluminum mirrors with a protected silver coating and operating at cryogenic temperatures. Ariel's primary telescope mirror is in fact a first in terms of size of the collecting area for this particular choice of substrate material, so the development of these technologies is pushing forward the state-of-the-art in metal mirrors manufacturing, providing novel heritage for future space missions.

The treatment of each topic covers both the review of the relevant scientific background and the specific engineering results in a general context, on top of explaining how the work contributed to the successful adoption of the Ariel mission, and its progress to the implementation phase. This further highlights the broad validity of the work presented in this dissertation.

At a personal level, the breadth of topics reflects the variety of competences required to properly support the Italian Telescope Team and complements perfectly, in the author's opinion, the multidisciplinary nature of the PhD program from CISAS and its several curriculum courses and seminars on Optics, space materials and constructions, and coatings. At the same time, the topics are naturally tied together not only from a programmatic point of view, but also by the underlying coherence of themes, as should be apparent by the treatment done in this dissertation.

Most of the work presented here reached a definite result or a specific project milestone, allowing a conclusive treatment of the topics from a scientific point of view. In the broader perspective of the mission however, as it progresses towards the end of Phase B2 and to the Preliminary Design Review (PDR), the author's contribution will continue in the following areas:

- tolerance and STOP analyses, since further iterations will be necessary as the mechanical design is refined (Chapter 3);
- support on the planning and execution of the telescope integration, assembly, alignment and verification activities (Chapter 3);
- further characterization of the manufacturing processes (stress releasing thermal treatments, diamond turning, optical surface polishing) for the M1 mirror prototypes and later for the M1 qualification and flight models (Chapter 4);
- characterization tests and measurements of coating of mirror prototypes and models (Chapter 5);
- coating radiation tests and ageing of coating on samples, and further refinement of the contamination and throughput models (Chapter 5).

From a managerial point of view, the author will also continue to provide his support to the Product Assurance / Quality Assurance tasks for the Telescope Assembly.

# Acknowledgments

The activities described in this dissertation have been funded under the Implementation Agreement n. 2018-22-HH.0 of the ASI-INAF Framework Agreement "Participation to the B1 study phase of the Ariel mission"), the ESA – CSL/INAF contract "Cryotesting of Ariel M1 mirror and coating process qualification" and the Implementation Agreement n. 2021-5-HH.0 of the ASI-INAF Framework Agreement "Italian Participation to Ariel mission phase B2/C".

I wish to thank the advisors Dr. Paola Zuppella and Dr. Vania Da Deppo for their guidance and support, Professors Emanuele Pace, Andrea Tozzi, Enzo Pascale and the rest of the Ariel team for their insights, Dr. Fabio Frassetto for the introduction at CNR–IFN. I'd also like to thank my family, Marco and Fabrizio for their inspiration.

This document was typeset in EB Garamond typeface with LaTeX $2_\varepsilon$, using a modified version of the Dissertate document syle[1]. Most figures were created with the Python graphing package Matplotlib [7], and open source graphic software Inkscape[2] and GIMP[3].

[1] https://github.com/suchow/Dissertate

[2] Inkscape: Open Source Scalable Vector Graphics Editor, https://inkscape.org/

[3] GNU Image Manipulation Program, https://www.gimp.org/

# Part II

# Papers and Conference Proceedings

# 1

# Preliminary analysis of ground-to-flight mechanical tolerances of the Ariel mission telescope

Paolo Chioetto[1 a,b,c], Andrea Tozzi[d], Anna Brucalassi[d], Debora Ferruzzi[d], Andrew Caldwell[f], Martin Caldwell[f], Fausto Cortecchia[e], Emiliano Diolaiti[e], Paul Eccleston[f], Elisa Guerriero[m], Matteo Lombini[e], Giuseppe Malaguti[e], Giuseppina Micela[g], Emanuele Pace[h], Enzo Pascale[i], Raffaele Piazzolla[k], Giampaolo Preti[h], Mario Salatti[k], Giovanna Tinetti[l], Elisabetta Tommasi[k], Paola Zuppella[c]

a CNR-Istituto di Fotonica e Nanotecnologie di Padova, Via Trasea 7, 35131 Padova, Italy
b Centro di Ateneo di Studi e Attività Spaziali "Giuseppe Colombo"- CISAS, Via Venezia 15, 35131 Padova, Italy
c INAF-Osservatorio Astronomico di Padova, Vicolo dell'Osservatorio 5, 35122 Padova, Italy
d INAF-Osservatorio Astrofisico di Arcetri, Largo E. Fermi 5, 50125 Firenze, Italy
e INAF-Osservatorio di Astrofisica e Scienza dello spazio di Bologna, Via Piero Gobetti 93/3, 40129 Bologna, Italy
f RAL Space, STFC Rutherford Appleton Laboratory, Didcot, Oxon, OX11 0QX, UK
g INAF-Osservatorio Astronomico di Palermo, Piazza del Parlamento 1, 90134 Palermo, Italy
h Dipartimento di Fisica ed Astronomia-Università degli Studi di Firenze, Largo E. Fermi 2, 50125 Firenze, Italy
i Dipartimento di Fisica, La Sapienza Università di Roma, Piazzale Aldo Moro 2, 00185 Roma, Italy
j INAF-IAPS, Via del Fosso del Cavaliere 100, I-00133 Roma, Italy
k ASI, Agenzia Spaziale Italiana, Via del Politecnico snc, Roma, Italy
l Department of Physics and Astronomy, University College London, Gower Street, London WC1E 6BT, UK

[1] Presenting and contact author.

m Dipartimento di Fisica e Chimica-Università degli Studi di Palermo, Via Archirafi 36, 90128 Palermo, Italy

### Abstract

Ariel (Atmospheric Remote-Sensing Infrared Exoplanet Large Survey) is the adopted M4 mission of ESA "Cosmic Vision" program. Its purpose is to conduct a survey of the atmospheres of known exoplanets through transit spectroscopy. Launch is scheduled for 2029. Ariel scientific payload consists of an off-axis, unobscured Cassegrain telescope feeding a set of photometers and spectrometers in the waveband between 0.5 and 7.8 μm, and operating at cryogenic temperatures.

The Ariel Telescope consists of a primary parabolic mirror with an elliptical aperture of 1.1 m of major axis, followed by a hyperbolic secondary, a parabolic recollimating tertiary and a flat folding mirror directing the output beam parallel to the optical bench. The secondary mirror is mounted on a roto-translating stage for adjustments during the mission.

Proper operation of the instruments prescribes a set of tolerances on the position and orientation of the telescope output beam: this needs to be verified against possible telescope misalignments as part of the ongoing Structural, Thermal, Optical and Performance Analysis.

A specific part of this analysis concerns the mechanical misalignments, in terms of rigid body movements of the mirrors, that may arise after ground alignment, and how they can be compensated in flight. The purpose is to derive the mechanical constraints that can be used for the design of the opto-mechanical mounting systems of the mirrors.

This paper describes the methodology and preliminary results of this analysis, and discusses future steps.

## 1.1 Introduction

Ariel is the M4 mission of ESA "Cosmic Vision" program, with the scientific purpose of carrying out a survey of the atmospheres of a large sample of known exoplanets. Officially adopted in 2020, Ariel is scheduled to launch in 2029 [5]. The payload consists of a set of spectrometers and photometers operating in the waveband 0.5–7.8 μm, fed by an afocal, unobscured Cassegrain-type telescope [1].

The Cassegrain is designed off-axis, with a large elliptical entrance pupil of 0.6 $m^2$ of area, followed by a tertiary recollimating mirror and a fourth folding mirror that bends the light beam towards the optical bench. The global optical coordinate reference (OPT) for the telescope has its center at the vertex of the parent parabola of M1, and the axes oriented as in Fig. 1.1. The secondary mirror is mounted on a mechanism (M2M) permitting regulation of tip/tilt and shift along the optical axis (focus) for fine adjustments during flight. The telescope has diffraction limited optical performance at the wavelength of 3 μm and on a 30″ Field of View (FoV). Telescope and instruments will operate at a temperature below 50 K.

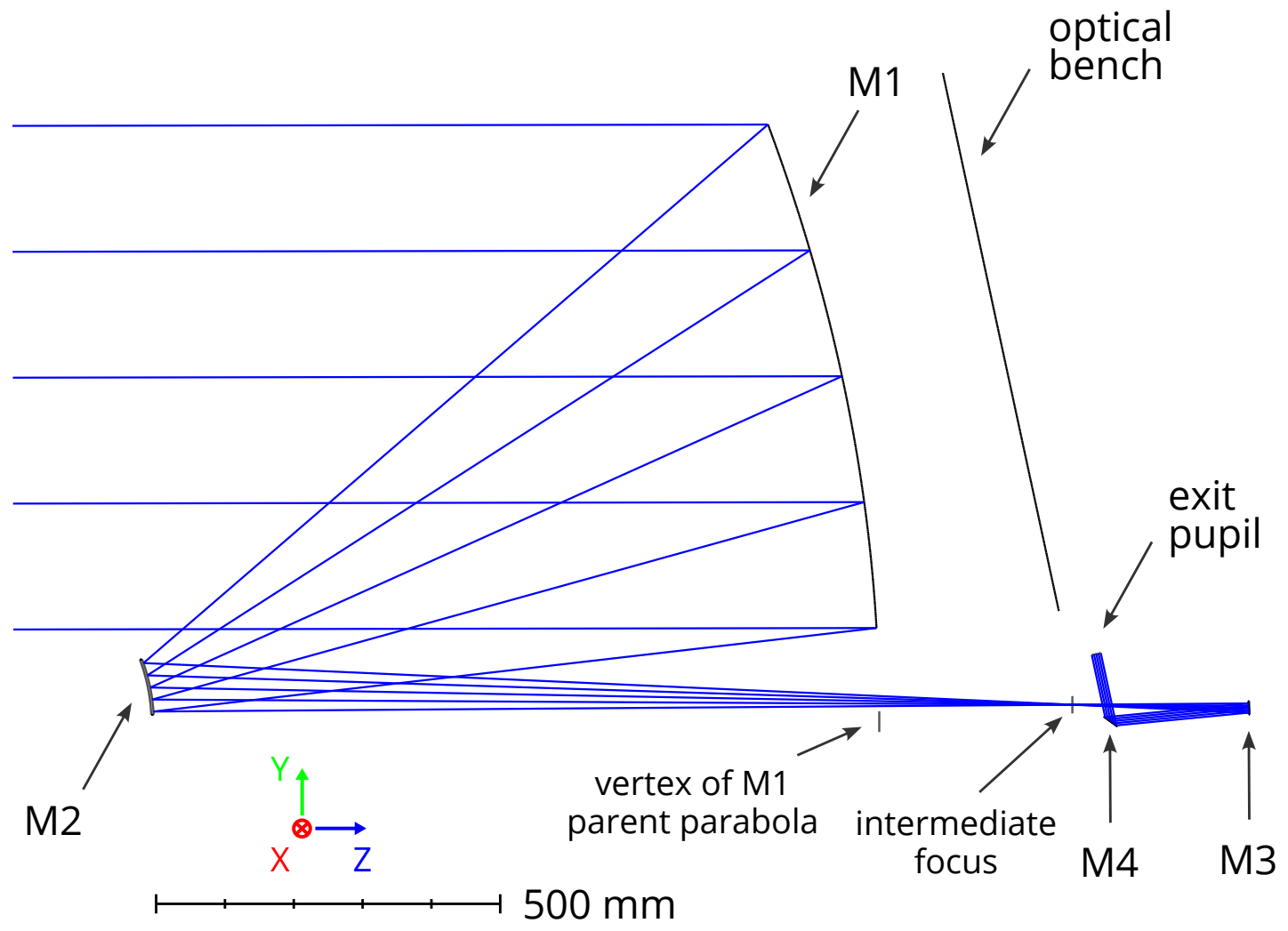

**Figure 1.1:** Ariel Telescope optical design.

The baseline material for the telescope mirrors and supporting structures, chosen on the criteria of athermalization, ease of construction and cost, is aluminum alloy 6061-T651. Aluminum has a relatively high coefficient of thermal expansion compared with other mirrors construction materials such as glass-ceramics, and a low stiffness, so a detailed thermal and termo-mechanical analysis is required to determine the extent of deformations induced at the operating temperature by thermal gradients and mechanical stresses, and their effects on optical performance. The analysis, known with the acronym STOP from the initials of Structural, Thermal, Optical Performance Analysis, consists of a series of cases with specific boundary conditions set in terms of temperature maps and mechanical loads.

The analysis presented here is complementary to the STOP Analysis, and is concerned with another possible source of optical performance degradation: the effect of rigid body motions of the mirrors with respect to their nominal position (misalignments), caused for example by dynamic loads (accelerations and vibrations) during launch. For this case, as opposed to thermo-elastic deformation, it is not always feasible to determine analytically the direction and extent of the motions, so a statistical approach is required. This is especially true for the analysis presented here since, at the current level of development, a detailed mechanical design of the mountings of the mirrors is not yet finalized.

The purpose of the analysis is twofold: the first, given a set of requirements on the optical performance and the position and orientation of the telescope collimated output beam, is to determine the maximum range of misalignments that can still be tolerated with respect to the requirements and the second, reversing the analysis, is to calculate the maximum effect on performance given the extent of the misalignments.

Optical performance is evaluated in terms of Enclosed Energy (EE) within a specific area of the Point Spread Function (PSF), at the exit pupil of the telescope. More specifically, since the footprint of the exit pupil is elliptical, as it is the image of the elliptical aperture of the primary mirror, what is actually computed is the percentage of energy within an ellipse. The size of the EE ellipse has been determined through simulation from the scientific requirements of the mission.

Position and orientation of the exit beams, on the other hand, are measured in reference to a plane that is orthogonal to the beam, and placed at the exit pupil of the telescope. The position is calculated at the intersection of the chief ray (the ray passing through the center of the entrance pupil) with the plane, while the orientation as the exit angle of the chief ray.

## 1.2 Simulation Setup

The starting point for the analysis presented here is the nominal telescope design at the operating temperature of 50 K.

Aim of the analysis is to determine the maximum misalignments that are still recoverable using the two compensation mechanisms available in flight: the secondary mirror M2, and a tilt of the line of sight (LoS) of the entire telescope.

The analysis is carried out with the aid of the optical simulation software Zemax OpticStudio©, and routines written by the author in Python. The amount of compensation is calculated through the optimization functionality of OpticStudio, employing a Merit Function (MF) based on Encircled Energy and maximum allowed displacements of the Telescope exit pupil (more on this in Section 1.2.3).

The steps used in the analysis are:

1. an initial inverse sensitivity analysis to determine the range of admissible misalignments for each element taken individually, based on the acceptable range of performance parameters (acceptance criteria);

2. a Monte Carlo simulation on the ranges identified in Step 1 to evaluate statistically the combined effect of multiple misalignments.

Tilts and shifts of each mirror are relative to the local reference frame as defined by the simulation software, and a center of rotation (CoR) (Figure 1.2).

For M1, M3 and M4 the CoR is arbitrarily positioned at the center of the optical surface, since the details of the mounting mechanism are still being finalized. For M2 the CoR of the M2M mechanism is known, so it is used also as center of rotation for the misalignments.

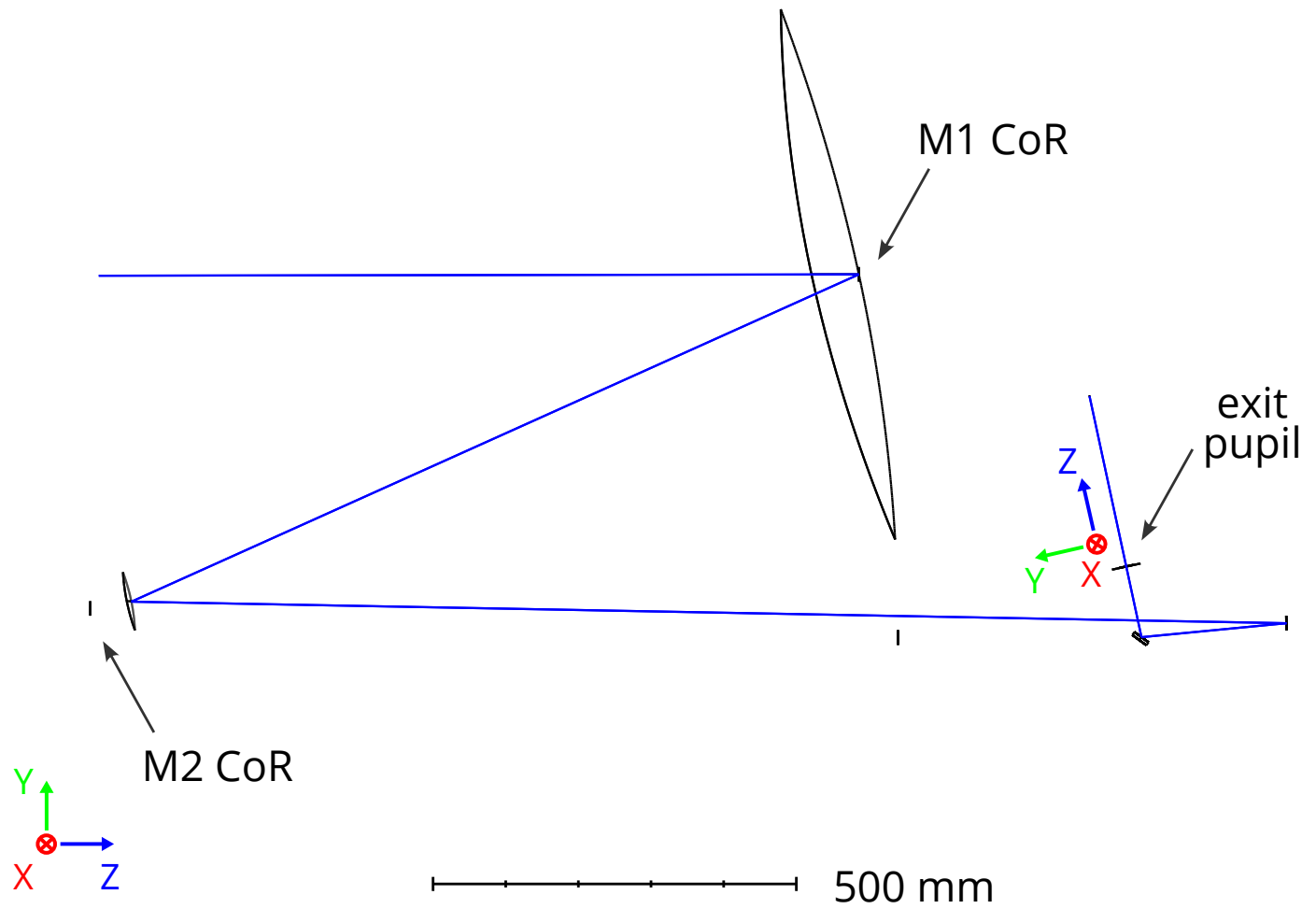

**Figure 1.2:** Diagram of the optical design used in the simulations, showing the chief ray (blue), global (left) and local (right, at the EXP surface) axes orientations, and Centers of Rotation (CoR) for M1 and M2.

### 1.2.1 Acceptance Criteria

The set of acceptance criteria, derived from telescope requirements, are the percentage of Enclosed Energy that is lost with respect to the nominal design and the exit beam shift and tilt at the nominal exit pupil. They are reported in Table 1.1, together with their knowledge error.

**Table 1.1:** Acceptable performance range for ground-to-flight misalignment errors. Corresponding TA requirement are also indicated.

| Parameter | Allowed Range | Knowledge Error |
|---|---|---|
| Lost EE | $\leq 5\,\%$ | – |
| Beam Tilt x/y | $\pm 45''$ | $\pm 15''$ |
| Beam Shift x/y | $\pm 60$ µm | $\pm 20$ µm |

The baseline requirement for Enclosed Energy at the wavelength of 0.55 µm is given for an ellipse with semi-axes $41''$ and $27''$, and imposes that at least 83.8% of the energy of the PSF at the telescope exit pupil be within the ellipse, for all fields within the telescope FoV.

If we consider the nominal telescope design, that is close to diffraction limit, the energy enclosed in the such ellipse is at least 96%, so the margin is 12.2%. The figure of 5%, used for this simulation, is an arbitrary allocation of this margin.

This is so because the ellipse from the requirement for radial Encircled Energy at 0.55 µm, as given above, is larger than the "diffraction limited" ellipse that has axes of 10.4″ and 6.9″, according to the approximated formula for the first minimum of the Airy disk with $\lambda$ = 0.55 µm:

$$\theta = 1.22\frac{\lambda}{D},$$

and with $D$ equal to the major and minor axes of the telescope entrance pupil (730 and 1100 mm), and taking into consideration the telescope angular magnification of 55.

Admissible tilts and shifts of the exit beam have instead been derived from alignment requirements of the instruments downstream from the telescope.

### 1.2.2 Compensators

At each step of the inverse sensitivity analysis and the Monte Carlo simulation, the perturbed system is re-optimized using the compensators within the nominal ranges specified in Table 1.2.

**Table 1.2:** Compensators rages for the ground-to-flight case.

| Compensator | Allowed Range |
|---|---|
| M2 Tip/Tilt | ±0.115° (414″) |
| M2 z-axis translation | ±350 µm |
| Line of Sight rotation (x/y axes) | – |

The Line of Sight (LoS) compensation is a rigid body rotation of the entire telescope, resulting in fact in a re-pointing to an off-axis field. It is easy to see how this compensation results in a tilt of the exit beam with a scale factor equal to the magnification of the telescope.

M2 compensations are applied wrt. the M2M movement origin, which is located behind the M2 mirror vertex, at a point with the following coordinates: +40.1 mm in the $y_{\mathrm{OPT}}$ direction from M2 vertex and -61.8 mm in the $z_{\mathrm{OPT}}$ direction from M2 vertex.

### 1.2.3 Optimization Function

OpticStudio local optimization functionality utilizes a proprietary version of the damped least squares algorithm to minimize a target function, known as Merit Function (MF), by varying a set of parameters [3, 4]. In our case the parameters correspond to the set of compensators specified in Section 1.2.2.

The MF is specified using various operands, and for both the inverse sensitivity analysis and the Monte Carlo simulation, consists of the following terms, all contributing equally.

1. An approximation of the lost Enclosed Energy (1 - EE) of the PSF at the exit pupil, taking the maximum value calculated for eight radial fields at the edge of the 30″ field of view of the telescope. The wavelength of 0.55 μm is considered.

2. The direction cosines of the chief ray at the exit pupil ("REAA" and "REAB" operands). These values are derived from the angle at which the incoming beam exits the surface that materializes the Exit Pupil.

3. The local $x$ and $y$ coordinates of the intersection of the chief ray for the on-axis field at the Exit Pupil ("REAX" and "REAY").

OpticStudio does not provide a way to calculate the fraction of energy enclosed in an elliptical aperture in the PSF. For this reason, an approximation is made by calculating the energy enclosed in the two circles inscribing and circumscribing the ellipse, and averaging the two, as illustrated in Figure 1.3 for a sample aberrated PSF. The approximation was found to be reasonably good for the intended purposes [2].

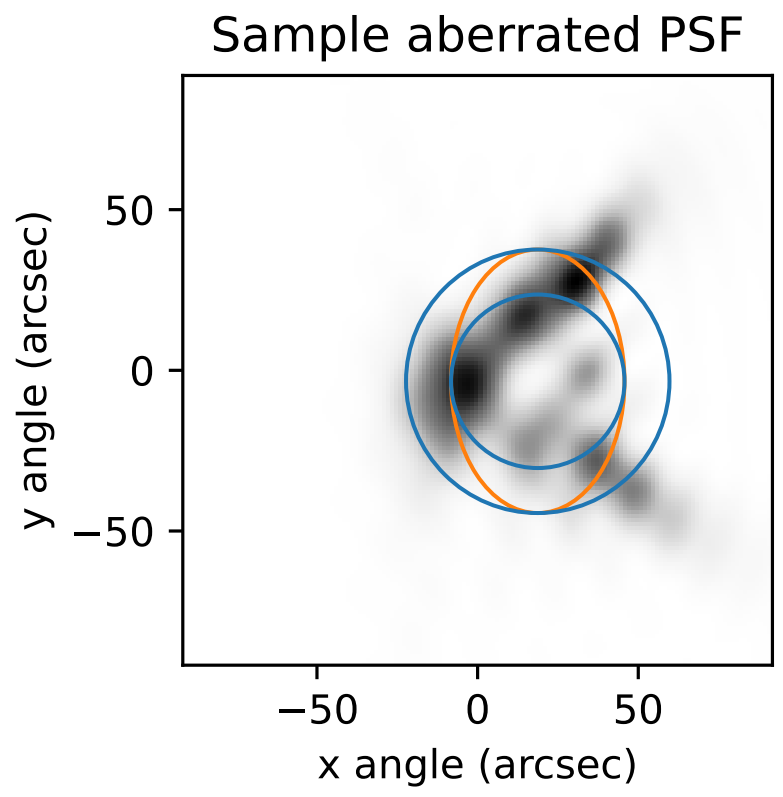

**Figure 1.3:** Sample aberrated PSF illustrating the approximated calculation of the Enclosed Energy on the circles inscribing and circumscribing an elliptic aperture.

The direction cosines of point 2 above ("REAA" and "REAB") are the cosines of the angles that the chief ray vector forms with the $x, y$ axes, so the relation with the beam angle is:

$$\theta_{\{x/y\}} = \pi/2 - \arccos(\{\text{REAA}/\text{REAB}\})$$

Each term is evaluated using "greater than" or "less than" operands ("OPGT" and "OPLT") against the mandated ranges of variation, so its contribution to the MF goes to zero as soon as the corresponding requirement is satisfied, and therefore the MF equals zero when all requirements are satisfied.

Table 1.3 summarizes the operands used.

The operands evaluating the position and orientation of the exit pupil must be employed on the surface that is located at the actual position of the exit pupil along the local $z$ axis, in the perturbed telescope design, which may move slightly, changing the calculation of the $x$ and $y$ intersection coordinates.

For this reason, the surface materializing the exit pupil in the nominal design is preceded by an auxiliary surface of variable thickness, and an additional optimization operand, "EXPP", is employed to make sure that the surface marked as "exit pupil" is indeed located at the exit pupil of the telescope, even in the perturbed design (in the paraxial approximation).

The Merit Function presented above is very computationally intensive because of the DENF operands, that require calculation of the PSF to determine the radius of encircled energy. For this reason, the final optimization of each configuration is preceded by a "pre-optimization" step using a less intensive Merit Function based on the root mean squared (RMS) size of the Spot Radius. This step can be seen as a way to bring the perturbed telescope close to the optimal compensation very quickly, before performing the last optimization step.

Since the final evaluation of the optimized case is performed on the proper merit function using DENF operands, the procedure is equivalent in terms of results.

Please also note that no global optimization algorithm has been applied in the procedure, so it is possible that some cases result non compliant when the MF does not converge to zero because it is stuck in a local minimum outside the acceptance criteria range. Some of the non compliant cases where therefore checked manually, but were rarely found to be false negatives.

**Table 1.3:** Operands used in the OpticStudio optimization Merit Function.

| | Operand Used | Explanation |
|---|---|---|
| Lost EE | DENF | Average of EE on circles inscribed and circumscribed on elliptical aperture |
| EXP Tilt x | REAA on EXP | Chief Ray angle (deg) with local x axis (parallel to global x) |
| EXP Tilt y | REAB on EXP | Chief Ray angle (deg) with local y axis (orthogonal to OB) |
| EXP Shift x | REAX on EXP | Chief Ray x shift (mm) |
| EXP Shift y | REAY on EXP | Chief Ray y shift (mm) |
| EXP Position | EXPP | Paraxial exit pupil position |

## 1.3 Results

The simulation setup described in the previous sections was applied to two scenarios, identified by the set of acceptance criteria utilized.

**Scenario 1** Nominal acceptance criteria defined in Paragraph 1.2.1. The analysis consists of an inverse sensitivity to determine the maximum perturbation ranges, a Monte Carlo simulation with the maximum perturbation ranges and finally a Monte Carlo with a set of perturbations with reduced ranges derived from industry expertise;

**Scenario 2** Inverse sensitivity analysis with a relaxed set of acceptance criteria.

### 1.3.1 Nominal Acceptance Criteria Scenario

#### Inverse Sensitivity Analysis

The goal of the inverse sensitivity analysis procedure is to determine the maximum range of perturbations that, applied one at a time, can still be compensated to achieve a telescope configuration that is compliant with requirements. The procedure consists of the following steps, applied to each parameter:

1. start from the unperturbed optical design;
2. perturb a single parameter and re-optimize the optical model using compensators, until the MF can converge to zero, meaning that the design has been compensated within requirement;
3. increase the magnitude of the perturbation until a value is reached beyond which the design can no longer be optimized, or the limits of the compensators are reached (this is done independently for the positive and negative extremes of the range).

In Table 1.4 we report the maximum and minimum values of the tolerance operands, and the compensation applied. The limits imposed on operands “REAA” and “REAB” are $2.182\times 10^{-4}$, calculated with the following formula:

$$\cos\left(\frac{\pi}{2} - \theta_{\{x/y\}}\right)$$

where $\theta_{x/y}$ are the components in $x$ and $y$ in radians of the chief ray angle wrt. the surface normal (equal to the $45''$ range limit of Table 1.1).

#### Monte Carlo Simulation with maximum perturbation ranges

A Monte Carlo simulation was then ran to consider the effect of all perturbations derived in the previous section, applied simultaneously. For each Monte Carlo case, all the parameters are assigned a random value from a uniform probability distribution in the perturbation interval (Table 1.4).

The goal of the simulation is to determine the percentage of compliant cases, i.e. cases that can be optimized to be within telescope requirements using the two steps optimization merit function described in paragraph 1.2.3.

This simulation was performed in order to get a sense of how sensitive the telescope design is in terms of the combination of more than one displacements, and in fact only 17 cases resulted viable out of the total run of 100.

### Monte Carlo Simulation with reduced perturbation ranges

The Monte Carlo simulation was then repeated using a set of reduced displacement ranges of ±10 µm for linear shifts and ±10″ for rotations.

The new set was devised based on feedback from the telescope prime contractor that the displacements can be greatly reduced from the ranges determined from the inverse sensitivity analysis, upon considerations that the primary mirror can be hold in position by pins, and that no specific challenges are foreseen for the mounting of the smaller mirrors.

Before proceeding with the actual Monte Carlo, the boundary cases of each reduced range was optimized separately, to determine the effect on the system. Results, presented in Table 1.5, show how each case can be compensated. Results are in line with those of Table 1.4, showing that smaller perturbations require smaller adjustments of the same compensators.

Then the Monte Carlo analysis was performed, assigning random values from uniform probability distributions in the perturbation intervals to all parameters.

With this new reduced set of ranges, 98% of the cases are recoverable with compensation (up from 17% in the previous case).

Table 1.6 shows the statistics of the results of the simulation, while the statistics on the values attained by the compensators after the optimizations are shown in Table 1.7. Note that in this case compensation does not require using the entire M2M range of motion, but 9% of the range for the shift along the optical axis and 63% of the range for the the tilts.

## 1.3.2 Relaxed Acceptance Criteria Scenario

This paragraph presents an additional simulation scenario with an arbitrary set of relaxed requirements on exit pupil position and orientation and encircled energy (Table 1.8). The purpose is to assess dependency of the tolerances on the requirements.

The analysis was performed on M1 only, as results for this mirror already show that relaxing the requirements does not lead to a significant improvement in the range of admissible perturbations.

The analysis seems to point to a strong dependence of the results to the limits in the range of motion of the M2M compensator.

Table 1.9 and Table 1.10 present the results of the inverse sensitivity analysis on the Relaxed Requirements scenario and the comparison with the Base Scenario, respectively.

**Table 1.4:** Maximum and minimum values for the tolerance operands for the nominal acceptance scenario. Cells highlighted in red color are values at the extreme of the acceptable range, while cells in yellow are close to the limit.

| | | | Compensators | | | | |
|---|---|---|---|---|---|---|---|
| Mirror | Parameter | Range | M2 Focus (mm) | M2 Tilt-x (deg) | M2 Tilt-y (deg) | LoS Tilt-x (deg) | LoS Tilt-y (deg) |
| M1 | TILT X | 0.0051 | 0.037 | 0.109 | 0.000 | 0.006 | 0.000 |
| | (range in deg) | -0.0052 | -0.036 | -0.109 | 0.000 | -0.005 | 0.000 |
| | TILT Y | 0.0053 | 0.000 | -0.001 | 0.115 | 0.000 | 0.006 |
| | (range in deg) | -0.0053 | 0.000 | -0.001 | -0.114 | 0.000 | -0.006 |
| | DEC X | 0.1030 | -0.002 | -0.006 | -0.115 | -0.001 | -0.011 |
| | (range in mm) | -0.1030 | 0.000 | -0.002 | 0.115 | 0.000 | 0.011 |
| | DEC Y | 0.0980 | 0.081 | 0.109 | 0.000 | 0.011 | 0.000 |
| | (range in mm) | -0.1010 | -0.082 | -0.109 | 0.000 | -0.011 | 0.000 |
| | MOVE Z | 0.3610 | 0.350 | -0.012 | 0.000 | -0.002 | 0.000 |
| | (range in mm) | -0.3630 | -0.350 | 0.021 | 0.000 | 0.003 | 0.000 |
| M2 | DEC X | 0.1030 | 0.000 | -0.003 | 0.115 | -0.001 | 0.011 |
| | (range in mm) | -0.1030 | 0.000 | -0.003 | -0.115 | 0.000 | -0.011 |
| | DEC Y | 0.1000 | -0.086 | -0.115 | 0.000 | -0.012 | 0.000 |
| | (range in mm) | -0.1030 | 0.086 | 0.115 | 0.001 | 0.011 | 0.000 |
| M3 | TILT X | 0.0069 | 0.009 | 0.022 | 0.000 | 0.004 | 0.000 |
| | (range in deg) | -0.0069 | -0.010 | -0.022 | 0.000 | -0.003 | 0.000 |
| | TILT Y | 0.0066 | 0.000 | 0.001 | 0.018 | 0.000 | 0.002 |
| | (range in deg) | -0.0066 | 0.000 | 0.000 | -0.018 | 0.000 | -0.002 |
| | DEC X | 0.0590 | 0.000 | 0.000 | -0.018 | 0.000 | -0.003 |
| | (range in mm) | -0.0580 | -0.001 | -0.003 | 0.018 | 0.000 | 0.003 |
| | DEC Y | 0.0610 | 0.008 | 0.021 | 0.007 | 0.003 | 0.001 |
| | (range in mm) | -0.0610 | -0.010 | -0.021 | -0.001 | -0.003 | 0.000 |
| | MOVE Z | 0.7600 | -0.001 | -0.019 | 0.000 | -0.003 | 0.000 |
| | (range in mm) | -0.7600 | 0.001 | 0.019 | 0.000 | 0.003 | 0.000 |
| M4 | TILT X | 0.0180 | -0.004 | -0.005 | 0.001 | -0.001 | 0.000 |
| | (range in deg) | -0.0200 | 0.008 | 0.017 | 0.000 | 0.003 | 0.000 |
| | TILT Y | 0.0290 | 0.000 | -0.003 | -0.014 | 0.000 | -0.002 |
| | (range in deg) | -0.0290 | 0.000 | 0.000 | 0.014 | 0.000 | 0.002 |
| | MOVE Z | 0.0480 | -0.009 | -0.023 | 0.000 | -0.003 | 0.000 |
| | (range in mm) | -0.0480 | 0.008 | 0.021 | -0.001 | 0.003 | 0.000 |

**Table 1.5:** Configuration of compensators that optimize each of the extremes of the reduced perturbation ranges of ±10 μm for linear shifts and ±10″ for rotations. M3 and M4 are not shown since their perturbations do not require reoptimization to comply with requirements.

| | | | Compensators | | | | |
|---|---|---|---|---|---|---|---|
| Mirror | Parameter | Range | M2 Focus (mm) | M2 Tilt-x (deg) | M2 Tilt-y (deg) | LoS Tilt-x (deg) | LoS Tilt-y (deg) |
| M1 | TILT X | 0.0028 | 0.025 | 0.070 | 0.000 | 0.004 | 0.000 |
| | (range in deg) | -0.0028 | -0.025 | -0.069 | 0.000 | -0.005 | 0.000 |
| | TILT Y | 0.0028 | 0.000 | 0.000 | 0.069 | 0.000 | 0.004 |
| | (range in deg) | -0.0028 | 0.000 | 0.000 | -0.069 | 0.000 | -0.004 |
| | DEC X | 0.0100 | 0.000 | 0.000 | -0.002 | 0.000 | 0.000 |
| | (range in mm) | -0.0100 | 0.000 | 0.000 | -0.001 | 0.000 | 0.000 |
| | DEC Y | 0.0100 | 0.001 | 0.001 | 0.002 | 0.000 | 0.000 |
| | (range in mm) | -0.0100 | -0.001 | -0.003 | 0.000 | 0.000 | 0.000 |
| | MOVE Z | 0.0100 | 0.010 | 0.000 | 0.000 | 0.000 | 0.000 |
| | (range in mm) | -0.0100 | -0.010 | 0.000 | 0.000 | 0.000 | 0.000 |
| M2 | DEC X | 0.0100 | 0.000 | 0.000 | 0.001 | 0.000 | 0.000 |
| | (range in mm) | -0.0100 | 0.000 | 0.000 | -0.001 | 0.000 | 0.000 |
| | DEC Y | 0.0100 | -0.020 | -0.020 | 0.000 | 0.000 | 0.000 |
| | (range in mm) | -0.0100 | 0.010 | 0.010 | 0.010 | 0.000 | 0.000 |

**Table 1.6:** Statistical description of Monte Carlo results (500 cases) with uniform parameters distribution. Absolute values of parameters are used for the statistics.

| | beam shift (mm) | | beam tilt (arcsec) | | EE lost |
|---|---|---|---|---|---|
| | $x$ | $y$ | $x$ | $y$ | |
| min | 0 | 0 | 0.1 | 0.4 | 4.15% |
| 50% | 0.021 | 0.023 | 45.0 | 45.0 | 4.18% |
| 75% | 0.034 | 0.039 | 45.0 | 45.0 | 4.20% |
| 90% | 0.048 | 0.051 | 45.0 | 45.0 | 4.23% |
| 95% | 0.055 | 0.057 | 45.0 | 45.0 | 4.66% |
| max | 0.061 | 0.065 | 45.3 | 46.4 | 6.54% |

**Table 1.7:** Statistics of compensators for the compliant cases from the Monte Carlo run. *EXP shift* is the shift along the optical axis of the paraxial exit pupil position. Tilts are in degrees, shifts in mm.

| | M2 | | | LoS | | EXP shift |
|---|---|---|---|---|---|---|
| | $z$ shift | $x$ tilt | $y$ tilt | $x$ tilt | $y$ tilt | |
| min | 0.0000 | 0.0002 | 0.0000 | -0.006 | 0.0000 | 0.0000 |
| 50% | 0.0127 | 0.0350 | 0.0372 | 0.0001 | 0.0021 | 0.0070 |
| 75% | 0.0205 | 0.0518 | 0.0523 | 0.0021 | 0.0031 | 0.0118 |
| 80% | 0.0226 | 0.0558 | 0.0550 | 0.0025 | 0.0034 | 0.0131 |
| 90% | 0.0270 | 0.0656 | 0.0647 | 0.0035 | 0.0041 | 0.0161 |
| 95% | 0.0318 | 0.0714 | 0.0719 | 0.0041 | 0.0047 | 0.0186 |
| max | 0.0416 | 0.0816 | 0.0902 | 0.0057 | 0.0065 | 0.0253 |

**Table 1.8:** Beam position and orientation and Enclosed Energy for the relaxed acceptance criteria scenario.

| Parameter | Required range | Relaxed range |
|---|---|---|
| Lost EE | $\leq$5 % | $\leq$7 % |
| Beam Tilt x | $\pm 45''$ | $\pm 150''$ |
| Beam Tilt y | $\pm 45''$ | $\pm 150''$ |
| Beam Shift x | $\pm$60 μm | $\pm$100 μm |
| Beam Shift y | $\pm$60 μm | $\pm$100 μm |

**Table 1.9:** Inverse sensitivity analysis results on the relaxed requirements scenario.

| Mirror | Parameter | Range | Compensators | | | | |
|---|---|---|---|---|---|---|---|
| | | | M2 Focus (mm) | M2 Tilt-x (deg) | M2 Tilt-y (deg) | LoS Tilt-x (deg) | LoS Tilt-y (deg) |
| M1 | TILT X | 0.0058 | 0.039 | 0.115 | 0.003 | 0.006 | 0.001 |
| | (range in deg) | -0.0059 | -0.039 | -0.115 | 0.001 | -0.005 | 0.001 |
| | TILT Y | 0.0058 | 0.002 | 0.003 | 0.115 | 0.001 | 0.005 |
| | (range in deg) | -0.0058 | 0.000 | -0.002 | -0.115 | 0.000 | -0.005 |
| | DEC X | 0.1110 | 0.000 | -0.002 | -0.115 | 0.000 | -0.011 |
| | (range in mm) | -0.1110 | 0.000 | -0.002 | 0.115 | -0.001 | 0.011 |
| | DEC Y | 0.1120 | 0.089 | 0.115 | -0.003 | 0.011 | -0.001 |
| | (range in mm) | -0.1150 | -0.089 | -0.115 | 0.000 | -0.011 | 0.000 |
| | MOVE Z | 0.3690 | 0.350 | -0.035 | -0.002 | -0.006 | -0.001 |
| | (range in mm) | -0.3690 | -0.350 | 0.035 | -0.002 | 0.006 | -0.001 |

**Table 1.10:** Comparison of inverse sensitivity results for the Base and Relaxed Requirements scenarios.

| Mirror | Parameter | Base Scenario | Relaxed Scenario |
|---|---|---|---|
| M1 | TILT X (deg) | 0.0051 | 0.0058 |
| | | -0.0052 | -0.0059 |
| | TILT Y (deg) | 0.0053 | 0.0058 |
| | | -0.0053 | -0.0058 |
| | DEC X (mm) | 0.1030 | 0.1110 |
| | | -0.1030 | -0.1110 |
| | DEC Y (mm) | 0.0980 | 0.1120 |
| | | -0.1010 | -0.1150 |
| | MOVE Z (mm) | 0.3610 | 0.3690 |
| | | -0.3630 | -0.3690 |

## 1.4 Conclusions and Next Steps

The analyses presented in this paper serve as a starting point to understand the range of misalignments that are recoverable in flight by the Ariel telescope, in terms of the requirements on alignment and optical performance of the exit beam, and the effects of the range of misalignments that are considered achievable by the manufacturer.

The two inverse sensitivity analyses of Sections 1.3.1 and 1.3.2 show that the telescope is very sensitive to small misalignments, especially tilts of M1: a tilt of 21″ around the $x$ or $y$ axis already produces a configuration that cannot be recovered even with the relaxed requirements on the exit pupil position and orientation.

Monte Carlo simulations show that combining all perturbations from the maximum perturbation ranges leads to a large percentage of non-recoverable cases.

Restricting the perturbation ranges to much smaller values that are deemed as realistic by the telescope manufacturing Prime Contractor, leads however to the vast majority of cases to be recoverable with the M2M using at most 63% of the range of motion for tilts (and a much smaller percentage for focus).

These analyses will need to be integrated with the results from STOP Analysis and with the expected margins of the on-ground alignment plan.

## Acknowledgments

This activity has been realized under the Implementation Agreement n. 2021-5-HH.0 of the Italian Space Agency (ASI) and the National Institute for Astrophysics (INAF) Framework Agreement "Italian Participation to Ariel mission phase B2/C".

## References

[1] V. Da Deppo et al. "The Optical Configuration of the Telescope for the ARIEL ESA Mission". In: *Proc. SPIE 10698, Space Telescopes and Instrumentation 2018: Optical, Infrared, and Millimeter Wave*. Aug. 21, 2018, 106984O. DOI: 10.1117/12.2313412.

[2] E. Diolaiti et al. *Telescope Assembly Zemax OpticStudio® Tools for Statistical Analysis of Optical Tolerances*. Ariel Payload Consortium Phase B2 Study ARIEL-INAF-PL-TN-015 Issue 1.0. 2021.

[3] *OpticStudio User Manual*. Zemax LLC, 2022.

[4] F. E. Sahin. "Open-Source Optimization Algorithms for Optical Design". In: *Optik* 178 (Feb. 2019), pp. 1016–1022. DOI: 10.1016/j.ijleo.2018.10.073.

[5] G. Tinetti et al. "A Chemical Survey of Exoplanets with ARIEL". In: *Experimental Astronomy* 46.1 (Nov. 2018), pp. 135–209. DOI: 10.1007/s10686-018-9598-x.

# 2

# Qualification of the thermal stabilization, polishing and coating procedures for the aluminum telescope mirrors of the Ariel mission

Paolo Chioetto[1 2a,b,c], Paola Zuppella[2a,c], Vania Da Deppo[a,c], Emanuele Pace[d], Gianluca Morgante[e], Luca Terenzi[e], Daniele Brienza[f], Nadia Missaglia[g], Giovanni Bianucci[g], Sebastiano Spinelli[g], Elisa Guerriero[j,g,i], Massimiliano Rossi[g], Catherine Grèzes-Besset[h], Colin Bondet[h], Grégory Chauveau[h], Caroline Porta[h], Giuseppe Malaguti[e], Giuseppina Micela[i], and the Ariel Team

a CNR–Istituto di Fotonica e Nanotecnologie di Padova, Via Trasea 7, 35131 Padova, Italy
b Centro di Ateneo di Studi e Attività Spaziali "Giuseppe Colombo"- CISAS, Via Venezia 15, 35131 Padova, Italy
c INAF–Osservatorio Astronomico di Padova, Vicolo dell'Osservatorio 5, 35122 Padova, Italy
d Dipartimento di Fisica ed Astronomia–Università degli Studi di Firenze, Largo E. Fermi 2, 50125 Firenze, Italy
e INAF–Osservatorio di Astrofisica e Scienza dello spazio di Bologna, Via Piero Gobetti 93/3, 40129 Bologna, Italy
f INAF–Istituto di Astrofisica e Planetologia Spaziali, Via Fosso del Cavaliere 100, 00133 Roma, Italy
g Media Lario S.r.l., Località Pascolo, 23842 Bosisio Parini (Lecco), Italy
h CILAS-ArianeGroup, Etablissement de Marseille, 600 avenue de la Roche Fourcade, Pôle ALPHA Sud–Z.I. Saint Mitre, 13400 Aubagne, France
i INAF–Osservatorio Astronomico di Palermo, Piazza del Parlamento 1, 90134 Palermo, Italy

[1] Corresponding author.
[2] The authors contributed equally to this work.

j Dipartimento di Fisica e Chimica–Università degli Studi di Palermo, Via Archirafi, 90128 Palermo, Italy

## Abstract

Ariel, the Atmospheric Remote-sensing Infrared Exoplanet Large-survey, was selected as the fourth medium-class mission in ESA's Cosmic Vision program. Ariel is based on a 1 m class telescope optimized for spectroscopy in the waveband between 1.95 and 7.8 micron and operating in cryogenic conditions.

Fabrication of the 1.1 m aluminum primary mirror for the Ariel telescope requires technological advances in the three areas of substrate thermal stabilization, optical surface polishing and coating. This article describes the qualification of the three procedures that have been set up and tested to demonstrate the readiness level of the technological processes employed.

Substrate thermal stabilization is required to avoid deformations of the optical surface during cool down of the telescope to the operating temperature below 50 K. Purpose of the process is to release internal stress in the substrate that can cause such shape deformations.

Polishing of large aluminum surfaces to optical quality is notoriously difficult due to softness of the material, and required setup and test of a specific polishing recipe capable of reducing residual surface shape errors while maintaining surface roughness below 10 nm RMS.

Finally, optical coating with protected silver must be qualified for environmental stability, particularly at cryogenic temperatures, and uniformity.

All processes described in this article have been applied to aluminum samples of up to 150 mm of diameter, leading the way to the planned final test on a full size demonstrator of the Ariel primary mirror.

## Declarations

### Funding

This activity has been realized under the Italian Space Agency (ASI) contract with the National Institute for Astrophysics (INAF) n. 2018-22-HH.0, and is partly funded under the ESA contract with Centre Spatial de Liège, Belgium (CSL) and INAF n. 4000126124/18/NL/BW.

### Conflicts of interest/Competing interests

The authors declare no conflict of interest.

### Data Availability

The data that support the findings of this study are available from the corresponding author, but restrictions apply to the availability of these data, which were used under license for the current study, and so are not publicly available.

Data are however available from the authors upon reasonable request and with permission of the companies and institutions involved in the study.

## 2.1 Introduction

The main design drivers of the telescope for the Ariel mission are a collecting area of at least 0.6 $m^2$, diffraction limited performance at a wavelength 3 µm on a Field of View of 30″, and an average throughput of 96 % in the operating waveband of 0.5 µm to 8 µm [5, 12]. The telescope will operate at a temperature below 50 K.

To guarantee the desired throughput while keeping the size of the primary mirror small, an unobscured Cassegrain design was chosen, leading to an off-axis parabolic primary mirror featuring an elliptical aperture with major and minor axes measuring 1100 mm and 768 mm respectively.

In terms of optical performance, the diffraction limit requirement imposes the total wavefront error at the telescope exit pupil to be below 200 nm RMS, of which 160 nm RMS have been assigned as the primary mirror tolerance budget.

Aluminum alloy 6061, in the T651 temper, was chosen as construction material for the mirrors substrates and most of the supporting structures of the telescope, based primarily on JWST MIRI heritage [9] and on consideration of manufacturability and cost [6].

Aluminum mirrors of such large size, operating at cryogenic temperatures in space, are however relatively untested, and present specific manufacturing challenges, related in particular to opto-mechanical stability of the substrate at cryogenic temperatures and polishing of the optical surface.

Aluminum is also prone to oxidation and its spectral reflectivity in the visible band does not lead to the required throughput, so a suitable protected coating with space heritage has been identified.

In order to demonstrate the viability of the manufacturing procedures selected for the primary mirror of Ariel, a specific qualification campaign on substrate thermal stabilization, optical surface polishing and coating was conducted on samples of Al 6061-T651. Results of the campaign will then be translated to a full size prototype of the primary mirror (PTM) that will demonstrate the readiness level of the technologies employed.

This paper reports the results of the qualification campaign, starting from a description of the samples used in the qualification, and then presenting the three procedures under test.

### 2.1.1 Samples description

Aluminum samples of three different sizes have been used in the qualification activities: disks of 150 mm of diameter and 19 mm of thickness have been employed for machining, polishing and thermal stabilization tests; disks of 50 mm of diameter and 10 mm of thickness have been used for setup of the polishing process and finally disks of 25 mm of diameter, 6 mm of thickness, have been used for the coating qualification. All sample types have been cut from the same plate of rolled Al6061-T651 used for the PTM.

Additional glass samples (NBK-7), 25 mm of diameter by 4 mm of thickness, have been used for profilometry of the coating depth.

The 150 mm disks, four in total, have been used to qualify the thermal stabilization and polishing procedures. They have been machined flat before initiating the thermal cycles of the thermal stabilization procedure described in 2.2, and later underwent a series of polishing runs on a Lamplan M8400 flat lapping machine.

The first disk to be processed, identified as "LTU-1", had been manufactured during Phase A of the Ariel mission in 2017. It was then used to set up the thermal stabilization procedure and test it for the first time. The three remaining disks have been used to confirm the results and qualify the procedure since the outcome on LTU-1 was deemed satisfactory.

All aluminum samples and the PTM have been procured by MediaLario[3].

## 2.2 Thermal Stabilization

Dimensional stability is one of the main issues of aluminum as substrate material for cryogenic mirrors, especially when a large aperture is required, as is the case of Ariel.

The purpose of a thermal stabilization procedure is to minimize residual substrate stresses that may be released during flight and final cool down of the telescope, causing an unpredictable variation of the shape of the optical surface of the mirrors that can ultimately affect optical performance.

The following paragraphs describe the sources of dimensional instability, identifying residual stress as the most prominent and actionable one, the stress release process that has been identified, adapted and tested, and the steps taken to verify compliance of the procedure with the program goals.

### 2.2.1 Sources of Dimensional Instability

There are four major types of dimensional instabilities that can affect optical performance of mirrors: temporal instability, thermal/mechanical hysteresis, thermal instability and other instabilities [1].

The first two are irreversible dimensional changes caused by the simple passage of time, or from changing mechanical or thermal environmental conditions. They are both caused by relaxation of residual stresses present in the material, and we expect them to be the main source of instability in aluminum mirrors.

[3]Media Lario S.r.l., Via al Pascolo, 23842 Bosisio Parini (LC), Italy

The thermal stabilization procedure described here was identified and set up to tackle these two sources of instability.

"Thermal instability" refers to changes that are independent of the environmental change path, such as an intrinsic inhomogeneity in the coefficient of thermal expansion in the material. This type of instability cannot be improved by stress release processes.

Finally, there are other sources of instability that are specific to the environmental change path, such as the rate of temperature change, but have been rarely observed in metals, so they have not specifically targeted by the risk mitigation effort.

### 2.2.2 Thermal Stabilization Procedure

T651 temper specifications already include a thermal/mechanical hardening and stabilization procedure performed by the aluminum plate supplier, and consisting of a sequence of "solution heat-treating", "artificially ageing" and "stress relieving by stretching" processes [2].

While this may be sufficient for less demanding applications, use as optical material requires further stress release cycles to minimize the possibility of surface shape variations [10].

Many stress release procedures are available in the literature. Based on a recommendation by the samples manufacturer, we decided to employ as baseline the one proposed by R. G. Ohl et al. from NASA/Goddard Space Flight Center and successfully applied to the *Infrared Multi-Object Spectrograph (IRMOS)* instrument [11].

The main reasons for choosing this procedure are the similarities between the two projects: same substrate material, observation wavelength in the IR and similar operating temperature (80 K).

The procedure, after adaptation according to the availability of cryotesting and manufacturing facilities, was first applied to LTU-1.

The test took place in 2019, and results were satisfactory, leading to a mirror that did not exhibit any change in optical surface shape, within the reproducibility error of the interferometer used, when measured at room temperature before and after the last verification thermal cycle. A detailed discussion and results of this test have been published elsewhere [4].

Based on the results, the procedure was then validated on the three additional 150 mm samples.

The procedure consists of a series of thermal cycles at high temperature followed by a second series of cold/hot thermal cycles (Table 2.1). Mirror machining and polishing phases are interspersed with the thermal cycles. A final thermal cycle serves as validation step to confirm that the mirror surface reached stability with the required optical shape given by the last polishing phase.

**Table 2.1:** Steps of the thermal stabilization procedure employed to minimize residual substrate stresses on mirrors.

| Stress Release Recipe Steps | |
|---|---|
| 1. | Thermal aging at 175 °C for 8 hours |
| 2. | Finish machining, leaving 1 mm of margin for SPDT/polishing |
| 3. | Age again at 175 °C for 8 hours |
| 4. | Perform three thermal cycles from −190 °C to 150 °C with rates not to exceed 1.7 °C/min |
| 5. | Repeat three thermal cycles as in Step 4 |
| 6. | Diamond turning/polishing |
| 7. | Repeat three thermal cycles as in Step 4 |
| 8. | Repeat three thermal cycles as in Step 4 (validation cycle) |

### 2.2.3 Verification Methods

To assess the effectivity of the procedure, the optical surface of the mirrors under test was measured at room temperature before and after each thermal cycle with either a Wyko 8600 Fizeau-type interferometer or a MPR 700 optical profilometer by MediaLario.

If the last thermal cycle of the procedure (validation cycle) had not produce any measurable difference in surface shape, within the reproducibility error of the instrument, the test was deemed as successful.

The comparison of two measurements at room temperature, instead of assessing surface form variation between room and operating temperature, was carried out under the assumption that any release of residual stress would produce a permanent variation in surface shape, and would therefore be detectable also after bringing the sample back at room temperature.

Final qualification of the procedure will then be completed on the prototype of the primary mirror (PTM) with the assessment of the optical surface shape variation between room temperature and operating temperature, validating the assumption stated above.

Surface roughness measurements have been performed with a Taylor Hobson CCI White Light Interferometer (WLI).

### 2.2.4 Results

The samples underwent the initial hot thermal cycles of the procedure alongside the PTM in a Nabertherm W 2200/A air circulation oven at TAG[4], and the cold/hot cycles at INAF-OAS Bologna[5], at their CryoWaves Lab for the cryogenic part of the cycle and in an Angelantoni CH250 climatic chamber for the hot part of the cycle. The validation cycle was performed in a thermal vacuum chamber at Criotec Impianti[6].

In the case of LTU-1, to bring the optical surface to its final shape, an actual fly-cutting process had been applied. For the other three samples, instead, to ease scheduling of the entire validation procedure, figuring was performed through heavy polishing on a LampPlan 8400 polishing machine at MediaLario, and final polishing was performed on a Zeeko IRP 1200 polishing machine, again at MediaLario.

The first measurements of surface error of the samples had been taken before and after the first cold/hot cycle with the MPR 700 optical profilometer (see Sironi et al. 2011, for a description of the instrument), since the low reflectivity of the unpolished surface was not measurable on the Wyko. Variation in SFE RMS was in the range 0.3–0.9 μm.

The following thermal cycle caused a smaller form variation in the range 0.2–0.6 μm, measured this time with the interferometer.

Final polishing was performed, as anticipated, on the Zeeko IRP1200 using an aggressive procedure in order to bring the form error to specification as much as possible to make form variation assessment more meaningful, while sacrificing surface roughness.

Results were satisfactory on the first two samples, achieving 101 nm RMS and 96 nm RMS of SFE respectively after several polishing runs. For the third sample, polishing had to stop before the required SFE was achieved since by simple visual assessment the optical quality of the surface was degrading so quickly that a serious concern on its interferometric measurability was raised. The sample was then left at 217 nm RMS SFE.

Surface roughness of the three samples, as measured on the Talysurf CCI at 10x and 50x magnification, was in the range 32–101 nm RMS.

Measurement of the samples using the Wyko interferometer proved anyhow problematic because of surface quality degradation. Eventually, surface shape variation was assessed by comparing a central circular region of 100 mm of diameter, and proved to be within the measurement error for the first two samples: 5 nm RMS and 1 nm RMS respectively, with an error of 10 nm (Figures 2.1 and 2.2). As a comparison, form variation for LTU1, after the last thermal cycle, was (6 ± 10) nm.

The third sample showed a larger variation: 26 nm RMS. This result, and the low reflectivity of the sample, prompted further evaluation of the quality of the interferometric measurement. Measurement error on this sample was then assessed to be in the order of 30 nm, higher than the measured difference itself. A summary of the measurements is presented in Table 2.2.

[4]TAG s.r.l., via Guglielmo Marconi 9, 23843 – Dolzago (LC) Italy

[5]INAF-Osservatorio di Astrofisica e Scienza dello spazio di Bologna, Via Piero Gobetti 93/3, 40129 Bologna, Italy

[6]Criotec Impianti SpA, Via Francesco Parigi 32/A, 10034 Chivasso (TO), Italy

**Table 2.2:** Summary of surface error measurements of the three samples before and after HT5 (validation step of the stress release procedure).

| | SFE RMS (nm) | | Difference (nm) |
|---|---|---|---|
| | before valid. cycle | after valid. cycle | |
| Sample 1 | 101 | 96 | 5 |
| Sample 2 | 77 | 76 | −1 |
| Sample 3 | 217 | 243 | 26 |

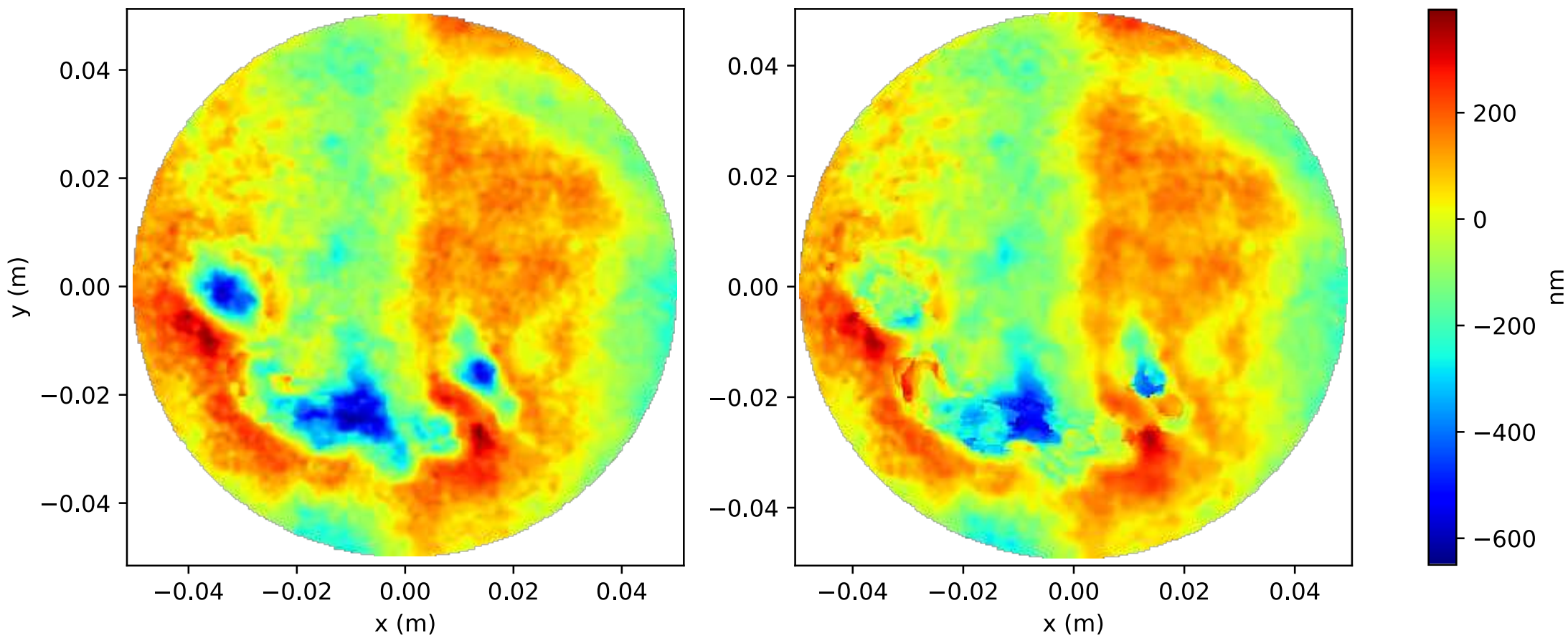

**Figure 2.1:** Comparison of surface error measurement of the first sample with the Wyko interferometer, before (left) and after (right) the validation thermal cycle.

## 2.3 Optical Surface Polishing

Optical surfaces are polished to reduce surface roughness, lowering scattering at the waveband of interest and improving reflectivity. Depending on the process, polishing can also achieve a sufficiently high material removal rate to affect surface shape and mitigate residual surface errors left from the previous machining steps.

Aluminum alloys are notoriously difficult to polish down to less than a few nanometers RMS of surface roughness, and for this reason they are used mostly for IR instruments that have less stringent requirements on surface finish [10]. For this reason, and based on the expertise and manufacturing capabilities of MediaLario, the requirement on surface roughness was set at 10 nm RMS for the qualification phase of the polishing procedure.

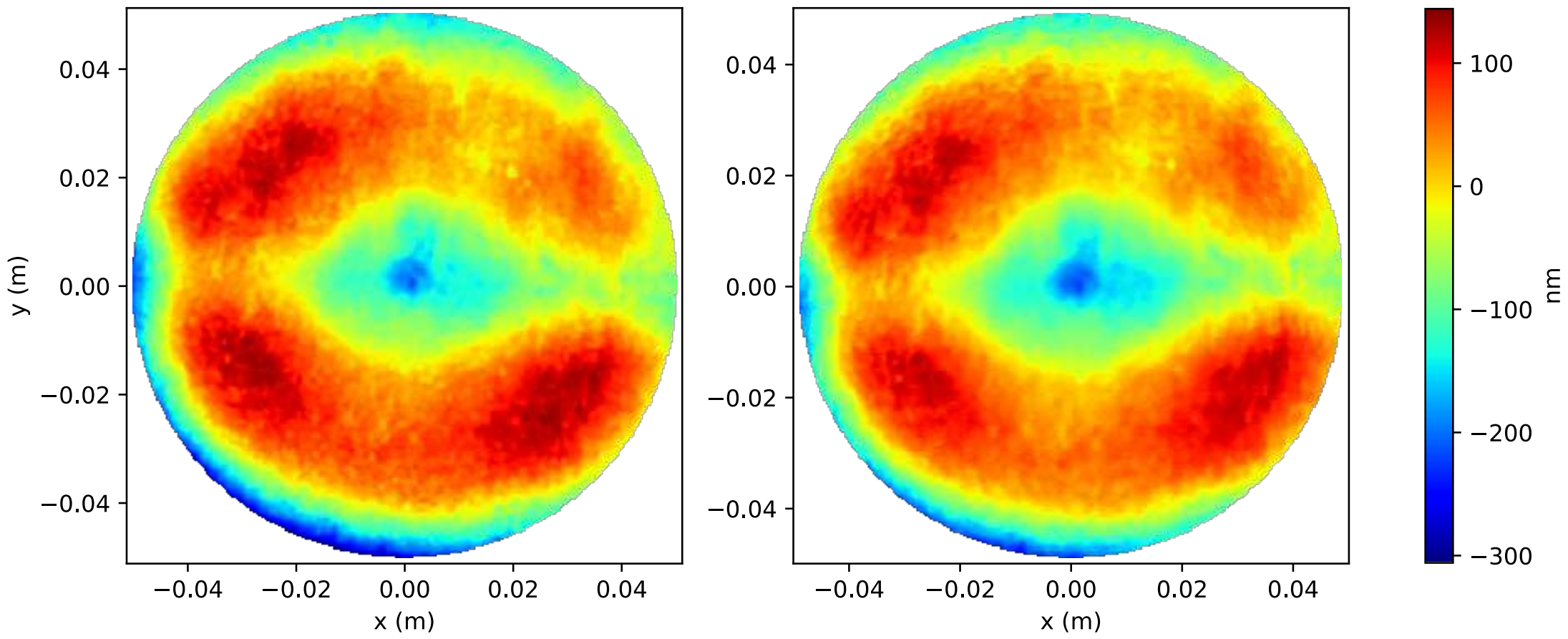

**Figure 2.2:** Comparison of surface error measurement of the second sample with the Wyko interferometer, before (left) and after (right) the validation thermal cycle.

Results of directly polishing aluminum alloys are heavily influenced by grain structure and orientation. In particular, polishing of Al6061-T651 in rolled plate form proved particularly difficult because of easy detachment of aggregates of heavy components (such as iron and magnesium) that produce micro-holes and scratches on the surface.

Set up and qualification of a suitable polishing procedure required several iterative steps. For comparison purposes, the tests involved also samples of Al6061-T651 in extruded form, and RSA6061-T6, a rapidly solidified Al alloy from RSP Technology.

The first step was carried out on a flat lapping machine, and consisted in screening existing polishing recipes and setting up a suitable combination of rotating speed, pressure, polishing slurry and pads. Once a promising combination was found, the second step had been transfer of the procedure on the robot polisher. The whole procedure was carried out by MediaLario.

### 2.3.1 Optical Surface Polishing Procedure

In order to set up the polishing procedure, several tests were carried out on a LamPlan M8400 flat lapping machine (Figure 2.3). The short machine setup time allowed for fast screening of different combinations of pads and slurries to determine the most promising one.

After a good candidate had been identified, the process was transferred to the Zeeko IRP 1200X deterministic polishing machine. The Zeeko consists of a rotating platform on which the piece to polish is secured, and a spindle mounted on a robotic arm (Figure 2.4). The spindle contacts the surface to be polished through an inflatable rubber head covered by a polishing cloth or pad. A lubricating liquid and/or an abrasive slurry are sprayed on the spindle while rotating [17].

The machine operates deterministically, using a surface map of the piece to be polished (usually an interferogram) to determine the optimal polishing path, spindle dwell time on the spots to be polished, rotation speed, angle and inflating pressure.

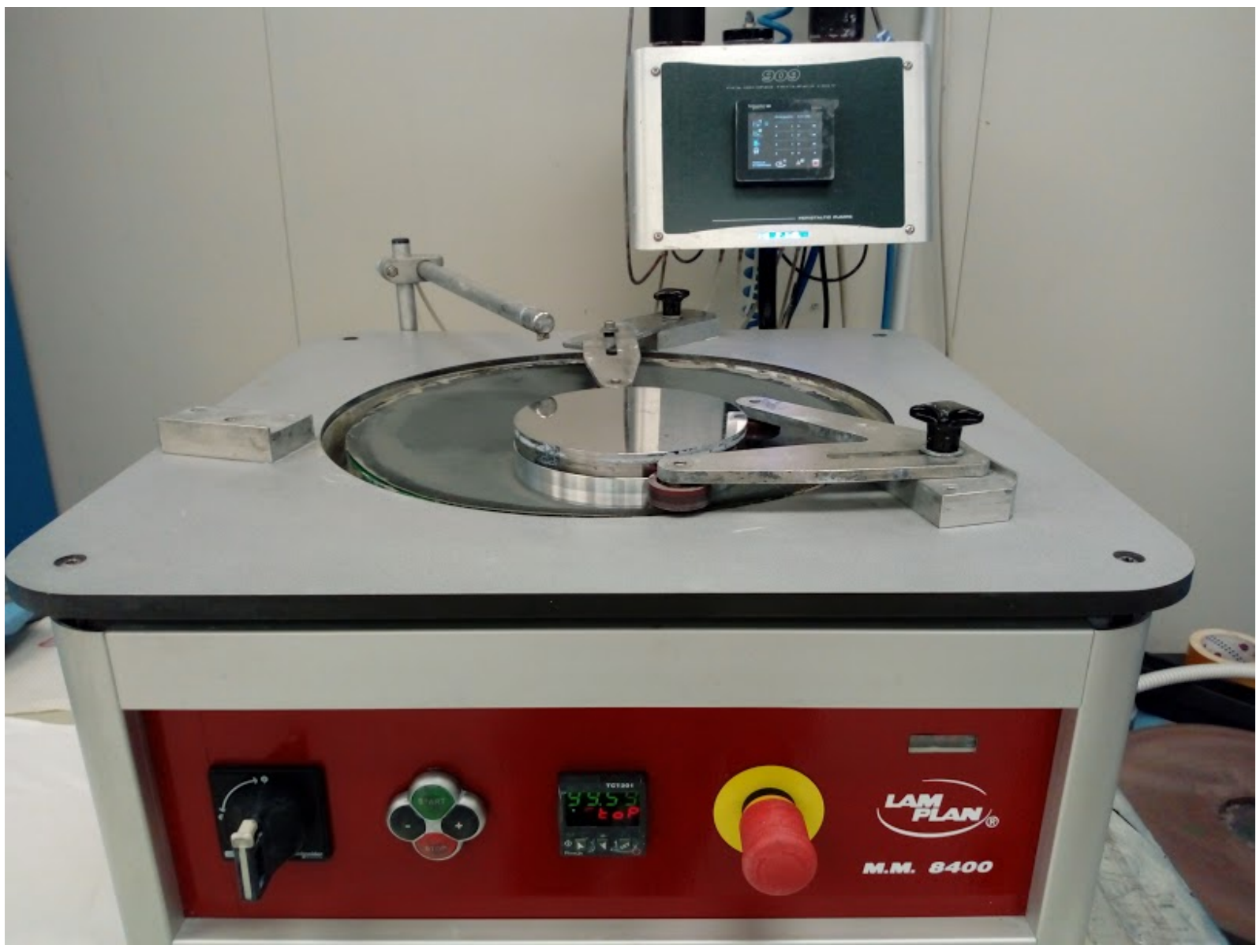

**Figure 2.3:** Picture of the LamPlanmm8400 flat lapping machine used by MediaLario to test different slurries and pads to set up the polishing procedure

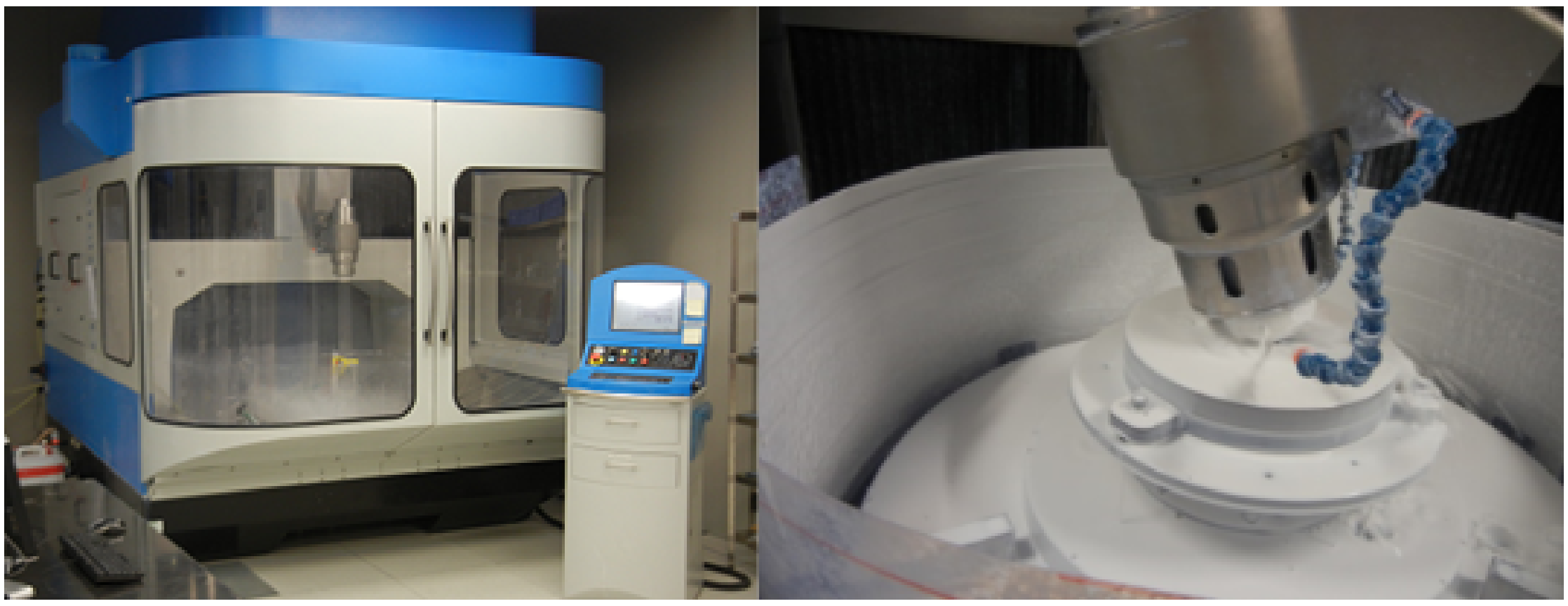

**Figure 2.4:** Pictures of the Zeeko IRP 1200X robotic polishing machine at MediaLario: ensemble view (left) and close up of the spindle while operating (right).

Small surface form errors can also be corrected in the process, up to a micron on NiP plated mirrors, according to MediaLario experience. Several polishing runs are usually required to achieve the desired shape and surface roughness.

### 2.3.2 Verification Methods

Surface roughness of the polished samples was measured using a Taylor Hobson Talysurf CCI optical profilometer and a Zygo White Light Interferometer (WLI). Measurement spots have been chosen to be representative of the whole surface.

Additional micrographs and SEM-EDX analyses have been performed on the samples during the tests to determine the nature of the imperfections and problematic morphological feature visible on the profilometric images.

In particular, 25 mm diameter samples have been encapsulated in epoxy adhesive and polished in order to perform further metallographic analyses.

Surface form error was measured with a Wyko 8600 Fizeau type interferometer.

### 2.3.3 Results

#### Reference results with aggressive polishing process

The first round of tests was performed on the LamPlanmm8400 flat lapping machine (see Section 2.3.1) on a 150 mm sample disk.

Purposes of these initial tests were to perform a baseline characterization of the results obtainable with the aggressive polishing process normally used by MediaLario on aluminum mirrors with an electroless nickel-phosphorous plating, to use as basis for assessing progress, and to study in details any issue than might had appeared.

Results were in fact unsatisfactory: visually, the surface of the sample appeared affected by opacity, and measurements on the Talysurf CCI confirmed the impression, showing that surface roughness was generally higher that the requirement of 10 nm RMS (Figure 2.5). In particular, the increased surface roughness seemed to be caused by localized defects, clearly visible at 50x magnification (Figure 2.5, center): the areas between defects has instead a roughness below 10 nm RMS (Figure 2.5, left).

Appearance of the defects is of holes on a generally even surface, leading to the hypothesis that the aggressive polishing process removed grains of material, probably aggregates of alloy solutes. Additional investigation efforts had then been made to characterize the localized defects, with the aim of guiding the polishing process development effort, using 25 mm aluminum samples.

Firstly, the 25 mm samples were measured with a Zygo WLI to confirm that surface morphology after polishing is equivalent to the one obtained on the 150 mm samples (Figure 2.6). The sample in 6061-T651 rolled plate clearly shows large hollowed structures, also found in the 150 mm samples.

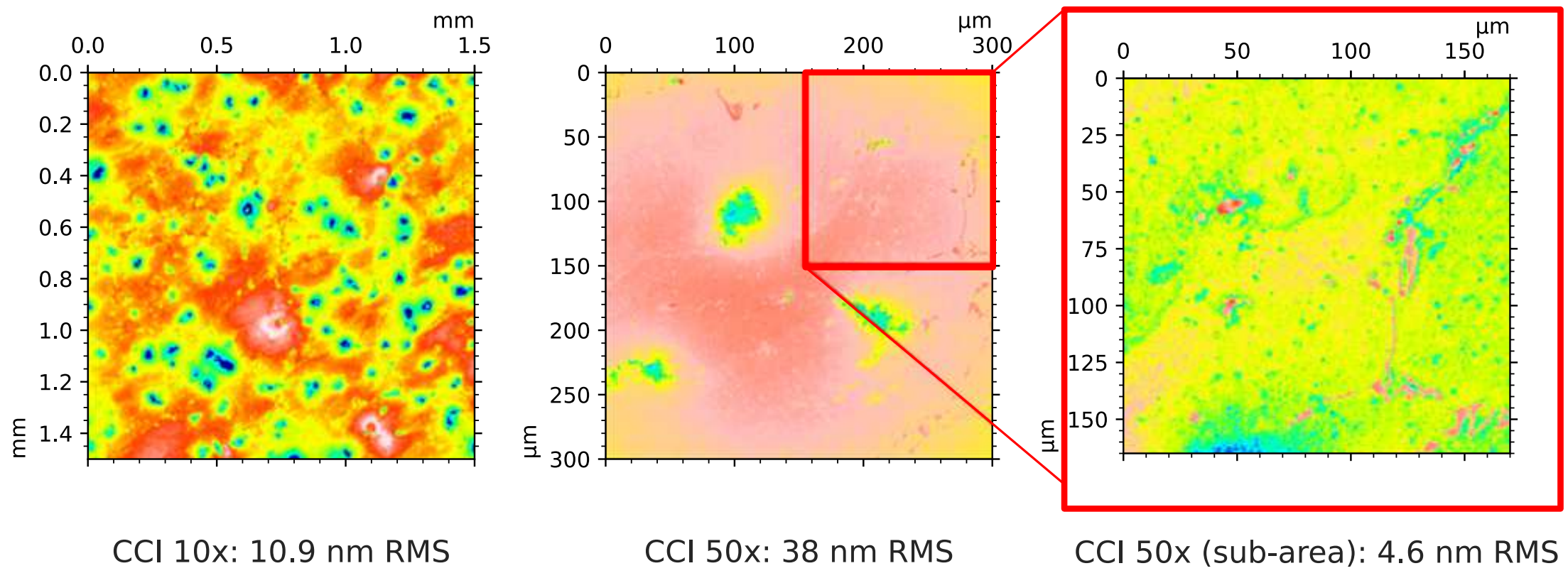

**Figure 2.5:** Representative roughness measurements of a 6061-T651 rolled plate 150 mm sample, with the aggressive polishing process, taken at different magnifications with the Talysurf CCI optical profilometer. The leftmost picture shows a specific area at 50x magnification with no significant surface defects.

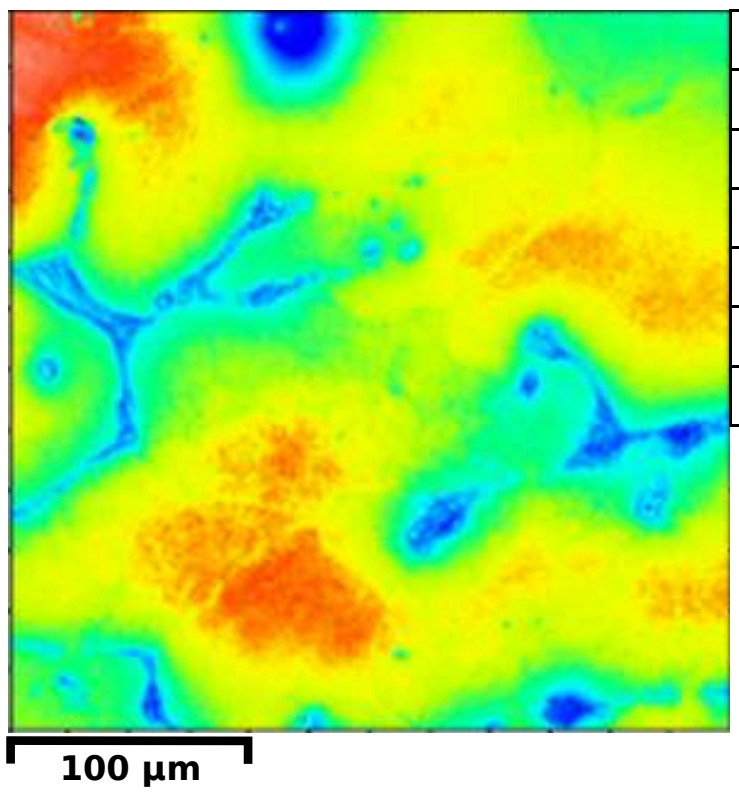

| **ISO 25178** | |
|---|---|
| **Height Parameters** | |
| **Sq** | 21.0 nm |
| **Sp** | 48.2 nm |
| **Sv** | 99.3 nm |
| **Sz** | 148.0 nm |
| **Sa** | 15.7 nm |

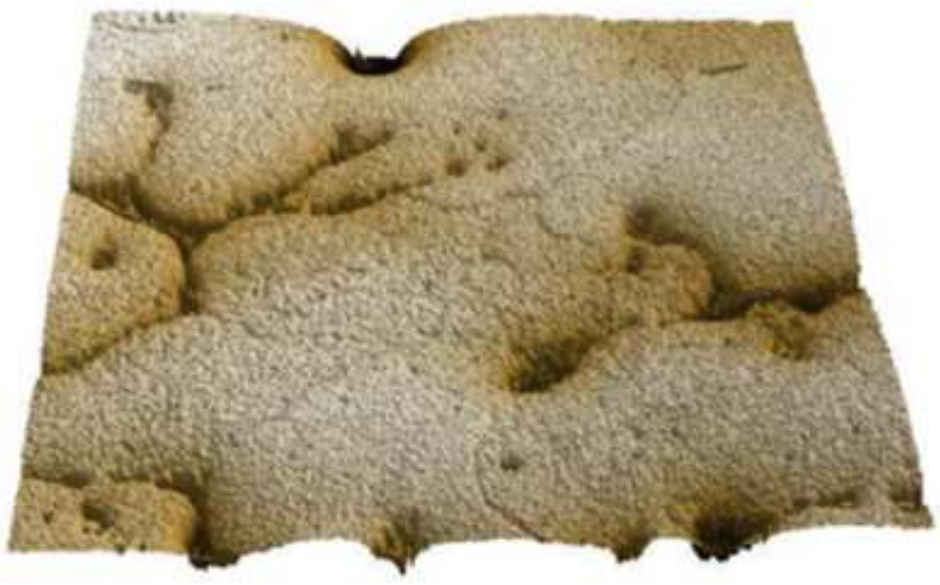

**Figure 2.6:** 25 mm Al sample after aggressive polishing measured with the Zygo WLI.

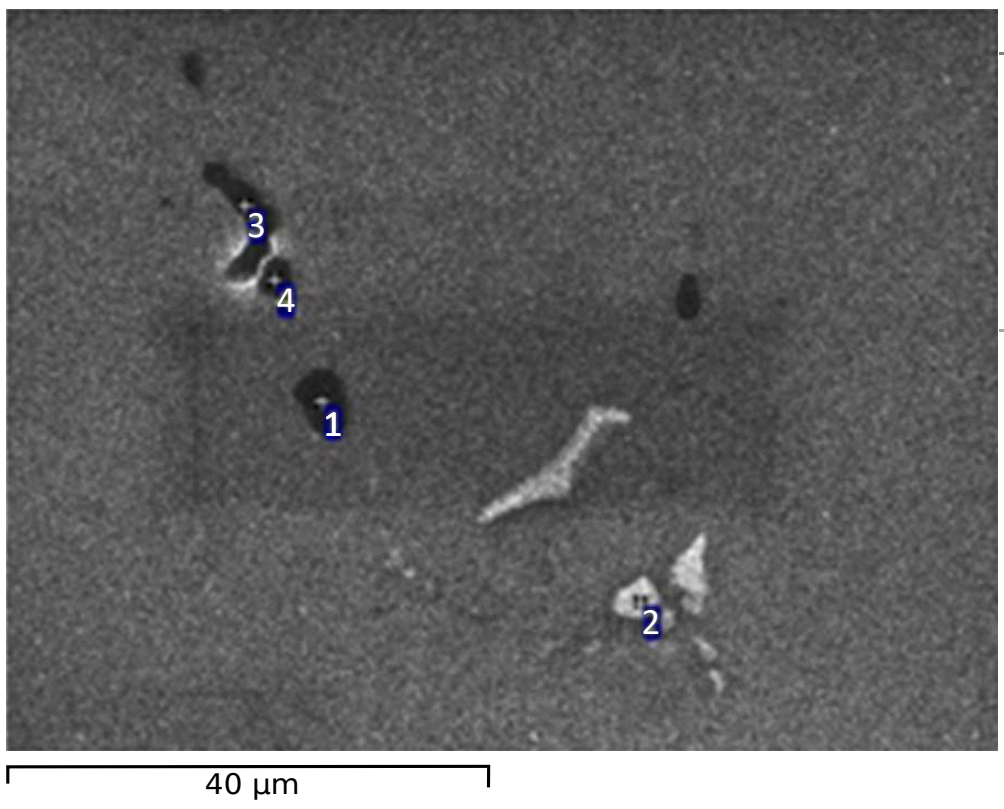

| *Spectrum* | O | Mg | Al | Si | Fe | Total |
|---|---|---|---|---|---|---|
| *1* | 15.15 | 19.27 | 43.36 | 22.22 | | 100.00 |
| *2* | | | 76.87 | | 23.13 | 100.00 |
| *3* | 14.10 | 18.92 | 48.32 | 18.66 | | 100.00 |
| *4* | 18.45 | | 66.68 | 14.87 | | 100.00 |

Processing option: all elements analyzed (normalized)
All results in weight%

**Figure 2.7:** SEM-EDX report of the analysis of one of the 25 mm sample in Al 6061-T651.

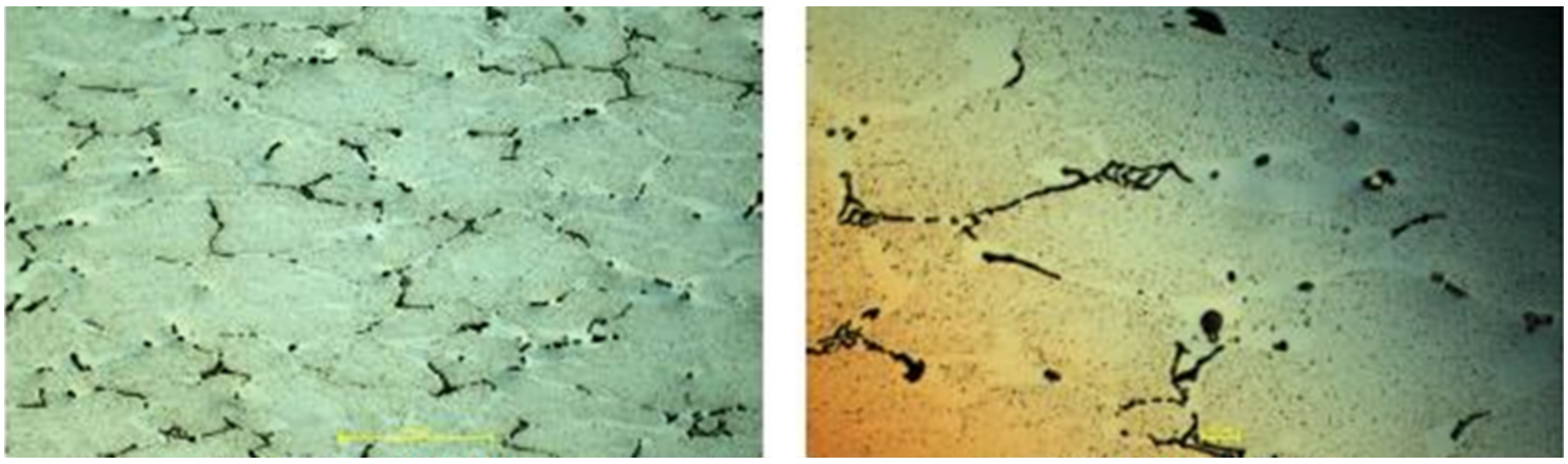

**Figure 2.8:** Optical micrograph of cross sections of a 25 mm sample subjected to aggressive polishing, at 200× magnification (left) and 500× magnification (right).

The samples were then analyzed with a Scanning Electron Microscope performing Energy-dispersive X-ray spectroscopy (SEM-EDX), as shown in Figure 2.7. The analysis identified two kinds of agglomerates: dark areas with generally high concentrations of Mg and Si, and light areas with high concentration of Fe.

Micrographs of the cross-section of the samples (Figure 2.8) were also taken, showing that the agglomerate structures are in fact uniformly distributed throughout the material.

### Final procedure development

The final polishing procedure was developed on the LamPlan, working on 50 mm flat samples of 6061-T651 rolled plate and using a different combination of polishing pad and slurry, and then transferred to the Zeeko IRP1200. The procedure consisted in three polishing phases, with each phase seeing a reduction in removal rate, lowering the pressure of the rotating head and speed at each step.

In the first phase, the 50 mm samples were first lapped to a flatness of approximately 1 µm RMS, while also removing machining marks. The second phase removed the deeper scratches

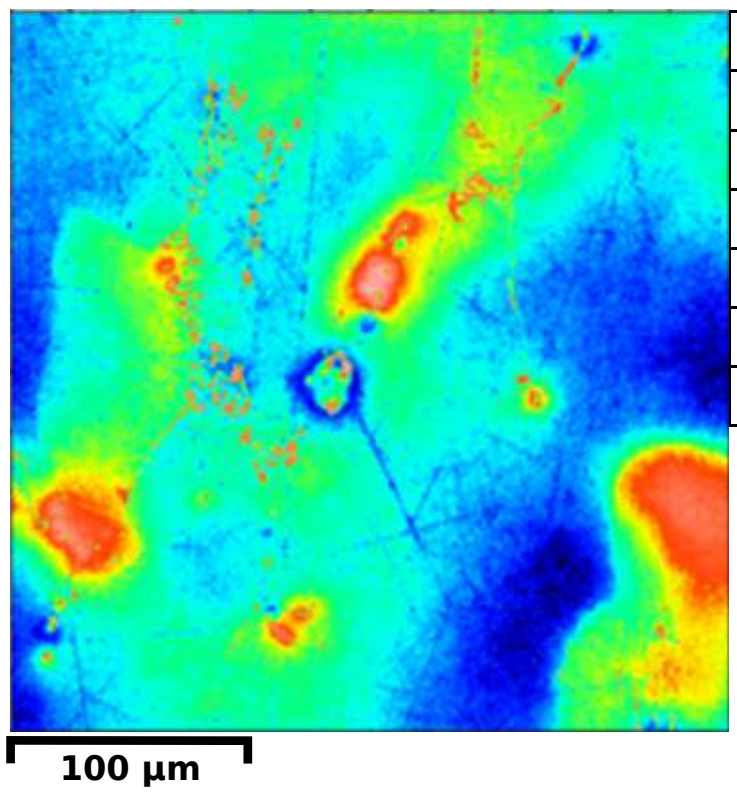

| ISO 25178 | |
|---|---|
| Height Parameters | |
| Sq | 3.2 nm |
| Sp | 16.7 nm |
| Sv | 5.5 nm |
| Sz | 22.2 nm |
| Sa | 2.3 nm |

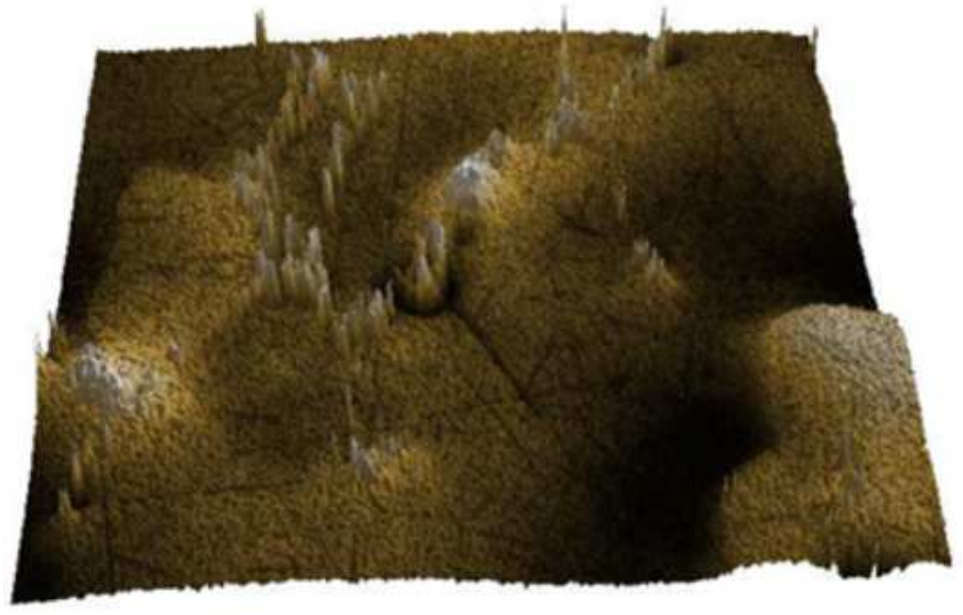

**Figure 2.9:** Surface roughness measurement on the Zygo WLI of a 50 mm diameter disk of Al6061-T651 rolled plate, polished on the LamPlan using the developed procedure

left from the first phase. These two phases used relatively standard tool parameters of rotational speed and pressure applied.

The third phase required more experimentation, as empirical observations highlighted a strong dependence of final roughness on pressure, but eventually results were satisfactory, with roughness in the range of 3–5 nm RMS. A sample measurement of the resulting surface is presented in Figure 2.9.

Adapting the identified procedure on the Zeeko required another phase of experimentation with different tool parameters, in particular spindle rotation speed and dwell time, using the same polishing slurry and pad materials identified on the LamPlan.

The final procedure was then tested on the LTU-1 aluminum sample, that had been previously machined flat on a fly-cutting tool. Two polishing runs on the Zeeko were then sufficient to bring the shape error from 220 RMS to 76 nm RMS, within specifications, and surface roughness to 12 nm RMS, with large areas within the requirement of 10 nm RMS (Figure 2.10).

Although the final results on LTU1 were not compliant with the specification of surface roughness, it was decided to proceed anyway with the final test on the PTM itself: the diffuse surface defects identified on the samples polished with the reference procedure were in fact mostly absent with the new process, the removal rate was enough to correct expected shape errors and the polishing runs could be kept within 24 hours of duration on the larger PTM surface, allowing reasonable process performance.

## 2.4 Optical Surface Coating

During Phase A, the Ariel Consortium decided to apply a protected silver coating to the aluminum mirrors of the telescope to protect them from oxidation and increase reflectivity. The choice of silver, as opposed to gold or aluminum was dictated by the throughput requirement

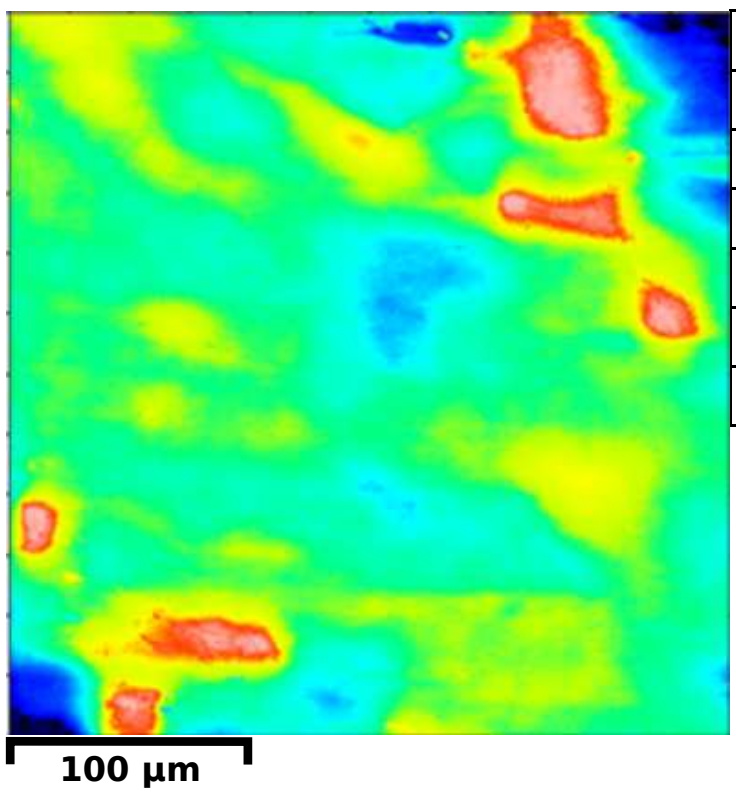

| ISO 25178 | |
|---|---|
| Height Parameters | |
| Sq | 12.4 nm |
| Sp | 54.8 nm |
| Sv | 32.7 nm |
| Sz | 87.5 nm |
| Sa | 8.29 nm |

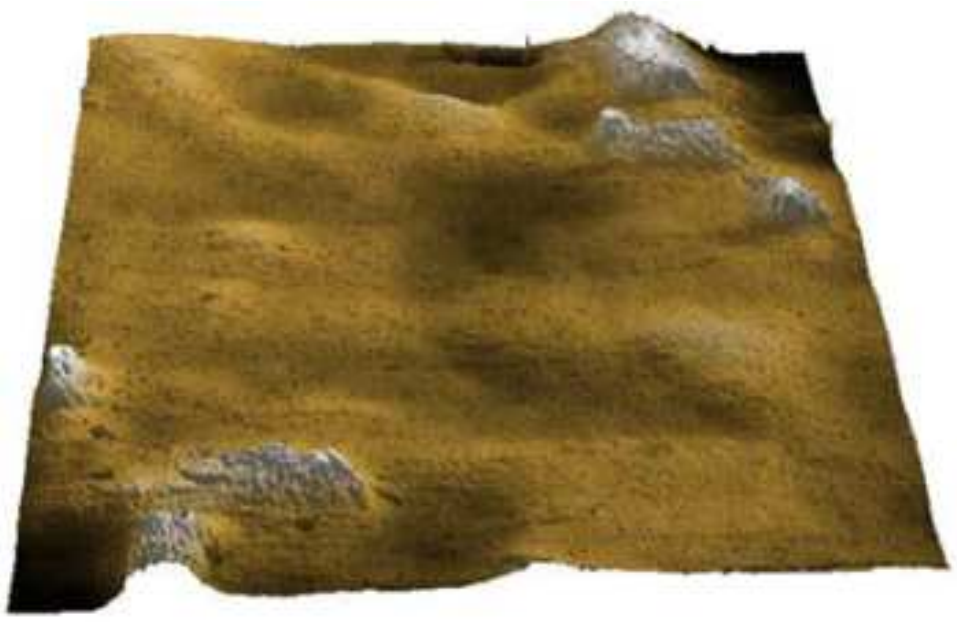

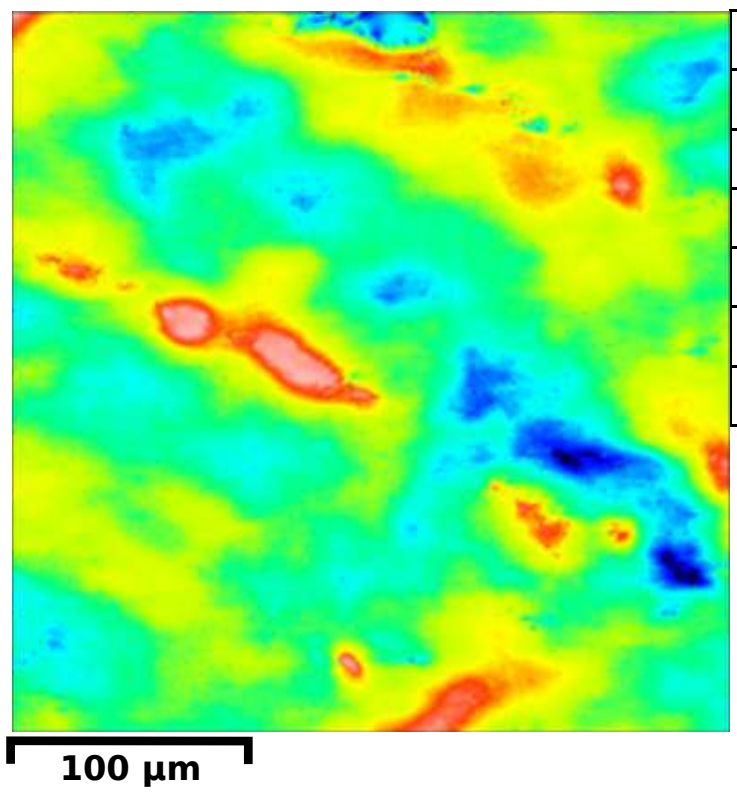

| ISO 25178 | |
|---|---|
| Height Parameters | |
| Sq | 8.36 nm |
| Sp | 36.1 nm |
| Sv | 19.7 nm |
| Sz | 55.8 nm |
| Sa | 6.5 nm |

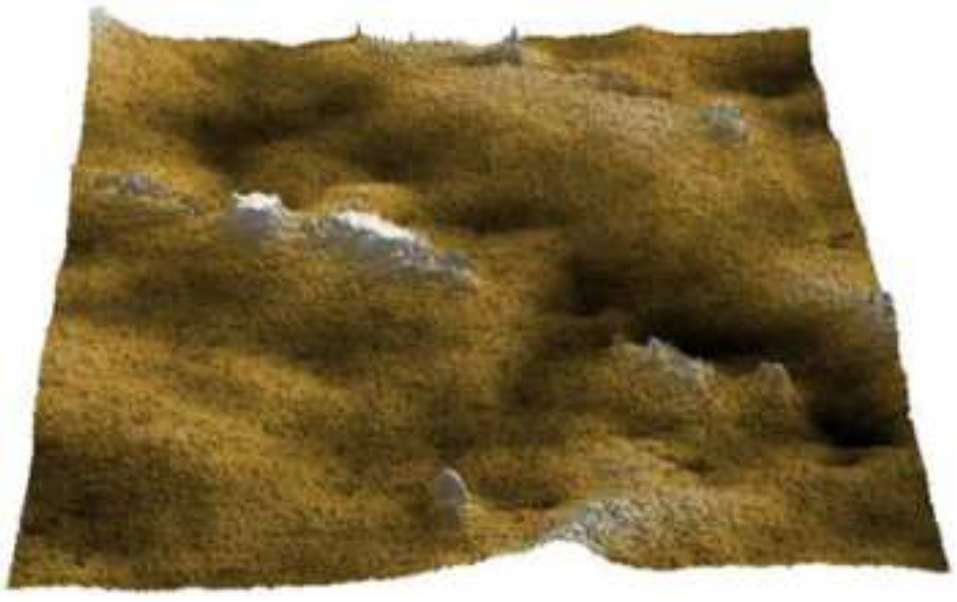

**Figure 2.10:** Representative surface roughness measurements on the Zygo WLI of the LTU1 Al sample, after applying the final procedure on the Zeeko polishing machine

at the operating waveband, with particular regard to the shorter wavelengths.

Protected silver on aluminum substrate has a strong heritage both for ground and space based telescopes, operating also at cryogenic temperatures [3, 14, 15].

A detailed qualification study was nonetheless required because of the large size of Ariel primary mirror, with particular emphasis on the possible issues caused by process deposition uniformity and CTE mismatch between the specific aluminum alloy used for the mirror substrate and silver.

Another area of concern was the environmental durability of silver, and the effectiveness of the protecting coating layer to avoid exposition and corrosion from common atmospheric pollutants [7, 16].

For the study, a protected silver coating with space heritage from CILAS[7] was selected and tested, mainly based on coating characteristics and capability and availability of the supplier of coating a 1.1 m diameter mirror within the timeframe of the mission.

The study, still ongoing at the time of writing, involves two main activities: a qualification of the coating on aluminum samples, and testing of the coating on the PTM itself. This paper describes the first activity.

### 2.4.1 Coating Process

CILAS coating process is based Physical Vapour Deposition. The coating platform consists in a large magnetron sputtering chamber capable of holding objects up to 2 m by 2 m of footprint, 0.4 m of thickness [8].

The tray holding the samples is able to move back and forth inside the chamber, allowing uniform deposition from the cathodes [13].

The protected silver coating consists of three layers: an adhesion layer in NiCr, less than 10 nm thick, the silver layer and a dielectric capping and protection layer. The actual coating composition and thickness is a trade secret. The total coating thickness is approximately 350 nm.

### 2.4.2 Verification Methods

A total of 30 samples of 25 mm of diameter were lined up for coating on the major and minor axis of an elliptically shaped sample holder with a curved surface modeled after the optical surface of the PTM (Figure 2.11), alternating 11 glass samples and 19 aluminum samples.

The 150 mm disks were instead coated lying flat on the coating tray outside of the sample holder. These were used exclusively to test stability of the coating after a series of cryogenic cycles.

Aluminum samples measured surface roughness was generally below 10 nm RMS, with two samples presenting the slightly higher values of 12.1 nm and 11.3 nm RMS. Surface rough-ness of glass samples had not been measured, but presumed to be less than that of aluminum samples.

[7] CILAS-ArianeGroup, 8 avenue Buffon, CS16319, 45063 Orleans CEDEX 2, France

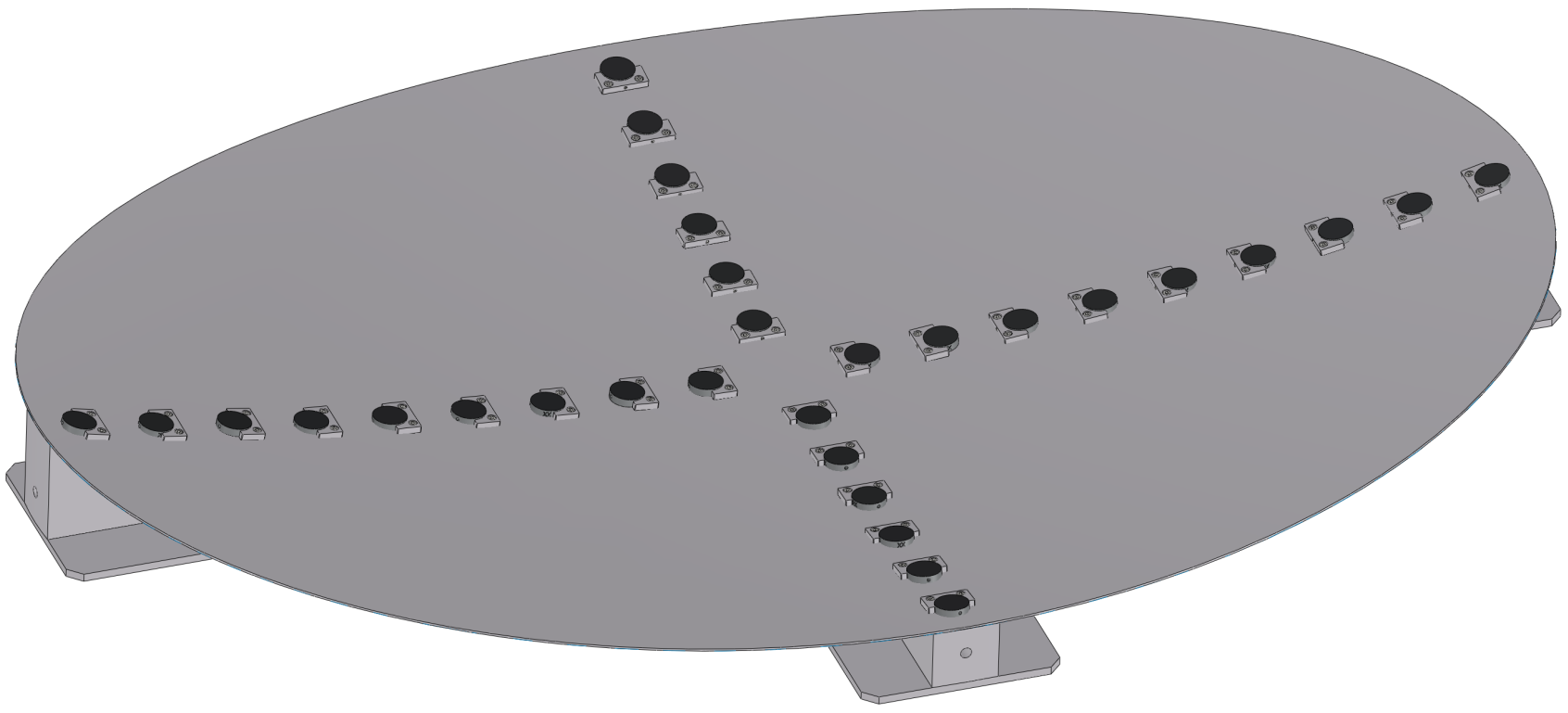

**Figure 2.11:** Drawing of the samples holder used for the coating deposition, shaped as the PTM optical surface.

After coating, the samples were subjected to a series of environmental (humidity and temperature cycling) and mechanical (adhesion, abrasion) tests to verify stability and performance. Spectral reflectivity, coating thickness and surface roughness have also been measured. A summary of the test specification and equipment used is presented in Table 2.3.

Visual inspections and relative reflectivity measurements were performed after each test step to identify possible degradation. Reflectivity measurements were limited by instrument availability to a waveband of 0.45–2.5 µm and incidence angles of 8 and 20 degrees, instead of covering the entire ARIEL operating range up to 8 µm and between 3 and 21 degrees AOI. Performance beyond 2.5 µm was positively assessed on a previous coating run, and repeating the measurement was not considered essential, as degradation is most likely to affect reflectivity at lower wavelengths; as for the AOI, results from reflectivity simulations indicated that the available setup would be sufficiently representative of the entire range.

Coating uniformity has also been assessed, both in terms of coating thickness and reflectivity at the lower end of the waveband of interest (500 nm). Glass samples were used for this measurement.

Further ageing tests have been planned to verify stability of the coating and effectiveness of the protection layer in normal storage conditions. These tests are however still ongoing and are not presented here.

### Test Success Criteria

Qualification tests were evaluated according to the following success criteria, applied to the aluminum samples only:

1. Reflectivity >90 % (goal >95 %) in the 0.5–2.5 µm waveband, best effort in the 0.45–

**Table 2.3:** Summary of test specifications and equipment used for coating qualification. All tests were performed at CILAS, unless otherwise indicated

| Test | Specifications | Testing Equipment |
|---|---|---|
| Relative Reflectivity | Wavelength range: 0.45–2.5 μm, AOI 8° and 20° | Perkin-Elmer Lambda 950, calibrated on results from absolute spectral measurements. |
| Visual Inspection | ISO 10110-7 | ISO compliant setup with 50 W halogen lamp |
| Adhesion | ISO 9211-4, Method 2 Severity 2 | ISO compliant cellophane tape |
| Humidity | ISO 9022-2<br>Method 12 Severity 06 but with a test duration of 24h:<br>90 % RH, 24 h, 55±3 °C (no condensation) | WEISS, WKL 64/70 climatic test chamber |
| Temperature cycling at ambient pressure | ISO 9022-2<br>T. range: -40 °C / +70 °C<br>T. change rate: 2°C/min<br>Dwell time: 15 min<br>Number of cycles: 30 | WEISS, WKL 64/70 climatic test chamber |
| Abrasion resistance | ISO 9211-4<br>Method 01 Severity 01 | 6 mm thick pad of clean, dry cheesecloth |
| Cryogenic cycling in vacuum | ECSS-Q-ST-70-04C<br>T. range: 54 K / 293 K<br>T. change rate: 5°C/min<br>Dwell time: 15 min<br>Vacuum: $<1 \times 10^{-4}$ mbar<br>Number of cycles: 10 | CryoWaves Lab at INAF-OAS Bologna |
| Coating thickness uniformity | Profilometry on float samples, coated with a mask to expose a ridge between coated surface and substrate | Alpha-Step® D-300 Stylus Profiler KLA Tencor |

0.50 μm waveband, angles of incidence 8 and 20 degrees.

2. No change in reflectivity, within the measurement reproducibility error, after the tests.

3. No visually detectable signs of degradation or delamination.

The specification on reflectivity is the result of a compromise between mission requirements and the expected performance of the coating, and is applicable to this phase of the qualification campaign only.

Uniformity of coating thickness and uniformity of reflectivity had not been included among the success criteria, but as further means to investigate possible failure, and to provide a baseline characterization of the coating.

### 2.4.3 Results

Spectral reflectivity was measured at 8 and 20 degrees of angle of incidence, for each sample, at wavelengths up to 2500 nm. The two measurements are identical within the instrument repeatability error, so the following considerations are valid regardless of the angle of incidence of the measurements.

Spectral reflectivity of the aluminum samples was generally above the requirement of 90 % for wavelengths greater than 500 nm, as illustrated in Figure 2.12, except for three of the outermost samples, positioned at the edge of three of the four "arms" of the sample holder of Figure 2.11. Variation in reflectivity at 500 nm is in the range 89.1–92.1 %.

Glass samples reflectivity is on average higher than aluminum samples by less than 1 %, probably due to better surface roughness. Reflectivity uniformity on the glass samples also follows the same trend of the aluminum samples, showing a variation in the range 88.6–93.6 % at 500 nm. The wider range may be explainable by the fact that three of the outermost positions in the holder are occupied by glass samples.

Coating thickness was determined on the glass samples by applying a mask during coating, and measuring the height of the ridge with a profilometer. Samples further away from the center of the holder showed up to 10 % higher thickness than the central ones, however no physical relation to reflectivity could be established, and the measurement will only be used for reference with further coating runs.

All mechanical and environmental tests were performed successfully, with no visible sign of delamination or degradation, nor impact on measured spectral reflectivity. A series of representative pictures of the visual appearance of the samples before and after coating and tests can be seen in Figure 2.13.

Results of the cryogenic cycles, especially on the larger 150 mm samples, are considered particularly important, since the specific combination of substrate and coating has not been tested at Ariel operating temperature of 50 K. The coating did not present any visible change in morphology nor changed its spectral reflectivity, leading to conclude that the exposure to high and cryogenic temperatures did no produce any short term degradation. Tape stripping tests were also repeated after the temperature cycles, showing that coating adhesion was not affected.

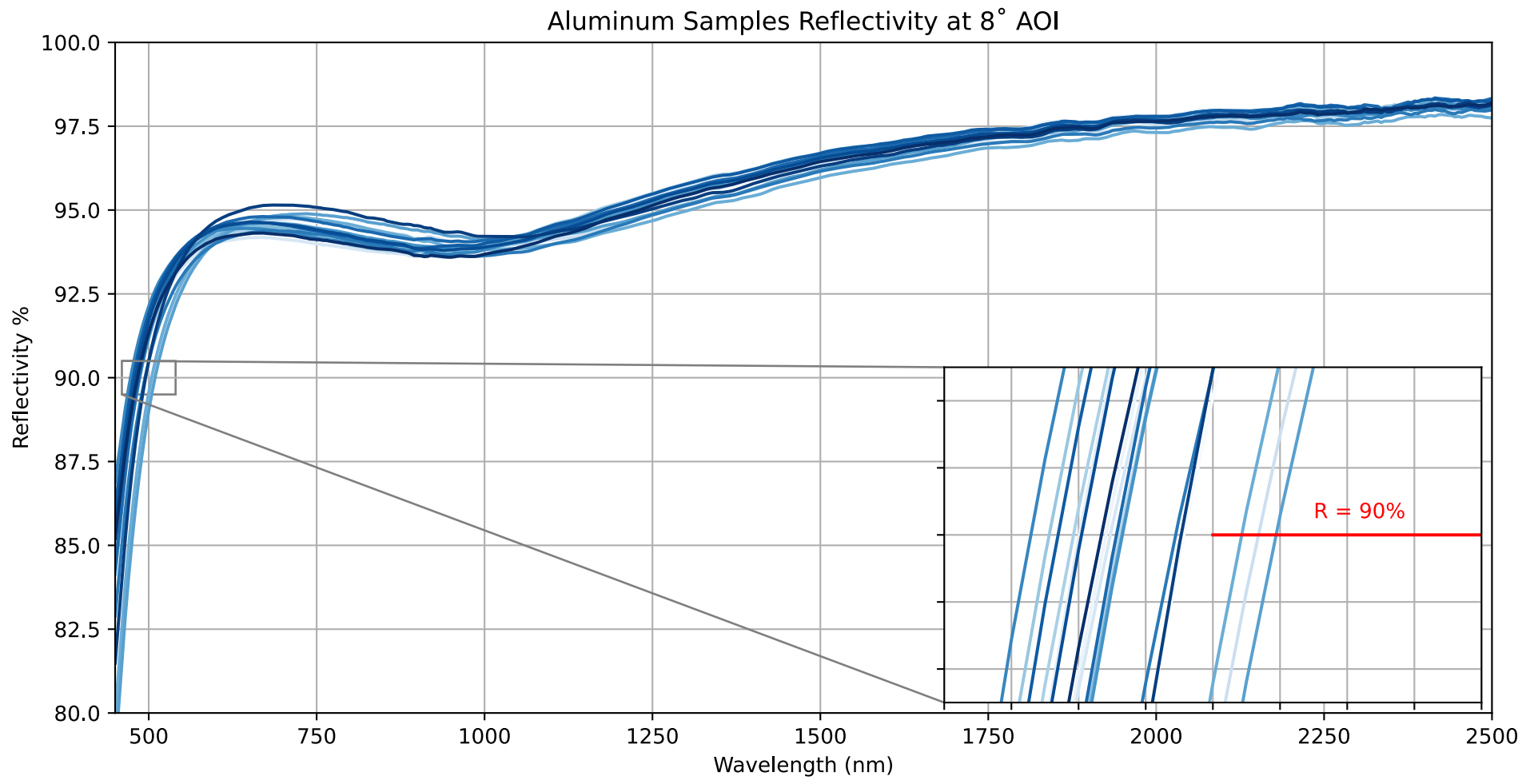

**Figure 2.12:** Reflectivity of all coated aluminum samples at 8 degrees angle of incidence, and comparison with the requirement of R >90 % at wavelengths >500 nm.

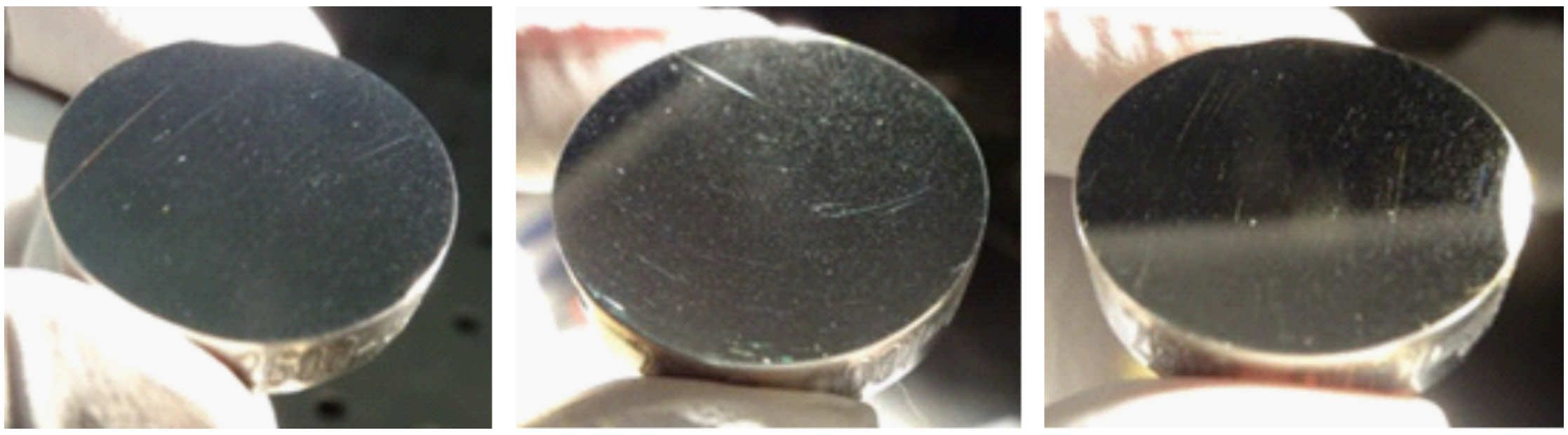

**Figure 2.13:** Representative pictures of one of the aluminum samples before coating (left), immediately after the coating (middle) and after the tests (right).

Finally, reflectivity measurements after the tests showed no variation within instrument accuracy wrt. the measurements taken before the tests.

## 2.5 Conclusions

The three core processes required to build Ariel telescope primary mirror, namely substrate thermal stabilization, optical surface polishing and coating were tested on samples of the same aluminum alloy foreseen for the mirror, with the purpose of assessing and improving the level of technological readiness.

Substrate thermal stabilization was successfully verified on two samples. A third sample was found not to be representative due to lower reflectivity affecting the measurement accuracy.

Polishing proved to be particularly difficult, requiring a very delicate and careful process, leading to longer execution times than expected, but eventually the procedure proved to be able to produce the desired results.

Finally, coating reflectivity, although at short wavelengths was slightly lower for the samples at the outer edges of the holder, was on average above specification and therefore compliant with the requirements. In view of the successful results, the team decided to proceed with the application of the coating to the full size demonstrator of the primary mirror (PTM).

## References

[1] A. Ahmad, ed. *Handbook of Optomechanical Engineering*. Boca Raton, Fla: CRC Press, 1997. 396 pp. ISBN: 978-0-8493-0133-9.

[2] ASM Handbook Committee, ed. *ASM Handbook, Volume 2 – Properties and Selection: Nonferrous Alloys and Special-Purpose Materials*. ASM International, Jan. 1, 1990. DOI: 10.31399/asm.hb.v02.9781627081627.

[3] M. Boccas et al. "Coating the 8-m Gemini Telescopes with Protected Silver". In: *Proc. SPIE 5494, Optical Fabrication, Metrology, and Material Advancements for Telescopes*. Sept. 24, 2004, p. 239. DOI: 10.1117/12.548809.

[4] P. Chioetto et al. "The Primary Mirror of the ARIEL Mission: Testing of a Modified Stress-Release Procedure for Al 6061 Cryogenic Opto-Mechanical Stability". In: *EPSC Abstracts*. Vol. 13. Sept. 15–20, 2019, p. 2. URL: https://meetingorganizer.copernicus.org/EPSC-DPS2019/EPSC-DPS2019-1625.pdf.

[5] V. Da Deppo et al. "An Afocal Telescope Configuration for the ESA ARIEL Mission". In: *CEAS Space Journal* 9.4 (Dec. 2017), pp. 379–398. DOI: 10.1007/s12567-017-0175-3.

[6] V. Da Deppo et al. *ARIEL Telescope Material Trade-Off*. Technical Paper ARIEL-INAF-PL-TN-004. 2017.

[7] K. A. Folgner et al. "Environmental Durability of Protected Silver Mirrors Prepared by Plasma Beam Sputtering". In: *Applied Optics* 56.4 (Feb. 1, 2017), p. C75. DOI: 10.1364/AO.56.000C75.

[8] C. Grèzes-Besset et al. "High Performance Silver Coating with PACA2M Magnetron Sputtering". In: *Proc. SPIE 11180, International Conference on Space Optics — ICSO 2018*. July 12, 2019, p. 1118083. DOI: 10.1117/12.2536210.

[9] G. Kroes et al. "MIRI-JWST Spectrometer Main Optics Opto-Mechanical Design and Prototyping". In: *Proc. SPIE 5877, Optomechanics 2005*. Aug. 18, 2005, 58770P. DOI: 10.1117/12.614784.

[10] T. Newswander et al. "Aluminum Alloy AA-6061 and RSA-6061 Heat Treatment for Large Mirror Applications". In: *Proc. SPIE 8837, Material Technologies and Applications to Optics, Structures, Components, and Sub-Systems, 883704*. Sept. 30, 2013, p. 883704. DOI: 10.1117/12.2024369.

[11] R. G. Ohl IV et al. "Comparison of Stress Relief Procedures for Cryogenic Aluminum Mirrors". In: *Proc. SPIE 4822, Cryogenic Optical Systems and Instruments IX*. Nov. 1, 2002, p. 51. DOI: 10.1117/12.451762.

[12] L. Puig et al. "The Phase A Study of the ESA M4 Mission Candidate ARIEL". In: *Experimental Astronomy* 46.1 (Nov. 2018), pp. 211–239. DOI: 10.1007/s10686-018-9604-3.

[13] I. Savin de Larclause et al. "PACA2m Magnetron Sputtering Silver Coating: A Solution for Very Big Mirror Dimensions". In: *Proc. SPIE 10563, International Conference on Space Optics — ICSO 2014*. Jan. 5, 2018, p. 1056308. DOI: 10.1117/12.2304237.

[14] M. Schürmann et al. "Manufacturing and Coating of Optical Components for the EnMAP Hyperspectral Imager". In: *Proc. SPIE 9912, Advances in Optical and Mechanical Technologies for Telescopes and Instrumentation II, 991230*. July 22, 2016. DOI: 10.1117/12.2232914.

[15] D. A. Sheikh. "Improved Silver Mirror Coating for Ground and Space-Based Astronomy". In: *Proc. SPIE 9912, Advances in Optical and Mechanical Technologies for Telescopes and Instrumentation II*. July 22, 2016, p. 991239. DOI: 10.1117/12.2234380.

[16] A. Sytchkova et al. "Optical Characterisation of Silver Mirrors Protected with Transparent Overcoats". In: *Proc. SPIE 10691, Advances in Optical Thin Films VI*. June 5, 2018, 106910N. DOI: 10.1117/12.2312077.

[17] D. D. Walker et al. "New Results from the Precessions Polishing Process Scaled to Larger Sizes". In: *Proc. SPIE 5494, Optical Fabrication, Metrology, and Material Advancements for Telescopes*. Sept. 24, 2004. DOI: 10.1117/12.553044.

# 3

# The primary mirror of the Ariel mission: cryotesting of aluminum mirror samples with protected silver coating

Paolo Chioetto[1 2a,b,c], Paola Zuppella[a,c], Vania Da Deppo[a,c], Emanuele Pace[d], Gianluca Morgante[e], Luca Terenzi[e], Daniele Brienza[f], Nadia Missaglia[g], Giovanni Bianucci[g], Sebastiano Spinelli[g], Elisa Guerriero[j,g,i], Massimiliano Rossi[g], Colin Bondet[h], Gregory Chauveau[h], Caroline Porta[h], Catherine Grezes-Besset[h], Giuseppe Malaguti[e], Giuseppina Micela[i], and the Ariel Team

a CNR-Istituto di Fotonica e Nanotecnologie di Padova, Via Trasea 7, 35131 Padova, Italy
b Centro di Ateneo di Studi e Attività Spaziali "Giuseppe Colombo"- CISAS, Via Venezia 15, 35131 Padova, Italy
c INAF-Osservatorio Astronomico di Padova, Vicolo dell'Osservatorio 5, 35122 Padova, Italy
d Dipartimento di Fisica ed Astronomia-Università degli Studi di Firenze, Largo E. Fermi 2, 50125 Firenze, Italy
e INAF-Osservatorio di Astrofisica e Scienza dello spazio di Bologna, Via Piero Gobetti 93/3, 40129 Bologna, Italy
f INAF-Istituto di Astrofisica e Planetologia Spaziali, Via Fosso del Cavaliere 100, 00133 Roma, Italy
g Media Lario S.r.l., Località Pascolo, 23842 Bosisio Parini (Lecco), Italy
h CILAS-ArianeGroup, Etablissement de Marseille, 600 avenue de la Roche Fourcade, Pôle ALPHA Sud - Z.I. Saint Mitre, 13400 Aubagne, France
i INAF-Osservatorio Astronomico di Palermo, Piazza del Parlamento 1, 90134 Palermo, Italy
j Dipartimento di Fisica e Chimica-Università degli Studi di Palermo, Via Archirafi 36, 90128 Palermo, Italy

[1] Presenting and contact author.
[2] The two authors contributed equally to the article.

## Abstract

Atmospheric Remote-Sensing Infrared Exoplanet Large Survey (Ariel) has been adopted as ESA "Cosmic Vision" M4 mission, with launch scheduled for 2029. Ariel is based on a 1 mclass telescope optimized for spectroscopy in the waveband between 1.95 and 7.8 µm, operating in cryogenic conditions in the range 40–50 K.

Aluminum has been chosen as baseline material for the telescope mirrors substrate, with a metallic coating to enhance reflectivity and protect from oxidation and corrosion.

As part of Phase B1, leading to SRR and eventually mission adoption, a protected silver coating with space heritage has been selected and will undergo a qualification process.

A fundamental part of this process is assuring the integrity of the coating layer and performance compliance in terms of reflectivity at the telescope operating temperature. To this purpose, a set of flat sample disks have been cut and polished from the same baseline aluminum alloy as the telescope mirror substrates, and the selected protected silver coating has been applied to them by magnetron sputtering.

The disks have then been subjected to a series of cryogenic temperature cycles to assess coating performance stability.

This study presents the results of visual inspection, reflectivity measurements and atomic force microscopy (AFM) on the sample disks before and after the cryogenic cycles.

## 3.1 Introduction

Ariel is ESA M4 mission, with the purpose of carrying out a survey of the atmospheres of known exoplanets. The main instruments consist in a set of spectrometers operating in the waveband between 0.5 µm and 8 µm.

Science requirements on the telescope can be summarized as a light collecting area of at least 0.6 square meters, diffraction limited performance at the wavelength of 3 µm on a 30″ Field of View, and an average throughput of 96 % [3, 10]. Telescope and instruments will operate at a temperature below 50 K.

These requirements led to an off-axis, unobscured Cassegrain telescope design. The primary mirror will have an elliptical aperture with dimensions of 1100 mm (major axis) and 768 mm (minor axis).

Aluminum alloy 6061-T651 has been proposed as construction material for mirrors substrates and supporting structures of the telescope, after a tradeoff study [2] considering manufacturability and cost and based on JWST MIRI heritage [8].

Bare aluminum is however prone to oxidation and its reflectivity in the visible portion of the operating waveband is not sufficient to reach the required throughput. For these reasons, early on in project development, the Ariel Consortium decided to apply a protected silver coating to the telescope mirrors. The choice of silver, as opposed to gold or a pure aluminum coating was dictated primarily by the throughput requirement.

Abundant literature exists on the heritage of silver-coated aluminum mirrors for ground and space telescope applications operating also at cryogenic temperatures [1, 13, 12].

The large collecting area of Ariel primary mirror however posed specific concerns on uniformity of the deposition process and possible issues caused by the difference in the coefficients of thermal expansion (CTE) between aluminum alloy 6061 and silver. Environmental durability of the coating, and the effectiveness of the protecting capping layer to avoid exposition and corrosion from common atmospheric pollutants was taken into account [4, 14].

For these reasons, the Ariel Consortium selected a specific protected silver coating with space heritage from CILAS[3] and devised a comprehensive qualification study to test its optical performance and durability on Al6061-T651 substrates.

The study, still ongoing at the time of writing, involves two main activities: a qualification of the coating on aluminum samples, and testing of the coating on a full size demonstrator of the primary mirror of the Ariel telescope (named PTM).

The first activity was further divided into two phases: a set of environmental and durability tests on samples coated while lying on a flat surface, and a follow up study with samples coated on top of a curved surface mimicking the optical surface of the PTM. This paper describes the cryogenic tests performed during the first phase and in particular the evaluation of coating performance and durability by means of adhesion tests, reflectivity measurements and Atomic Force Microscopy (AFM) metrology.

## 3.2 Materials and Processes

### 3.2.1 Samples Description

The coating qualification campaign is performed on samples of Al 6061-T651 in rolled plate form, the same aluminum alloy and forge currently foreseen for Ariel Telescope mirrors and supporting structure. Two of the samples used have been subjected to the cryogenic thermal cycles described below and tested for signs of degradation.

The samples, obtained from the same metal plate from which the PTM substrate had been cut, are shaped as 6 mm thick disks with a diameter of 25 mm. Figure 3.1 shows one of the samples being held for visual inspection before the coating run.

The samples have been procured, polished and cleaned by MediaLario[4] before delivery to CILAS for coating.

Roughness of the optical surface was measured with a Taylor Hobson CCI White Light Interferometer with magnifications 10× and 50×. All samples were within the 10 nm RMS specification.

[3]CILAS-ArianeGroup, Etablissement de Marseille, 600 avenue de la Roche Fourcade, Pôle ALPHA Sud - Z.I. Saint Mitre, 13400 Aubagne, France

[4]Media Lario S.r.l., Via al Pascolo, 23842 Bosisio Parini (LC), Italy

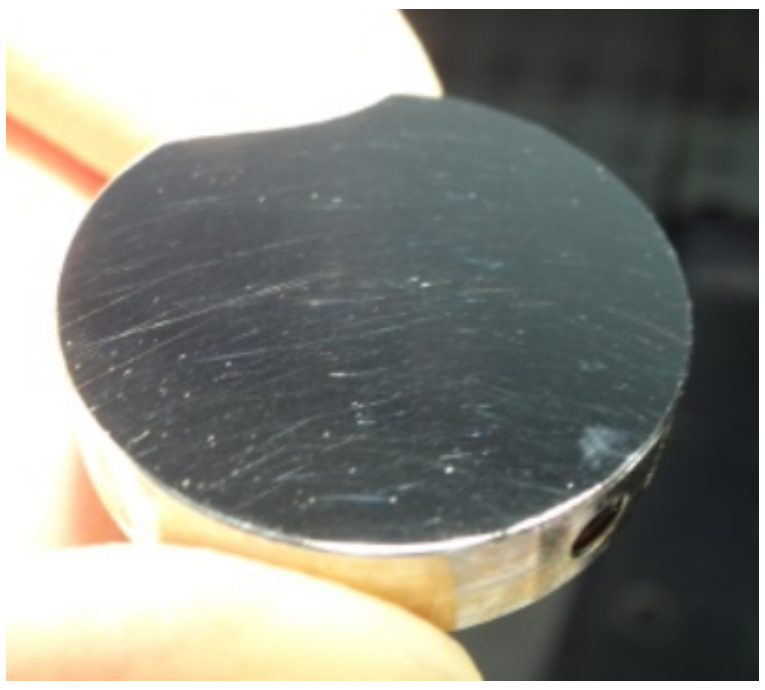

**Figure 3.1:** Picture of an aluminum sample used for the qualification during a visual inspection before being coated.

### 3.2.2 Coating Process

CILAS in-line coating process is based on physical vapor deposition. The coating platform consists in a large magnetron sputtering chamber where a tray is slid back and forth beneath a set of cathodes, in order to minimize deposition inhomogeneities [11].

The process is suited to optical substrates up to 2 m by 2 m of footprint and 0.4 m of thickness [5].

The protected silver coating tested during this qualification campaign is on average 350 nm thick, with a thickness uniformity measured at 10 %, and consists of three layers: a NiCr adhesion layer of less than 10 nm of thickness, the silver layer, and a dielectric capping and protection layer. The actual thicknesses and composition of the capping layer are covered by industrial trade secret.

### 3.2.3 Cryogenic Thermal Cycling Procedure

The procedure consists in a series of 10 cryogenic cycles under vacuum, between room temperature and 54 K. The test took place at the Blue Barrel facility in the CryoWaves Laboratory at INAF OASBo in Bologna[5].

Details of the cycles are presented in Table 3.1.

Samples were attached to copper straps with aluminum tape on the non-optical side.

## 3.3 Verification Methods

Purpose of the test is to make an initial assessment of stability of the coating on the aluminum substrate both in terms of spectral reflectivity and adhesion, after subjecting a set of samples to cryogenic temperatures close to those the Ariel telescope will operate at. For the test to be considered successful, the following three criteria have been established:

[5] INAF-Osservatorio di Astrofisica e Scienza dello spazio di Bologna, Via Piero Gobetti 93/3, 40129 Bologna, Italy

**Table 3.1:** Details of the planned cryogenic thermal cycles procedure.

| | |
|---|---|
| ECSS reference standard | ECSS-Q-ST-70-04-C |
| Start Temperature | 20 °C |
| Cycle temperatures | 54–293 K |
| Temperature change rate | <5 K/min (goal 1 K/min) |
| Dwell time | 15 minutes |
| Vacuum | $<1 \times 10^{-4}$ mbar |
| Number of cycles | 10 |

**Criterion 1:** the coating must appear unaltered, i.e. no obvious delamination or degradation must be apparent upon visual inspection.

**Criterion 2:** relative reflectivity should not change, within the experimental measurement error.

**Criterion 3:** the adhesion tests after the cryogenic cycles must be successful.

The samples were subjected to an adhesion test also before the cryogenic cycles, to make sure that a failure of Criterion 3 would be imputable to the test itself, and not to an intrinsic defect of the deposition process.

In addition to the three criteria, AFM images were taken before and after the test. AFM scans could further qualify a “fail” test result, analyzing microscopically any alteration detected through visual inspection and/or reflectivity measurements.

The following paragraphs describe the details of the verification methods employed.

### 3.3.1 Relative Reflectivity Measurements

Relative reflectivity measurements were taken at the Institute for photonics and nanotechnologies of the National Research Council in Padova[6] with a custom built setup.

Reproducibility of the measurements in the worst case has been determined to be ±3.8 % over the 400–1000 nm wavelength range and ±1.2 % over the 500–900 nm wavelength range.

Data analysis and visualization have been performed in Python with Matplotlib [6].

### 3.3.2 Adhesion Tests

Adhesion tests were performed before and after the cryogenic cycles at MediaLario using Kapton® tape strips, following ISO Standard 9211-4, Method 02, Severity 02 [7].

Before the tests, the samples had been cleaned with an acetone-based solvent and optical wipes.

[6] CNR-IFN Padova, Via Trasea 7, 35131 Padova, Italy

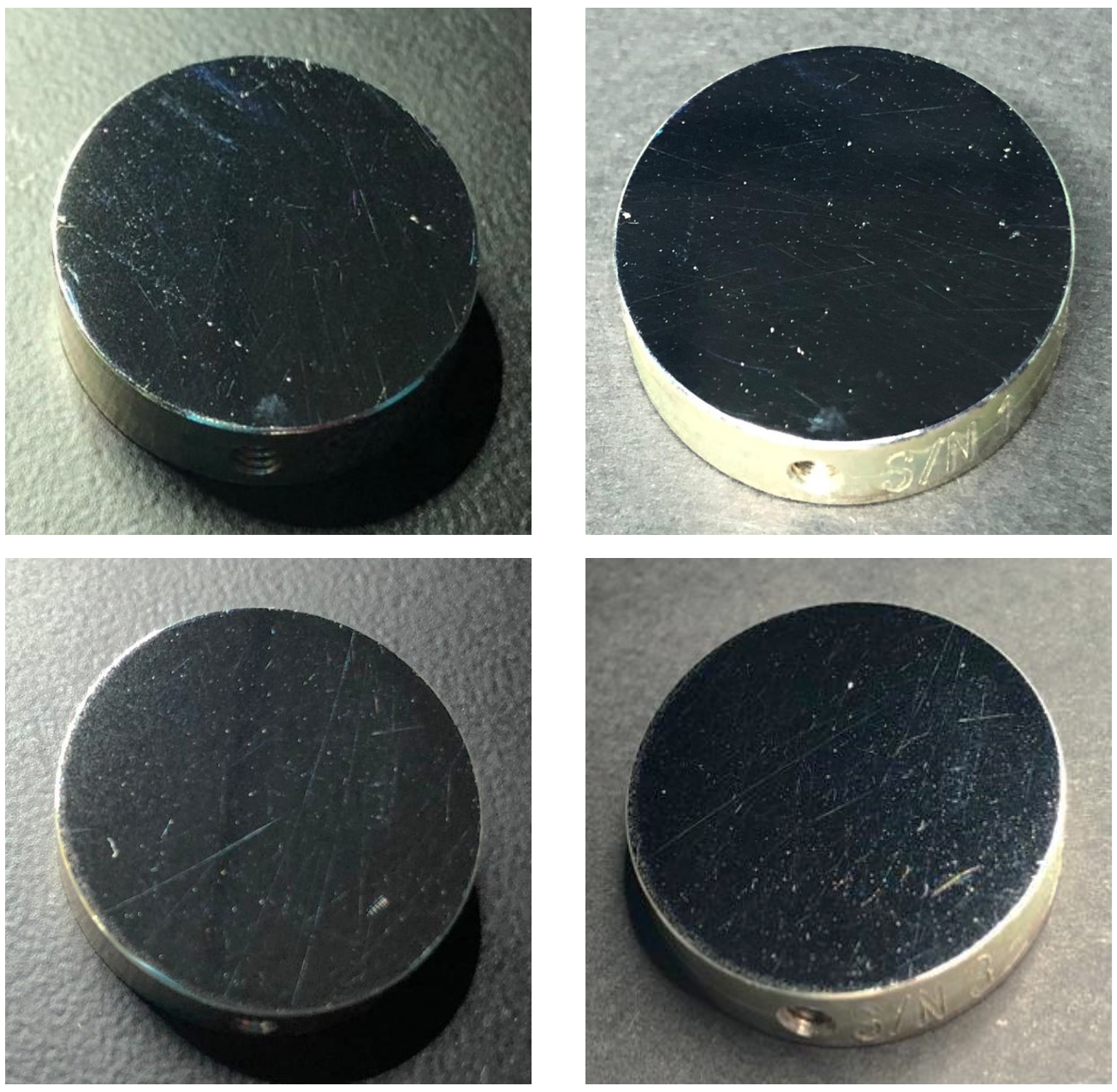

**Figure 3.2:** Pictures of the samples taken before (left) and after (right) the cryogenic cycles. The spot marked with a red oval is a reflection artifact from the lamp used for the inspection, and not a surface blemish.

### 3.3.3 Atomic Force Microscopy

AFM scans were taken with a Park System XE-Series 70 microscope in non-contact mode and processed with Gwyddion [9].

## 3.4 Test Results

### 3.4.1 Visual Inspection

Figure 3.2 shows each sample before and after performing the cryogenic cycles.

Although some minor scratches and slight shadowing and halos appeared after the cycles, especially on the first sample, the overall surface appearance of both samples presented no discernible signs of coating degradation, cracks or visible signs of delamination.

Some residual traces of adhesive from the adhesion tests were visible despite the cleaning.

### 3.4.2 Reflectivity Measurements

Figure 3.3 illustrates reflectance measurements of the two samples before and after the cryogenic cycles. Reflectance figures before cryogenic cycles are the average of four different measurements taken in two consecutive days, while post-cycles figures are the average of two mea-surements, also taken during two days.

The difference in reflectance measurements taken before and after the cryogenic cycles is less than 1 %, well within the reproducibility error of the setup for both mirror samples.

### 3.4.3 Adhesion Test

Adhesion tests performed before and after cryo-testing were both successful: no visible signs of delamination appeared, nor traces of coating were spotted on the tape used for the test.

### 3.4.4 Atomic Force Microscopy

AFM measurements of representative areas of the two samples are presented in Figure 3.4 and 3.5. As discussed in Section 3, the scans were performed to provide a qualitative assessment of surface morphology variations.

Since the cryogenic cycles did not produce any visible signs of surface degradation nor delamination, the sampling location for the AFM had been chosen to be reasonably representative of the central area, where reflectivity was also measured, without aiming at specific surface blemishes or scratches. Moreover, measurements before and after cryo-testing did not image the exact same portion of surface.

A comparison of the AFM images does not indicate the appearance of new topological structures of relevance: most features are attributable to scratches and dents that were equally present before and after the cryogenic cycles.

The white areas in relief are likely caused by residuals of tape glue from the adhesion tests that resisted cleaning. These were in fact visible upon careful examination of the surface area affected by the test.

RMS roughness measurements also do not appear to change significantly before and after the cycles.

## 3.5 Conclusions

Two Al6061-T651 disks, deposited with a protected silver coating with space heritage from CILAS, have been subjected to a series of 10 cryogenic cycles between room temperature and 50 K, the operating temperature of the Ariel telescope.

Results of visual inspections, adhesion tests and reflectivity measurements showed no alteration in appearance imputable to deterioration or delamination of the coating, nor a degradation in optical performance in the waveband 400–1000 nm.

These results were considered satisfactory and led to the second phase of the qualification campaign, consisting of a battery of tests on equivalent samples coated while lying on a curved surface in the shape of the primary mirror of the telescope.

Further measurements on the samples will be repeated periodically to assess possible aging deterioration.

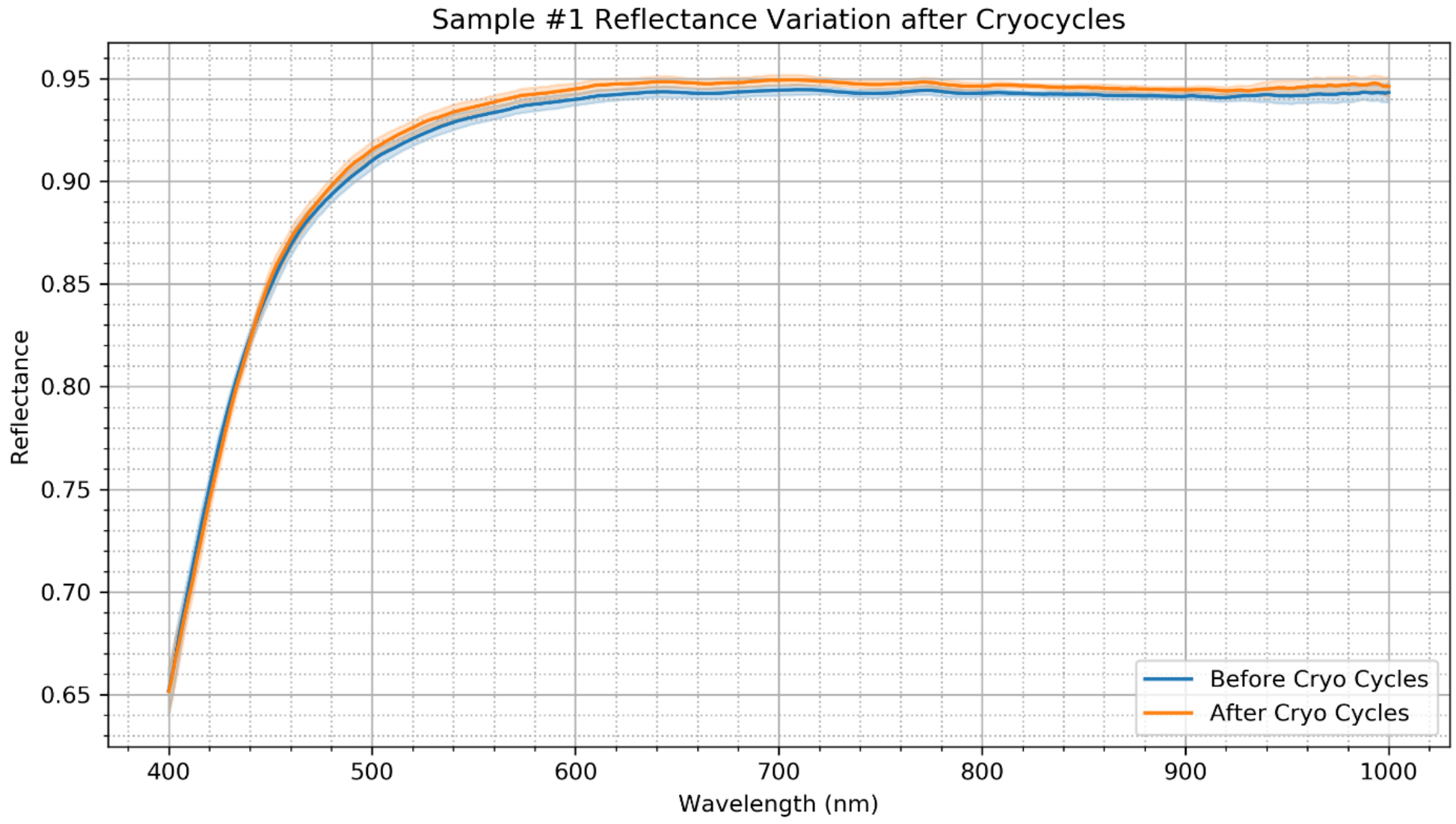

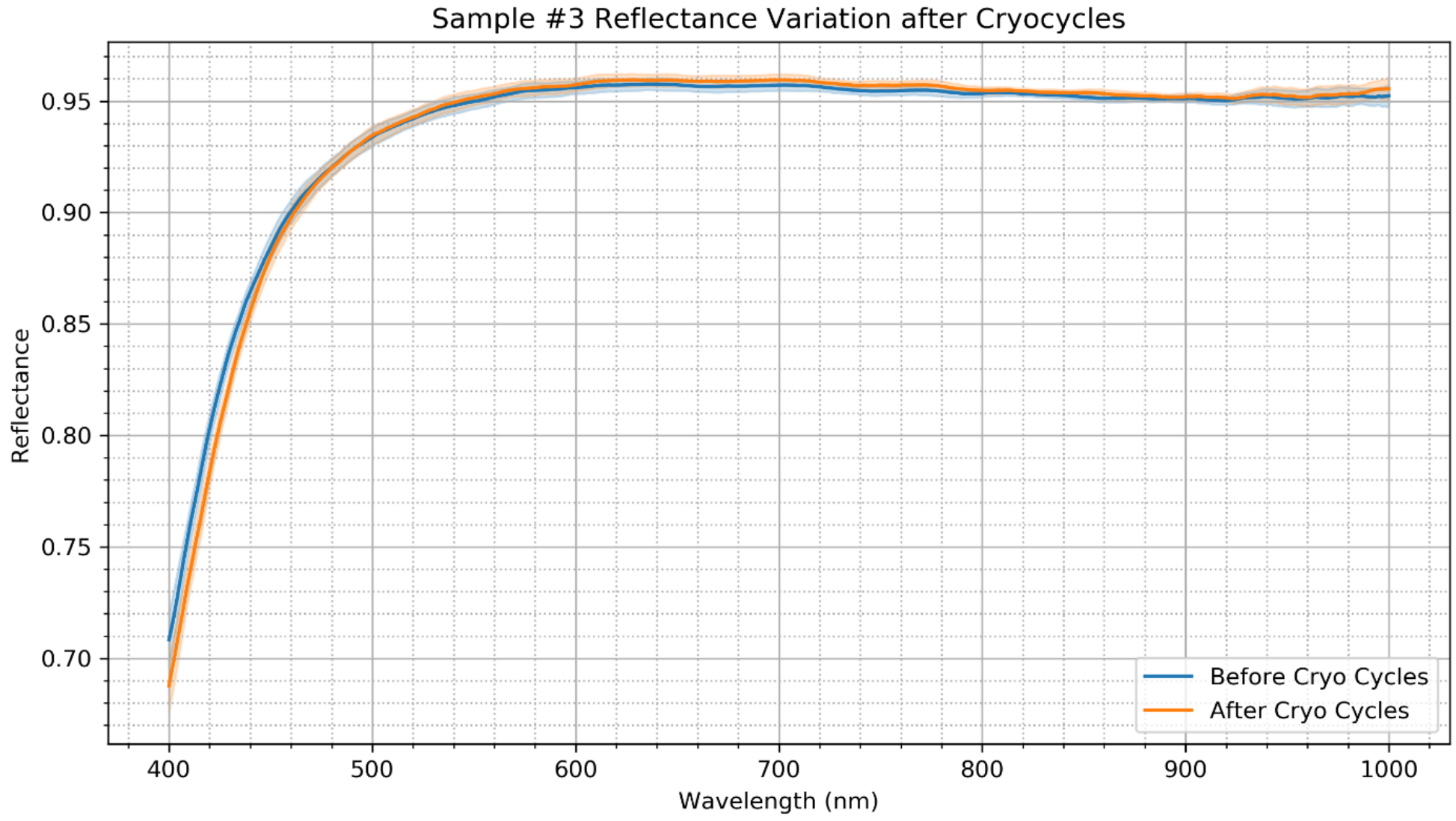

**Figure 3.3:** Reflectance measurements of the two samples before (blue) and after (orange) the cryogenic cycles. The colored areas show reproducibility of the measurements. Graphs have been smoothed with a Savitzky–Golay convolution filter.

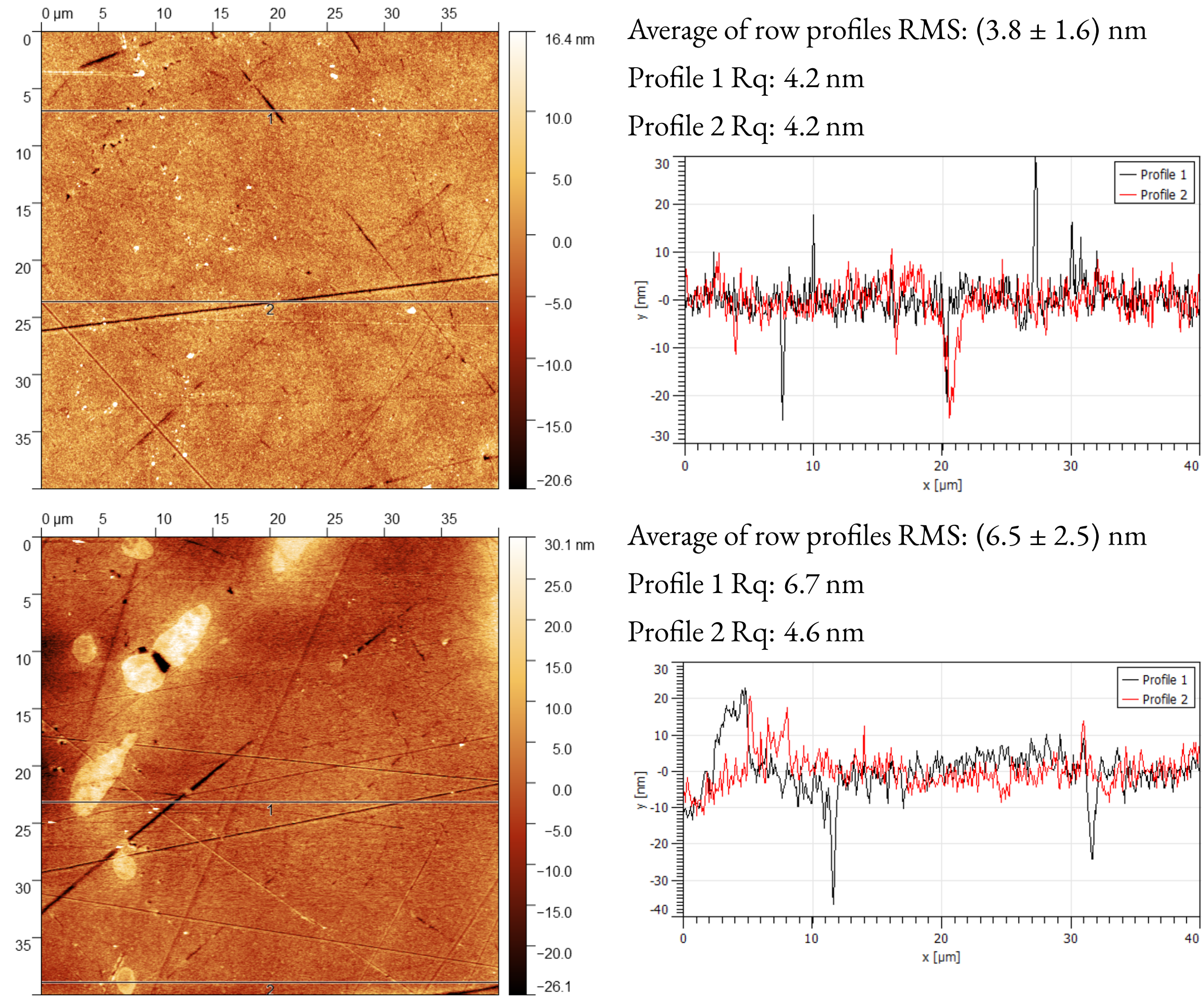

**Figure 3.4:** On the left, AFM images of different areas of the first sample before (top) and after (bottom) the cryogenic cycles. On the right, roughness parameters and row profiles.

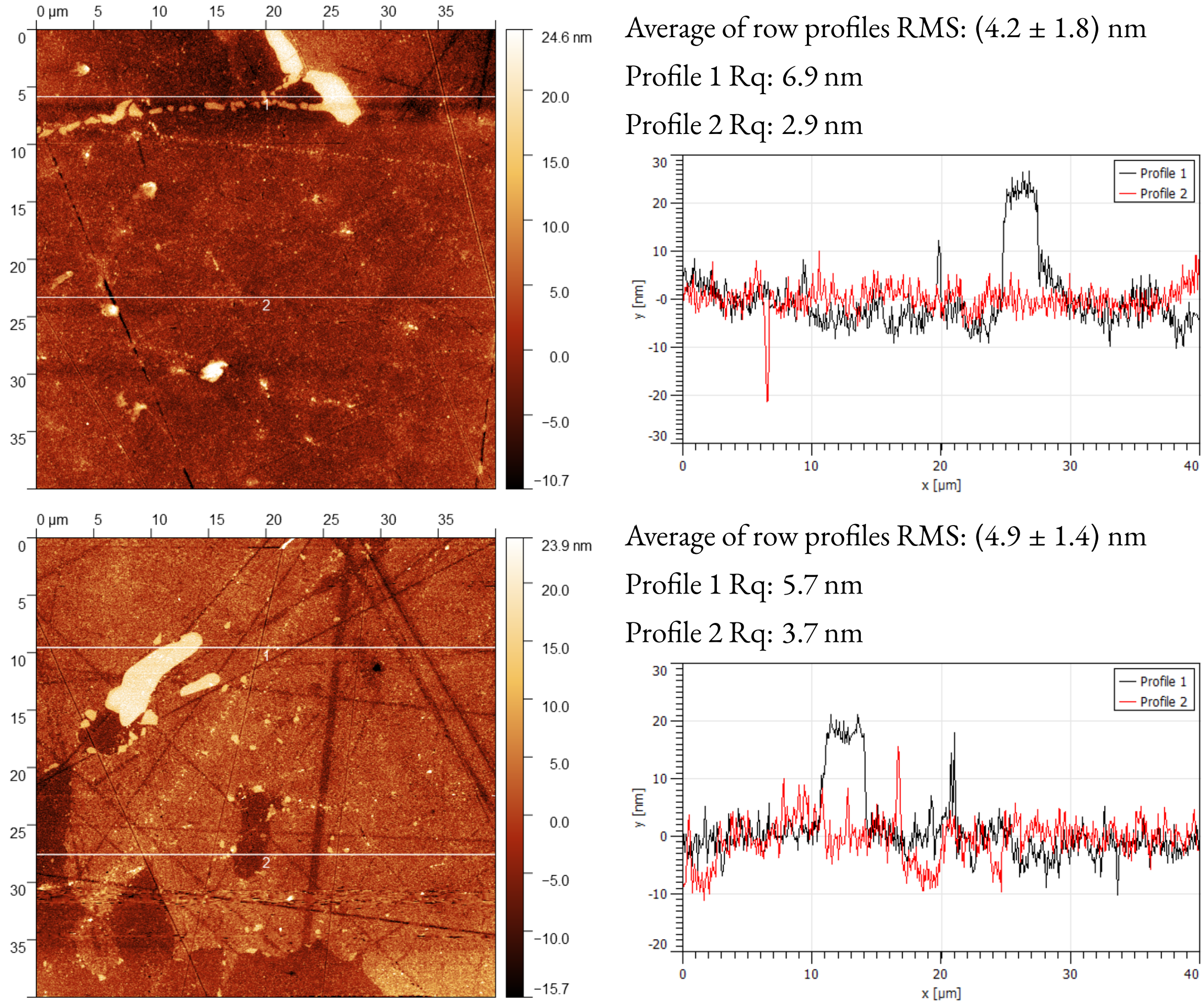

**Figure 3.5:** On the left, AFM images of different areas of the second sample before (top) and after (bottom) the cryogenic cycles. On the right, roughness parameters and row profiles.

## Acknowledgements

This activity has been realized under the Italian Space Agency (ASI) contract with the National Institute for Astrophysics (INAF) n. 2018-22-HH.0, and is partly funded under the ESA contract with Centre Spatial de Liège, Belgium (CSL) and INAF n. 4000126124/18/NL/BW.

## References

[1] M. Boccas et al. "Coating the 8-m Gemini Telescopes with Protected Silver". In: *Proc. SPIE 5494, Optical Fabrication, Metrology, and Material Advancements for Telescopes*. Sept. 24, 2004, p. 239. DOI: 10.1117/12.548809.

[2] V. Da Deppo et al. *ARIEL Telescope Material Trade-Off*. Technical Paper ARIEL-INAF-PL-TN-004. 2017.

[3] V. Da Deppo et al. "The Afocal Telescope Optical Design and Tolerance Analysis for the ESA ARIEL Mission". In: *Proc. SPIE 10590, International Optical Design Conference 2017*. Nov. 27, 2017, 105901P. DOI: 10.1117/12.2292299.

[4] K. A. Folgner et al. "Environmental Durability of Protected Silver Mirrors Prepared by Plasma Beam Sputtering". In: *Applied Optics* 56.4 (Feb. 1, 2017), p. C75. DOI: 10.1364/AO.56.000C75.

[5] C. Grèzes-Besset et al. "High Performance Silver Coating with PACA2M Magnetron Sputtering". In: *Proc. SPIE 11180, International Conference on Space Optics — ICSO 2018*. July 12, 2019, p. 1118083. DOI: 10.1117/12.2536210.

[6] J. D. Hunter. "Matplotlib: A 2D Graphics Environment". In: *Computing in Science & Engineering* 9.3 (2007), pp. 90–95. DOI: 10.1109/MCSE.2007.55.

[7] International Organization for Standardization. *ISO 9211-4:2022 Optics and Photonics — Optical Coatings — Part 4: Specific Test Methods: Abrasion, Adhesion and Resistance to Water*. Vernier, Geneva, Switzerland: International Organization for Standardization, 2022.

[8] G. Kroes et al. "MIRI-JWST Spectrometer Main Optics Opto-Mechanical Design and Prototyping". In: *Proc. SPIE 5877, Optomechanics 2005*. Aug. 18, 2005, 58770P. DOI: 10.1117/12.614784.

[9] D. Nečas and P. Klapetek. "Gwyddion: An Open-Source Software for SPM Data Analysis". In: *Open Physics* 10.1 (Jan. 1, 2012). DOI: 10.2478/s11534-011-0096-2.

[10] L. Puig et al. "The Phase A Study of the ESA M4 Mission Candidate ARIEL". In: *Experimental Astronomy* 46.1 (Nov. 2018), pp. 211–239. DOI: 10.1007/s10686-018-9604-3.

[11] I. Savin de Larclause et al. "PACA2m Magnetron Sputtering Silver Coating: A Solution for Very Big Mirror Dimensions". In: *Proc. SPIE 10563, International Conference on Space Optics — ICSO 2014*. Jan. 5, 2018, p. 1056308. DOI: 10.1117/12.2304237.

[12] M. Schürmann et al. "High-Reflective Coatings for Ground and Space Based Applications". In: *Proc. SPIE 10563, International Conference on Space Optics — ICSO 2014*. Nov. 17, 2017, p. 105630M. DOI: 10.1117/12.2304172.

[13] D. A. Sheikh. "Improved Silver Mirror Coating for Ground and Space-Based Astronomy". In: *Proc. SPIE 9912, Advances in Optical and Mechanical Technologies for Telescopes and Instrumentation II*. July 22, 2016, p. 991239. DOI: 10.1117/12.2234380.

[14] A. Sytchkova et al. "Optical Characterisation of Silver Mirrors Protected with Transparent Overcoats". In: *Proc. SPIE 10691, Advances in Optical Thin Films VI*. June 5, 2018, 106910N. DOI: 10.1117/12.2312077.

# 4

# Test of protected silver coating on aluminum samples of Ariel main telescope mirror substrate material

Paolo Chioetto[1 2 a,b,c], Paola Zuppella[a,c], Vania Da Deppo[a,c], Simone Nordera[a], Emanuele Pace[d], Andrea Tozzi[e], Gianluca Morgante[f], Luca Terenzi[f], Daniele Brienza[g], Nadia Missaglia[h], Giovanni Bianucci[h], Sebastiano Spinelli[h], Elisa Guerriero[j,k,h], Massimiliano Rossi[h], Gabriele Grisoni[h], Colin Bondet[i], Gregory Chauveau[i], Caroline Porta[i], Catherine Grezes-Besset[i], Giuseppe Malaguti[f], Giuseppina Micela[j]

a CNR-Istituto di Fotonica e Nanotecnologie di Padova, Via Trasea 7, 35131 Padova, Italy
b Centro di Ateneo di Studi e Attività Spaziali "Giuseppe Colombo"- CISAS, Via Venezia 15, 35131 Padova, Italy
c INAF-Osservatorio Astronomico di Padova, Vicolo dell'Osservatorio 5, 35122 Padova, Italy
d Dipartimento di Fisica ed Astronomia-Università degli Studi di Firenze, Largo E. Fermi 2, 50125 Firenze, Italy
e INAF-Osservatorio Astrofisico di Arcetri, Largo E. Fermi 5, 50125 Firenze, Italy
f INAF-Osservatorio di Astrofisica e Scienza dello spazio di Bologna, Via Piero Gobetti 93/3, 40129 Bologna, Italy
g INAF-Istituto di Astrofisica e Planetologia Spaziali, Via Fosso del Cavaliere 100, 00133 Roma, Italy
h Media Lario S.r.l., Località Pascolo, 23842 Bosisio Parini (Lecco), Italy
i CILAS-ArianeGroup, Etablissement de Marseille, 600 avenue de la Roche Fourcade, Pôle ALPHA Sud - Z.I. Saint Mitre, 13400 Aubagne, France
j INAF-Osservatorio Astronomico di Palermo, Piazza del Parlamento 1, 90134 Palermo, Italy

[1]paolo.chioetto@pd.ifn.cnr.it
[2]The two authors contributed equally to the article.

k Dipartimento di Fisica e Chimica-Università degli Studi di Palermo, Via Archirafi 36, 90128 Palermo, Italy

## Abstract

Ariel (Atmospheric Remote-Sensing Infrared Exoplanet Large Survey) has been adopted as the M4 mission for ESA "Cosmic Vision" program. Launch is scheduled for 2029.

ARIEL will study exoplanet atmospheres through transit spectroscopy with a 1 mclass telescope optimized in the waveband between 1.95 and 7.8 µm and operating in cryogenic conditions in the temperature range 40–50 K.

Aluminum alloy 6061, in the T651 temper, was chosen as baseline material for telescope mirror substrates and supporting structures, following a trade-off study. To improve mirrors reflectivity within the operating waveband and to protect the aluminum surface from oxidation, a protected silver coating with space heritage was selected and underwent a qualification campaign during Phase B1 of the mission, with the goal of demonstrating a sufficient level of technology maturity.

The qualification campaign consisted of two phases: a first set of durability and environmental tests conducted on a first batch of coated aluminum samples, followed by a set of verification tests performed on a second batch of samples coated alongside a full-size demonstrator of Ariel telescope primary mirror.

This study presents the results of the verification tests, consisting of environmental (humidity and temperature cycling) tests and chemical/mechanical (abrasion, adhesion, cleaning) tests performed on the samples, and abrasion tests performed on the demonstrator, by means of visual inspections and reflectivity measurements.

## 4.1 Introduction

Ariel has been recently adopted as ESA Cosmic Vision Program M4 mission. In its 4-year nominal mission, Ariel will conduct a survey of known exoplanets to characterize their atmospheres through transit spectroscopy in the wavelength band between 0.5 µm and 8 µm.

The Ariel telescope is based on an off-axis, unobscured Cassegrain design with an elliptical primary mirror with an aperture of 1100 mm (major axis) and 768 mm (minor axis) and a light collecting area of approximately 0.6 square meters. Telescope performance is diffraction limited at the wavelength of 3 µm on a 30″ Field of View. The required average telescope throughput is 96 % [3, 9]. Telescope and instruments will operate at a temperature below 50 K.

Following the heritage of the JWST MIRI instrument [8], aluminum alloy 6061-T651 has been chosen for mirrors substrates and supporting structures of the telescope, after a trade-off study [4] on manufacturability and cost.

To protect the mirrors and to improve their reflectivity in the visible section of the operating waveband, a protected silver coating with space heritage from CILAS[3] was chosen as baseline.

Although several examples of cryogenic silver-coated aluminum mirrors are found in literature [1, 12, 11], the large size of the primary mirror and its curvature raised concerns on the uniformity of the deposition process and stability of the coating.

An initial study was therefore devised to test optical performance and durability on Al6061-T651 substrates, consisting of a qualification campaign on coated aluminum samples [2], and a verification test on additional samples and on a full-scale demonstrator of Ariel primary mirror denoted PTM.

This paper describes the verification tests, in particular the evaluation of coating performance and durability by means of adhesion tests, reflectivity measurements and environmental tests.

## 4.2 Materials and Processes

### 4.2.1 Items Under Test

The coating verification tests were performed on the following items coated together in the same run:

1. the PTM (Figure 4.1), a spherical mirror with a radius of curvature of 2401 mmand an elliptical optical aperture of 1100 x 730 mm cut from a rolled plate of Al 6061-T651, the current baseline for Ariel telescope mirrors and supporting structure;

2. 6 aluminum samples, shaped as disks 6 mmthick 25 mmin diameter, obtained from the same plate from which the PTM substrate had been cut (Figure 4.2 shows one of the samples being held for visual inspection before the coating run);

3. 2 sets of glass samples measuring 25 mmand 48 mmin diameter respectively to serve as reference for spectral reflectivity and for the profilometry measurements.

The roughness of the optical surface on the aluminum samples was measured with a Taylor Hobson CCI White Light Interferometer with magnifications 10× and 50×. All samples measured roughness was within the 10 nm RMS specification.

The surface roughness of the PTM, measured at 10 different locations on its optical surface, was instead in the range 21.5–26.9 nm RMS.

The PTM and aluminum samples had been manufactured by MediaLario[4].

[3]CILAS-ArianeGroup, Etablissement de Marseille, 600 avenue de la Roche Fourcade, Pôle ALPHA Sud - Z.I. Saint Mitre, 13400 Aubagne, France

[4]Media Lario S.r.l., Via al Pascolo, 23842 Bosisio Parini (LC), Italy

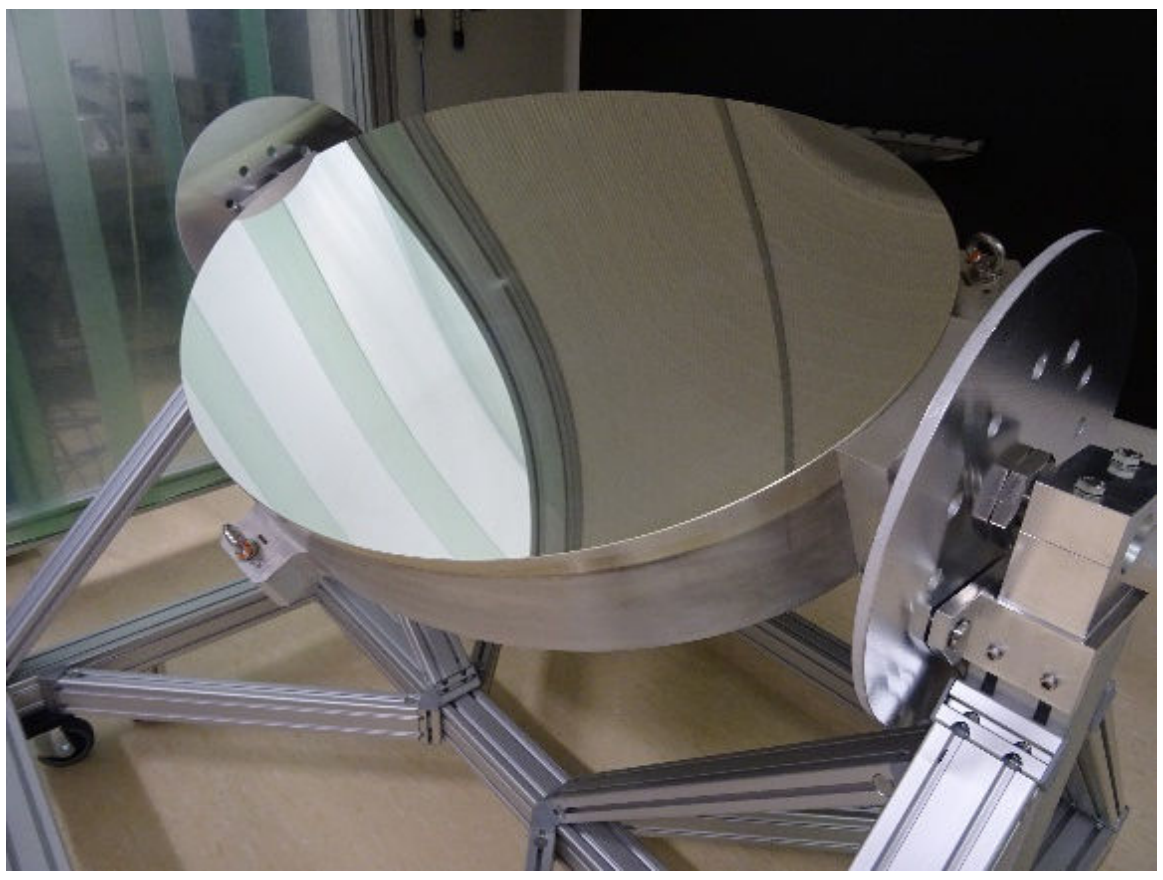

**Figure 4.1:** Picture of the PTM mirror mounted on its transport and handling trolley before coating (courtesy of CILAS).

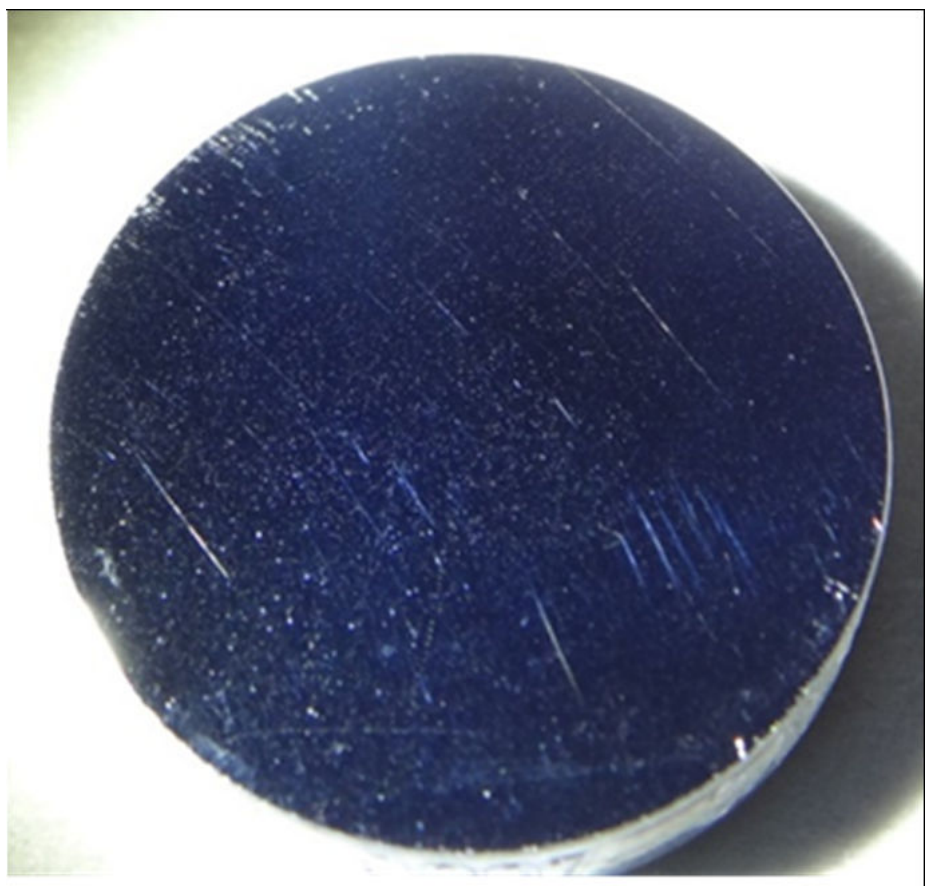

**Figure 4.2:** Picture of an aluminum sample used for the qualification during a visual inspection before being coated (courtesy of CILAS).

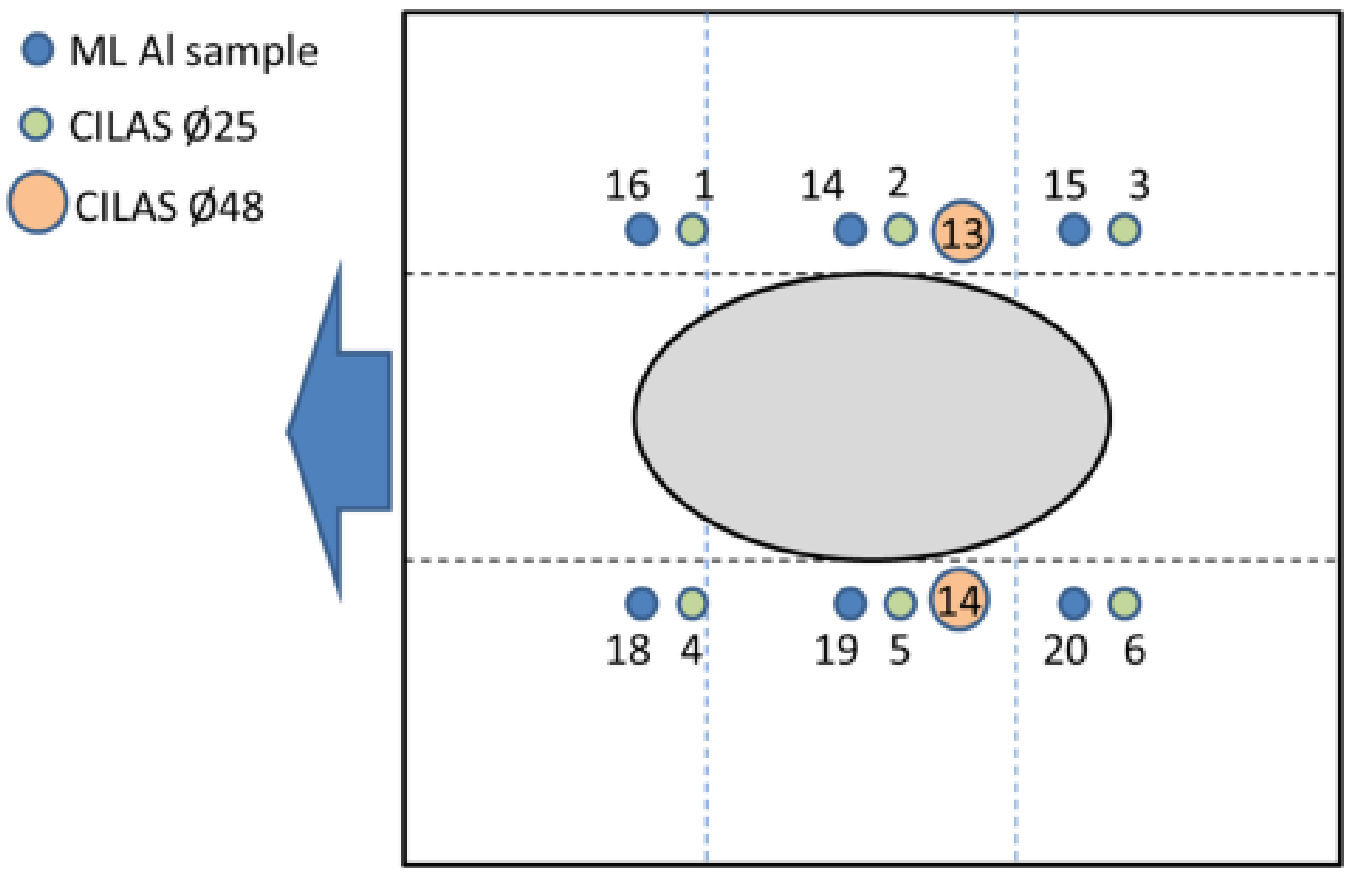

**Figure 4.3:** Layout of the items to be coated on the tray of the coating chamber. The blue arrow indicates the direction of insertion into the chamber.

### 4.2.2 Coating Process

The coating process used by CILAS for this batch is based on physical vapor deposition. The items to be coated are laid on a tray that slides inside a large magnetron sputtering chamber, and moved back and forth beneath a set of cathodes, to minimize deposition inhomogeneities [10]. The process is suited to optical substrates up to 2 m by 2 m of footprint and 0.4 m of thickness [5].

The protected silver coating employed for the study consists of three layers: a NiCr adhesion layer of less than 10 nm of thickness, the silver layer, and a dielectric capping and protection layer. The process had already been qualified on aluminum samples and produces a coating layer measuring on average 350 nm of thickness, with 10 % uniformity.

The layout of the items on the coating tray is shown in Figure4.3.

## 4.3 Verification Methods

The tests described in this paper have the purpose of verifying that the performance and durability of the protected silver coating deposited on the PTM are consistent with the results obtained during the qualification phase [2].

Four of the witness samples deposited together with the PTM were therefore subjected to a series of humidity, temperature, cleaning, abrasion and adhesion tests, evaluated by visual inspection and assessment of the variation of relative spectral reflectivity measurements taken before and after the set of tests.

Additionally, the coating thickness was measured on one of the glass samples, and the durability of the coating on the PTM was assessed through adhesion testing.

The following paragraphs describe the details of the verification methods employed.

### 4.3.1 Visual Inspection

Visual inspection was performed at CILAS on the optical area of each sample according to ISO standard 9211-4:2012 [7]. Inspection of the PTM was performed both by CILAS and MediaLario.

### 4.3.2 Relative Reflectivity Measurements

Relative reflectivity measurements were performed at CILAS with a Perkin-Elmer Lambda 950 spectrophotometer with a reflectometry accessory in the waveband 500–2500 nm and with an accuracy of ±0.6 % from 500 nm to 890 nm and ±1 % above 890 nm.

### 4.3.3 Humidity Test

The test was realized at CILAS according to ISO 9211-3:2008 [6] Test #5 (damp heat). The samples were exposed to a 90 % humidity environment at 55 °C (±3 °C) for 24 hours with a maximum temperature slope of 2 °C/minute. No condensation was observed on the coated surfaces during the test.

### 4.3.4 Temperature Cycling Test

The samples were subjected to 30 cycles between -40 °C and 70 °C with a maximum temperature slope of 2 °C/minute and a dwell time of 15 minutes. The test was performed at CILAS.

### 4.3.5 Cleaning Test

The test was performed at CILAS applying their standard cleaning procedure based on ethanol and acetone solutions, using an optical wipe. The test was repeated 5 times.

### 4.3.6 Abrasion Test

Test realized at CILAS on one sample, according to ISO 9211-4:2012 [7], Test Method 01, Severity Level 01.

### 4.3.7 Adhesion Test

Adhesion tests were performed following ISO Standard 9211-4, Method 02, Severity 02 [7].

Tests on samples were performed at CILAS with a cellophane tape, while the test on the optical surface of the PTM was performed at MediaLario using Kapton® tape strips.

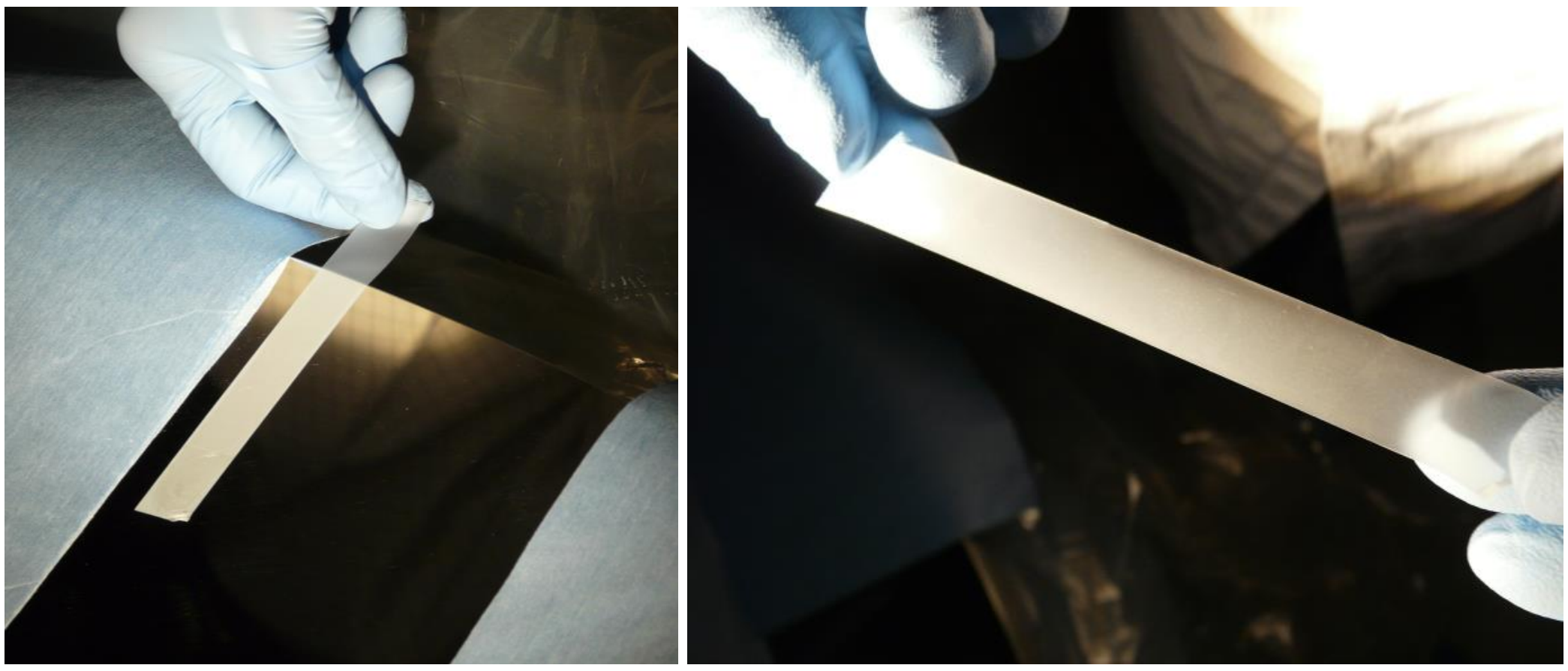

**Figure 4.4:** Adhesion test on the PTM mirror. On the left, the tape positioned perpendicular to the edge of the mirror right before lift-off. On the right, the sticky side of the tape after lift-off, showing no detached coating particles (pictures courtesy of MediaLario).

## 4.4 Test Results

### 4.4.1 Environmental and Mechanical Tests

One aluminum sample was subjected to the whole sequence of chemical/mechanical tests (cleanability, abrasion and adhesion), while three other samples underwent the environmental set of tests (humidity and temperature cycling) and later the adhesion test. The two remaining aluminum samples have been kept as references for aging.

Results were satisfactory and in line with the outcomes from the coating qualification phase: no discernible signs of coating degradation nor delamination were apparent.

Reflectivity measurements did not highlight any change in performance either, as described in paragraph 4.2.

Adhesion tests on the coated PTM were performed on two separate areas of the mirror, and were also successful with no sign of coating degradation nor visually detectable traces of the coating on the tape strip (Figure 4.4).

### 4.4.2 Reflectivity Measurements

Figure 4.5 illustrates reflectivity measurements of the samples that underwent mechanical (#18, 19, 20) and cleaning (#16) tests. Variations between measurements of the same sample are within the accuracy of the measurement instrument.

### 4.4.3 Profilometry

Glass sample number 14 was partially covered with a mask during coating, and the depth of the resulting ridge was measured with a profilometer in two points. The resulting coating thicknesses were 342 nm and 345 nm compatible with measurements taken during the qualification coating run at the same position in the coating chamber of 348 nm.

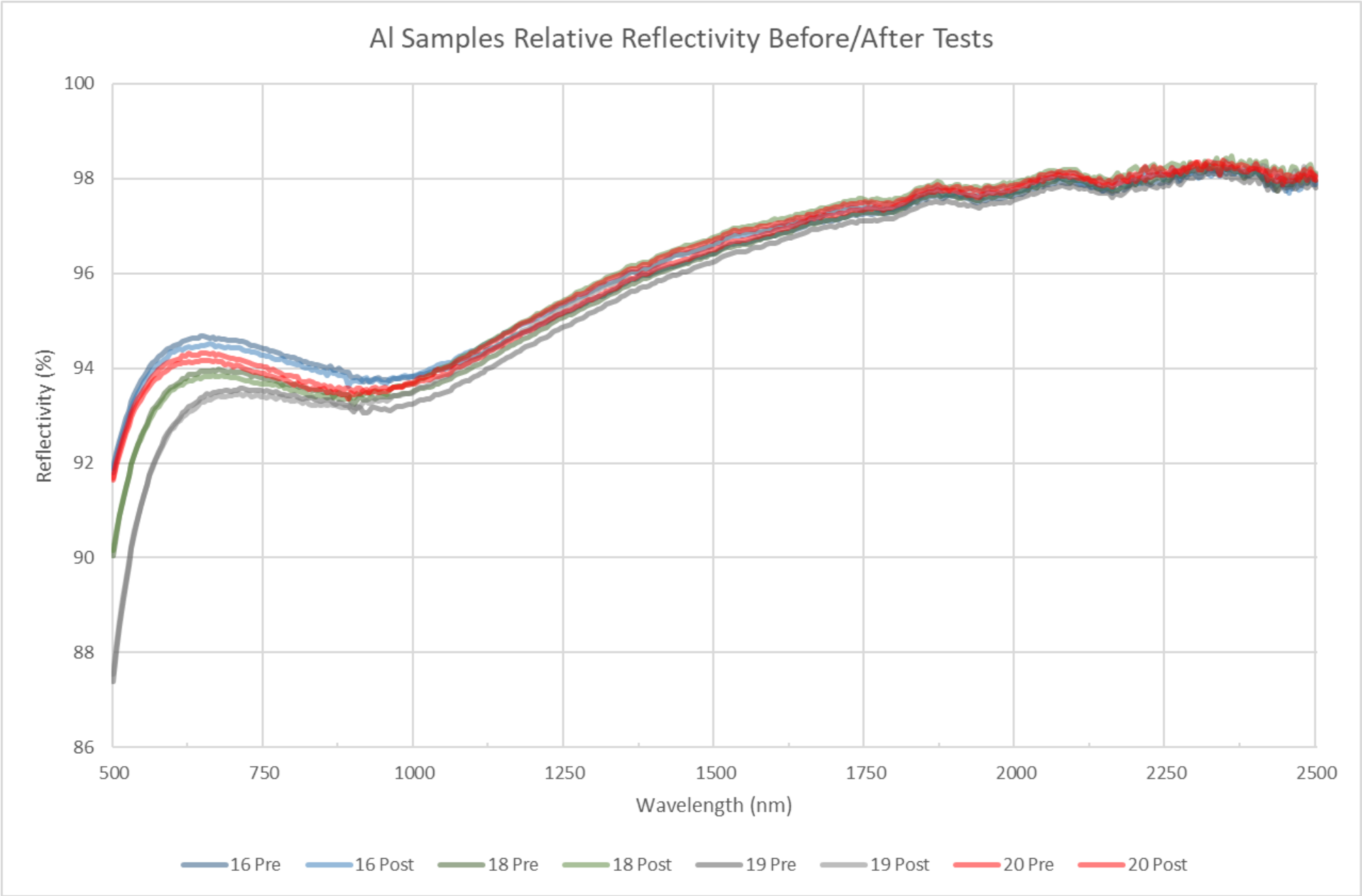

**Figure 4.5:** Reflectivity measurements of samples 16, 18, 19 and 20 before ("Pre" in the legend) and after ("Post" in the legend) the tests, showing no significant change.

## 4.5 Conclusions

Following the successful qualification of the selected protected silver coating for the aluminum mirrors of the Ariel telescope, a coating test run was performed on the full size demonstrator of the telescope primary mirror (PTM), together with 6 aluminum samples, to further confirm performance and durability of the coating.

Results of adhesion, abrasion, cleaning, humidity and thermal cycling tests performed on the samples and verified by visual inspections and reflectivity measurements showed no alteration in appearance imputable to deterioration or delamination of the coating, nor a degradation in optical performance in the waveband 500–2500 nm.

Adhesion tests performed on the optical surface of the PTM further confirmed the durability of the coating.

These results were considered satisfactory and led to a successful termination of the coating qualification campaign.

Further measurements on the samples will be repeated periodically to assess possible aging deterioration.

### Acknowledgements

This activity has been realized under the Italian Space Agency (ASI) contract with the National Institute for Astrophysics (INAF) n. 2018-22-HH.0, and is partly funded under the ESA contract with Centre Spatial de Liège, Belgium (CSL) and INAF n. 4000126124/18/NL/BW.

### References

[1] M. Boccas et al. "Coating the 8-m Gemini Telescopes with Protected Silver". In: *Proc. SPIE 5494, Optical Fabrication, Metrology, and Material Advancements for Telescopes*. Sept. 24, 2004, p. 239. doi: 10.1117/12.548809.

[2] P. Chioetto et al. "The Primary Mirror of the Ariel Mission: Cryotesting of Aluminum Mirror Samples with Protected Silver Coating". In: *Proc. SPIE 11451, Advances in Optical and Mechanical Technologies for Telescopes and Instrumentation IV*. Dec. 13, 2020, 114511A. doi: 10.1117/12.2562548.

[3] V. Da Deppo et al. "An Afocal Telescope Configuration for the ESA ARIEL Mission". In: *CEAS Space Journal* 9.4 (Dec. 2017), pp. 379–398. doi: 10.1007/s12567-017-0175-3.

[4] V. Da Deppo et al. *ARIEL Telescope Material Trade-Off*. Technical Paper ARIEL-INAF-PL-TN-004. 2017.

[5] C. Grèzes-Besset et al. "High Performance Silver Coating with PACA2M Magnetron Sputtering". In: *Proc. SPIE 11180, International Conference on Space Optics — ICSO 2018*. July 12, 2019, p. 1118083. doi: 10.1117/12.2536210.

[6] International Organization for Standardization. *ISO 9211-3:2008 Optics and Photonics — Optical Coatings — Part 3: Environmental Durability*. Vernier, Geneva, Switzerland: International Organization for Standardization, 2008.

[7] International Organization for Standardization. *ISO 9211-4:2022 Optics and Photonics — Optical Coatings — Part 4: Specific Test Methods: Abrasion, Adhesion and Resistance to Water*. Vernier, Geneva, Switzerland: International Organization for Standardization, 2022.

[8] G. Kroes et al. "MIRI-JWST Spectrometer Main Optics Opto-Mechanical Design and Prototyping". In: *Proc. SPIE 5877, Optomechanics 2005*. Aug. 18, 2005, 58770P. doi: 10.1117/12.614784.

[9] L. Puig et al. "The Phase A Study of the ESA M4 Mission Candidate ARIEL". In: *Experimental Astronomy* 46.1 (Nov. 2018), pp. 211–239. doi: 10.1007/s10686-018-9604-3.

[10] I. Savin de Larclause et al. “PACA2m Magnetron Sputtering Silver Coating: A Solution for Very Big Mirror Dimensions”. In: *Proc. SPIE 10563, International Conference on Space Optics — ICSO 2014*. Jan. 5, 2018, p. 1056308. DOI: 10.1117/12.2304237.

[11] M. Schürmann et al. “High-Reflective Coatings for Ground and Space Based Applications”. In: *Proc. SPIE 10563, International Conference on Space Optics — ICSO 2014*. Nov. 17, 2017, p. 105630M. DOI: 10.1117/12.2304172.

[12] D. A. Sheikh. “Improved Silver Mirror Coating for Ground and Space-Based Astronomy”. In: *Proc. SPIE 9912, Advances in Optical and Mechanical Technologies for Telescopes and Instrumentation II*. July 22, 2016, p. 991239. DOI: 10.1117/12.2234380.

# 5

# Long term durability of protected silver coating for the mirrors of Ariel mission telescope

Paolo Chioetto[1 a,b,c], Daniele Brienza[d], Rodolfo Canestrari[e], Paul Eccleston[f], Elisa Guerriero[g,e], Giuseppe Malaguti[h], Giuseppina Micela[e], Emanuele Pace[i], Enzo Pascale[j], Raffaele Piazzolla[d], Giampaolo Preti[i], Mario Salatti[d], Giovanna Tinetti[k], Elisabetta Tommasi[d], Andrea Tozzi[l], Paola Zuppella[a,c]

a CNR–Istituto di Fotonica e Nanotecnologie di Padova, Via Trasea 7, 35131 Padova, Italy
b Centro di Ateneo di Studi e Attività Spaziali "Giuseppe Colombo"–CISAS, Via Venezia 15, 35131 Padova, Italy
c INAF–Osservatorio Astronomico di Padova, Vicolo dell'Osservatorio 5, 35122 Padova, Italy
d ASI, Agenzia Spaziale Italiana, Via del Politecnico snc, Roma, Italy
e INAF–Osservatorio Astronomico di Palermo, Piazza del Parlamento 1, 90134 Palermo, Italy
f RAL Space, STFC Rutherford Appleton Laboratory, Didcot, Oxon, OX11 0QX, UK
g Dipartimento di Fisica e Chimica, Università degli Studi di Palermo, Via Archirafi 36, 90128 Palermo, Italy
h INAF–Osservatorio di Astrofisica e Scienza dello spazio di Bologna, Via Piero Gobetti 93/3, 40129 Bologna, Italy
i Dipartimento di Fisica ed Astronomia, Università degli Studi di Firenze, Largo E. Fermi 2, 50125 Firenze, Italy
j Dipartimento di Fisica, La Sapienza Università di Roma, Piazzale Aldo Moro 2, 00185 Roma, Italy
k Department of Physics and Astronomy, University College London, Gower Street, London WC1E 6BT, UK
l INAF–Osservatorio Astrofisico di Arcetri, Largo E. Fermi 5, 50125 Firenze, Italy

[1] Presenting and contact author.

ABSTRACT

Ariel (Atmospheric Remote-sensing Infrared Exoplanet Large survey) is the fourth medium-size mission in ESA "Cosmic Vision" program. It is scheduled to launch in 2029. Ariel will conduct spectroscopic and photometric observations of a large sample of known exoplanets to survey their atmospheres with the transit method.

Ariel is based on a 1 m class telescope designed for the visible and near infrared spectrum, but optimized specifically for spectroscopy in the waveband between 1.95 and 7.8 μm. Telescope and instruments will be operating in cryogenic conditions in the range 40–50 K.

The telescope mirrors will be manufactured in aluminum 6061, with a protected silver coating deposited onto the optical surface to enhance reflectivity and prevent oxidation and corrosion.

During the preliminary definition phase of the development work, leading to mission adoption, a silver coating with space heritage was selected and underwent a qualification process on disc-shaped samples of the mirrors substrate material. The samples were deposited through magnetron sputtering and then subjected to a battery of tests, including environmental durability tests, accelerated aging, cryogenic tests and mechanical resistance tests. Further to the qualification, the samples have been stored in cleanroom (ISO6) conditions and periodically re-examined and measured to detect any sign of coating degradation.

The test program, still ongoing at the time of writing this article, consists of visual inspection with a high intensity lamp, spectral reflectance measurements and Atomic Force Microscopy (AFM) evaluation of nanometric surface features.

The goal is to ensure stability of the optical performance, in terms of coating reflectance, during a time span comparable to the period that the actual mirrors of the telescope will spend in average cleanroom conditions.

This study presents the interim results after three years of storage.

## 5.1 Introduction

Ariel is the fourth medium-class mission under development in the framework of ESA "Cosmic Vision" Program. It was adopted in 2020 and launch is planned for 2029. During its 4 years of nominal mission duration, Ariel will conduct a survey of known exoplanets to characterize their atmospheres through transit spectroscopy and photometry in the waveband between 0.5 μm and 7.8 μm [15].

The Ariel telescope is based on an off-axis, unobscured Cassegrain design with an elliptical primary mirror with an aperture of 1100 mm (major axis) and 768 mm (minor axis), corresponding to a light collecting area of approximately 0.6 $m^2$. The telescope was designed to be diffraction limited at the wavelength of 3 μm on a 30″ field of view [6]. The required average telescope throughput is 96 % [1]. Telescope and instruments will operate at a temperature below 50 K.

Following the heritage of the JWST MIRI instrument [12], aluminum alloy 6061 in the T651 forge has been chosen for mirrors substrates and supporting structures of the telescope, after a trade-off study [5] on manufacturability and cost.

To comply with throughput requirements, particularly in the visible portion of the operating waveband, and to protect the aluminum substrate from oxidation, the Consortium decided to apply a protected silver coating to the telescope mirrors.

The coating, from CILAS[2] is qualified for space, but needed to be subjected to additional qualification tests to assess performance at the Ariel telescope operating temperature of <50 K and because of the large size of the primary mirror, raising possible concerns on the uniformity of the deposition process and stability of the coating.

An initial study was therefore devised to test optical performance and durability on Al6061-T651 substrates, consisting of a qualification campaign on coated aluminum samples [2, 4], and a verification test on additional samples and on a full-scale demonstrator of the Ariel primary mirror denoted PTM [3].

After the successful completion of the qualification, the samples have been kept in storage in a ISO6 cleanroom facility, and are being periodically re-examined to detect signs of functional (reflectance) or visual deterioration due to oxidation or delamination.

Silver coatings, although protected by a capping layer, are in fact particularly sensitive to damaging from contact with humidity, sulfur and chlorine pollutants, normally present even in the controlled atmosphere of a cleanroom [9]. It is therefore necessary to ensure that the coating will not deteriorate in the time span from deposition on the Ariel telescope mirrors to launch.

## 5.2 Materials and Processes

### 5.2.1 Samples Description

The coating qualification campaign was performed on several samples of Al 6061-T651 in rolled plate form, the same aluminum alloy and forge currently foreseen for Ariel Telescope mirrors and supporting structure.

The samples are shaped as discs of 25 mm of diameter and 6 mm thick (Figure 5.1 shows one of the samples being held for visual inspection before the coating run). The samples were procured, polished and cleaned by MediaLario[3] before delivery to CILAS for coating. Roughness of the optical surface was measured with a Taylor Hobson CCI White Light Interferometer on areas measuring 1.5 mm × 1.5 mm and 300 µm × 300 µm. All samples were within the 10 nm RMS specification.

[2] CILAS-ArianeGroup, Etablissement de Marseille, 600 avenue de la Roche Fourcade, Pôle ALPHA Sud - Z.I. Saint Mitre, 13400 Aubagne, France

[3] Media Lario S.r.l., Via al Pascolo, 23842 Bosisio Parini (LC), Italy

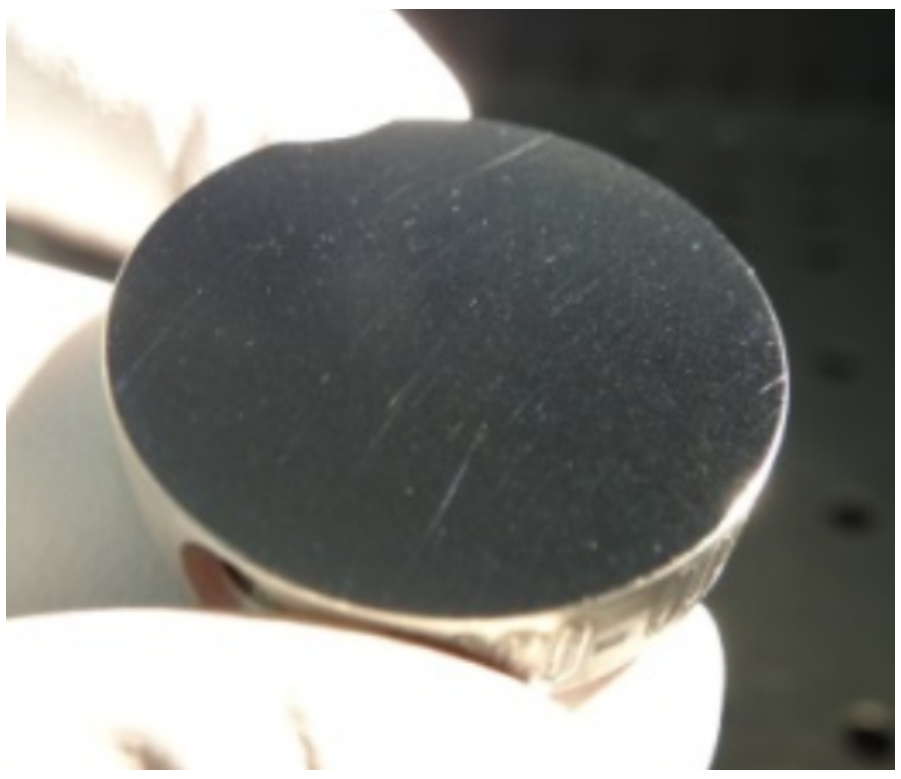

**Figure 5.1:** Picture of one of the aluminum samples during a visual inspection before coating.

The qualification campaign (in fact, a *delta*-qualification), consisted in a series of environmental and mechanical resistance tests to ensure that the coating will reach the end of life of the instrument without significant performance degradation, following the European Cooperation for Space Standardization (ECSS) Q-ST-70-17C standard [7]. A brief description of the set of tests is reported in Table 5.1. The qualification program was overall successful [2].

**Table 5.1:** Summary of specifications of the coating qualification tests performed on the samples.

| Test | Specifications |
|---|---|
| Adhesion | ISO 9211-4, Method 2 Severity 2 |
| Humidity | ISO 9022-2 Method 12 Severity 06, 24 h test duration, 90 % RH, 55±3 °C (no condensation) |
| Temperature cycling at ambient pressure | ISO 9022-2, 30 cycles, T. range −40–70 °C |
| Abrasion resistance | ISO 9211-4, Method 01 Severity 01 |
| Cryogenic cycling in vacuum | ECSS-Q-ST-70-04C, 10 cycles, T. range: 54–293 K, Vacuum: $<1 \times 10^{-4}$ mbar |

After the qualification campaing, a subset of the samples was retrieved by the authors and stored in airtight containers in a cleanroom environment. The risk of exposure to excessive humidity is further minimized by placing silica gel dessicant bags inside the containers. The samples are being re-examined periodically in a normal laboratory environment.

Table 5.2 identifies the samples being monitored, coming from two subsequent coating de-position runs: a first test run, performed on April 3rd, 2019, and the actual qualification run, on December 12th, 2019. Both runs were conducted with the nominal coating procedure and produced equivalent results. The set of treatments to which each sample was subjected during the qualification is also reported in the table.

**Table 5.2:** List of samples under test, with coating deposition date and qualification tests performed.

| Coating Run | Sample | Humidity | Thermal | Adhesion | Cryotest | Cleaning | Abrasion |
|---|---|---|---|---|---|---|---|
| 03/04/2019 | SN01 | | | ✓ | ✓ | ✓ | |
| | SN12 | ✓ | | | | | |
| 12/12/2019 | SN-01M | | | | | | |
| | SN-02M | | | ✓ | | | |
| | SN-04M | ✓ | | ✓ | | | |
| | SN-05M | | ✓ | ✓ | | | |
| | SN-06M | | | ✓ | ✓ | | |
| | SN-07M | ✓ | ✓ | ✓ | ✓ | | |
| | SN-08M | ✓ | ✓ | ✓ | ✓ | ✓ | ✓ |
| | SN-09M | | | ✓ | | ✓ | ✓ |
| | SN-10M | | | ✓ | | ✓ | ✓ |

All verification and measurements described in this paper have been performed on all samples listed Table 5.2, but for the reminder of the treatment we will focus on Samples SN01 and SN12, that have been in storage for a longer period, Sample SN06M since it was already examined in details in a previous work [4] and Sample SN08M because it was subjected to the entire set of qualification tests.

### 5.2.2 Coating Process

The coating process employed by CILAS for the samples is based on physical vapor deposition. The coating platform consists in a large magnetron sputtering chamber. Samples are inserted on a tray sliding beneath a set of cathodes [14]. The process is suited to optical substrates up to 2 m by 2 m of footprint and 0.4 m of thickness [10].

The protected silver coating described in this paper is on average 350 nm thick, with a thickness uniformity measured at 10 %.

The structure of the stack is illustrated in Figure 5.2 and consists of at least three layers: a NiCr adhesion layer of less than 10 nm of thickness, the reflecting silver layer, and a dielectric capping and protection layer. An additional intermediate adhesion layer may be present between the silver and the capping layer. The actual layers thicknesses and composition of the coating cannot be disclosed due to business confidentiality.

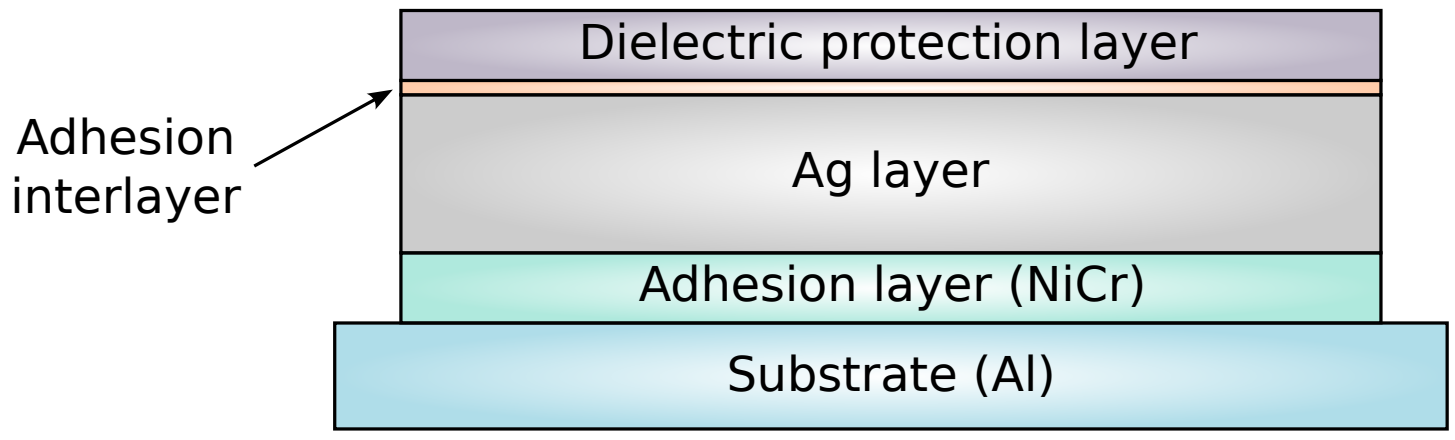

**Figure 5.2:** Indicative structure of the multilayer coating stack. The total coating thickness is approximately 350 nm.

## 5.3 Verification Methods

The ECSS Q-ST-70-17C standard [7] defines a set of verification and acceptance criteria to be performed on coatings to assess their performance and compliance with requirements.

More specifically, after each test, the coating shall present:

1. no visual degradation;
2. no delamination or adherence loss;
3. thickness according to requirements;
4. performance measurements compliant with coating specifications.

The following paragraphs describe the assessment methods employed to verify compliance of the samples under test according to the list above, except for item 3 (thickness) since it was already verified during qualification and it is not expected to be affected by storage.

### 5.3.1 Relative Reflectance Measurements

Immediately after coating deposition, the reflectance of all samples was measured by the manufacturer with a Perkin-Elmer Lambda 950 spectrophotometer with a reflectometry accessory, in the waveband 500–2500 nm. Accuracy, as reported in the instrument datasheet, is ±0.6 % from 500 nm to 890 nm and ±1 % above 890 nm.

Subsequent relative reflectance measurements were taken at the Institute for Photonics and Nanotechnologies of the National Research Council (CNR–IFN) in Padova with a custom built setup. Reproducibility of the measurements in the worst case has been determined to be better than ±1.2 % over the 500–900 nm wavelength range, and better than ±3.8 % over the 400–1000 nm wavelength range.

In order to cross-calibrate the two setups, which appear to have a measurement bias between each other (see Figure 5.5 in the *Results* section), we used one of the earliest measurement of Sample SN01 as reference point, since it presented the shortest temporal gap between measurements.

### 5.3.2 Visual Inspection

The ECSS standard suggests following Annex C of ISO 9211-4:2022 for Visual Inspection (VI), that mandates the use of two cool white 15 W lamps positioned directly above the sample, and to look at the sample against a black matte background at a distance of ≤45 cm and at a near grazing angle [11]. The use of optical micrographs is suggested only in case of suspected degradation, to further qualify it. In our case it was employed to look for signs of oxidation, as explained in Section 5.4.1.

Darkfield imaging using a compact digital camera (Canon IXUS 220 HS) and a custom built LED lighting setup was also employed to highlight the presence of light scattering defects.

### 5.3.3 Atomic Force Microscopy

Besides optical imaging techniques, we employed an Atomic Force Microscope (AFM) for a qualitative analysis of surface topography and to measure surface roughness.

AFM scans were taken with a Park System[4] XE-Series 70 microscope in non-contact mode and processed with the Gwyddion open source software [13] (the processing pipeline consisted in removal of low spacial frequencies by fitting and subtraction of an $x, y$ third order polynomial surface, rows alignment using “median” as statistic and “scars removal”).

## 5.4 Results

### 5.4.1 Visual Inspection

The optical surface of all samples appeared visually unaltered after the storage period, as exemplified in Figure 5.3: no discernible signs of coating degradation, such as cracks, blistering or change in color/iridescence, nor other visible signs of delamination could be spotted. In particular, we could not detect signs of oxidation developing from surface grains, as described for example by Folgner *et al.* in their studies of protected silver coatings exposed to mixed flowing gas [9, 8].

Apart from scratches and occasional dust particles, the most prominent features on the optical surface continued to be the glue residues from adhesion tests, especially evident on sample SN01.

It is worth noting that the uncoated back and lateral surfaces of the samples do exhibit a slight brownish coloration and faint whitish areas, compatible with the oxidation of bare aluminum, so the level of exposure to pollutants seems to be at least sufficient to cause slight degradation of the substrate.

Darkfield photographs also show a mostly uniformly dark optical surface, indicative of low scattering (Figure 5.4).

[4]Park Systems Inc., KANC 15F, Gwanggyo-ro 109, Suwon 16229, Korea

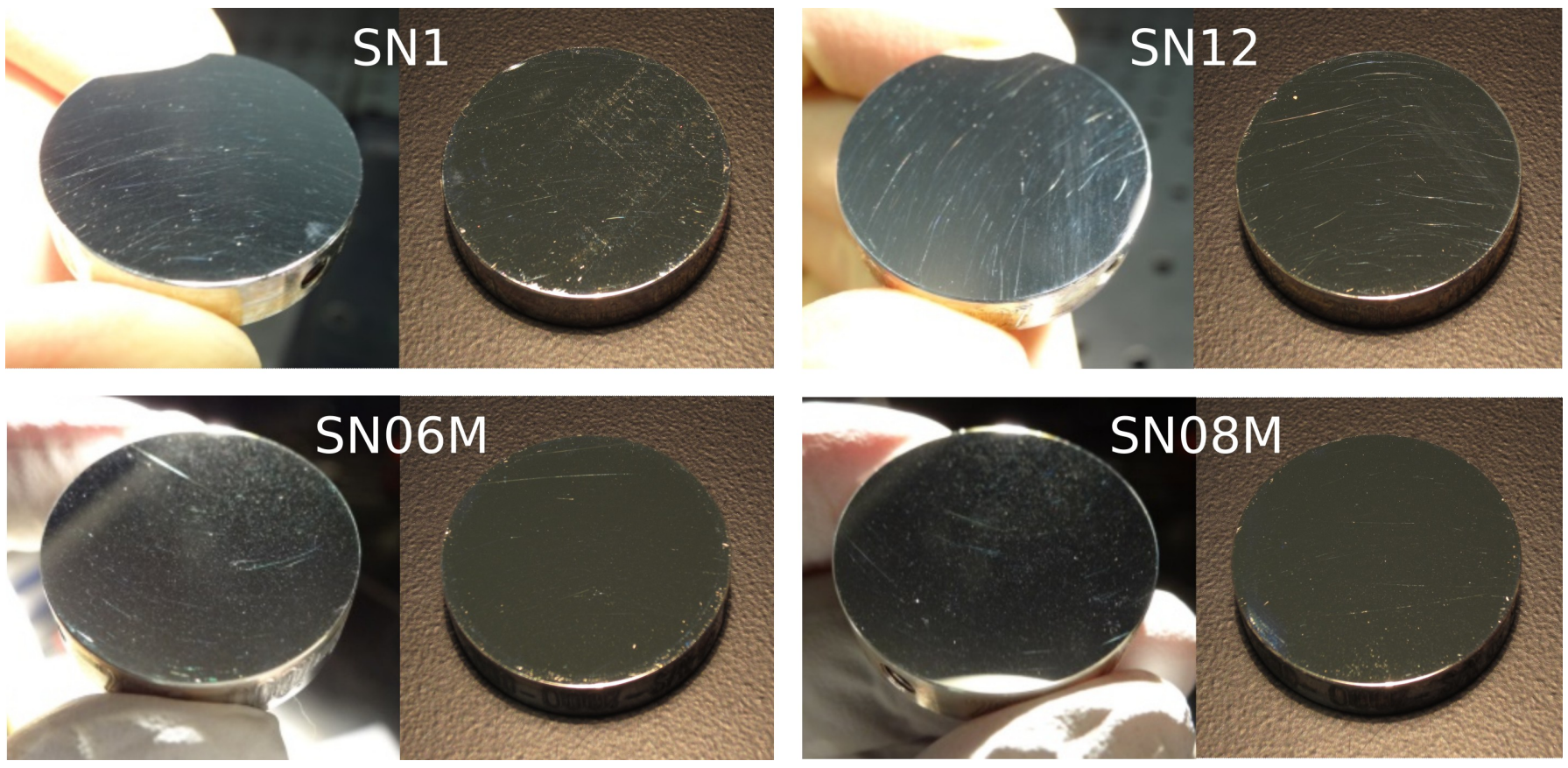

**Figure 5.3:** Couples of photographs of each sample immediately after coating (left), and after the storage period (right). Note that the orientation of the sample is not consistent in each couple.

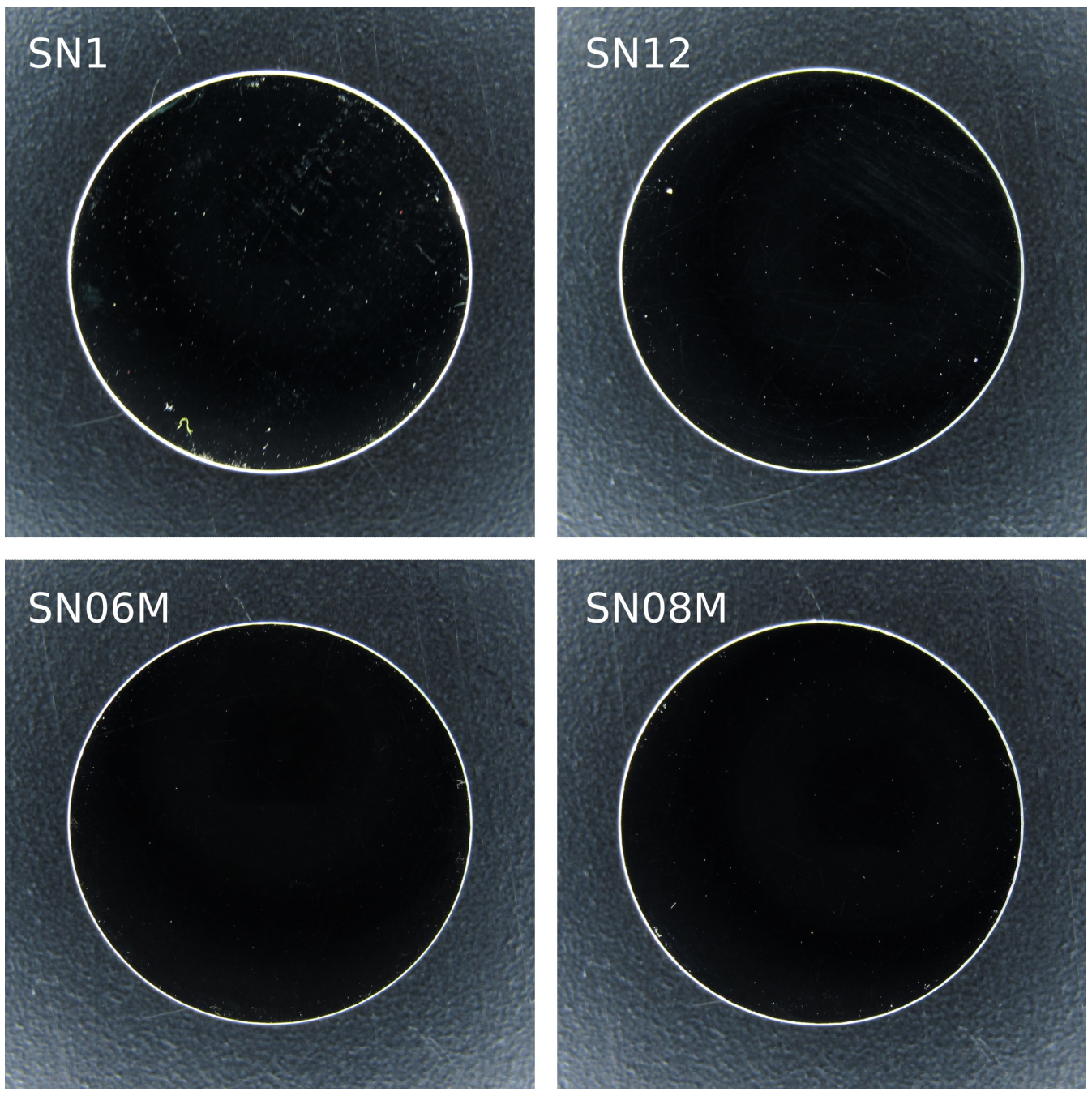

**Figure 5.4:** Darkfield photographs of the four samples after the storage period. Orientation of the samples is consistent with each right image of Figure 5.3.

### 5.4.2 Reflectance Measurements

Figure 5.5 shows the results of the reflectance measurements performed on the samples at different points in time. Samples reflectance was measured after coating and after qualification tests by the manufacturer (solid gray line), who found no degradation in performance [2]. One of the samples (SN01) was also measured by the authors early on during the qualification campaign, so this measurement could be used as reference for the cross-calibration of the two setups (dashed gray line).

Eventually two comprehensive measurement campaigns could be performed six months apart (in October 2021 and April 2022, orange and blue lines in the plots).

Considering the difference between the initial measurement of the in-house setup and the manufacturer's setup, and noting the estimated repeatability error, the results do not highlight a significant change in reflectance of the samples.

More precise measurements may be performed in the future to further confirm this preliminary result.

### 5.4.3 Atomic Force Microscopy

Two sets of AFM scans of representative areas of samples SN01 and SN06M are presented in Figures 5.6 and 5.7. For each set, the first scan (on top) was performed at the beginning of the storage period, after the sample underwent the qualification tests, and the second one (at the bottom), in August 2022.

As discussed in Section 5.3.3, the scans were performed to provide a qualitative assessment of surface morphology variations and to measure surface roughness. Since the time in storage did not produce any visible signs of surface degradation nor delamination, the sampling location for the AFM was chosen to be reasonably representative of the central area, where reflectance was also measured, without aiming at specific surface blemishes or scratches. Please also note that measurements of the same sample do not image the exact same portion of surface.

A comparison of the AFM images does not indicate the appearance of new topological structures of relevance: most features are attributable to scratches and dents that were equally present before storage. The white areas in relief on SN01 are likely the residues of tape adhesive from the adhesion tests. These were in fact visible upon careful examination of the surface area affected by the test. Anecdotally, AFM scans of sample SN01 did result in frequent tip pollution that required replacement, possibly because of the residues.

RMS roughness measurements also do not appear to change significantly before and after the cycles.

## 5.5 Conclusions and Next Steps

In the framework of the coating qualification program for the mirrors of the Ariel mission telescope, a series of samples of the mirrors substrate material, Al6061-T651, were tested and are currently being kept in storage and re-examined periodically for signs of degradation.

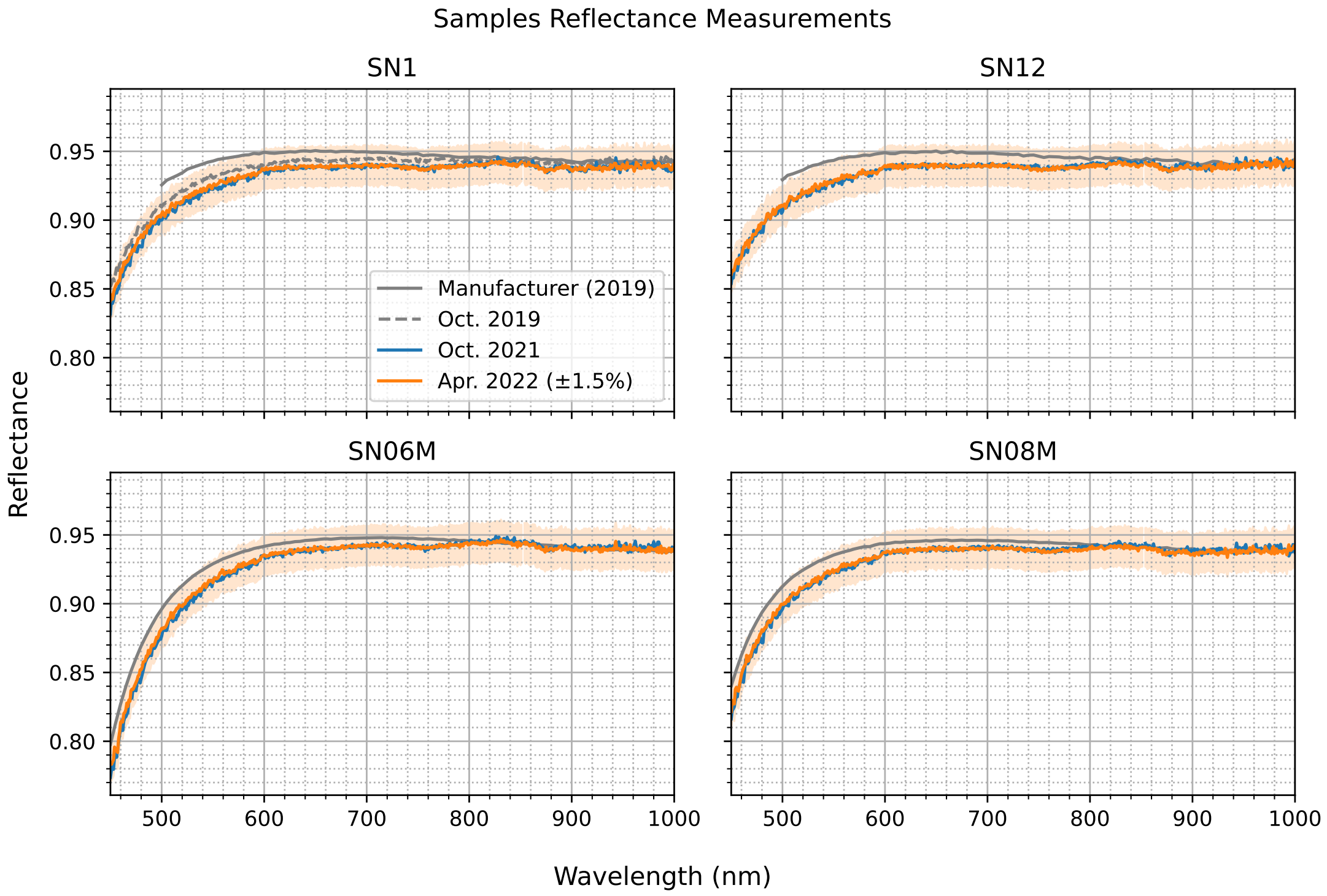

**Figure 5.5:** Reflectance measurements of four coated samples performed with two different setups, by the manufacturer and by the authors. One comparison measurement of sample SN1 (dashed gray line) was performed sufficiently close in time to the manufacturer's measurement to be useful to assess the cross-calibration of the two setups, which appear to have a bias.

After three years, results of visual inspections (both with direct and darkfield illumination) and reflectance measurements showed no alteration in appearance imputable to deterioration or delamination of the coating, nor a significant degradation in optical performance in the waveband 400–1000 nm, according to preliminary measurements. Additional AFM scans of the samples showed no qualitative morphology variations nor an increase in surface roughness.

Further testing will be repeated in the future to confirm coating stability under cleanroom environmental conditions, equivalent or worse to those foreseen for Ariel telescope mirrors. Additional operational environment tests are also planned for the near future, in particular radiation testing with an ion bombardment program simulating Ariel L2 operating orbital environment.

## Acknowledgments

This activity has been realized under the Implementation Agreement n. 2021-5-HH.0 of the Italian Space Agency (ASI) and the National Institute for Astrophysics (INAF) Framework Agreement "Italian Participation to Ariel mission phase B2/C" and was partly funded under the ESA contract with Centre Spatial de Liège, Belgium (CSL) and INAF n. 4000126124/18/NL/BW.

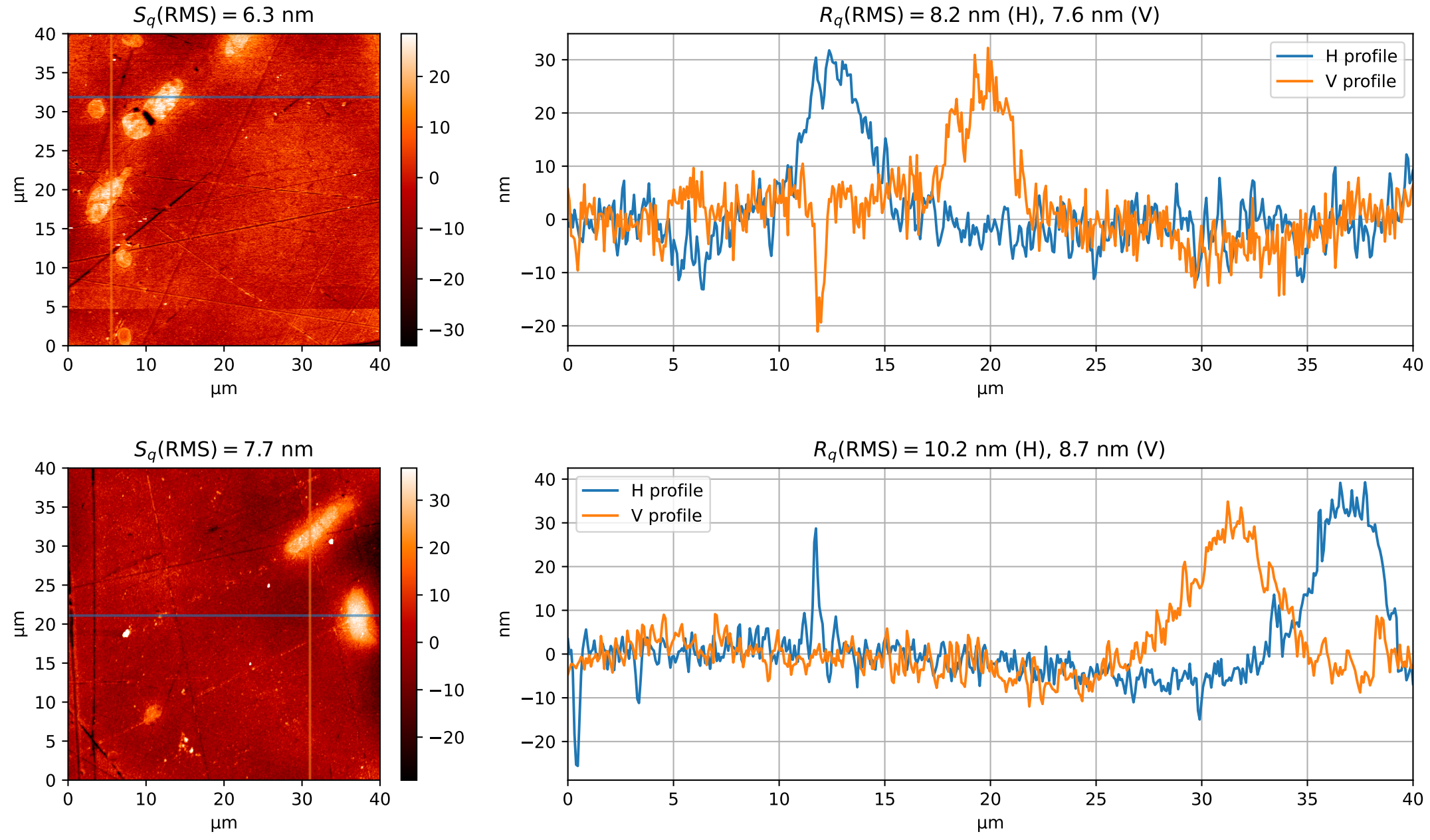

**Figure 5.6:** AFM scans (left) of sample SN01 before (top) and after (bottom) the storage period. For each scan, two orthogonal line profiles are also provided (plots on the right). The two scanned areas do not necessarily coincide.

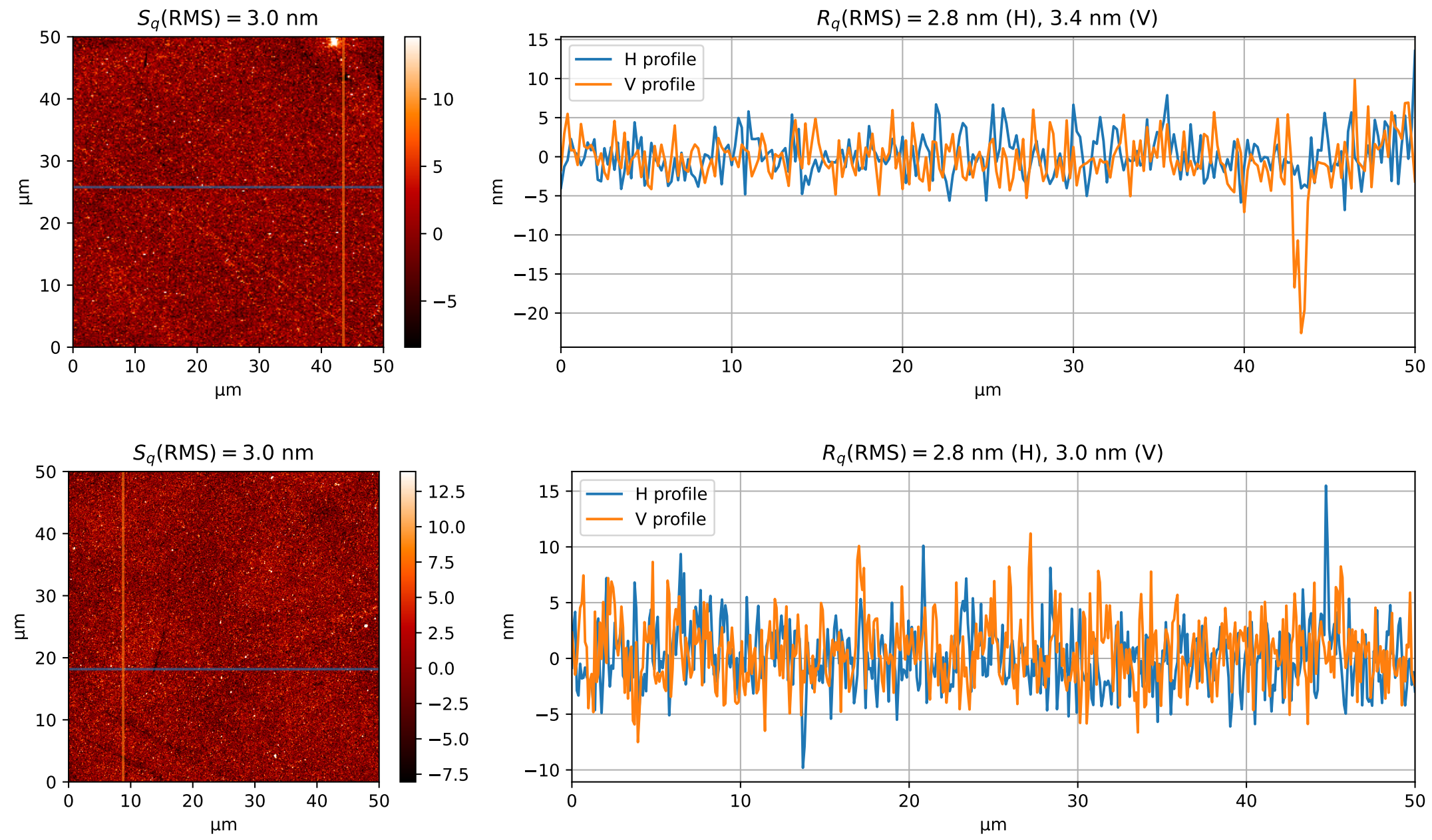

**Figure 5.7:** AFM scans (left) of sample SN06M before (top) and after (bottom) the storage period.

## References

[1] P. Chioetto et al. “Initial Estimation of the Effects of Coating Dishomogeneities, Surface Roughness and Contamination on the Mirrors of Ariel Mission Telescope”. In: *Proc. SPIE 11871, Optical Design and Engineering VIII*. Oct. 4, 2021, 118710N. DOI: 10.1117/12.2603768.

[2] P. Chioetto et al. “Qualification of the Thermal Stabilization, Polishing and Coating Procedures for the Aluminum Telescope Mirrors of the ARIEL Mission”. In: *Experimental Astronomy* 53 (Apr. 19, 2022), pp. 885–904. DOI: 10.1007/s10686-022-09852-x.

[3] P. Chioetto et al. “Test of Protected Silver Coating on Aluminum Samples of ARIEL Main Telescope Mirror Substrate Material”. In: *Proc. SPIE 11852, International Conference on Space Optics — ICSO 2020*. June 11, 2021, p. 118524L. DOI: 10.1117/12.2599794.

[4] P. Chioetto et al. “The Primary Mirror of the Ariel Mission: Cryotesting of Aluminum Mirror Samples with Protected Silver Coating”. In: *Proc. SPIE 11451, Advances in Optical and Mechanical Technologies for Telescopes and Instrumentation IV*. Dec. 13, 2020, 114511A. DOI: 10.1117/12.2562548.

[5] V. Da Deppo et al. *ARIEL Telescope Material Trade-Off*. Technical Paper ARIEL-INAF-PL-TN-004. 2017.

[6] V. Da Deppo et al. “The Optical Configuration of the Telescope for the ARIEL ESA Mission”. In: *Proc. SPIE 10698, Space Telescopes and Instrumentation 2018: Optical, Infrared, and Millimeter Wave*. Aug. 21, 2018, 106984O. DOI: 10.1117/12.2313412.

[7] European Cooperation for Space Standardization. *ECSS-Q-ST-70-17C – Durability Testing of Coatings (1 February 2018)*. European Space Agency, Feb. 1, 2018.

[8] K. A. Folgner et al. “Development and Growth of Corrosion Features on Protected Silver Mirrors during Accelerated Environmental Exposure”. In: *Applied Optics* 59.5 (Feb. 10, 2020), A187. DOI: 10.1364/AO.375891.

[9] K. A. Folgner et al. “Environmental Durability of Protected Silver Mirrors Prepared by Plasma Beam Sputtering”. In: *Applied Optics* 56.4 (Feb. 1, 2017), p. C75. DOI: 10.1364/AO.56.000C75.

[10] C. Grèzes-Besset et al. “High Performance Silver Coating with PACA2M Magnetron Sputtering”. In: *Proc. SPIE 11180, International Conference on Space Optics — ICSO 2018*. July 12, 2019, p. 1118083. DOI: 10.1117/12.2536210.

[11] International Organization for Standardization. *ISO 9211-4:2022 Optics and Photonics — Optical Coatings — Part 4: Specific Test Methods: Abrasion, Adhesion and Resistance to Water*. Vernier, Geneva, Switzerland: International Organization for Standardization, 2022.

[12] G. Kroes et al. "MIRI-JWST Spectrometer Main Optics Opto-Mechanical Design and Prototyping". In: *Proc. SPIE 5877, Optomechanics 2005*. Aug. 18, 2005, 58770P. DOI: 10.1117/12.614784.

[13] D. Nečas and P. Klapetek. "Gwyddion: An Open-Source Software for SPM Data Analysis". In: *Open Physics* 10.1 (Jan. 1, 2012). DOI: 10.2478/s11534-011-0096-2.

[14] I. Savin de Larclause et al. "PACA2m Magnetron Sputtering Silver Coating: A Solution for Very Big Mirror Dimensions". In: *Proc. SPIE 10563, International Conference on Space Optics — ICSO 2014*. Jan. 5, 2018, p. 1056308. DOI: 10.1117/12.2304237.

[15] G. Tinetti et al. "A Chemical Survey of Exoplanets with ARIEL". In: *Experimental Astronomy* 46.1 (Nov. 2018), pp. 135–209. DOI: 10.1007/s10686-018-9598-x.

# 6

# Initial estimation of the effects of coating dishomogeneities, surface roughness and contamination on the mirrors of Ariel mission telescope

Paolo Chioetto[1a,b,c], Enzo Pascale[d], Paola Zuppella[a,c], Emanuele Pace[e], Andrea Tozzi[f], Giuseppe Malaguti[g], Giuseppina Micela[h], and the Ariel Team

a CNR-Istituto di Fotonica e Nanotecnologie di Padova, Via Trasea 7, 35131 Padova, Italy
b Centro di Ateneo di Studi e Attività Spaziali "Giuseppe Colombo"- CISAS, Via Venezia 15, 35131 Padova, Italy
c INAF–Osservatorio Astronomico di Padova, Vicolo dell'Osservatorio 5, 35122 Padova, Italy
d Dipartimento di Fisica, La Sapienza Università di Roma, Piazzale Aldo Moro 2, 00185 Roma, Italy
e Dipartimento di Fisica ed Astronomia–Università degli Studi di Firenze, Largo E. Fermi 2, 50125 Firenze, Italy
f INAF–Osservatorio Astrofisico di Arcetri, Largo E. Fermi 5, 50125 Firenze, Italy
g INAF–Osservatorio di Astrofisica e Scienza dello spazio di Bologna, Via Piero Gobetti 93/3, 40129 Bologna, Italy
h INAF–Osservatorio Astronomico di Palermo, Piazza del Parlamento 1, 90134 Palermo, Italy

## Abstract

Ariel (Atmospheric Remote-Sensing Infrared Exoplanet Large Survey) is ESA "Cosmic Vision" M4 adopted mission, to survey exoplanet atmospheres through transit spectroscopy.

[1]Contact and presenting author.

Launch is scheduled for 2029. Ariel scientific payload consists of a 1-m class, all-aluminum telescope feeding a set of photometers and spectrometers in the waveband between 0.5 and 7.8 μm. The operating temperature is below 50 K.

To improve reflectivity and to prevent degradation of the optical surface of the telescope mirrors, a protected silver coating with space heritage has been selected and qualified.

The wide operating waveband of the instruments poses some concerns on the achievable throughput. In particular, technological limitations on large aluminum mirrors manufacturing and coating demand an estimation of the overall throughput achievable by the telescope for the entire 4-year scientific duration of the mission.

The starting point for the estimation is the spectral reflectivity of the coated mirrors, as measured on samples as part of the coating qualification campaign and considering the effects of deposition uniformity. On top of this, throughput losses caused by scattering of light from surface roughness, particulate contamination and cosmetic defects, and absorption from molecular contamination have been modelled or estimated from the available literature on the subject.

This work presents an initial estimation of the overall spectral throughput of the telescope, in order to assess compliance with scientific requirements and identify areas of concern.

## 6.1 Introduction

In 2020 ESA formally adopted Ariel as the fourth "M" class mission of its "Cosmic Vision" program. Ariel is currently in the implementation phase.

Ariel's purpose is to carry out a survey of the atmospheres of known exoplanets. The Payload consists of a telescope feeding a collimated beam to a set of spectrometers operating in the waveband between 2 μm and 8 μm and a fine guidance/photometer/VIS-NIR spectrometer operating in the visible to near infrared spectrum.

The design chosen for the Telescope (Figure 6.1) is an off-axis, unobscured Cassegrain followed by a tertiary recollimating mirror and a fourth folding mirror that bends the beam towards the optical bench (hereafter designated as M1, M2, M3 and M4).

Scientific requirements and preliminary design and performance considerations led to the following constraints: at least 0.6 square meters for the light collecting area of the Telescope, diffraction limited optical performance at the wavelength of 3 μm and on a 30″ Field of View and an average throughput of 96 % [3, 9].

The optical aperture of the primary mirror was then designed as an ellipse with dimensions of 1100 mm (major axis) and 730 mm (minor axis). The remaining mirrors are much smaller in aperture: 110 mm × 80 mm, 28 mm × 20 mm and the folding mirror has a diameter of 24 mm.

Telescope and instruments will operate at a temperature below 50 K.

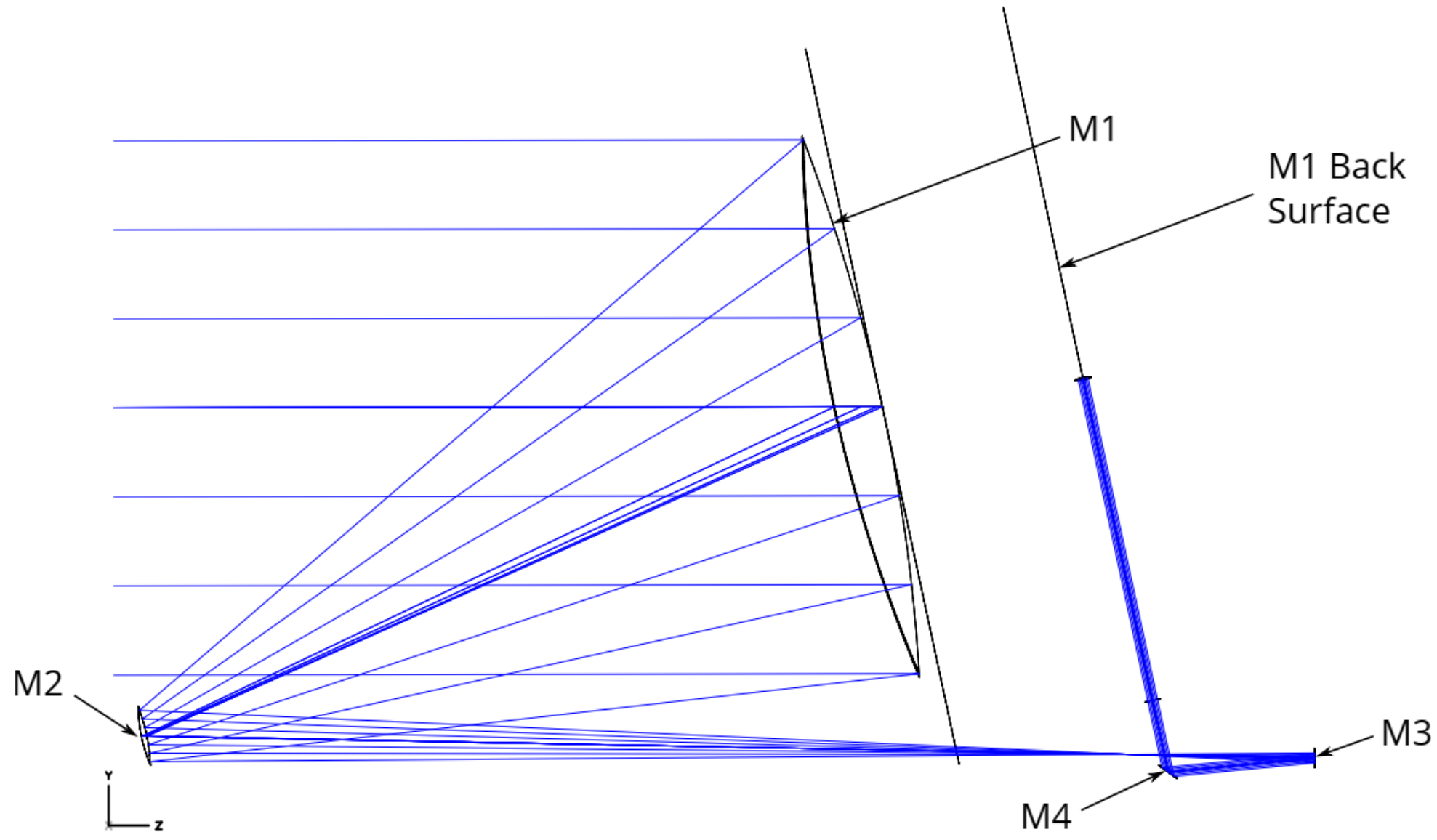

**Figure 6.1:** Optical model of Ariel Telescope.

Aluminum alloy 6061-T651 was chosen as construction material for mirrors substrates and supporting structures of the telescope to achieve athermalization under isothermal conditions. A protected silver coating with space heritage from CILAS[2] was then selected and qualified, and will be applied to the optical mirror surface to avoid oxidation and to improve reflectivity.

While heritage of silver-coated aluminum mirrors for ground and space telescope applications is well established, including instruments operating at cryogenic temperatures [1, 11, 10], the large collecting area of Ariel primary mirror poses some specific manufacturing challenges, in particular the achievable optical surface roughness and uniformity of the coating process, that directly affect mirror reflectivity, and therefore Telescope throughput.

During Phase A, a preliminary estimation of the Telescope throughput at the end of the nominal scientific mission was made using catalog reflectivity data, in order to prepare an initial budget at payload level. After the successful qualification of the optical coating, the estimation was then revised to include measured reflectivity data from the actual coating and better estimations of the effects of surface roughness, particulate and molecular contamination and cosmetic defects, either through modelling or from the available literature on the subject.

This paper illustrate the data and procedures employed to calculate the revised throughput estimation.

[2]CILAS-ArianeGroup, Etablissement de Marseille, 600 avenue de la Roche Fourcade, Pôle ALPHA Sud - Z.I. Saint Mitre, 13400 Aubagne, France

## 6.2 Materials and Methods

### 6.2.1 End-of-life Telescope Throughput Estimation

The end-of-life (EOL) Telescope throughput is define as the end-to-end throughput of the Telescope at the end of the scientific mission, and is calculated as the product of the EOL reflectivity of the four mirrors.

The EOL reflectivity of each mirror is estimated as the product of the following components:

1. baseline mirror reflectivity, from coating qualification measurements on samples;
2. AOI coefficient, calculated to adjust the baseline reflectivity to the variation of coating reflectivity wrt. the angle of incidence of the incoming light;
3. reflectivity loss coefficient from scattering due to surface roughness;
4. reflectivity loss coefficient from scattering due to particulate contamination;
5. reflectivity loss coefficient from absorption due to molecular contamination and water ice;
6. reflectivity loss coefficient from scattering due to cosmetic defects.

The following paragraphs describe each of the component and the method by which it was estimated.

### 6.2.2 Baseline Reflectivity

The first component of the estimated EOL mirror reflectivity is the baseline reflectivity, calculated from actual measurements on samples employed to qualify the selected optical coating. To obtain a single averaged spectral reflectivity value for each mirror, samples surface roughness and reflectivity dependence on the position inside the coating chamber need to be taken into consideration, as explained in the following paragraphs.

Each of the four Telescope mirrors has different requirements in terms of maximum acceptable surface roughness. For this reason, baseline reflectivity was calculated to be indicative of a perfectly smooth mirror, and the effects of different levels of roughness was then added, corresponding to the worst case scenario for each mirror.

Coating qualification was performed on 19 polished samples made of Al 6061-T651 and 11 glass (N-BK7) samples [2].

The measured surface roughness of the Al samples was variable, with an average RMS of 7.35 nm and 5.33 nm, measured respectively on areas of 1.5 mm × 1.5 mm and 300 µm × 300 µm. Surface roughness of the glass samples was not measured before coating, but can be presumed consistently lower than 1 nm RMS and thus negligible at the wavelengths under consideration.

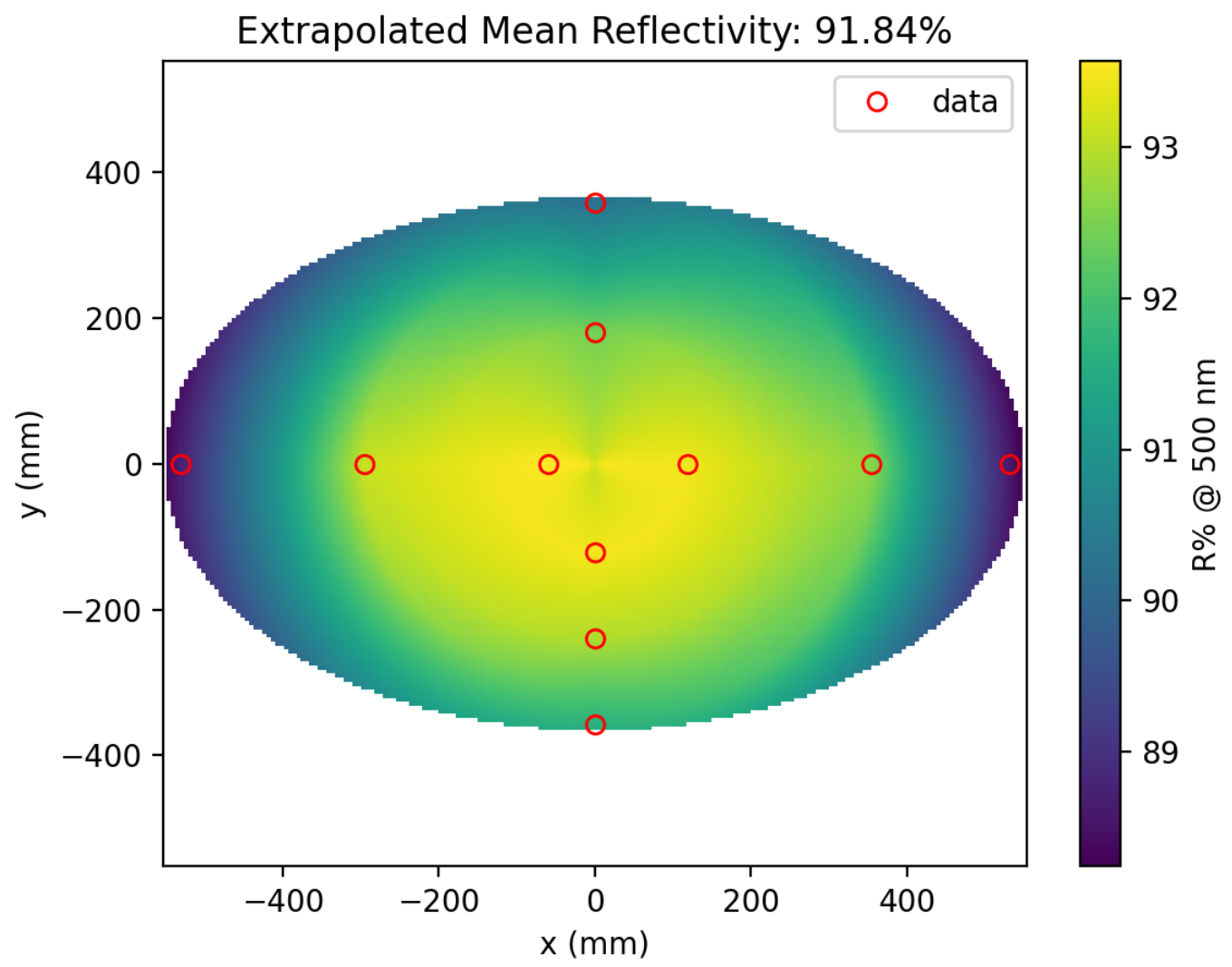

**Figure 6.2:** M1 extrapolated mean reflectivity at 500 nm wavelength from the positions of glass samples (red circles) in the coating chamber.

Therefore, in order to model the effects of surface roughness on the baseline reflectivity, we could have either used measurements on the glass samples directly, or adjusted the measurements of each Al sample subtracting the effect of the specific level of surface roughness. For simplicity, and to avoid having to subtract and then add back the effect of roughness on the samples, we chose to use the measurements on glass samples as the starting point.

Coating reflectivity was also found to vary throughout the mirror surface, given the large size of the primary mirror of the Telescope, as anyway anticipated by the coating manufacturer. For this reason, the qualification samples were coated while lying on the axes of an elliptical surface shaped roughly like the Telescope primary mirror, in order to study the correlation between position in the coating chamber and reflectivity.

To determine a suitable averaging strategy, we first attempted a two-dimensional interpolation of the measurements, starting from the samples positions on the axes of the elliptical footprint of the mirror. Reflectivity was first linearly extrapolated on the axes, and then along concentric circles to provide an average figure representative of the whole mirror. The process was repeated at several wavelengths (see Figure 6.2 for the results of this procedure at a wavelength of 500 nm).

The result proved equivalent to a simple averaging of reflectivity from all glass samples, so the latter was used in the calculations.

For the other three mirrors, given the smaller apertures, the average reflectivity from the two central samples was used directly.

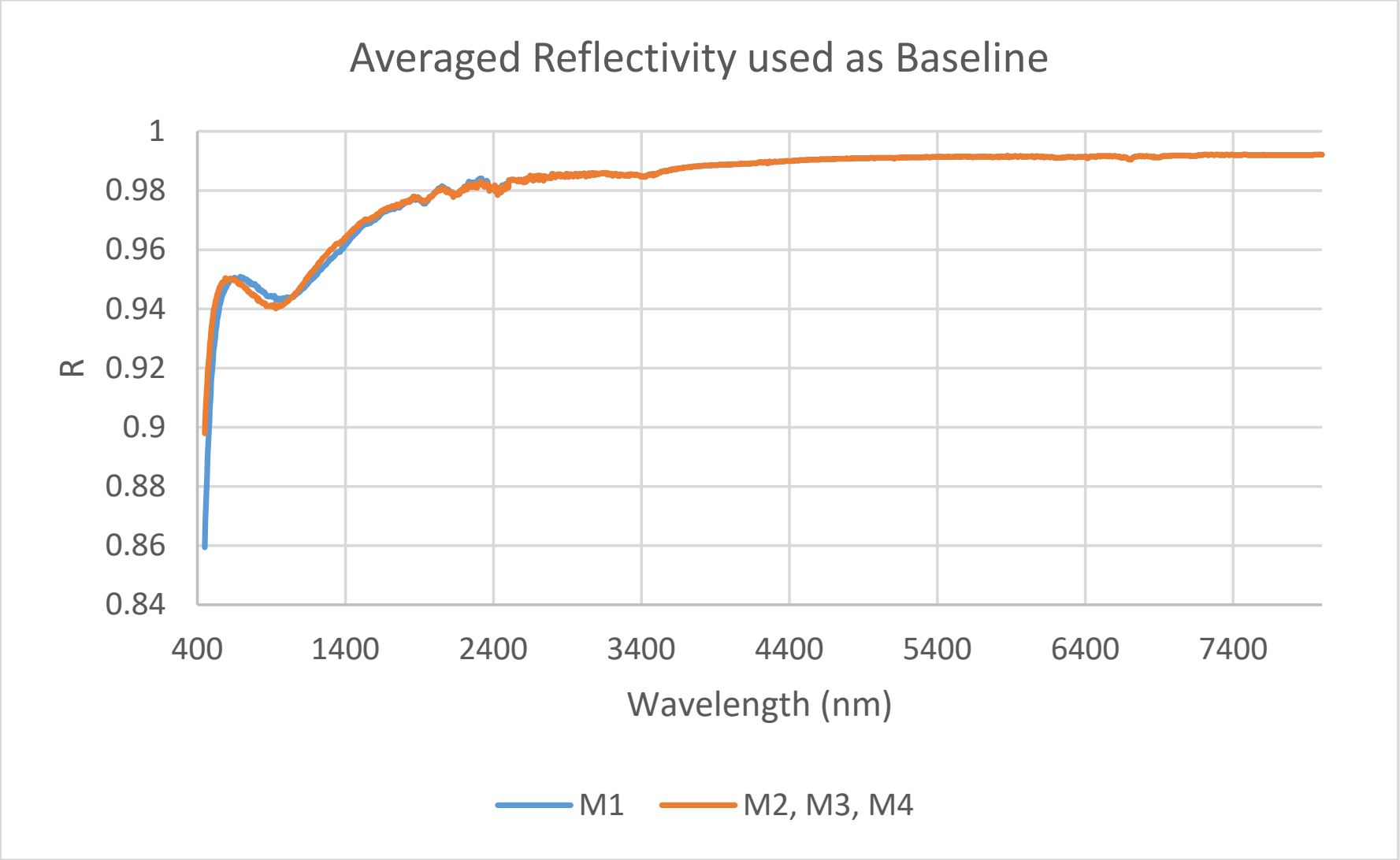

**Figure 6.3:** Averaged reflectivity of glass samples used as baseline for the Telescope throughput calculations. M1 is the Telescope primary mirror, M2–4 are the other mirrors.

The chart in Figure 6.3 below shows the results of the averaging process, for the primary and the other mirrors.

### 6.2.3 AOI Coefficient

As reflectivity depends on the angle of incidence (AOI) of incoming light, the baseline reflectivity needed to be adjusted to be representative of the range of angles expected for each mirror from the optical design of the Telescope.

For M1–3, the adjustment coefficient was determined from measurements of the coating qualification samples at 8 and 20 degrees AOI, an acceptable compromise between the extremes of the range of AOI (approximately 3-21 degrees) and the constraints of the measurement apparatus.

Unfortunately, only aluminum samples were measured both at 8 and 20 degrees AOI, while glass samples were only measured at 8 degrees. Assuming that the effect of AOI on reflectivity is mostly independent on the mirror substrate, the variation in reflectivity determined on Al samples was the applied to the measurements of glass samples according to the formula below (Equation 6.1), to obtain their average reflectivity wrt. AOI.

$$R_i^{glass} = \left\langle \frac{R_j^{Al,8} + R_j^{Al,20}}{2} / R_j^{Al,8} \right\rangle R_i^{glass,8} \tag{6.1}$$

$R_i^{glass}$ is the estimated average spectral reflectivity of the i-th glass sample, $R_i^{glass,8}$ the mea-

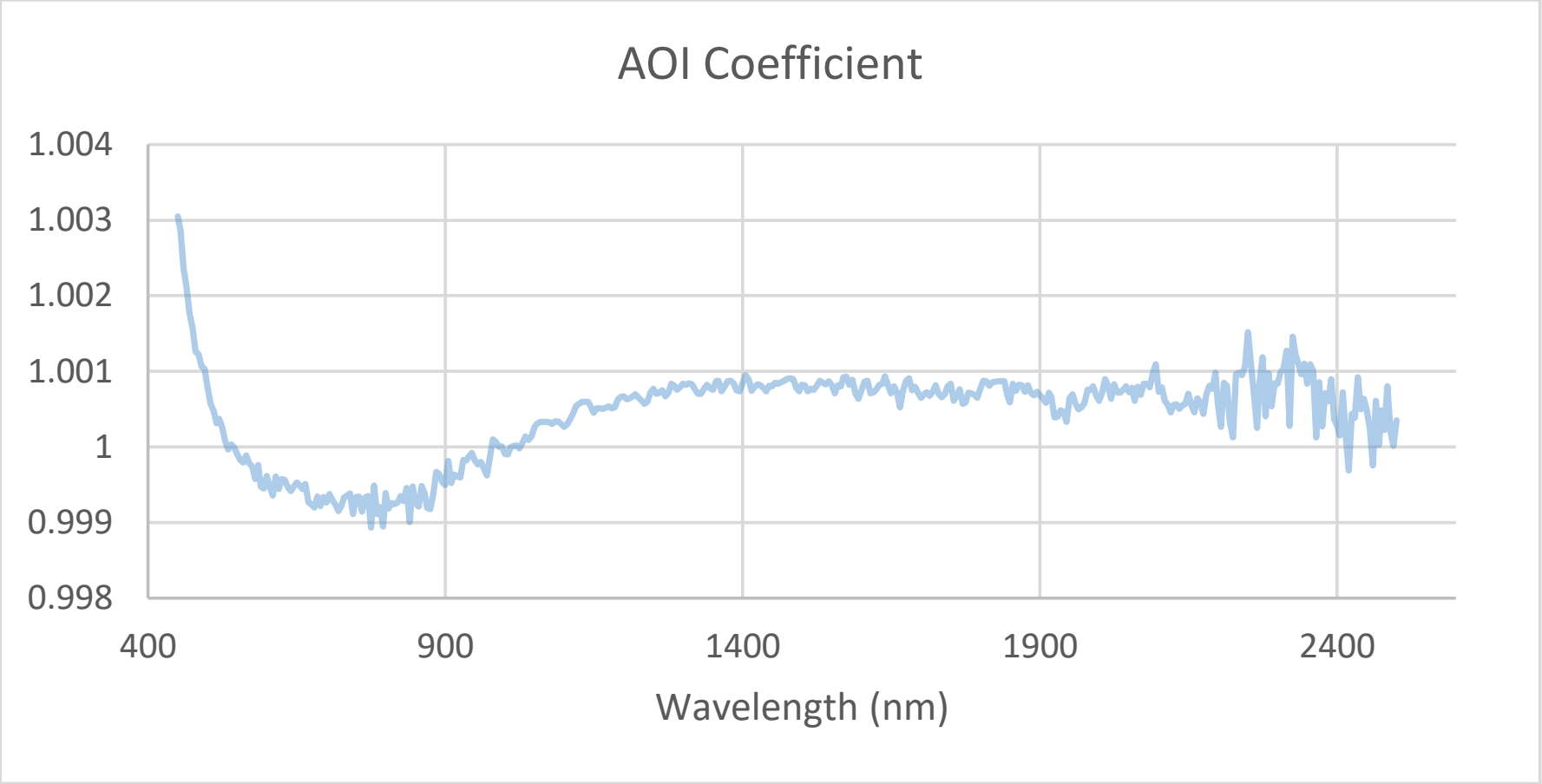

**Figure 6.4:** Average of the ratios of reflectivity measured for each Al sample at 8 and 20 degrees AOI, to be applied to the baseline reflectivity from glass samples (measured only at 8 degrees AOI).

sured reflectivity at 8 deg AOI of the same sample, and $R_j^{Al,\,8}$ is the measured reflectivity of the j-th aluminum sample at 8 degrees AOI. The average ratio, together with the range of variation, is shown in Figure 6.4.

For M4 (the flat mirror folding the Telescope collimated output beam), reflectivity measurements at the 48 degrees AOI were not available due to measurement apparatus limitations. Proper thin film interference simulations could not be setup either since knowledge of the exact composition and thickness of the coating capping layer are not available.

A set of tentative simulations with varying thicknesses of a protective SiO2 layer were anyway performed, and showed small variations in reflectivity, both with positive and negative signs, always below 1 %. This value was then used as worst case adjustment coefficient.

### 6.2.4 Microroughness Scattering Model

Throughput loss due to microroughness has been modelled with the Harvey-Shack Total Integrated Scatter (TIS) formula (Equation 6.2 below):

$$\mathrm{TIS} = 1 - \exp\left[-\left(\frac{4\pi\sigma cos\theta_i}{\lambda}\right)^2\right] \tag{6.2}$$

with the values in Table 6.1 below and $\lambda$ in Ariel operating wavelength range.

The surface roughness values used here have been agreed with the mirrors manufacturer and have been considered reasonably achievable given the foreseen manufacturing processes. The angles of incidence used are the average of the maximum and minimum AOI for incoming rays for each mirror.

**Table 6.1:** AOI and roughness values used in roughness scattering model.

| | M1 | M2 | M3 | M4 |
|---|---|---|---|---|
| Angle of incidence $\theta_i$ (deg) | 10 | 10 | 10 | 48 |
| Roughness (nm RMS) | 10 | 2 | 2 | 2 |

### 6.2.5 Particulate Contamination Scattering Model

Losses from particulate contamination scattering have been modelled according to Mie scatter theory.

The starting point is the assumption of an end-of-life cleanliness level of 400 per IEST-STD-CC1246D [5] for the optical surfaces of all elements of the Ariel Telescope and instruments, a value considered achievable with careful contamination control. This leads to a Total Integrated Scatter of 0.2 % at a wavelength of 632 nm [4]. This figure is used as a worst case scenario for the remainder of the estimation.

In order to allow for a looser cleanliness requirement on the primary and secondary mirrors of the Telescope, requiring a longer and more complex manufacturing and integration process, the TIS figure was then arbitrarily increased to a value of 0.55 %, and the figure for the remaining optical elements adjusted to 0.1 %, so that budgeted scattering loss for the entire Ariel optical train, consisting of 15 items, remains the same.

### 6.2.6 Molecular and Ice Absorption Model

Losses from molecular contamination, depicted in Figure 6.5, derive from thin film interference calculations performed for the James Webb Telescope for a 20 nm layer of amorphous hydrocarbons from cleanroom environment exposure [6]. The assumption was considered compatible the molecular contamination requirement for Ariel Telescope primary and secondary mirrors (2000 ng/cm$^2$) and used as a worst case scenario for the remaining mirrors.

An additional layer of 20 nm of water ice, deposited by outgassing during flight, was then added, taking the worst case scenario analyzed by the JWST team.

The wavelength range of the original data from JWST literature (800–5000 nm) has been extended for Ariel using constant extrapolation.

### 6.2.7 Cosmetic (Scratch/Dig) Scattering Model

The potential loss of reflectance caused by cosmetic defects was estimated numerically integrating formulas 6.3 and 6.4 here below, derived from the scattering models presented by Peterson [8] for cosmetic defects characterized according to the MIL-PRF-13830B “scratch and dig” standard [7].

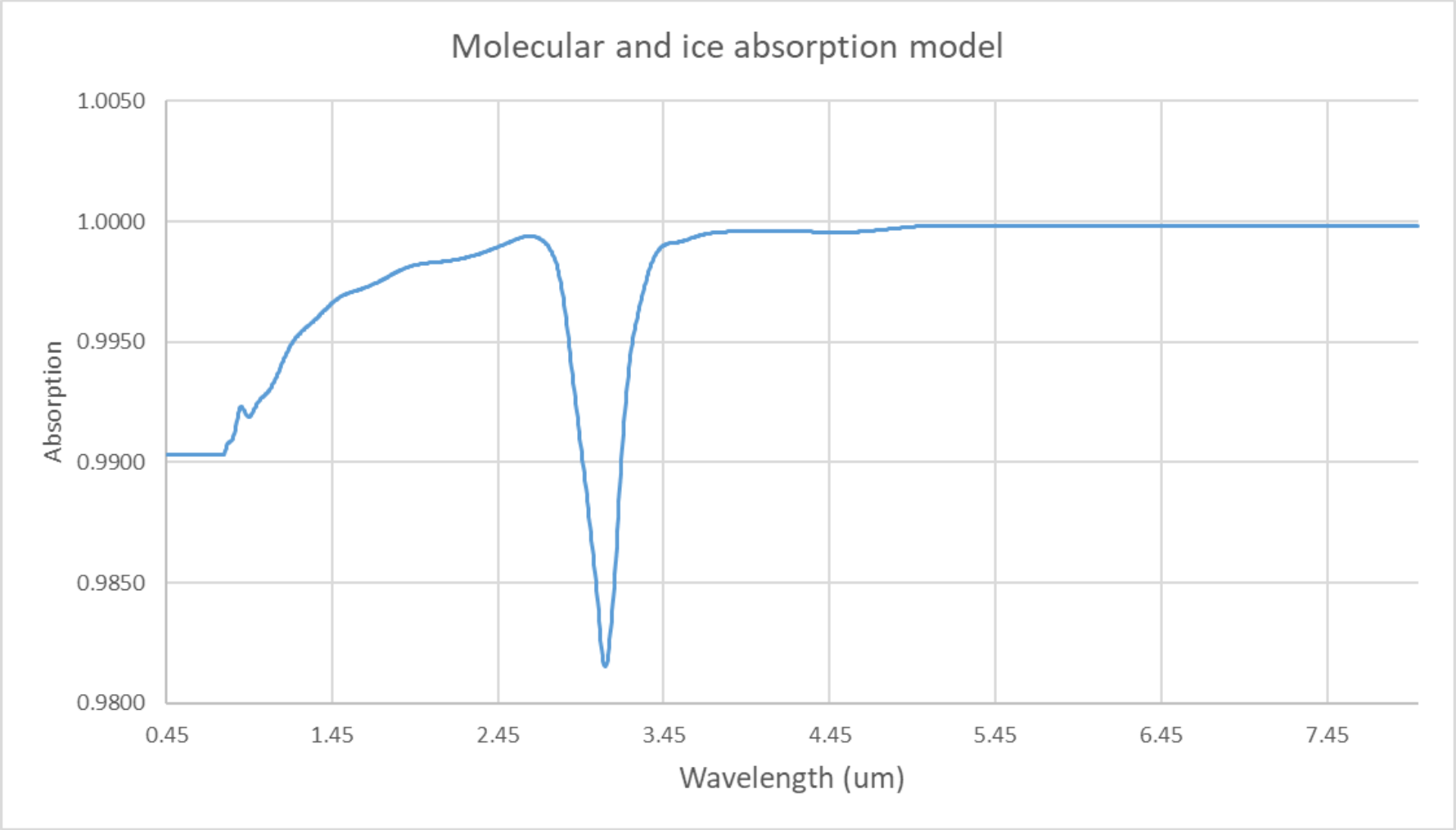

**Figure 6.5:** Molecular and ice absorption curve used in the model.

$\mathrm{BSDF_s}$ and $\mathrm{BSDF_d}$ are the Bidirectional Scattering Distribution Function for scratches and digs, and:

- $N_d$ and $N_s$ are the defects densities;
- $d$ is the average diameter of digs;
- $\lambda$ is the wavelength;
- $w$ the width of the typical scratch;
- $l$ is the average scratch length;
- $\theta$ is the angle of scatter;
- $l_s$ and $l_d$ are functions of $\lambda$ and the defects dimensions.

$$BSDF_s = N_s \frac{wl}{\pi}\left[1 + \frac{\pi wl}{\lambda^2}\left(1 + \frac{\sin^2\theta}{l_s^2}\right)^{-\frac{3}{2}}\right] \tag{6.3}$$

$$BSDF_d = \frac{1}{4} N_d d^2 \left[1 + \frac{\pi^2 d^2}{4\lambda^2}\left(1 + \frac{\sin^2\theta}{l_d^2}\right)^{-\frac{3}{2}}\right] \tag{6.4}$$

For the estimation, a cosmetic quality of 60/40 scratch/dig was assumed, which is generally considered to be an "acceptable" level for scientific applications, and therefore an achievable worst case scenario for the Telescope.

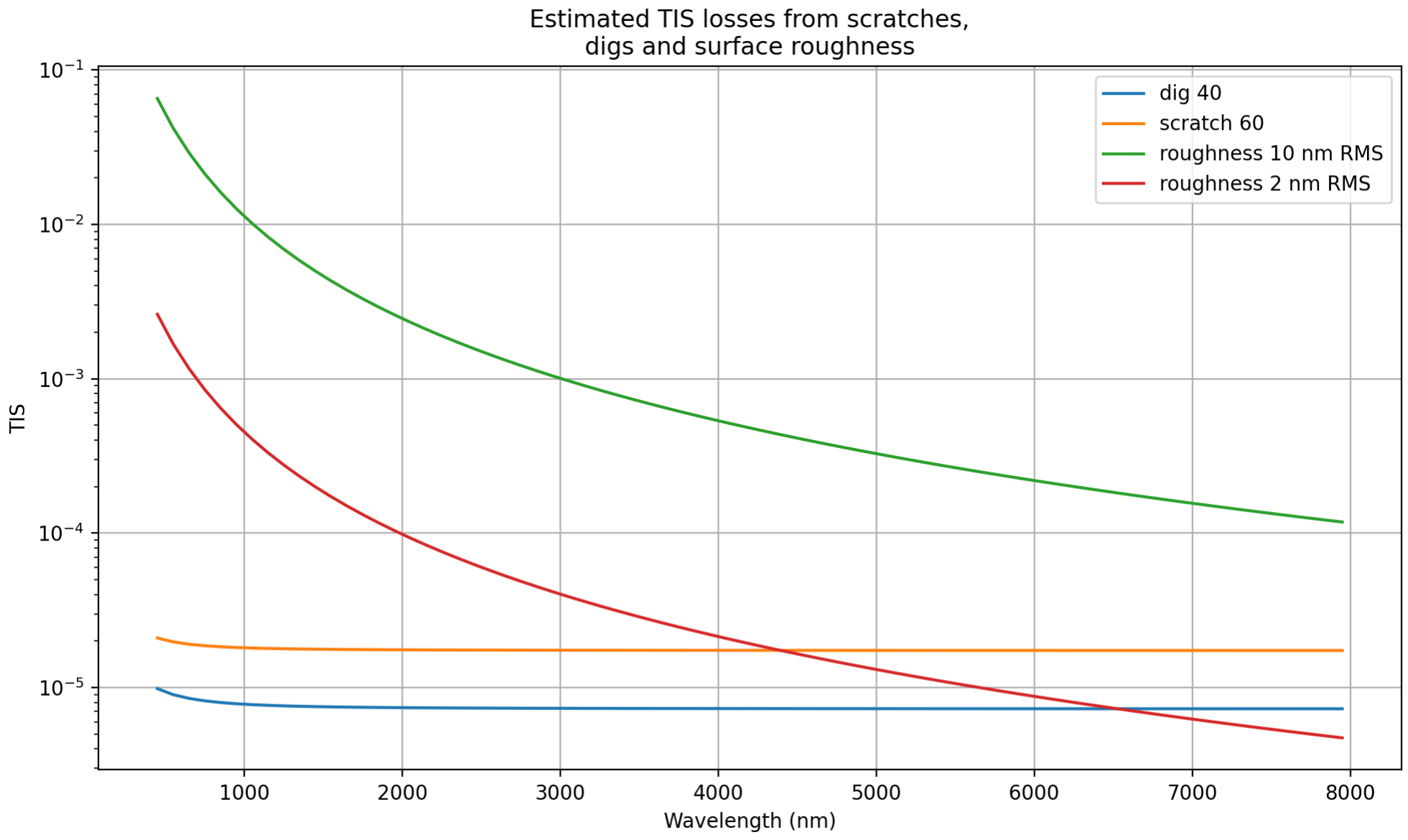

**Figure 6.6:** Comparison of estimated reflectivity losses due to cosmetic defects on the optical surface of the mirrors, and surface roughness (10 nm RMS for M1, 2 nm RMS for M2–4).

The figure "60" refers to the visibility of the scratch, as determined by visual comparison of the defects with a set of standard scratches of calibrated brightness. For this analysis, the number is also used as the actual width of the scratches. A further assumption was made that scratches would be on average 5 mm of length, as the standard does not specify it. The figure "40" for digs is instead the diameter of digs in tenths of millimeters.

The TIS calculated from these models results at least two orders of magnitude smaller than the one expected from surface roughness (Figure 6.6) at lower wavelengths, a result in accordance with the general considerations found in the literature on the subject. For this reason, it has not been included in the final throughput calculation.

It is however important to note that a cosmetic specification on the mirror is nonetheless important to limit straylight effects, as these may not be negligible.

## 6.3 Results

Figure 6.7 below shows the estimated end-of-life reflectivity of each mirror (brown line), calculated from the baseline spectral reflectivity, multiplied by each of the coefficients presented in the previous paragraphs.

The EOL reflectivity figures from each mirror are then multiplied together to derive the final Telescope EOL throughput, shown in the chart of Figure 6.8, together with Ariel throughput requirements (horizontal segments) for each instrument operational wavelength range.

For each range, the requirement is expressed as an average (dotted line) and/or a minimum throughput (solid line). The reason is that Ariel photometric instruments, all operating below 1.1 µm of wavelength, mandate only a requirement on the average throughput of the Telescope in the range, while spectrometric instruments, operating above 1.1 µm, require both a minimum and an average throughput.

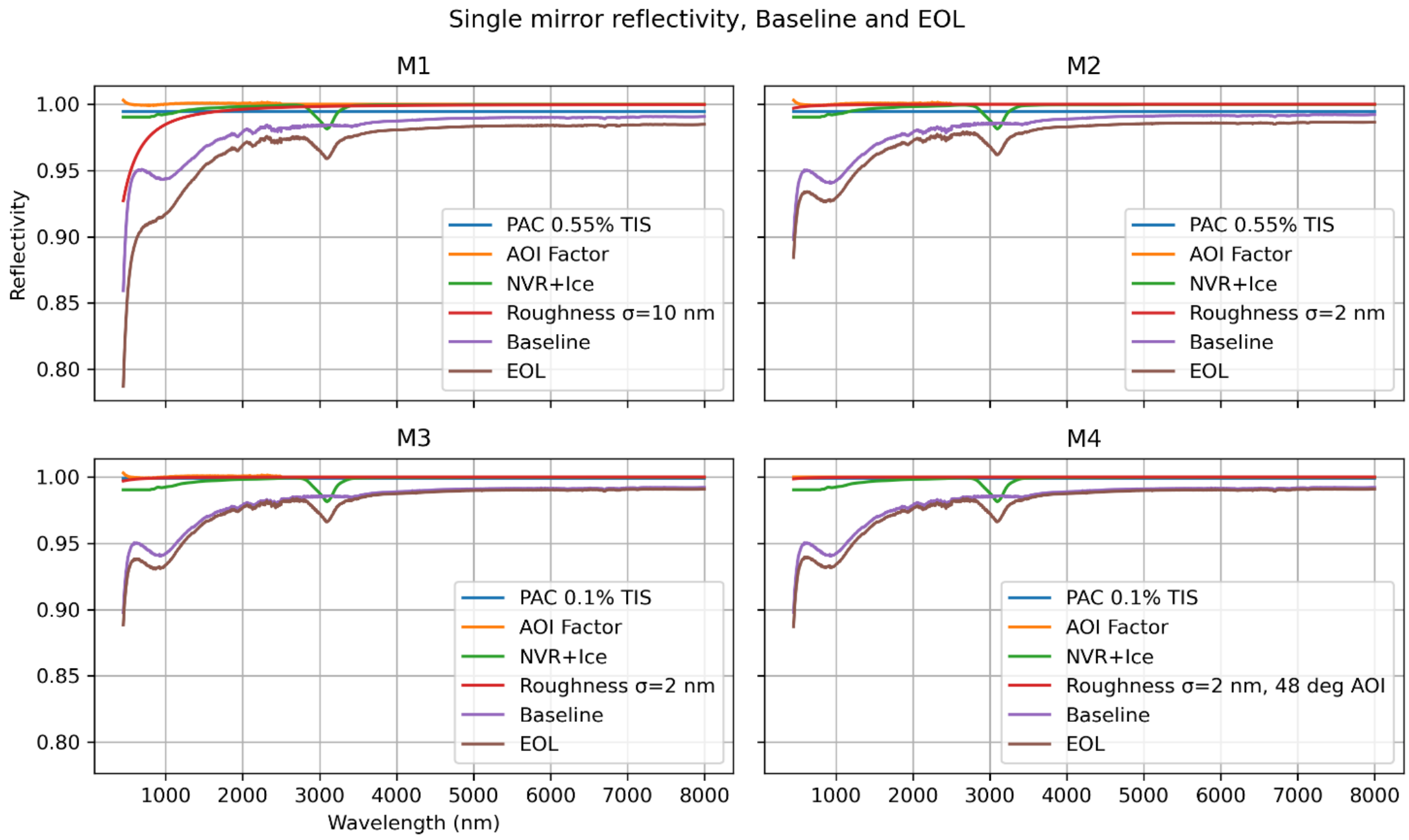

**Figure 6.7:** Final estimation of the EOL spectral reflectivity of each Telescope mirror. "PAC" refers to particulate contamination loss, while "Baseline" refers to the baseline reflectivity defined in paragraph 6.2.2.

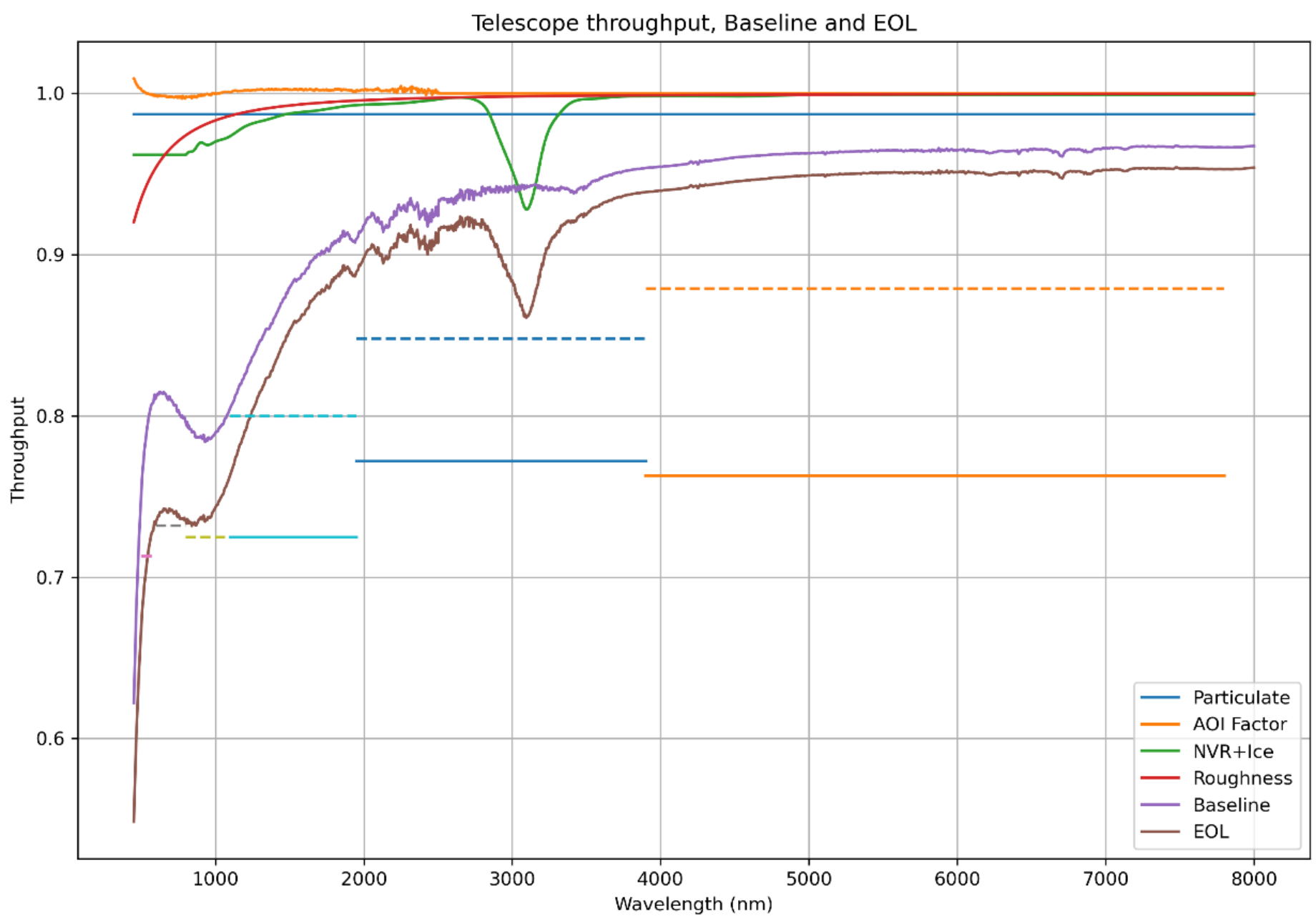

**Figure 6.8:** EOL Telescope throughput compared with Ariel requirements (horizontal segments) at each instrument wavelength range (see paragraph 4 for the discussion of this result). Each curve is the combined effect on all four mirrors. Dotted lines mark the required average Telescope throughput for the range, while solid lines indicate the required minimum.

To be compliant, the Telescope throughput curve has to be entirely above the minimum requirement, and the average throughput for each range has to be above the average requirement.

## 6.4 Discussion

The main purpose of the model developed in this paper is to reassess compliance of the Telescope throughput to requirements using the newly available reflectivity measurements on samples of the same material and coated with the same process envisioned for the flight mirrors.

Wherever assumptions and simplifications have been made, a worst case approach has been taken to ensure that the end result would not overestimate throughput.

Ariel payload instruments are both photometers, for which a requirement on the average throughput over the operating waveband is sufficient, and spectrometers, for which a requirement on the average and on the minimum throughput are both necessary.

As it is apparent from Figure 8 of the previous chapter, while minimum throughput requirements are amply met, the estimation affords very little margin towards average requirements for wavelengths in the ranges between 0.5 and 1.1 µm: less than 2.2 % of the required average throughput.

Of the various components that constitute the loss in throughput at these wavelength ranges, nearly half is due to scattering from surface roughness, and it is mostly imputable to M1.

## 6.5 Conclusions

This paper presented an updated throughput model for the Ariel Telescope, based on mirrors reflectivity measurements from the coating qualification campaign performed in Phase B1 of the Payload study and the updated requirements on surface particulate contamination. A detailed explanation for the use of glass coated samples as the basis for reflectivity measurements, as opposed to aluminium samples, is also provided.

Finally, a preliminary comparison against requirements is discussed, showing compliance but with no safety margin at lower wavelengths.

## Acknowledgements

This activity has been realized under the Italian Space Agency (ASI) contract with the National Institute for Astrophysics (INAF) n. 2018-22-HH.0, and is partly funded under the ESA contract with Centre Spatial de Liège, Belgium (CSL) and INAF n. 4000126124/18/NL/BW.

## References

[1] M. Boccas et al. “Coating the 8-m Gemini Telescopes with Protected Silver”. In: *Proc. SPIE 5494, Optical Fabrication, Metrology, and Material Advancements for Telescopes*. Sept. 24, 2004, p. 239. doi: 10.1117/12.548809.

[2] P. Chioetto et al. “Test of Protected Silver Coating on Aluminum Samples of ARIEL Main Telescope Mirror Substrate Material”. In: *Proc. SPIE 11852, International Conference on Space Optics — ICSO 2020*. June 11, 2021, p. 118524L. doi: 10.1117/12.2599794.

[3] V. Da Deppo et al. "An Afocal Telescope Configuration for the ESA ARIEL Mission". In: *CEAS Space Journal* 9.4 (Dec. 2017), pp. 379–398. DOI: 10.1007/s12567-017-0175-3.

[4] E. Fest. *Stray Light Analysis and Control*. SPIE Digital Library. SPIE Press, 2013. ISBN: 978-0-8194-9325-5.

[5] *IEST-STD-CC1246 Revision E: Product Cleanliness Levels – Applications, Requirements, and Determination*. Institute of Environmental Sciences and Technology, Feb. 2013.

[6] P. A. Lightsey et al. "Optical Transmission for the James Webb Space Telescope". In: *Proc. SPIE 8442, Space Telescopes and Instrumentation 2012: Optical, Infrared, and Millimeter Wave*. Aug. 22, 2012, 84423A. DOI: 10.1117/12.924841.

[7] *MIL-PRF-13830, Revision B - OPTICAL COMPONENTS FOR FIRE CONTROL INSTRUMENTS; GENERAL SPECIFICATION GOVERNING THE MANUFACTURE, ASSEMBLY, AND INSPECTION OF.* Military and Government Specs & Standards (Naval Publications and Form Center) (NPFC), Jan. 9, 1997.

[8] G. L. Peterson. "A BRDF Model for Scratches and Digs". In: *Proc. SPIE 8495, Reflection, Scattering, and Diffraction from Surfaces III*. Oct. 15, 2012, 84950G. DOI: 10.1117/12.930860.

[9] L. Puig et al. "The Phase A Study of the ESA M4 Mission Candidate ARIEL". In: *Experimental Astronomy* 46.1 (Nov. 2018), pp. 211–239. DOI: 10.1007/s10686-018-9604-3.

[10] M. Schürmann et al. "High-Reflective Coatings for Ground and Space Based Applications". In: *Proc. SPIE 10563, International Conference on Space Optics — ICSO 2014*. Nov. 17, 2017, p. 105630M. DOI: 10.1117/12.2304172.

[11] D. A. Sheikh. "Improved Silver Mirror Coating for Ground and Space-Based Astronomy". In: *Proc. SPIE 9912, Advances in Optical and Mechanical Technologies for Telescopes and Instrumentation II*. July 22, 2016, p. 991239. DOI: 10.1117/12.2234380.